\documentclass[reprint,amsmath,amssymb,prl,superscriptaddress]{revtex4-2}
\usepackage{graphicx}
\usepackage{dcolumn}
\usepackage{bm}
\usepackage[mathlines]{lineno}
\RequirePackage[l2tabu,orthodox]{nag}
\usepackage{siunitx}

\usepackage{charter,microtype,BOONDOX-cal}
\DeclareMathAlphabet{\mathcal}{OMS}{cmsy}{m}{n}
\DeclareSymbolFont{largesymbols}{OMX}{cmex}{m}{n}
\usepackage{amsthm,amsbsy,amstext,marvosym,amsfonts}
\usepackage[bookmarks=true,colorlinks,linkcolor=blue,urlcolor=blue,citecolor=blue,plainpages=false,pdfpagelabels,final,breaklinks=true
]{hyperref}
\DeclareMathOperator{\sech}{sech}
\usepackage[utf8]{inputenc}
\usepackage[T1]{fontenc}
\usepackage{array,float,tabularx,booktabs,multirow,lipsum}
\usepackage{siunitx,xcolor}
\usepackage[version=4]{mhchem}
\graphicspath{{figs/}{figsgaoerb/}} 

\usepackage{framed}
\usepackage{color}
\definecolor{shadecolor}{rgb}{0.92,0.92,0.92}

\usepackage{lastpage}
\usepackage{fancyhdr}
\usepackage{autobreak}
\allowdisplaybreaks

\usepackage{ragged2e}

\usepackage{geometry}
\usepackage{tikz,xcolor,hyperref}
\definecolor{lime}{HTML}{A6CE39}
\DeclareRobustCommand{\orcidicon}{
\begin{tikzpicture}
\draw[lime, fill=lime] (0,0)
circle[radius=0.14]
node[white]{{\fontfamily{qag}\selectfont \tiny \.{I}D}};
\end{tikzpicture}
\hspace{-2mm}}
\foreach \x in {A, ..., Z}{
\expandafter\xdef\csname orcid\x\endcsname{\noexpand\href{https://orcid.org/\csname orcidauthor\x\endcsname}{\noexpand\orcidicon}}}

\begin{document}

\preprint{APS/123-QED}

\title{Finite Temperature Dynamics of Spin Solitons with Applications in Thermocouples and Refrigerators}

\author{Chaofan Gong\hspace{-2mm}\orcidA{}} 
\affiliation{waiting for revisions}

\begin{abstract}
The exploitation of spin Berry phases to generate emergent fields for producing miniaturized and high-quality inductors has enjoyed considerable popularity among proponents of quantum technologies [\href{https://www.nature.com/articles/d41586-020-02721-7}{Nature 586, 202 (2020)}].
Inspired by this breakthrough, we extend its mechanism to spin thermoelectrics by probing responses of ferrimagnetic domain walls (DWs) to thermal gradients.
Similarly, voltages here stem from DW-spin collective motion, in contrast to normal electron transport phenomena.
We further develop finite-temperature dynamics to investigate thermoelectric figures of merit and attribute corresponding quantum superiority to ultrafast spin  evolution of ferrimagnetism with tunable non-Abelian phases.
We propose a more likely cause of DW motion towards hot or cold regions (contrary to conclusions of previous reports) and verify existence of efficient magnon-momentum transfers.
These findings deepen our understanding of heat-driven DW kinetics and suggest profitable new directions in an emerging realm of spincaloritronics.
\end{abstract}

\maketitle
\pagestyle{fancy}

\textbf{Introductions.}
After-heat in our life is ubiquitous.
It occurs in human bodies, industrial products, solar beams, electronic components, and geothermal collections.
Heat is the most widely existing form of energy, and our domestic electricity mainly comes from thermal power generation.
Conventional thermocouples utilize Seebeck, Nernst, and Thomson effects to convert waste heat into electricity, and we can also conversely harness Peltier effects for refrigeration.
As Moore's law reaches saturation, these thermoelectric elements are encountering significant obstacles, shifting our attention to their quantum counterparts.
In quantum physics, spins are excellent candidates, and they have been utilized with striking success in memories \cite{parkin2008Magnetic} and computations \cite{luo2020Currentdriven}.

Recently, the further revolution \cite{yokouchi2020Emergent} in which spin Berry phases are used to generate electromagnetic fields has enabled inductors to be fabricated, and these quantum inductors are surprisingly effective even above room temperature \cite{kitaori2021emergenta}.
It has also been discovered that large Berry curvature in magnetic semi-metals \cite{xiang2020large} and antiferromagnets \cite{roychowdhury2022large,you2022anomalous,xu2022observation,lin2022evidence,chen2021anomalous,ikhlas2017large} will enhance thermoelectric transports, indicating that combinations of spins and thermoelectronics are becoming new trends \cite{uchida2021transverse,kuroyama2022realtime,krzysteczko2015domain,niemann2016thermoelectric}.
When spins condense into various textures, their Berry phases will form scattering potentials hindering heat transport \cite{lee2022magnon}, thus effectively improving thermoelectric figures of merit.
These results all testify remarkable efficiency of spin emergent fields, and making full use of this emergent electromagnetism to design advanced topology-quantum-materials \cite{keimer2017physics} is a grand challenge \cite{tokura2022quantum} in the realm of contemporary science-technology, inspiring us to employ thermal gradients to excite these fields for inducing electricity.
Intuitively, thermal gradients will stimulate magnon flows (i.e. spin Seebeck and Nernst effects \cite{uchida2008observation,cheng2016Spin}) and entropy increases \cite{raimondo2022temperaturegradientdriven}, boosting spin collective motion to make Berry phases (serving as ``magnetic vector potential'') change with spacetime to generate spin-motive force for power generation.

Thermal gradients furnish driving methods which are more productive \cite{slonczewski2010Initiation,matsuo2018spin}, convenient \cite{choi2015Thermal}, and low-loss \cite{uchida2008observation} than currents and magnetic fields, leading to pollution-free and recyclable spincaloritronics \cite{bauer2012Spin}.
Meanwhile, thermal gradients also offer very effective ways for producing \cite{wang2020thermal} and stabilizing \cite{mazo-zuluaga2016controlling} DWs.
Beyond applications in spin batteries \cite{yang2009universal}, heat-driven DW dynamics can realize green refrigerators \cite{kovalev2012thermomagnonic,kovalev2010magnetocaloritronic,kovalev2009Thermoelectric}, chaos generators \cite{shen2021Signal}, and random-number devices \cite{zazvorka2019Thermal}.
Compared with techniques of magnetic susceptibilities, neutron scatterings, and magneto-optical microscopes, thermal transport measurements can detect magnetic orders more operatively, especially for antiferromagnetism \cite{fernandezscarioni2021thermoelectric,bartell2015Tabletop,katsura2010Theory} without net magnetic moments.

To deepen our understanding of responses of DWs to thermal gradients, we naturally choose ferrimagnets with Dzyaloshinskii-Moriya interactions (DMIs).
Ferrimagnets have advantages in strong quantum fluctuations \cite{chernyshev2015strong}, long spin-coherence lengths \cite{yu2019Long}, low damping factors \cite{kim2019Low}, tunable spin densities \cite{okuno2019Spintransfer}, high electrical sensitivities \cite{siddiqui2018CurrentInduced}, and enormous emergent fields \cite{surgers2014large,park2020numerical}.
DMIs preferably suppress the degree of freedom of DW precession \cite{okuno2020Magnetic,yu2018polarizationselective} and expedite Zeeman couplings and spin stagger \cite{dasgupta2021zeeman}, benefitting translational motion to attain ultrafast spin dynamics \cite{kim2017fast} so that relativistic effects \cite{caretta2020Relativistic} cannot be bypassed.
These phenomena essentially originate from ferrimagnetic non-Abelian Berry phases \cite{cheng2012Electron} producing extra spin nutation \cite{mondal2021spin}, in contrast to ferromagnetic Abelian phases.

Nevertheless, under thermal gradients, physical mechanisms behind Berry-phase-dominated DW dynamics still remain elusive.
As mentioned in Ref. \cite{pacchioni2020heat}, it is unclear what forces are exerted on DWs, and what is the predominant factor that causes DWs to move to hot or cold regions.
Recent experiments observe that ferromagnetic DWs can move to hot \cite{yu2021realspace,jiang2013direct,tolley2015generation,torrejon2012unidirectional,qin2022dynamics} or cold \cite{shokr2019steeringa,wang2020thermal,jen1986thermal,qin2022dynamics} regions, individually attributed to magnon spin torque \cite{yan2011allmagnonic,qin2022dynamics} or magnon momentum transfers \cite{shokr2019steeringa}/ thermal diffusion \cite{wang2020thermal,qin2022dynamics}/ Nernst-Ettingshausen effects \cite{jen1986thermal} (DWs we refer to here include skyrmions as well).
We perceive that these conclusions are not complete because they not quantitatively analyze various factors and neglect entropy-increase processes \cite{schlickeiser2014role,wang2016thermally,sukhov2016swift} that have significant impacts on DW velocities.
DW entropy relates to Heisenberg interactions \cite{donges2020unveiling,hinzke2011domain,kim2015landaulifshitz, schlickeiser2014role,selzer2016inertiafree,wang2014thermodynamic,yan2015thermodynamic}, DMIs \cite{wang2021rectilinear,gorshkov2022dmigradientdriven,yershov2020domain}, anisotropy interactions \cite{shen2018dynamics,sukhov2016swift,tomasello2018chiral}, and Zeeman couplings \cite{raimondo2022temperaturegradientdriven} and can promote DW migration towards hot or cold ends depending on positive or negative signs of the above interactions.
In addition, thermal gradients will stimulate excessive dipole fields \cite{moretti2017domain}, making ferromagnetic DWs move to hot regions.
Here we revisit these problems in ferrimagnetism generalized to ferromagnetism and antiferromagnetism and conclude DW migration to hot regions in other experiments due primarily not to magnon spin-transfer torque but to entropy and Dzyaloshinskii-Moriya \cite{kim2019tunable} vector potential. 

On the other hand, interactions between magnons and DWs also need to be solved.
Several articles in Refs. \cite{selzer2016inertiafree,kovalev2014skyrmionic,kovalev2012thermomagnonic,yan2011allmagnonic,bayer2005phase,hertel2004domainwall,zhao2020spin} assume that DWs are non-scattering solitons having fully-transmitted Pöschl-Teller potential \cite{kiriushcheva1998scattering}.
A small number of reports think that wide DWs \cite{yan2015thermodynamic} and high-frequency magnons \cite{yan2015thermodynamic,kim2012interaction} are conducive to scatterings. 
Quite a few papers ruminate that narrow DWs and low-frequency magnons will have noticeable reflection \cite{ross2020propagation,yu2018polarizationselective,qaiumzadeh2018controlling,hamalainen2018control,chang2018ferromagnetic,lan2017antiferromagnetic,borys2016spinwave,wang2015magnondriven,schroeter2015scattering,pirro2015experimental,iwasaki2014theory,hata2014spinwaveinduced,bogdan2014spin,wang2013spinwave,wang2013analytical,wang2013magnonic,wang2012domain,macke2010transmission,yuan2006quantum,liu1979spin,mochizuki2014thermally,shen2020driving,schutte2014magnonskyrmion,laliena2022scattering,sterk2021green,faridi2022atomicscale,daniels2019topological,zhang2018spinwavedriven,shen2018motion,lee2022magnon,kim2022interaction} and even inelastic scatterings \cite{dadoenkova2019inelastic} and resonances \cite{hata2014spinwaveinduced,wang2013spinwave,iwasaki2014theory,janutka2013resonance,kim2012interaction,seo2011magnetic,han2009magnetic,jin2022spinwave,gruszecki2022influence}, and Refs. \cite{iwasaki2014theory,hu2022micromagnetic} further claim that scatterings will provoke the strongest momentum transfer when magnon wavelengths are close to DW widths. 
From our viewpoint, various factors (comprising DW configurations \cite{hamalainen2018control,chang2018ferromagnetic,borys2016spinwave,pirro2015experimental,seo2011magnetic,lan2021skew,lee2022magnon}, magnon polarization \cite{lan2017antiferromagnetic,faridi2022atomicscale,qaiumzadeh2018controlling}, temperature, magnetic fields \cite{bogdan2014spin}, exchange interactions, DMIs \cite{liu2022spinwavedriven,wang2022allmagnonic,yu2018polarizationselective,borys2016spinwave,wang2015magnondriven,lan2021skew,zhang2019thermala,gruszecki2022influence,qaiumzadeh2018controlling}, easy \cite{wang2015magnondriven,bogdan2014spin,wang2013analytical,gruszecki2022influence} $\&$ hard \cite{laliena2022scattering,qaiumzadeh2018controlling,wang2015magnondriven,janutka2013resonance} axis anisotropy, damping \cite{sterk2021green}, and pinning \cite{zhao2020spin}) affecting interactions between magnons and DWs can be attributed to Lagrangian alterations that vary DW potential and magnon frequencies to affect scatterings, providing intuitive-overall comprehension for magnon-DW interactions and demonstrating that DWs are excellent magnon regulators.
This permits us to better devise logic-gates \cite{yu2020magnetic,daniels2019topological}, filters \cite{hamalainen2018control,chang2018ferromagnetic}, polarizers \cite{lan2017antiferromagnetic}, phase-retarders \cite{ye2021magnetically}, fibers \cite{yu2016magnetic}, interferers \cite{hertel2004domainwall,moon2013control}, Stern-Gerlach generators \cite{wang2022allmagnonic}, Goos-Hänchen producers \cite{laliena2022magnonic,wang2019gooshanchen}, and focalizers \cite{bao2020offaxial} via magnons with the above regulable parameters.

Ultimately, the extrapolation \cite{donges2020unveiling} that ferrimagnetic DWs migrate to hot ends below Walker-breakdown limits and above angular-momentum compensation points (in other circumstances, they shift to cold ends) is also incomplete because it dismisses substantial momentum transfers so as to engender anomalous divergence of physical quantities owing to exploiting  zero-temperature equations to dissect finite-temperature cases  \cite{schlickeiser2012temperature,stanciu2006ultrafast,binder2006magnetization,caretta2018fast}.
Ref. \cite{donges2020unveiling} recognizes that magnons are entirely transmissive, but it misses a essential that magnons will transmit momentum via redshifts \cite{kim2014propulsion} after passing through DWs apart from angular-momentum transfers.
We demonstrate in this study that mutual scatterings of magnons and emergent fields is sufficiently robust to make DWs overcome entropy forces and migrate to cold ends.
By considering influences of various forces on DW, we can forecast their behaviors more accurately, avoiding information losses that DW velocities are slowed down or even reversed due to environmental heat.

\begin{figure}[tp]
\includegraphics[clip=true,width=1.0\columnwidth]{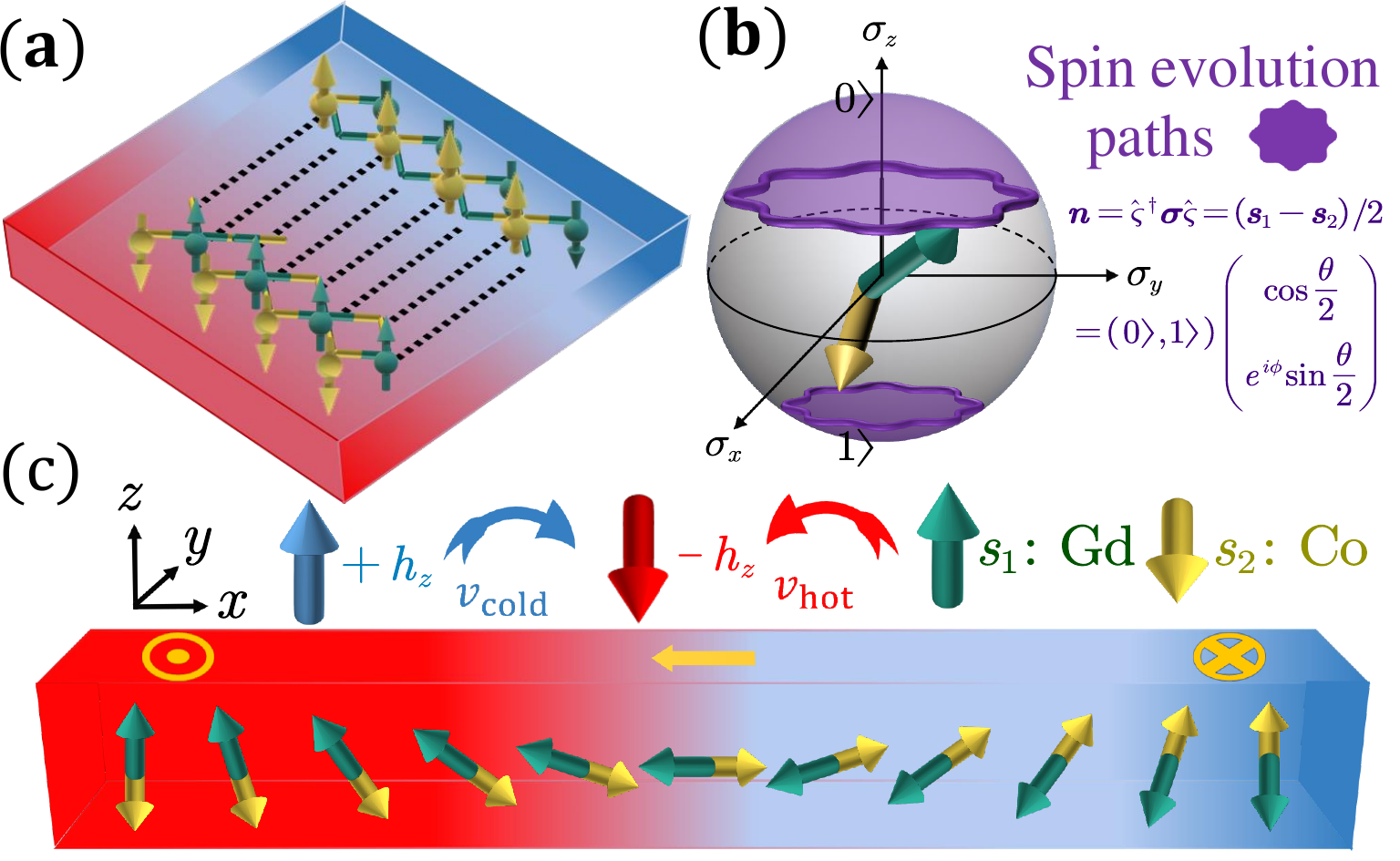}
\caption{
(a) Thermal gradients excite magnon flows in a ferrimagnetism composed by Co and Gd.
(b) Spin staggering evolution on a Bloch sphere constitutes non-Abelian Berry phases.
(c) Left-handed Néel DWs (topological charges equal one, i.e. $q=1$) driven by magnon spin torque, magnon momentum transfers, magnetic fields, and thermodynamic entropy forces can move to hot or cold regions under different parameters.
}
\label{2}
\end{figure}

\textbf{{Finite-temperature dynamics dominated by Berry phases.}} 
Thermal gradients will excite magnons and electrons [Fig. \ref{2}(a)]. 
Magnons exert spin-angular-momentum and momentum to DWs, while electrons impart angular momentum to DWs by spin transfer  torque and spin orbit torque (here electron momentum transfers \cite{tatara2004theory} are very small \cite{gong2022dynamics}).
Under thermal gradients, even accounting for charge pumping \cite{abbout2018cooperative}, DW velocities caused by magnons are much larger ($\sim 10^3$) than that of electron spin-transfer torque \cite{tatara2015thermala,shokr2019steeringa,torrejon2012unidirectional}.
Due to the existence of the non-magnetic metal layer Pt in our experiment, electrons also have spin-orbit torque, which is two times (proportional to ratios of DW widths to film thicknesses \cite{shen2020currentinduced}) greater than spin-transfer torque but still less than magnon influences (spin-orbit torque likewise enhance magnon propagations \cite{woo2017control}).  
In the experiment, we apply a current $1\times 10^9\,\,\mathrm{A}\cdot \mathrm{m}^{-2}$ corresponding to a thermal gradient $1\times 10^7\,\,\mathrm{K}\cdot \mathrm{m}^{-1}$ (Seebeck effects), and DWs can not overcome pinning to start moving. 
However, in insulators \cite{jiang2013direct,yu2021realspace} where only magnons are present, DW motion is observed with a small thermal gradient $\sim 1\times 10^3\,\,\mathrm{K}\cdot \mathrm{m}^{-1}$.
These experimental phenomena further show that electron flows excited by thermal gradients are inoperative in DW motion.
Additionally, thermal diffusion here is ineffective on account of a large cross-sectional area \cite{kim2015thermophoresis}.

Synthesizing the above analysis, we consider the magnon adiabatic and diabatic spin torque  $\varOmega_s$ and $\beta _{s}$, magnon redshift $\beta _r$, magnon momentum transfer $\beta _{\mathcal{R}}$, entropy effect $\beta _T$, vertical magnetic field $ h_z $, antiferromagnetic-coupling exchange interaction $\epsilon _{\mathrm{ne}}$, DMI $\epsilon _{\mathrm{dm}}$, easy axis anisotropy $\epsilon _{\mathrm{an}}$, hard axis anisotropy $\epsilon _{\mathrm{ah}}(m)$ including the thermally induced dipole interaction $\epsilon _{\mathrm{dip}}(m)$, and pinning potential $\hat{V}_{\mathrm{p}}$ and compose the DW Lagrangian (the dimension is $\mathrm{J}$) by a spinor \cite{lifshitzs1982course} [Fig. \ref{2}(b)] $\hat{\varsigma}_{j}^{\dagger}=\left( e^{-i\phi _j}\cos \frac{\theta _j}{2},\sin \frac{\theta _j}{2} \right)$ (since the spin $\boldsymbol{s}_j$ of each lattice point $j$ is quantized so that $\left( \theta _j,\phi _j \right)$ can only take a series of discrete values, for concision, we will adopt Einstein's summation convention to omit these subscripts):
\begin{gather}
\hat{\mathcal{L}}_{\mathrm{w}}=\rho _nn\left| \hat{\varsigma}^{\dagger}\left( -i\partial _{\tau}-\mathcal{J} _{\mathrm{m}} \right) \hat{\varsigma} \right|^2+2 n\hat{\varsigma}^{\dagger}\left( -i \rho _s \partial _{\tau}-\mathcal{J} _{\mathrm{m}} \right) \hat{\varsigma}-
\nonumber \\  
2\epsilon _{\mathrm{ne}}\left| a\left( -i \nabla -\boldsymbol{\mathcal{A} _r} \right) \hat{\varsigma} \right|^2+\frac{a^2}{2} ( \epsilon _{\mathrm{an}}-\left| \zeta \right| ) [ 1- \left( \hat{\varsigma}^{\dagger}\sigma _z\hat{\varsigma} \right) ^2 ]
\nonumber \\   
-(\epsilon _{\mathrm{ah}}+\epsilon _{\mathrm{dip}})a^2\left( \hat{\varsigma}^{\dagger}\sigma _x\hat{\varsigma} \right) ^2+nmh_z\hat{\varsigma}^{\dagger}\sigma _z\hat{\varsigma}+\hat{V}_{\mathrm{p}}
\label{1}
\end{gather}
with the dynamic coupling $\mathcal{J} _{\mathrm{m}}=\left( 1-\mathcal{R} \right) \rho _s\varOmega _s-\left[ \left( 1-\mathcal{R} \right) \left( \beta _s-\beta _r \right) -\mathcal{R} \beta _{\mathcal{R}}+\beta _T \right] \hat{\varsigma}^{\dagger}\boldsymbol{\sigma }\hat{\varsigma}\times ({\boldsymbol{j}_{\mathrm{m}}}/{\hbar}) \cdot ( \nabla -\frac{\epsilon _{\mathrm{dm}}}{\epsilon _{\mathrm{ne}}}\boldsymbol{e}_i )$ between magnons and Berry phases, magnon reflectivity $\mathcal{R}$, relative spin density $\rho _s=s/n$,  polar/azimuth angle $ \theta /\phi  $, Pauli matrix $\boldsymbol{\sigma }=\left( \sigma _x,\sigma _y,\sigma _z \right) $, magnon current $\boldsymbol{j}_{\mathrm{m}}$, reduced Planck constant $\hbar$, unit vector $\boldsymbol{e}_i=\left( e_x,e_y,e_z \right) $, spin inertia $\rho _n= n/\epsilon _{\mathrm{ne}}$, Néel vector $\boldsymbol{n}=\left( \boldsymbol{s}_1-\boldsymbol{s}_2 \right) /2= \hat{\varsigma}^{\dagger}\boldsymbol{\sigma }\hat{\varsigma}$, Co/Gd spin vector $ \boldsymbol{s}_1/\boldsymbol{s}_2 $, net spin density $ \boldsymbol{s}=\boldsymbol{s}_1+\boldsymbol{s}_2 $, Co/Gd magnetization $ M_1/M_2 $, vector lengths $n$ and $s$, time scale $ \tau  $, lattice constant $a$, DW Berry phase $\boldsymbol{\mathcal{A} _{\mu}} =\left( \mathcal{A} _{\tau}, \boldsymbol{\mathcal{A} _r} \right) =( -i\rho _s\hat{\varsigma}^{\dagger}\partial _{\tau}\hat{\varsigma}-\rho _n( \hat{\varsigma}^{\dagger}\partial _{\tau}\hat{\varsigma} ) ^2,-i\rho _s\hat{\varsigma}^{\dagger}\partial _{\boldsymbol{r}}\hat{\varsigma}-i \zeta \boldsymbol{\sigma } ) $, relative net magnetization $m={{\left( \gamma _1s_1-\gamma _2s_2 \right)}/{\nu n}}$, Co/Gd gyromagnetic ratio $ \gamma _1/\gamma _2 $, vacuum permeability $\nu$, and DMI characteristic scale $\zeta =\epsilon _{\mathrm{dm}}/\epsilon _{\mathrm{ne}}$.
The above interactions result in left-handed Néel DWs [Fig. \ref{2}(c)], and their dynamic processes under magnons are shown in Figs. \ref{3}(a)-(c).

\begin{figure}[tp]
\includegraphics[clip=true,width=1.0\columnwidth]{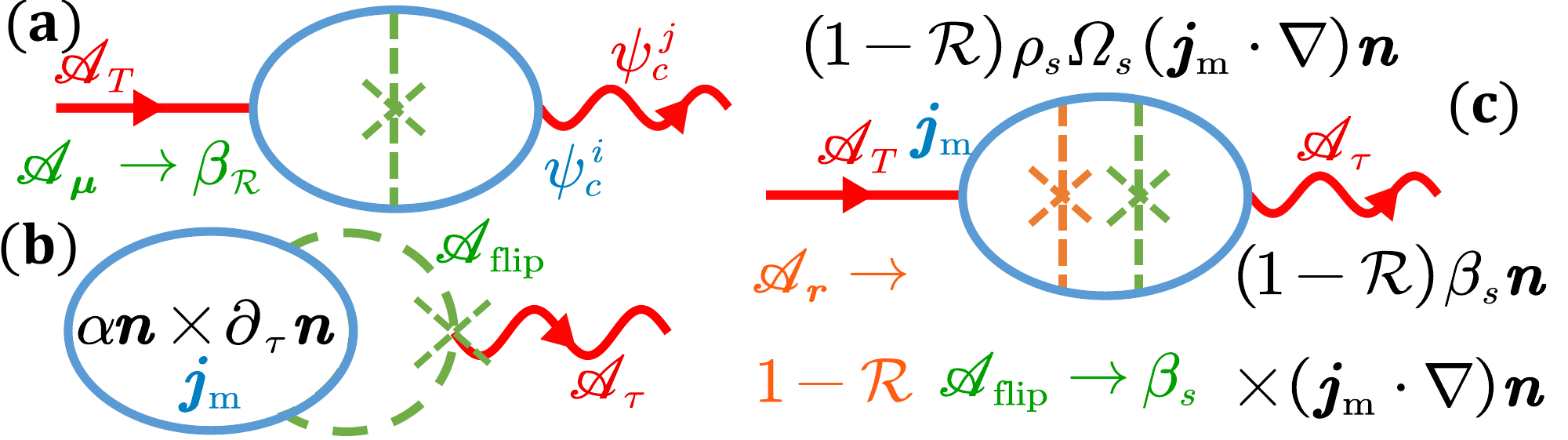}
\caption{
(a) Temperature-gradient (equivalent to a thermal vector potential $\mathcal{A} _{T}$) excited magnons are reflected ($\beta _{\mathcal{R}}$) by the DW Berry phase $\mathcal{A} _{\boldsymbol{\mu }}$.
Considering more general inelastic scatterings, magnon incident ($\psi _{c}^{i}$) and reflective ($\psi _{c}^{j}$) states are different.
(b) The spin flip scattering $\mathcal{A} _{\mathrm{flip}}$ produces the lateral damping $\alpha $, hindering a time-dependent variation $\mathcal{A} _{\tau}$ of Berry phases motivated by a magnon flow $j_{\mathrm{m}}$.
(c) When magnons interacting with an emergent field $\mathcal{A} _{\boldsymbol{r}}$ pass through ($1-\mathcal{R}$) DWs, spin torque including adiabatic and diabatic terms is produced, in which the diabatic term $\beta _s$ also comes from the spin flip scattering $\mathcal{A} _{\mathrm{flip}}$.
}
\label{3}
\end{figure}

\begin{figure}[tp]
\includegraphics[clip=true,width=1.0\columnwidth]{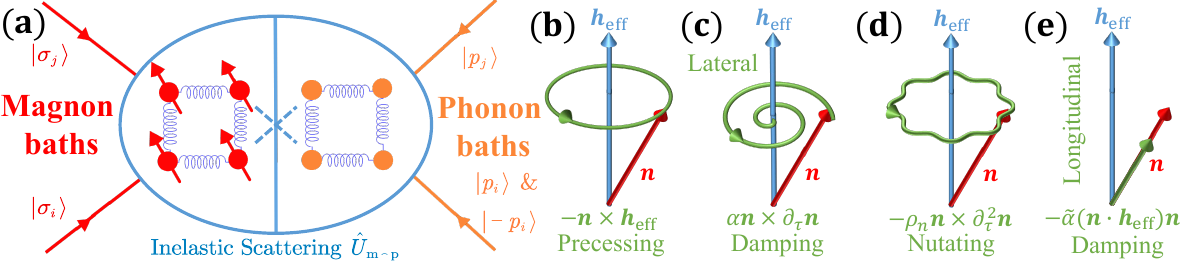}
\caption{
Magnon and phonon thermal baths at finite temperature will induce non-Markov and Markov  friction (a), making spin  evolution have extra longitudinal damping (e) in addition to precession (b), lateral damping (c), and nutation (d).
The motion (d) is the embodiment of ferrimagnetic non-Abelian phases, which does not exist in ferromagnetism.
The motion (e) causes magnetization to decrease with temperature and is the reason for the existence of Néel and Curie temperature.
}
\label{4}
\end{figure}

$\hat{\varsigma}^{\dagger}$ and $\hat{\varsigma}$ represent DW creation and annihilation operators with the number $N=\hat{\varsigma}^{\dagger}\hat{\varsigma}$ of DWs.
Other advantages of expressing Eq. \eqref{1} by spinors are that they avoid the Berry-phase-evoked uncertainty \cite{wen2010quantum} $2j\pi$ (i.e. $\int_0^{\tau}{\mathcal{A} _{\tau}d\tau}=-\int_0^{\tau}{[ i\rho _s\hat{\varsigma}^{\dagger}\partial _{\tau}\hat{\varsigma}+\rho _n\left( \hat{\varsigma}^{\dagger}\partial _{\tau}\hat{\varsigma} \right) ^2 ] d\tau}=-i\rho _s\left[ \phi \left( \tau \right) -\phi \left( 0 \right) +2j\pi \right] +\cdots $ with the integer $j$) and non-physical phenomenon that Lagrangian relies on choices \cite{tchernyshyov2015conserved} of gauge potential.
The first term in $\hat{\mathcal{L}}_{\mathrm{w}}$ signifies spin nutation \cite{mondal2021spin} bringing about DW mass and ultrafast dynamics [Fig. \ref{4}(d)].

At the finite temperature $T$, spins are coupled to phonon-and-magnon thermal baths [Fig. \ref{4}(a)], making fluctuation-dissipation terms appear so that magnetization of each lattice point decreases with temperature \cite{nieves2016selfconsistent}. 
The dissipative Lagrangian can be denoted as $\hat{\mathcal{L}}_{\mathrm{d}}=\frac{1}{2}\alpha n\rho _n\left( \partial _{\tau}\hat{\varsigma}^{\dagger}\boldsymbol{\sigma }\hat{\varsigma}+\hat{\varsigma}^{\dagger}\boldsymbol{\sigma }\partial _{\tau}\hat{\varsigma} \right) ^2+\frac{1}{2}\tilde{\alpha}U_{\mathrm{w}}\left( \hat{\varsigma}^{\dagger}\boldsymbol{\sigma }\hat{\varsigma} \right) ^2$ [$U_{\mathrm{w}}$ is the potential energy in Eq. \eqref{1}], where the damping parameter $\alpha$ contains the Markov $\alpha_{\mathrm{p}}$, non-Markov \cite{kim2018magnoninduced} $\int_{-\infty}^{\tau}{[ \alpha_{\mathrm{m}} ( \tau -\tau ^{'} )] \frac{\partial \tau ^{'}}{\partial \tau}}d\tau ^{'}$, and DW structure scattering-relation \cite{akosa2016phenomenology,akosa2017intrinsic,hals2009intrinsica} $\alpha _0\sqrt{{{\epsilon _{\mathrm{ne}}}/{\epsilon _{\mathrm{an}}}}}\boldsymbol{n}\cdot \left( \nabla \times \boldsymbol{n} \right)$ (the strength parameter $\alpha _0$) parts, and the longitudinal relaxation $\tilde{\alpha}$ evinces spin and magnetization vector lengths decreasing with temperature \cite{nieves2016selfconsistent} [Figs. \ref{4}(b)-(e)].
Spins in Markov processes are connected to phonons and in Berry-phase-induced non-Markov procedures are linked to magnons, and the homographic energy-fluctuation is $\left< \delta \mathcal{H} \left( t \right) \right> =\int_0^{\infty}{d\omega \frac{\rho _{n}^{2}\hbar \omega \cos \left( \omega \tau \right)}{2\tanh \left( \hbar \omega /2k_BT \right)}}\left[ \int_0^{\infty}{d\omega \exp \left( -\omega \tau \right) \alpha _{\mathrm{m}}}+4\pi \left( \alpha _{\mathrm{p}}-\tilde{\alpha} \right) \right]$ with the magnon frequency $\omega$ and Boltzmann constant $k_B$.
These phonon baths will also trigger the longitudinal-spin-length fluctuation $\left< \delta \boldsymbol{n}_{\parallel}\left( t \right) \right>  =\int_0^{\infty}{d\omega \frac{\tilde{\alpha}\rho _n\cos \left( \omega \tau \right)}{2\tanh \left( \hbar \omega /2k_BT \right)}}$.
Atomic spin dynamics simulations \cite{weissenhofer2021skyrmion} also verify these additional friction-damping effects due to temperature.

\begin{figure}[tp]
\includegraphics[clip=true,width=1.0\columnwidth]{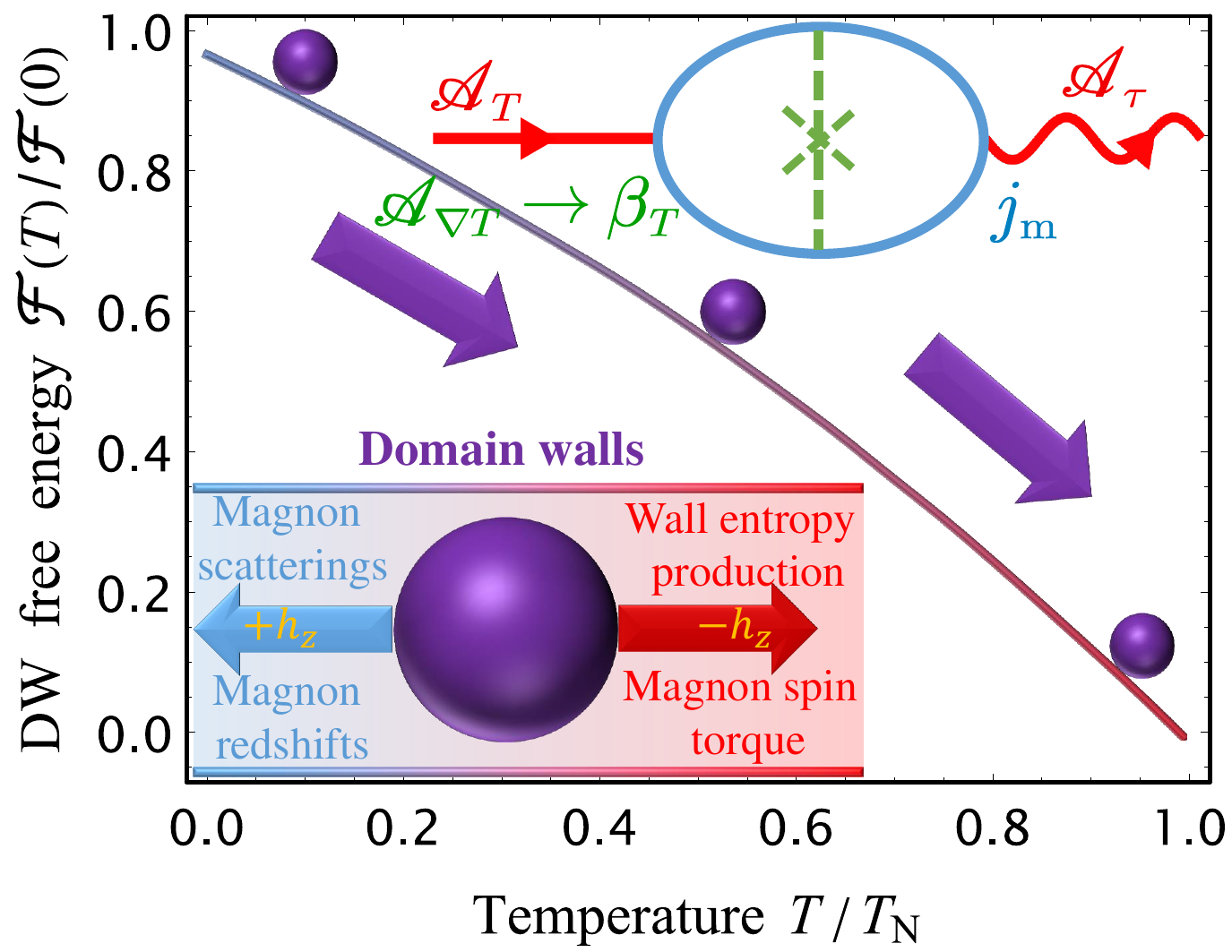}
\caption{
Of factors affecting DW velocities, entropy forces make DWs move to hot regions, while magnons and magnetic fields can make them move to hot (magnon spin torque and $+h_z$) or cold (magnon scatterings, redshifts, and $-h_z$) ends.
Entropy increases are tantamount to free energy reductions, and free energy decreasing with temperature always makes DWs move towards hot regions like playing slides (the Néel temperature is $T_{\mathrm{N}}$).
Here we equate DW entropy generation as a gauge field $\mathcal{A} _{\nabla T}$ and then write it as the diabatic strength parameter $\beta _T$ in Eqs. \eqref{6}-\eqref{7}.
}
\label{5}
\end{figure}

\textbf{{Entropy forces acting on DWs.}} 
To simplify problems, we equate entropy increase efficacy \cite{selzer2016inertiafree} under thermal gradients to forces acting on DWs \cite{gong2022dynamics}, avoiding complexity of dealing with non-equilibrium path integrals.
Entropy generation is equivalent to free-energy diminutions, which can be derived from Eq. \eqref{1}, i.e.
\begin{gather}
\{ 1+\alpha \left[ n\left( T \right) /n\left( 0 \right) \right] ^2 \} ^{-1}\varDelta \mathcal{F} =4\left\{ \epsilon _{\mathrm{ne}}\left[ \epsilon _{\mathrm{an}}+ \right. \right. 
\nonumber \\
\left. \left. \left( \epsilon _{\mathrm{ah}}+\epsilon _{\mathrm{dip}} \right) \cos ^2\phi \right] \right\} ^{1/2}-\pi \epsilon _{\mathrm{dm}}\cos \phi +nmh_z/a,
\label{2}
\end{gather}
and the homologous force $F=\frac{d\mathcal{F}}{dx}=\frac{d\mathcal{F}}{dT}\nabla T$ can be paraphrased as an intensity quantity $\beta _T=-\frac{2\hbar}{nk_B}\frac{\partial \mathcal{F}}{\partial T}$ in diabatic spin torque.
For convenience, we unify the thermally induced dipole field $\epsilon _{\mathrm{dip}}=\frac{1}{2}\frac{\nu aL_z}{L_z+\varDelta} \int{\frac{\partial \left( M_1-M_2 \right)}{\partial T}\frac{\partial T}{\partial x}dx}$ to hard axis anisotropy with the DW with $\varDelta $ and film thickness $L_z$.
Here these thermodynamic factors always make DWs tend to hot ends since all parameters decrease with temperature [Fig. \ref{5}].
This conclusion is also applicable to ferromagnetic and antiferromagnetic systems because their energy ground states can be mapped to each other via $\boldsymbol{s}_j \leftrightharpoons  \left( -1 \right) ^j\boldsymbol{s}_j$ so that they have the same thermodynamic feature.
The force caused by each interaction (containing the Heisenberg interaction \cite{hinzke2011domain,kim2015landaulifshitz, schlickeiser2014role,wang2014thermodynamic}, DMI \cite{hinzke2011domain,kim2015landaulifshitz,schlickeiser2014role,wang2014thermodynamic}, and magnetic anisotropy \cite{gorshkov2022dmigradientdriven,wang2016thermally,yershov2020domain}) in Eq. \eqref{2} is appreciable, confirming that thermal gradients are efficient driving methods.

\begin{figure}[tp]
\includegraphics[clip=true,width=1.0\columnwidth]{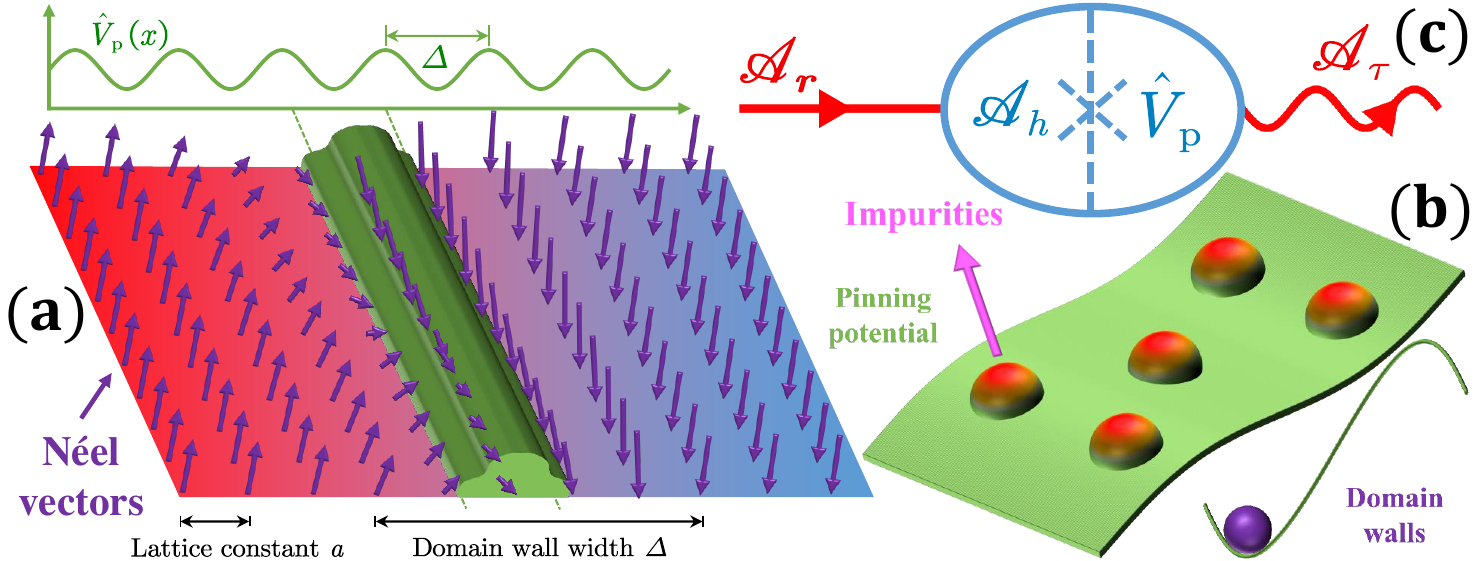}
\caption{
Impurities break the translational and rotational symmetry of DWs (a), deform DWs (a), and make them move in a periodic potential $\hat{V}_{\mathrm{p}}$ (b).
Through quantum tunneling and thermal crawling, DWs cross irregular pinning (a), in much the same way as cars cross speed humps (b). 
(c) An easy-axis field $\mathcal{A} _h$ can resist pinning and even depin DWs ($\boldsymbol{\mathcal{A} _{{r }}}\rightarrow \mathcal{A} _{\tau}$).
}
\label{6}
\end{figure}

\textbf{{Pinning potential.}} 
Pinning comprises both internal and external components.
Internal components \cite{kim2013twobarrier} come from the hard-axis anisotropy $\epsilon _{\mathrm{ah}}\,\,\&\,\ \epsilon _{\mathrm{dip}}$ and DMI $\epsilon _{\mathrm{dm}}$, while external components \cite{kohno2006microscopic} originate from the translational and rotational symmetry-breakings stimulated by system disorders [Figs. \ref{6}(a)-(b)].  
Pining is inherent \cite{litzius2017skyrmion}, dilates with film thicknesses \cite{mougin2007domain}, mediates Gilbert-damping and diabatic-torque production \cite{kohno2007gauge}, and can be counterbalanced by magnetic fields along easy axes \cite{jen1986thermal} [Fig. \ref{6}(c)].
Internal pinning caused by hard axes will be demonstrated in Eq. \eqref{6}, and here we delineate the external pinning potential as $\hat{V}_{\mathrm{p}}=\epsilon _{\mathrm{px}}\exp \left( i\frac{2\pi}{\varDelta}\hat{\varsigma}^{\dagger}\boldsymbol{\sigma }\hat{\varsigma}\cdot \boldsymbol{r}_i \right) \delta \left( \boldsymbol{r}-\boldsymbol{r}_i \right) +\epsilon _{\mathrm{p\phi}}\hat{\varsigma}^{\dagger}\boldsymbol{\sigma }\hat{\varsigma}\cdot \boldsymbol{s}_i\delta \left( \boldsymbol{r}-\tilde{\boldsymbol{r}}_i \right) =\epsilon _{\mathrm{px}}\left( 1-\cos \frac{x}{\varDelta} \right) +\epsilon _{\mathrm{p\phi}}\left( 1-\cos \phi \right)$ with impurities impeding translation (elastic impurities: densities $\epsilon _{\mathrm{px}}\sim \exp \left( -\varDelta /a \right)$ and positions $\boldsymbol{r}_i$) and rotation (magnetic impurities: densities $\epsilon _{\mathrm{p\phi}}$, positions $\boldsymbol{\tilde{r}}_i$, and spins $\boldsymbol{s}_i$) of DWs.
When DW widths approach to characteristic lengths of impurities, pinning is strongest \cite{gruber2022skyrmion}; while $\varDelta$ is away from these characteristic lengths, broad DWs are more facile to cross pinning \cite{berges2022sizedependent} owing to the slackly-variable $\sin x/\varDelta $ and small $\epsilon _{\mathrm{px}}$.
Eq. \eqref{1} asserts that pinning will diminish DW energy \cite{buijnsters2014motion}, tantamount to curtailing DW widths to enhance magnon reflection \cite{ross2020propagation}.
From another perspective, magnetic impurities can correspond to hard-axis anisotropy [see subsequent Eq. \eqref{6}], availing magnon reflection by shrinking the DW width $\varDelta ^2=\frac{4\epsilon _{\mathrm{ne}}\left( 4\epsilon _{\mathrm{an}}-\epsilon _{\mathrm{dm}}^{2}/\epsilon _{\mathrm{ne}} \right) ^{-1}}{1+\epsilon _{\mathrm{ah}}\sin ^2\phi /\left[ \epsilon _{\mathrm{an}}-\nu \left( M_1-M_2 \right) ^2/2 \right]}$. 
A micromagnetic simulation \cite{wang2015magnondriven} still visualizes benefit of hard-axis anisotropy to magnon reflection.
Elastic external pinning will affect scaling laws and pinning strength in the DW creeping-motion solution \eqref{7}, while magnetic-external and internal pinning only determines intensity terms (see Ref. \cite{jue2016chirala}).

\textbf{{Interactions between magnons and DWs.}} 
For magnons, we regard a spin oscillation $\delta \hat{\varsigma}^{\dagger}=( \hat{\psi}_{\delta}^{\phi},\hat{\psi}_{\delta}^{\theta} ) $ and define the magnon two-component field $\hat{\psi}_{c}^{\dagger}=( \hat{\psi}_{\delta}^{\theta}-i\hat{\psi}_{\delta}^{\phi},\hat{\psi}_{\delta}^{\theta}+i\hat{\psi}_{\delta}^{\phi})$, where left ($c=-$) or right ($c=+$) chirality is $\hat{\psi}_{l}^{\dagger}=\,\,\hat{\psi}_{\delta}^{\theta}+i\hat{\psi}_{\delta}^{\phi}$ or $\hat{\psi}_{r}^{\dagger}=\,\,\hat{\psi}_{\delta}^{\theta}-i\hat{\psi}_{\delta}^{\phi}$ [Fig. \ref{7}(a)].
Substituting the above expressions into Eq. \eqref{1}, we acquire the magnon Lagrangian $\hat{\mathcal{L}}_{\mathrm{m}}=\rho _n n| \hat{\psi}_{c}^{\dagger}\left( -i\partial _{\tau}-c\mathcal{A} _{\tau} \right) \hat{\psi}_c |^2+2 n\hat{\psi}_{c}^{\dagger}  \nonumber \left( -i \rho _s \partial _{\tau}-c\mathcal{A} _{\tau} \right) \hat{\psi}_c-\mathcal{H} _{\mathrm{m}}$ with the Hamilton $\mathcal{H} _{\mathrm{m}}=\frac{\hbar ^2}{2\mathcal{M} _{\mathrm{m}}}\left( -i\nabla -\mathrm{c} \boldsymbol{\mathcal{A} _r}-\boldsymbol{\mathcal{A}} _T\right) ^2+\mathcal{F} \exp \left( -\frac{\varDelta}{a} \right) \sin ^2\frac{x}{\varDelta}+nmh_z+\frac{\epsilon _{\mathrm{dm}}^{2}a^2}{2\epsilon _{\mathrm{ne}}}-c\epsilon _{\mathrm{ah}}a^2+\frac{n}{4\rho _n}+\epsilon _{\mathrm{an}}a^2\left( 1-2 \sech ^2\frac{x}{\varDelta} \right)$, magnon mass $\mathcal{M} _{\mathrm{m}}={{4\hbar \rho _n}/{a^2}}$, and thermal vector potential $\boldsymbol{\mathcal{A}} _T=\nabla T/T$.
The magnon flow interacting with DWs in Eq. \eqref{1} is generated by thermal gradients, i.e. $\boldsymbol{j}_{\mathrm{m}}=\delta \mathcal{L} _{\mathrm{m}}/\delta \boldsymbol{\mathcal{A}} _T$.
By the variational deductions $ \hat{\mathcal{L}} _{\mathrm{m}}/\hat{\psi}_{c}^{\dagger} $ and $ \hat{\mathcal{L}} _{\mathrm{m}}/\hat{\psi}_c $, we obtain the magnon-density ($ \hat{\varXi}_c=\hat{\psi}_{c}^{\dagger}\hat{\psi}_c $) evolution equation 
\begin{gather}
\partial _{\tau}\hat{\varXi}_c=\frac{i}{\hbar}[\hat{\varXi}_c,\mathcal{H} _{\mathrm{m}}]+i\left( \rho _n\partial _{\tau}^{2}-\alpha \partial _{\tau} \right) \hat{\varXi}_c+2\frac{\tilde{\alpha}}{\hbar}( \hat{\varXi}_cU_{\mathrm{w}}) \hat{\varXi}_c.
\label{3}
\end{gather}
When magnons interplay with DWs in more comprehensive inelastic processes \cite{landau1981course} [Fig. \ref{7}(b)], we divide the magnon field into the incident and scattered parts, i.e. $\hat{\psi}_{\mathrm{c}}^{\dagger}=\frac{\tanh x-ik}{1-ik} \exp{[i\left( kx-\omega \tau \right)]}+\frac{\tanh \ell +ik}{1+ik}\frac{S\left( \phi \right)}{\sqrt{-i\ell}}\mathcal{I} \left( \phi \right) \exp{[-i\left( k\ell -\omega \tau \right)]}$ with the wavenumber $k$, frequency $\omega$, elastic scattering amplitude $S\left( \phi \right)$, inelastic scattering trajectory $\mathcal{I} \left( \phi \right)$, and deflection path $\ell =x\sec \phi$.
Thus, we define the reflectance as $\mathcal{R} =\frac{1}{\varDelta}\,\,\int_{\frac{\pi}{2}}^{\frac{3\pi}{2}}{S^{\dagger}\left( \phi \right) S\left( \phi \right) \mathrm{Re}\mathcal{I} \left( \phi \right) d\phi}$ with the total cross-section $\varDelta =\int_0^{2\pi}{S^{\dagger}\left( \phi \right) S\left( \phi \right) \mathrm{Re}\mathcal{I} \left( \phi \right) d\phi}=\frac{4}{k}\sum_{\ell =-\infty}^{+\infty}{\sin ^2\vartheta _l}$, Aharonov-Bohm phase shift $\vartheta _l$ \cite{bayer2005phase,hertel2004domainwall}, and partial wave $l$.

$\mathcal{I} \left( \phi \right) =1$ infers complete elastic scatterings, $\mathcal{I} \left( \phi \right) =0$ means that magnons are absorbed by DWs, and $\mathcal{I} \left( \phi \right) >1$ means that the number of reflected particles is increased by $\mathcal{I} \left( \phi \right) -1$ times by releasing magnons from DWs.
Subsequently, by the optical theorem \cite{landau1981course}, we extract the situation of complete transmission magnons
\begin{gather}
\mathrm{Im}S\left( 0 \right) \mathrm{Re}\mathcal{I} \left( 0 \right) =\sqrt{\frac{2}{k\pi}}\sum_{\ell =-\infty}^{+\infty}{\sin ^2\vartheta _l}=\sqrt{\frac{k}{8\pi}}\varDelta ,
\label{4}
\end{gather}
evincing that the verdict of broad DWs and high-frequency magnons facilitating transmissions in the introductions is accurate.

\begin{figure}[tp]
\includegraphics[clip=true,width=1.0\columnwidth]{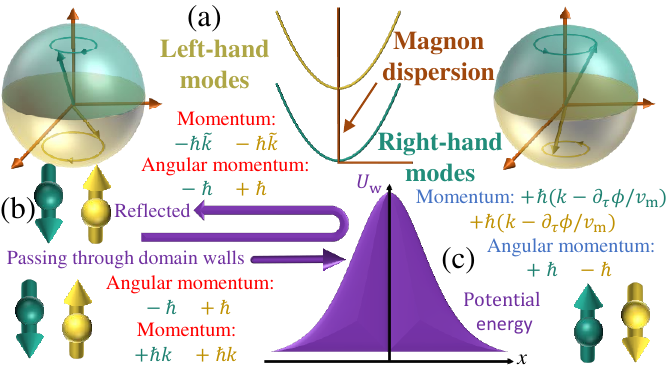}
\caption{
(a) Ferrimagnetic magnons having left and right chiral modes.
(b) Angular momentum of magnons reflected by DWs does not change, but their momentum varies $\hbar ( \tilde{k}+k ) $, and there is $\tilde{k}=k$ when scatterings are elastic.
(c) Angular-momentum and momentum of magnons transmitting through DWs change $2\hbar$ (spin torque) and $2\hbar \partial _{\tau}\phi /v_{\mathrm{m}}$ (redshifts), respectively, and it is accompanied by diabatic spin dislocations and redshifts when passing through DWs.
Magnon energy of left and right polarization is different and is a function of $\rho _s$.
When the spin density $\rho _s$ is equal to 1, magnons are all right-handed (ferromagnetism); when it is equal to 0, magnon left and right polarization is degenerate (antiferromagnetism).
Boson flows depend on choices of gauge potential \cite{wen2010quantum}, which is the physical mechanism by which we can regulate magnon scatterings via magnetic fields ($\mathcal{A} _h$), temperature ($\boldsymbol{\mathcal{A}} _T$), and DWs ($\boldsymbol{\mathcal{A} _{{\mu }}}$).
}
\label{7}
\end{figure}

Distinctively, when magnon energy approaches to DWs (DW quantized energy levels are given in Refs. \cite{su1994temperature,derras-chouk2022dynamics}), resonances will emerge, evoking scattering amplitudes to increase significantly.
The energy eigenvalue is plurality popularized as $E_{\mathrm{m}}=E_0-\frac{1}{2}i\mathcal{I}\varGamma$, where the physical implication of $\varGamma$ can be interpreted as the resonance width.
Supplanting this complex energy into Eq. \eqref{3}, the phase shift will have an excessive term $\delta \vartheta _l=-\mathrm{arc}\tan \frac{\mathcal{I} \varGamma}{2\left( E-E_0 \right)}$ with a new resonance amplitude
\begin{gather}
S_{\varGamma}\left( \phi \right) =-\frac{\mathcal{I} \varGamma \left( 2l+1 \right) /2^ll!\sqrt{k}}{\left[ 2\left( E_{\mathrm{m}}-E_0 \right) +i\varGamma \right]}\frac{d^l\left( \cos ^2\phi -1 \right) ^l}{\left( d\cos \phi \right) ^l}e^{2i\left( \delta \vartheta _l \right)}
\label{5}
\end{gather}
with $\varGamma =\frac{\hbar}{\alpha a}[ \frac{\epsilon _{\mathrm{ne}}}{n}k+\frac{\epsilon _{\mathrm{dm}}( \omega -{\epsilon _{\mathrm{dm}}}k/{n} )}{\epsilon _{\mathrm{an}}+2\epsilon _{\mathrm{ah}}+2\epsilon _{\mathrm{ne}}k^2}]$.
Compared with ferromagnetism, ferrimagnetism has additional nutation resonance modes \cite{mondal2021spin}, leading to stronger momentum transfers and weaker adiabatic spin torque.

After passing through (redshifts \cite{kim2014propulsion}) or being reflected by (scatterings \cite{macke2010transmission,dasgupta2018energymomentum}) DWs, magnons momentum transfers will always occur [Fig. \ref{7}(b)].
We obtain the momentum-transfer intensity owing to scatterings by the variation $\beta _{\mathcal{R}}=-\int{\frac{\delta \mathcal{L} _{\mathrm{w}}}{\hbar \delta \left( \partial _{\tau}\hat{\varsigma}^{\dagger} \right)}\nabla \hat{\varsigma}^{\dagger}\cdot d\boldsymbol{r}}\sim \varDelta k$.
For redshifts, their presence is because the wavenumber alters ${{2\partial _{\tau}\phi}/{v_{\mathrm{m}}}}$ when magnons pass through DWs.
Analogous to the magnon wavenumber variation $2k$ after being scattered, we define the redshift intensity as $\beta _r=\varDelta {{\partial _{\tau}\phi}/{v_{\mathrm{m}}}}$ [Fig. \ref{7}(c)].
Ultimately, the magnon adiabatic or diabatic torque intensity \cite{wang2013spinwave} is $\varOmega _s=1$ or $\beta _s=\alpha$.

\textbf{{Transition to observation representations.}} 
In actual observations, we should embody the above microcosmic imagery in measurable representations, of which one of the most straightforward is the DW velocity $\partial _{\tau}x=v$.
From Eq. \eqref{1}, we realize that the coefficient before the magnetic anisotropy $\epsilon _{\mathrm{an}}$ is positive at a small DMI $\epsilon _{\mathrm{dm}}$, i.e. $\epsilon _{\mathrm{ne}}\epsilon _{\mathrm{an}}>|\epsilon _{\mathrm{dm}}|$, constructing Néel-type DWs.
For Néel DWs, Eq. \eqref{1} subsists a kink-soliton solution $\boldsymbol{n}=\left( \sech \frac{x}{\varDelta}\cos \zeta x,\sech \frac{x}{\varDelta}\sin \zeta x,q\tanh \frac{x}{\varDelta} \right)$ with the topological charge $q=\frac{1}{\pi}\int_{-\infty}^{\infty}{\nabla \theta dx}= 1$, and the DW width is $\varDelta =\sqrt{{{\epsilon _{\mathrm{ne}}}/{\epsilon _{\mathrm{an}}}}}$ in desirable-undisturbed status.
At the new representation $\left( x,\phi \right)$, through a variational operation $\frac{d}{dt}\frac{\delta \hat{\mathcal{L}}_{\mathrm{w}}}{\delta \left( \partial _{\tau}\hat{\varsigma}^{\dagger} \right)}-\frac{\delta}{\delta \hat{\varsigma}^{\dagger}}( \hat{\mathcal{L}}_{\mathrm{w}}+\hat{\mathcal{L}}_{\mathrm{d}} ) =-\frac{\delta \hat{\mathcal{L}}_{\mathrm{d}}}{\delta \left( \rho _n\partial _{\tau}\hat{\varsigma}^{\dagger} \right)}$ with $\partial _{\tau}\hat{\varsigma}^{\dagger}=\partial _{\tau}\theta \left( x \right) \partial _{\theta \left( x \right)}\hat{\varsigma}^{\dagger}+\partial _{\tau}\phi \left( x \right) \partial _{\phi \left( x \right)}\hat{\varsigma}^{\dagger}$ ($\partial _{\tau}\boldsymbol{n}=-v\nabla \boldsymbol{n}+\hat{z}\times \boldsymbol{n}\partial _{\tau}\phi$), we deduce the DW motion equations 
\begin{gather}
\frac{\rho _s+\alpha \zeta \varDelta}{q+\zeta ^2\varDelta ^2}\varDelta \partial _{\tau}\phi +\alpha \partial _{\tau}x+\rho _n\partial _{\tau}^{2}x
\nonumber \\
=-v_{\mathrm{px}}\sin \frac{x}{\varDelta}+\frac{m\varDelta h_z}{\left( q+\zeta ^2\varDelta ^2 \right)}+\bar{\beta}v_{\mathrm{m}}
\nonumber
\end{gather}
and
\begin{gather}
\left( \rho _sq-\alpha \zeta \varDelta \right) \partial _{\tau}x-\alpha \varDelta \partial _{\tau}\phi -\rho _n\varDelta \partial _{\tau}^{2}\phi =v_{\mathrm{d}}\sin \phi -
\nonumber \\
(v_{\mathrm{a}}+v_{\mathrm{dip}}+v_{\mathrm{p\phi}})\frac{q\pi \zeta \sin 2\phi}{\sinh \left( \pi \zeta \varDelta \right)}-\left[ \left( 1-\mathcal{R} \right) q\rho _s\varOmega _s+\bar{\beta}\zeta \varDelta \right] v_{\mathrm{m}}
\label{6}
\end{gather}
with the reciprocal relation $\left[ x,\phi \right] ={\hbar a}/{2n}$, $v_{\mathrm{px}} \sim \epsilon _{\mathrm{px}}$, $v_{\mathrm{a}}\sim\epsilon _{\mathrm{ah}}$, $v_{\mathrm{p\phi}}\sim\epsilon _{\mathrm{p\phi}}$, $v_{\mathrm{dip}}\sim \epsilon _{\mathrm{dip}}$, $v_{\mathrm{d}}\sim\epsilon _{\mathrm{dm}}$, and $\bar{\beta}=\mathcal{R} \beta _{\mathcal{R}}-\left( 1-\mathcal{R} \right) \left( \beta _s-\beta _r \right) -\beta _T$.
At small external forces (below Walker-breakdown limits \cite{thiaville2005micromagnetic}), $\partial _{\tau}\phi =0$ and $\phi =\pi$ hold true.
On the premise of driving forces being weaker than pinning, DWs are crawling \cite{jeudy2016universal}, and the solutions combined with Walker limits are (we write the solution of Eq. \eqref{6} as the form of Arrhenius's law \cite{eltschka2010nonadiabatic,kim2013method})
\begin{gather}
v=\bar{v}_{\mathrm{p}}\sin \left( \omega _{\mathrm{w}}\tau \right) +\left\{ \frac{m\varDelta h_z}{\bar{\alpha}\left( q+\zeta ^2\varDelta ^2 \right)}\exp \left( -\left| \frac{h_{\mathrm{p}}}{h_z} \right|^{\chi _h} \right) \right. +
\nonumber \\
\frac{\bar{\beta}_{\phi}}{\bar{\alpha}}v_{\mathrm{m}}\left. \exp \left( -\left| \frac{\nabla T_{\mathrm{p}}}{\nabla T} \right|^{\chi _{\nabla T}} \right) \right\} \left( 1-e^{-\frac{\tau}{\bar{\tau}}} \right) \exp \left( -\left| \frac{T_{\mathrm{p}}}{T} \right|^{\chi _T} \right) 
\label{7}
\end{gather}
\maketitle
with the pinning strength parameters $h_{\mathrm{p}}\,\,\left( \chi _h \right)$, $\nabla T_{\mathrm{p}}\,\,\left( \chi _{\nabla T} \right)$, and $T_{\mathrm{p}}\,\,\left( \chi _T \right)$, relaxation time $\bar{\tau}\approx \bar{\alpha}/(\rho _n+\alpha )$, DW angular frequency (it is zero below Walker limits) $\omega _{\mathrm{w}}={\left( v_{0}^{2}-v_{\mathrm{p}}^{2} \right) ^{1/2}/{[ \left( \rho _sq-\alpha \zeta \varDelta \right) \frac{\left( \rho _s+\alpha \zeta \varDelta \right)}{\alpha \left( q+\zeta ^2\varDelta ^2 \right)}+\alpha ] \varDelta}}$, $v_0=[ \left( 1-\mathcal{R} \right) q\rho _s\varOmega _s+\bar{\beta}\zeta \varDelta +\frac{\bar{\beta}}{\alpha}\left( \rho _sq-\alpha \zeta \varDelta \right) ] v_{\mathrm{m}}+\frac{\left( \rho _sq-\alpha \zeta \varDelta \right) m\varDelta h_z}{\alpha \left( q+\zeta ^2\varDelta ^2 \right)}-\frac{\rho _sq-\alpha \zeta \varDelta}{\alpha}v_{\mathrm{px}}\sin \frac{x_0}{\varDelta}$, $v_{\mathrm{p}}=\frac{v_{\mathrm{d}}}{2\cos \phi_0}-\frac{q\pi \zeta (v_{\mathrm{a}}+v_{\mathrm{dip}}+v_{\mathrm{p\phi}})}{\sinh \left( \pi \zeta \varDelta \right)}$, $\bar{\alpha}\bar{v}_{\mathrm{p}}=\frac{\rho _s+\alpha \zeta \varDelta}{q+\zeta ^2\varDelta ^2}\frac{v_{\mathrm{d}}}{\alpha 2\cos \phi_0}-\frac{\rho _s+\alpha \zeta \varDelta}{q+\zeta ^2\varDelta ^2}(\frac{v_{\mathrm{a}}+v_{\mathrm{dip}}+v_{\mathrm{p\phi}}}{\alpha})\frac{q\pi \zeta}{\sinh \left( \pi \zeta \varDelta \right)}-v_{\mathrm{px}}\frac{\sin {{x_0}/{\varDelta}}}{\sin 2\phi_0}$, $\bar{\alpha}=\alpha +\frac{\rho _s+\alpha \zeta \varDelta}{q+\zeta ^2\varDelta ^2}\left( \frac{\rho _sq}{\alpha}-\zeta \varDelta \right)$, $\bar{\beta}_{\phi}=\bar{\beta}-\frac{\rho _s+\alpha \zeta \varDelta}{\left( q+\zeta ^2\varDelta ^2 \right) \alpha}\left[ \left( 1-\mathcal{R} \right) q\rho _s\varOmega _s+\bar{\beta}\zeta \varDelta \right]$, and the solution of unknown quantities $\left( x_0,\phi_0 \right)$ is in Supplementary Materials.
$h_{\mathrm{p}}$ and $\nabla T_{\mathrm{p}}$ are depinned by external driving forces \cite{jeudy2016universal}, and $T_{\mathrm{p}}$ is depinned by thermal activation \cite{litzius2020role} that increases DW energy to make them surmount pinning cushier.
Eq. \eqref{7} signifies that adiabatic spin torque only works above Walker-breakdown points, and DWs will be entirely depinned when $h_z\gg h_{\mathrm{p}}$, $\nabla T\gg \nabla T_{\mathrm{p}}$, or $T\gg \,T_{\mathrm{p}}$.

\textbf{{Thermal spin-motive force.}}  
Moving DWs will spawn spin locomotion, making ``Berry phases acting as magnetic vector potential'' vary with time to energize voltages \cite{yang2009universal}, which can also be regarded as inverse effects \cite{freimuth2015direct,zhang2020spin,reiss2022finitefrequency} of current-induced spin transfer torque and spin orbit torque [Fig. \ref{8}(a)].
Spontaneously, if we wield thermal gradients to compel DW motion, the concepts of thermoelectric generation \cite{bauer2010nanoscale} and electrical refrigeration \cite{kovalev2012thermomagnonic} will be appeared.
Hereby, the widely adopted thermoelectric figure of merit \cite{wong2012spin} expending to spin-motive force can be comparably expounded as $\mathcal{E} _{\mathrm{TE}}={{{\mathcal{S}}^2 T}/{\varrho \kappa}}$ with the resistance $\varrho$, Seebeck coefficient $\mathcal{S}$, and heat conductivity $\kappa$.
DW functions are to furnish spin voltages to intensify Seebeck coefficients and reduce thermal conductance by reflecting magnons \cite{huang2015influence} or consuming DW entropy \cite{yan2015thermodynamic}.
Here we give the expression of its ferrimagnetic counterpart as
\begin{gather}
\mathcal{V} _{\mathrm{w}}=\frac{\hbar}{e}pN\int{\left[ \left( \rho _s+\rho _{\mathrm{SOT}}\sin \phi \right) -\left( \beta _{\mathrm{STT}}+\beta _{\mathrm{SOT}}\sin \phi \right) \boldsymbol{n} \right.}
\nonumber \\
\left. \times \boldsymbol{\mathcal{A} _{\mu }} \right] \cdot d\boldsymbol{r}=-\frac{\hbar pN}{e\varDelta}\left\{ \left( \beta _{\mathrm{STT}}+\beta _{\mathrm{SOT}}\sin \phi +q\varDelta \zeta  \right) v \right. 
\nonumber \\
\left. +\left[ q-\left( \beta _{\mathrm{STT}}+\beta _{\mathrm{SOT}}\sin \phi \right) \varDelta \zeta  \right] \varDelta \partial _{\tau}\phi \right\} ,
\label{8}
\end{gather}
and the DW-devoted Seebeck coefficient can be portrayed as $\mathcal{S} =-\frac{\mathcal{V} _{\mathrm{w}}}{L\nabla T}=\frac{\hbar pN}{e\varDelta L\nabla T}(\beta _{\mathrm{STT}}+\beta _{\mathrm{SOT}}\sin\phi )v$ with the system length $L$, diabatic spin-torque intension $\beta _{\mathrm{STT}}$, effective spin-orbit-torque adiabatic or diabatic intension $\rho _{\mathrm{SOT}}$ or $\beta _{\mathrm{SOT}}$ \cite{boulle2014current}, spin polarization $p$, DW number $N$, and charge $e$.
The current corresponding to Eq. \eqref{8} is $j_{\mathrm{e}}=\frac{e}{2\pi}\mathrm{ImTr}[ \frac{\partial S_{\varGamma}}{\partial \tau}S_{\varGamma}^{\dagger} ] $, and then we can define the resistance including DW contributions \cite{saitoh2004currentinduced,tatara1997resistivity,tatara2000domain,tatara2001domain} as $\varrho_{\mathrm{STT}}=\mathcal{V} _{\mathrm{w}}/j_{\mathrm{e}}$, while the Pt resistance $\varrho _{\mathrm{SOT}}$ is only born of electron-phonon scatterings.
For another, as thermotransport mediums, magnons herein surpass phonons and electrons \cite{huang2015influence} (magnons will scatter electrons and phonons \cite{lucassen2011spintransfer}, making them inactive \cite{tian2022ultraweak}, especially for $k_BT \sim \epsilon _{\mathrm{an}}a^2$), and this thermal conductivity can be expressed as $\kappa _{\mathrm{m}}=\frac{\hbar}{4\pi TL}\mathrm{Tr[}\frac{\partial S_{\varGamma}}{\partial \tau}\frac{\partial S_{\varGamma}^{\dagger}}{\partial \tau}]$ by the scattering matrix of Eq. \eqref{5}. 
For the non-magnetic metal layer Pt, the thermal conductivity $\kappa_\mathrm{e}$ is dominated by electrons evoked by the inverse spin-orbit torque.
Perusing some pertinent calculations \cite{tatara1997resistivity,tatara2000domain,tatara2001domain,yan2015thermodynamic,tatara2015thermala,kovalev2012thermomagnonic}, we garner the thermoelectric figure of merit with adopting $q=1$ [Fig. \ref{8}(a)]
\begin{gather}
\mathcal{E} _{\mathrm{TE}}=\frac{2\sqrt{k_BT+nmh_z}}{3n\varDelta L\sqrt{\epsilon _{\mathrm{ne}}}\left( \alpha \pi e \right) ^2}\left\{ \left( \frac{\tilde{\beta}_{\mathrm{STT}}}{\sqrt{\varrho _{\mathrm{STT}}}}+\frac{\kappa _{\mathrm{m}}}{\kappa _{\mathrm{e}}}\frac{\beta _{\mathrm{SOT}}\sin \phi}{\sqrt{\varrho _{\mathrm{SOT}}}} \right) \right. 
\nonumber \\
\left. \left[ (\bar{\beta}+\varDelta \zeta )e^{-\left| \frac{\nabla T_{\mathrm{p}}}{\nabla T} \right|^{\chi _{\nabla T}}}+\frac{m\varDelta h_z}{v_{\mathrm{m}}}e^{-\left| \frac{h_{\mathrm{p}}}{h_z} \right|^{\chi _h}} \right] a\hbar pNe^{-\left| \frac{T_{\mathrm{p}}}{T} \right|^{\chi _T}} \right\} ^2
\label{9}
\end{gather}
with $\tilde{\beta}_{\mathrm{STT}}=\beta _{\mathrm{STT}}+\frac{q\rho _s}{\rho _s+\alpha}+\varDelta \frac{\epsilon _{\mathrm{dm}}}{\epsilon _{\mathrm{ne}}}\frac{\left( q-\beta _{\mathrm{STT}}-\beta _{\mathrm{SOT}}\sin \phi \right) \rho _s+q\alpha}{\rho _s+\alpha}$.
This thermoelectric figure of merit can also characterize the spin-Peltier refrigeration utility: moving DWs will dissipate energy by inelastic scatterings of releasing magnons, i.e. $\mathcal{E} _{\mathrm{Peltier}}=\varrho \kappa T\mathcal{E} _{\mathrm{TE}}$ [Fig. \ref{8}(b)].
By contrast, the thermoelectric figure of merit of ferromagnetic insulator DWs in Ref. \cite{kovalev2012thermomagnonic} is the magnetic Seebeck effect \cite{brechet2013evidence}, i.e. the competence of thermal gradients to convert into magnon flows or effective magnetic fields.

\begin{figure}[tp]
\includegraphics[clip=true,width=1.0\columnwidth]{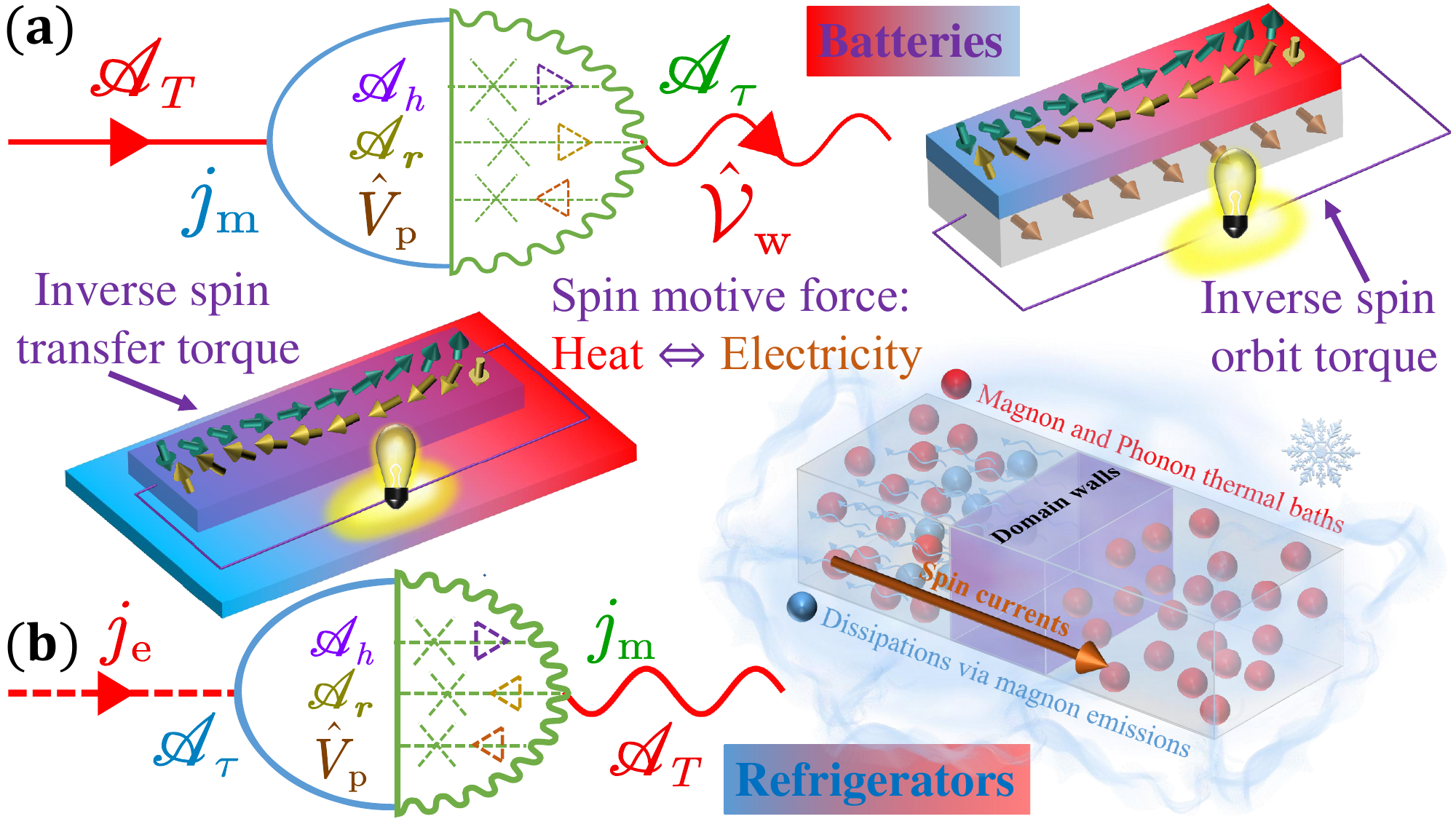}
\caption{
(a) The thermal-gradient-excited magnon flow $j_{\mathrm{m}}$ interacts with the DW $\mathcal{A} _{\boldsymbol{r}}$ under the magnetic field $\mathcal{A} _h$ and pinning impurity $\hat{V}_{\mathrm{p}}$ to convert the time-varying Berry phase $\mathcal{A} _{\tau}$ to the spin-motive force $\hat{\mathcal{V}}_{\mathrm{w}}$.
(b) The inverse effect of thermal spin-motive force, i.e. current-driven DW motion for cooling, where energy is dissipated by releasing magnons through DWs subjected to particle-bath friction.
}
\label{8}
\end{figure}

\textbf{{DW dynamics under thermal gradients.}}  
Initially, we discuss the necessity of finite-temperature  correction.
Figs. \ref{9}(a)-(b) reveal that ferrimagnetism is more sensitive to temperature than ferromagnetism and antiferromagnetism.
At the magnetization compensation temperature $T_m$, systems are immune to external magnetic fields ($m=0$) but have limited spin angular momentum ($\rho _s\ne 0$). 
At the angular momentum compensation temperature $T_s$, angular momentum disappears ($\rho _s = 0$), but there is limited magnetization ($m\ne0$). 
It is these individual features that allow ferrimagnetism to combine the advantages ($m\ne0$ and $ \rho _s = 0$) of ferromagnetism ($m\ne0$ and $\rho _s\ne 0$) and antiferromagnetism ($m=0$ and $\rho _s = 0$).
To further illustrate effects of temperature on dynamics, we rewrite Eq. \eqref{1} to $\rho _s\partial _{\tau}\boldsymbol{n}=-\boldsymbol{n}\times \boldsymbol{h}_{\mathrm{eff}}+\alpha \boldsymbol{n}\times \partial _{\tau}\boldsymbol{n}-\rho _n\boldsymbol{n}\times \partial _{\tau}^{2}\boldsymbol{n}-\tilde{\alpha}\left( \boldsymbol{n}\cdot \boldsymbol{h}_{\mathrm{eff}} \right) \boldsymbol{n}$ ($\boldsymbol{h}_{\mathrm{eff}} \ [ \mathrm{s}^{-1}] $ stands for system energy in a form of effective magnetic fields).
Equivalently, we have $\rho _s\boldsymbol{n}\cdot \partial _{\tau}\boldsymbol{n}=\boldsymbol{n}\cdot \boldsymbol{n}\times \left( -\boldsymbol{h}_{\mathrm{eff}}+\alpha \partial _{\tau}\boldsymbol{n}-\rho _n\partial _{\tau}^{2}\boldsymbol{n} \right) -\tilde{\alpha}\left( \boldsymbol{n}\cdot \boldsymbol{h}_{\mathrm{eff}} \right) \boldsymbol{n}^2\Rightarrow \frac{\rho _s}{2}\partial _{\tau}\boldsymbol{n}^2=-\tilde{\alpha}\left( \boldsymbol{n}\cdot \boldsymbol{h}_{\mathrm{eff}} \right) \boldsymbol{n}^2 \Rightarrow \left| \boldsymbol{n} \right|\sim \exp \left( -\int{\boldsymbol{n}\cdot \boldsymbol{h}_{\mathrm{eff}}\left( T \right) \tilde{\alpha}\left( \rho _s,T \right) dT} \right) $, certifying that the role of temperature enables the hitherto-widely-used zero-temperature spin-dynamics equation (i.e. $\tilde{\alpha}=0$) to be defeated.
Because the restriction $\partial _{\tau}\boldsymbol{n}^2=0$ in the zero-temperature equation ($\tilde{\alpha}=0$) leads to spin-vector-length conservation, a misconception will emerge that magnetization still exists even at Néel or Curie temperature.
Given this idea, there is no room for the consideration of the compensation points $T_m$ and $T_s$.
As a result, the previous studies \cite{caretta2018fast,kim2017fast,donges2020unveiling} which adopted the zero-temperature equation to analyze ferrimagnetism will produce a widely differing consequence, with the infinite divergence of the gyromagnetic ratios and damping parameters.
In fact, extending to finite temperature is a vital requirement \cite{nikitin2022thermal} in examining DW-entropy influences because it essentially derives from the spin vector length reducing with temperature, i.e. $\partial _{\tau}\left| \boldsymbol{n} \right|=\partial _T\left| \boldsymbol{n} \right|\nabla Tv$, and all interaction parameters are proportional to $\left( \partial _T\left| \boldsymbol{n} \right| \right) ^{\eta}$ with the respective power relation $\eta$.

\begin{figure}[tp]
\includegraphics[clip=true,width=1.0\columnwidth]{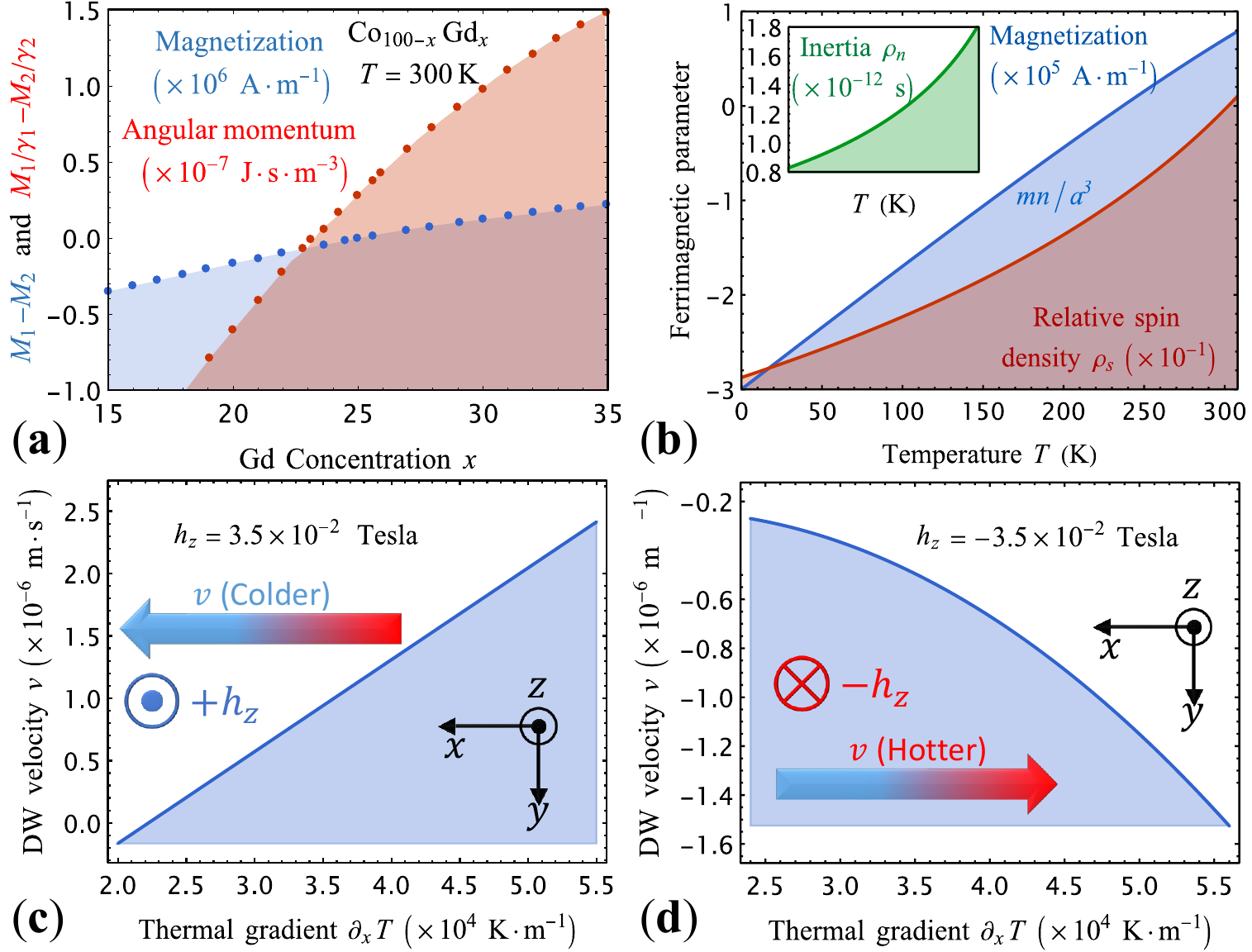}
\caption{
Below Walker-breakdown limits + At crawling intervals.
(a) Magnitudes of magnetization and angular-momentun vary with Gd contents at $T=300\mathrm{K}$ (experimental data selected from Ref. \cite{cai2020ultrafast}). 
(b) Parameters characterizing ferrimagnetic natures vary with temperature at $\mathrm{Co}:\mathrm{Gd}=80:20$.
(c) DW velocities vary with thermal gradients.
(d) DW velocities vary with thermal gradients when magnetic fields in (c) are reversed.
}
\label{9}
\end{figure}

Then, we will demonstrate that magnon momentum transfers (cold) can be strong enough to exceed magnon spin-transfer torque (hot) and entropy forces (hot) and are thereby capable of propelling DWs to cold regions.
Firstly, we add a magnetic field in the $+z$ direction to make DWs move towards cold regions and detect that DW velocities increase with thermal gradients [Fig. \ref{9}(c)].
Nevertheless, this increasing trend can not be solely explained as the role of momentum because magnetic-field-driven DW motion with thermal activation [Eq. \eqref{7}] can also evoke the similar result \cite{mazo-zuluaga2016controlling}.
We thereafter flip magnetic fields and find that, while velocities still increase with thermal gradients [Fig. \ref{9}(d)], they do so to a smaller extent than that depicted in Fig. \ref{9}(c).
This unfolds that the contribution of magnetic fields to velocities is larger than that of thermal gradients so that velocity magnitudes keep increasing with thermal gradients due to depinning by thermal activation.
Basically, magnon momentum transfers can be embodied in the following two facets: 
(1) the existence of magnon momentum transfers engenders velocities in Fig. \ref{10}(a) greater than that in Fig. \ref{10}(b);
(2) the nonlinearity of velocities and thermal gradients in Fig. \ref{9}(d) is stronger than that in Fig. \ref{9}(c), indicating that DWs in Fig. \ref{9}(d) suffer lesser forces due to their motion directions opposite to that contributed by magnons.
Consequently, magnon momentum-transfer proportions are the difference between velocities in Figs. \ref{9}(c)-(d) or Figs. \ref{10}(a)-(b).
Ref. \cite{khoshlahni2019ultrafasta} supports our conclusion via atomic-spin-dynamics simulations, i.e. finding that small radius skyrmions will move to cold regions at low damping.
This is a result of momentum transfers because scatterings will stand above thermal diffusion under low damping and small radiuses.

Identically, DW motion towards cold ends reported in Ref. \cite{jen1986thermal} is also the combined effect of magnetic fields and momentum transfers rather than the Nernst-Ettingshausen mechanism.
From this viewpoint, Ref. \cite{shokr2019steeringa} can not well prove the existence of magnon momentum transfers because it does not exclude influences of magnetic fields.
Actually, the phenomenon \cite{mochizuki2014thermally} that skyrmions rotate around the disk center can not also justify the existence of momentum transfers because the entropy force $F$ and magnon scatterings can cause the same velocity in the $y$-direction, i.e. $v_{y}^{\mathrm{sky}}=\frac{4\pi q\left( 4\pi \alpha v_{\mathrm{m}}+F \right)}{\left( \alpha \int{\partial _x\boldsymbol{n}\cdot \partial _y\boldsymbol{n}dxdy} \right) ^2+16\pi ^2q^2}$.
In particular, the phenomenon \cite{wang2020thermal} of skyrmion migration to cold ends is because a large number of skyrmions produced in high-temperature regions will repel each other, and this repulsive force can push skyrmions to cold regions (relevant simulation and numerical-comparison results are shown in Refs. \cite{lin2014aca,gong2022dynamics}).
A recent experiment \cite{qin2022dynamics} subsequently reports that skyrmion motion directions will be reversed from cold to hot regions when thermal gradients continue to increase, consistent with the conclusion of Refs. \cite{tatara2015thermala,gong2022dynamics} (since many factors are not considered, Ref. \cite{wang2022domainwalla} only draws a conclusion that DW velocities increase with temperature under a fixed thermal gradient).
When $\varDelta k=1$, momentum transfers and resonances enhanced by DMIs are strongest due to the scattering intensity $\beta _{\mathcal{R}}\sim \varDelta k$ and scattering rate $\mathcal{R} \sim {{1}/{\varDelta k}}$ [Fig. \ref{10}(c)], resembling to the cases \cite{iwasaki2014theory,hu2022micromagnetic} of magnons scattered by skyrmions.
Scattering intensity of DWs to left and right chiral magnons is different \cite{faridi2022atomicscalea}, and the relative number of two chiral magnons is affected by the spin density $\rho _s$, but its influence is not as great as the above factors.
Large thermal gradients will enhance magnon frequencies and excite soliton inner modes (i.e. $\partial _{\tau}\phi \ne 0$), facilitating spin torque to move towards hot regions [Fig. \ref{10}(d)].

\begin{figure}[tp]
\includegraphics[clip=true,width=1.0\columnwidth]{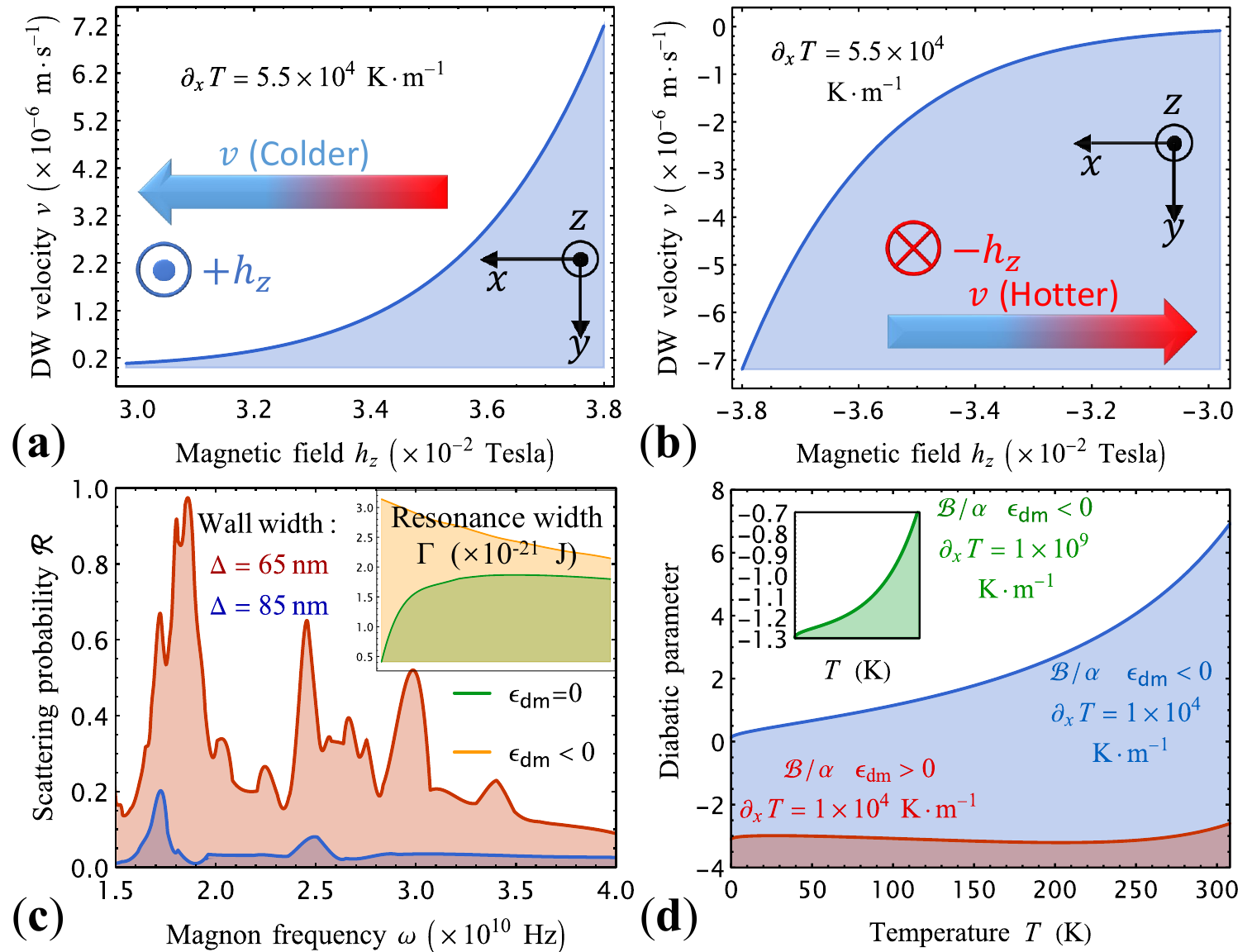}
\caption{
(a)-(b) DW velocities vary with magnetic fields below Walker-breakdown limits (in crawling intervals).
(c) Influences of magnon frequencies and DW widths on scatterings.
(d) The total diabatic parameter $\bar{\beta}_{\phi}/\bar{\alpha}$ including magnon momentum, magnon angular momentum, entropy effects, and DMIs varies with temperature under different thermal gradients (below Walker-breakdown limits).
Positive or negative values of this parameter represent DW motion towards cold or hot regions, respectively.
}
\label{10}
\end{figure}

We go on to show that the principal cause of DW migration to hot regions is not magnon spin torque but DMI vector potential and entropy forces.
When thermal gradients are so massive that DWs are above Walker-breakdown limits ($\partial _{\tau}\phi \ne 0$), adiabatic spin torque and redshifts will work [Eq. \eqref{7}].
Whereas large thermal gradients are likely to accompany by global high temperature, this will simultaneously increase magnon wavenumbers and DW widths, reducing scattering rates [Eq. \eqref{4} and Fig. \ref{10}(c)].
Because redshifts-induced momentum transfers are proportional to DW angular frequencies, their effects are to increase velocities by a factor of $\frac{\left( 1-\mathcal{R} \right) \rho _s}{\left( \rho _s-1+\mathcal{R} \right) \rho _s+\alpha ^2}$ but not to impact on velocity directions, i.e., we can eliminate $\beta _r\sim \partial _{\tau}\phi $ when we solve Eq. \eqref{6} [see Eq. \eqref{7}].
Another effect of the Walker-breakdown occurrence is that it can enhance the easy-axis anisotropy $\epsilon _{\mathrm{an}}+\rho _nn\left( \boldsymbol{n}\times \boldsymbol{e}_z\partial _{\tau}\phi /a \right) ^2$ to increase magnon reflectivity, but  its comprehensive influence will still push DWs toward hot regions when $\rho _s$ is considerable [Fig. \ref{10}(d)].

However, if we reverse the DMI sign, DW motion directions can be reversed [Fig. \ref{10}(d)].
DMIs resemble dipole fields \cite{thiaville2012dynamics}, inducing hard-axis anisotropy and jointly twisting the DW width as \cite{tretiakov2010current} $\varDelta ^2=\frac{4\epsilon _{\mathrm{ne}}\left( 4\epsilon _{\mathrm{an}}-\epsilon _{\mathrm{dm}}^{2}/\epsilon _{\mathrm{ne}} \right) ^{-1}}{1+\epsilon _{\mathrm{ah}}\sin ^2\phi /\left[ \epsilon _{\mathrm{an}}-\nu \left( M_1-M_2 \right) ^2/2 \right]}$.
Generally, DMIs have the following functions:
(1) decreasing magnon reflectivity by increasing DW widths and causing additional scattering potential [Eqs. \eqref{3}-\eqref{4}]; 
(2) devoting an extra entropy force that always makes DWs trend towards cold regions because $\pi \epsilon _{\mathrm{dm}}\cos \phi >0$ in Eq. \eqref{2} ($\epsilon _{\mathrm{dm}}>0\Rightarrow \phi =0$ and $\epsilon _{\mathrm{dm}}<0\Rightarrow \phi =\pi $) (but here it doesn't exceed effects of Heisenberg and anisotropic interactions, see Fig. \ref{5}); 
(3) attracting additional DMI magnon flows to make DWs move towards hot ($\frac{\left( 1-\mathcal{R} \right) q\rho _s\varOmega _s+\bar{\beta}\zeta \varDelta}{\alpha \zeta \varDelta -\rho _sq-\alpha ^2\frac{q+\zeta ^2\varDelta ^2}{\rho _s+\alpha \zeta \varDelta}}<0$) or cold ($\frac{\left( 1-\mathcal{R} \right) q\rho _s\varOmega _s+\bar{\beta}\zeta \varDelta}{\alpha \zeta \varDelta -\rho _sq-\alpha ^2\frac{q+\zeta ^2\varDelta ^2}{\rho _s+\alpha \zeta \varDelta}}>0$) regions [$v\sim \frac{\left( 1-\mathcal{R} \right) q\rho _s\varOmega _s+\bar{\beta}\zeta \varDelta}{\alpha \zeta \varDelta -\rho _sq-\alpha ^2\frac{q+\zeta ^2\varDelta ^2}{\rho _s+\alpha \zeta \varDelta}} v_{\mathrm{m}}$ in Eqs. \eqref{6}-\eqref{7}];
(4) enhancing \cite{chen2021chiral} amplitudes by a negative sign or attenuating \cite{chen2021chiral} amplitudes by a positive sign [the resonance width in Eq. \eqref{5}];
(5) intensifying ferrimagnetic properties of translation and rotation freedom-degree couplings and making relaxation time not always positive [Eqs. \eqref{6}-\eqref{7}] (especially when DMIs are relatively large, DWs can be accelerated throughout the whole movement process).
In these competitive relationships here, the third effect is dominant: contributions of magnons to DW velocity directions are inverse when DMI signs are reversed [Fig. \ref{10}(d)].
Ferromagnetic micromagnetic-simulations \cite{wang2015magnondriven,wang2016thermally} and ferromagnetic and antiferromagnetic micromagnetic-calculations \cite{liang2019motion} bear out our contention, and the experiment \cite{han2019mutuala} also observe that a positive DMI sign will make DWs move towards magnon sources (hot ends). 

The observed tendency \cite{yu2021realspace} of DW motion to hot regions is therefore due to influences of vector potential caused by DMIs on magnon scatterings.
Differently, the reference \cite{torrejon2012unidirectional} supposes that magnon spin torque pull DWs towards hot regions because DWs still move when thermal gradients are gradually reduced to zero.
While thermal gradients are zero, magnon flows will not disappear immediately due to finite diffusion lengths so they can continue to drive DWs.
Nevertheless, in the absence of magnons, only DW inertia can also make DWs slide for certain distances so the above phenomenon \cite{torrejon2012unidirectional} should be interpreted as the coalescent consequence of magnons and entropy forces.
One of the pivotal factors is that spin torque is smaller than entropy forces, and its magnitudes caused by exchange interactions alone can be larger than magnon torque \cite{schlickeiser2014role,wang2016thermally}.
In the experiment \cite{jiang2013direct}, there are no DMIs but hard-axis anisotropy stabilizing Bloch DWs.
Analogously, this does not prove that magnon spin torque make DWs move towards hot regions because more significant entropy forces are not excluded.
Summing up, phenomena of DWs moving towards hot regions observed by them are actually caused by entropy forces because
magnon amplitudes and their reflection on DWs are reduced when DMI signs are reversed or DMIs are equal to zero ($-\rightarrow + / 0$).

Eventually, the investigation \cite{donges2020unveiling} of ferrimagnetic DW dynamics under thermal gradients cannot explain our experimental phenomena, for reasons which we have explained in our introductions.
Distinctively, by micromagnetic simulation, Ref. \cite{kong2021dynamics} finds skyrmions moving to cold ends with spiral backgrounds \cite{knapman2021currentinduced} induced by large DMIs, because these backgrounds will provide boundary forces conducive to motion towards cold regions.
The article \cite{lee2022magnon} indicates that a structure called skyrmion-DWs can have Rosen-Morse potential with total reflection, and DWs and easily-excited gapless magnons formed in systems with second-order magnetic anisotropy also have ascendant reflectivity \cite{kim2022interaction}.
As we mentioned earlier, DW-structure changes and variations of various interactions are the same, which all modify system Lagrangian and thus affect magnon reflectivity.
If DMI strength becomes large enough, DWs will have additional rotational degrees of freedom \cite{boulle2013domain}, but our conclusion is still valid because this can be eliminated by $\phi$, equivalent to multiplying DW velocities by a coefficient.

\textbf{{Power generation and refrigeration by heat-spin reciprocities.}} 
Spin motive force depends on DW velocities under thermal gradients [Eq. \eqref{8}], which embodies change speeds of Berry phases, and having tremendous impacts on the changing rates is pinning [Figs. \ref{11}(a)-(b)].
Among them, interference of magnetic impurities on DW motion is weaker than that of elastic impurities because only adiabatic spin torque can immediately contribute to velocities through $\phi$ [Fig. \ref{11}(a)].
Moreover, DWs propelled by thermal gradients have a merit in depinning, i.e., they can preferably surmount impurities due to thermal activation [Figs. \ref{9}(c)-(d)] (antiferromagnetic couplings will strengthen this effect \cite{dohi2022enhanced}, compared with ferromagnetic DWs \cite{jen1982dragging}).
Another quantity having conspicuous influences on velocities is the relative spin density $\rho _s$, and velocities will accomplish the maximum at $ \rho _s =  0$ [Figs. \ref{11}(b)-(c)], i.e., systems are completely antiferromagnetic couplings.
Nonetheless, once systems are in these states, DWs will not oscillate and cannot output alternating voltages. 
This requires us to sacrifice some efficiency to swap a frequency $\omega _{\mathrm{w}}$ controled by thermal gradients, magnetic fields, and temperature [Eqs. \eqref{7}-\eqref{8}, and Fig. \ref{11}(c)].

In the previous ferromagnetic vortex experiment \cite{tanabe2012spinmotivea}, magnitudes ($ \sim \mathrm{\mu V} $) and frequencies ($\sim$ ns) of spin-motive force are all considerable, and these will be further consolidated in ferrimagnetism.
In any event, commercial levels ($\mathcal{E} _{\mathrm{TE}}\geqslant 3$) can be fulfilled [Fig. \ref{11}(d)] if we can manufacture devices with smoother surfaces \cite{kim2013twobarrier} and thinner films \cite{mougin2007domain} to achieve tiny external and internal pinning (thinner thicknesses and smaller lengths-widths can reduce internal pinning caused by dipole fields and numbers of foreign impurities, respectively).
Considering the DW count $N$ and pinning proportional to the cross-sectional area $A$, we find $\mathcal{E} _{\mathrm{TE}}\sim {{N^2}/{\varDelta V}}$ with the device volume $V$.
In small or large volume, there is $V\ll 1$ or $N\gg 1$, enunciating that spin motive force can demonstrate superior performances at any size.

\begin{figure}[tp]
\includegraphics[clip=true,width=1.0\columnwidth]{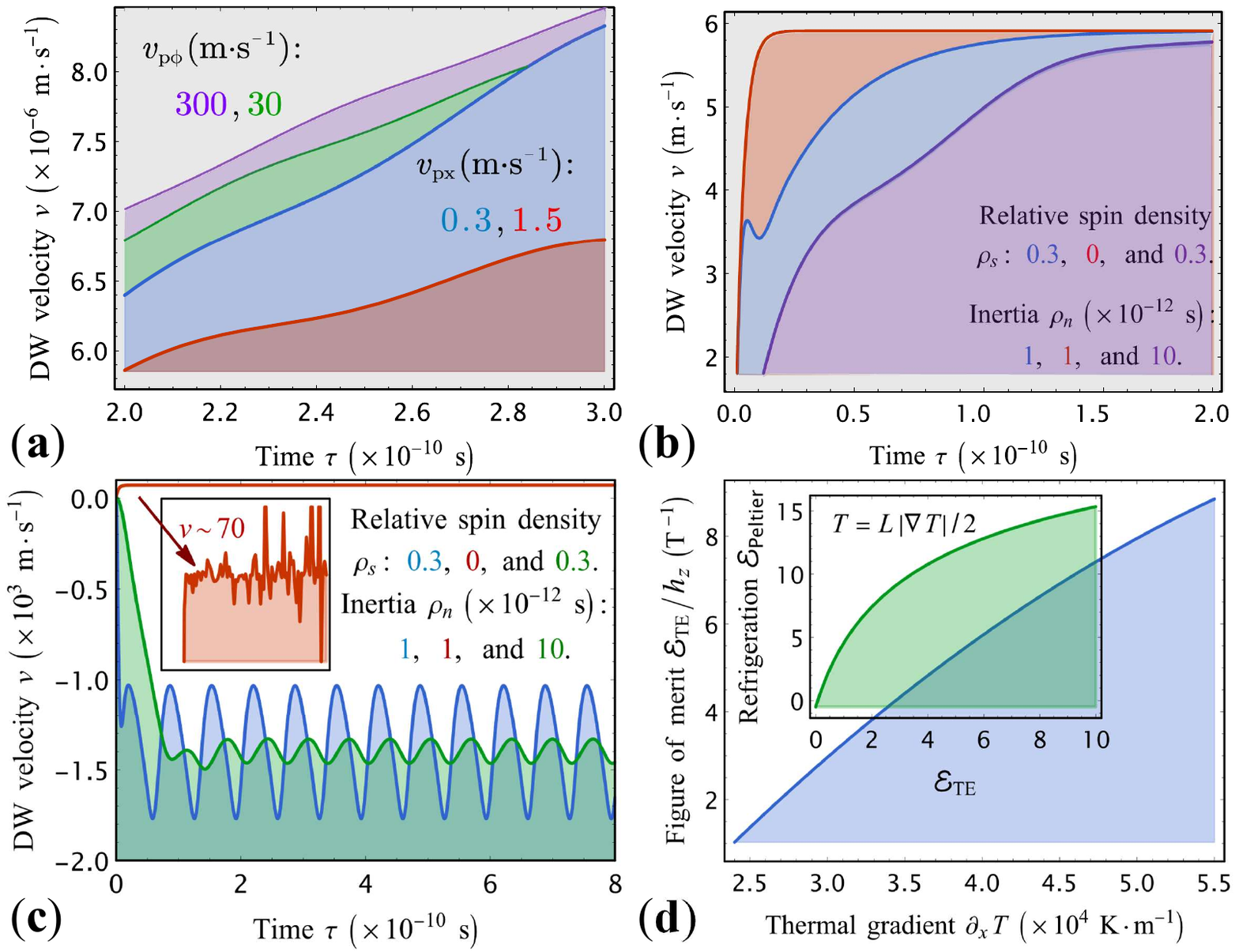}
\caption{
(a) Influences of pinning strength on DW velocities below Walker-breakdown limits (in crawling intervals).
(b)-(c) Influences of ferrimagnetic characteristic parameters on DW velocities below(b) or above(c) Walker-breakdown limits (in depinning intervals).
(d) The influences of magnetic fields and temperature on thermoelectric figures of merit.
}
\label{11}
\end{figure}

Fig. \ref{11}(d) proclaims that magnetic fields and temperature can raise thermoelectric figures of merit, which they do by affecting magnons and DWs.
As heat transport mediums \cite{agrawal2013direct,huang2015influence,an2013unidirectional,yan2013magnon}, low-energy magnons will be completely reflected.
At significant thermal gradients, high-energy magnons with small DMIs are transmissive but will lose their energy when passing through precession DWs due to redshifts.
We can also adjust DMI intensities so that magnons of various wavelengths are reflected \cite{zhang2019thermala}.
Hence, the existence of DWs will invariably reduce thermal conductance and improve electrical conductance, breaking the confinement of the Wiedemann-Franz law like Ref. \cite{lee2017anomalously}.
Magnetic fields and temperature can regulate magnon energy and can be usefully employed to depin DWs to ulteriorly enhance thermoelectric figures of merit [Eq. \eqref{9}].
Even if there are phenomenal thermal gradients (high energy magnons), we can still wield bigger magnetic fields \cite{moon2013control,an2013unidirectional,qiu2022tunable}, DMIs \cite{zhang2019thermala,qiu2022tunable}, and DW structure potential \cite{lee2022magnon} to make magnons be completely reflected.
Furthermore, magnons can also form non-dissipative and non-thermal-conductive superfluids \cite{flebus2016twofluid} in a wide range of temperature and magnetic fields to increase electrical conductance and reduce thermal conductance concurrently and scatter with electrons to shape drag effects \cite{boona2016research,yamaguchi2019microscopic,lucassen2011spintransfer,miura2012microscopic} for amplifying the diabatic factor $\beta _{\mathrm{STT}}$.

Spincaloritronics devices overcome the difficulty of thermoelectric-conversion disappearances due to electron-hole compensation, avert the constraint of the Wiedemann-Franz law constraining concurrent optimization of electrical and thermal conductance, and refine complex material stacks \cite{liu2019topologically}.
Analogously, magnetic topological semimetals \cite{xiang2020large} and topological insulators \cite{kasahara2018majorana,wei2016minimum,liang2016maximizing,xu2014enhanced,zahid2010thermoelectric,goyal2010mechanicallyexfoliated} can also retain these excellences and fulfil better thermoelectric figures of merit.
Magnetic semimetals have strong Berry curvature, while topological insulators have topological  embodiment in reciprocal  space (associated with real  space of DWs).
By contrast, DWs have advantages in miniaturization, low-energy consumption, and high-temperature resistance.
All these features are ascribed to huge emergence of spin Berry phases \cite{kitaori2021emergenta,ye1999berrya,hamamoto2015quantized} [Fig. \ref{12}], and generated electrons with spin degrees of freedom have greater chemical potential \cite{wang2003spin} to enhance charge accumulations.

The inverse effect of spin thermoelectricity, i.e. cooling, it stems from the fact that DWs subjected to frictional forces will dissipate energy by emitting magnons.
Different from the spin Peltier effect \cite{flipse2014observationa,miura2012microscopic} of only exchanging energy of magnons and phonons, the existence of DWs makes magnons affected by Berry phases generate non-Markovian  friction so that environments and systems are more closely related.

In practical applications, mutually perpendicular temperature gradients and voltage directions can better achieve separations of electrons and heat flows to break the Wiedemann-Franz law. 
We can replace DWs with vortex walls \cite{yang2010topological} or skyrmions \cite{yamane2019skyrmiongenerateda} to implement this transverse thermoelectric effect, i.e., the relationship between the direction of voltages and velocities is $\mathcal{V} _{\mathrm{w}}\boldsymbol{e}_y\sim v\boldsymbol{e}_x$ in Eq. \eqref{8}.

No experiments characterize influences of various factors on thermoelectric figures of merit in detail but only estimate the order of magnitudes in Fig. \ref{10}(d) based on existing measurement results.
For this realm, we still need to investigate effects of DWs with different widths, magnons with different frequencies, and static and alternating magnetic fields in any direction on thermal and electrical conductance.

\begin{figure}[tp]
\includegraphics[clip=true,width=1.0\columnwidth]{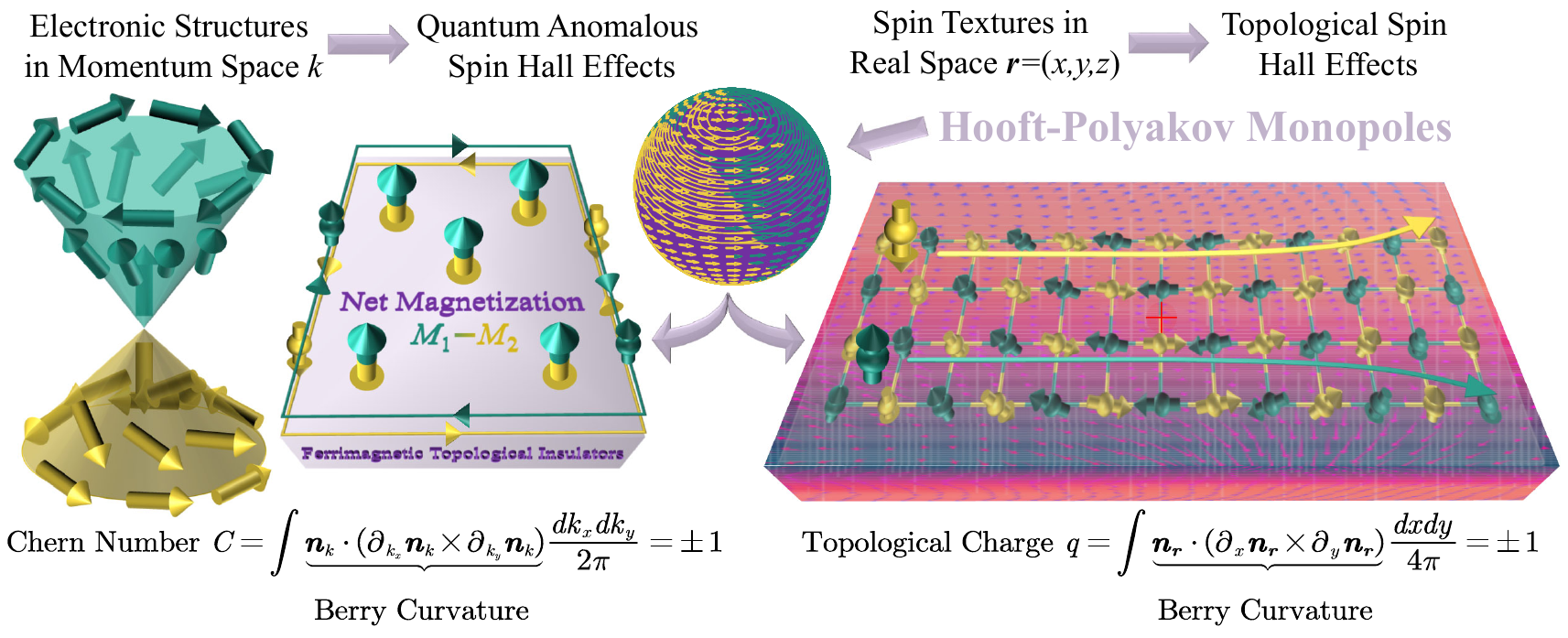}
\caption{
Different manifestations of non-Abelian Berry phases generated by Hooft-Polyakov magnetic monopoles in momentum and real space.
}
\label{12}
\end{figure}

\textbf{{Conclusions.}} 
We probe DW dynamics under thermal gradients in ferrimagnetism generalized to ferromagnetism and antiferromagnetism and analyze magnon momentum transfers, shedding light on why DWs can move to hot (magnon spin torque and entropy forces) or cold (magnon momentum transfers) regions in experiments.
We show that magnon momentum transfers are efficient driving forces and is much stronger than spin torque.
We uncover that time-varying spin Berry phases induced by thermal gradients can elicit a commercial-level thermoelectric figure of merit that will maintain advantages in diminutive-volume elements and at high temperature.
We hope that different forms of Berry phases can be more widely used in practical applications \cite{smejkal2018topological}, such as topological insulators, topological semi-metals, axion insulators, topological superconductors, and topological solitons [Fig. \ref{12}].
Our work indicates ways forward for green spincaloritronics and topological quantum materials in the post-Moore era.\\

\noindent\textbf{{Appendix}}\\
Using path integrals directly will cause integrand functions having classical-action forms, contrary to quantum essence of spins.
In spin motion equations, generalized velocities rather than acceleration are proportional to generalized forces, and generalized forces are perpendicular to velocity directions. 
The above consideration shows that Hamilton's principle cannot describe spin motion, and we resort to coherent state formalism here to obtain complete spin action and paths.
We write the spin coherent state $\left| \boldsymbol{s} \right> $ as
\begin{gather}
\left| \boldsymbol{s} \right> =\left| \hat{\varsigma} \right> =\left( \begin{array}{c}
	e^{-i\phi}\cos \frac{\theta}{2}\\
	\sin \frac{\theta}{2}\\
\end{array} \right) \,\, \mathrm{with}  \int{\frac{1}{2\pi}\left| \boldsymbol{s} \right> \left< \boldsymbol{s} \right|}d^2\boldsymbol{s}=\left( \begin{matrix}
	1&		0\\
	0&		1\\
\end{matrix} \right).
\label{10}
\end{gather}
Inserting $\int{|\boldsymbol{s}\rangle \langle \boldsymbol{s}|\frac{d^2\boldsymbol{s}}{2\pi}}$ into $\left< \boldsymbol{s}_j\mid \boldsymbol{s}_k \right> $, we can obtain the complete path integral
\begin{gather}
\left< \boldsymbol{s}_j\mid \boldsymbol{s}_k \right> =\int{\prod_{i=1}^{\infty}{\left< \boldsymbol{s}\left( \tau \right) \mid \boldsymbol{s}\left( \tau _{\infty} \right) \right>}}\cdots \left< \boldsymbol{s}\left( \tau _2 \right) \mid \boldsymbol{s}\left( \tau _1 \right) \right> 
\nonumber \\ 
\times \left< \boldsymbol{s}\left( \tau _1 \right) \mid \boldsymbol{s}\left( 0 \right) \right> \frac{d^2\boldsymbol{s}\left( \tau _i \right)}{2\pi}.
\label{11}
\end{gather}
Because of 
\begin{gather}
\left< \boldsymbol{s}\left( \delta \tau \right) \mid \boldsymbol{s}\left( 0 \right) \right> =\hat{\varsigma}^{\dagger}\hat{\varsigma}-\hat{\varsigma}^{\dagger}\left( \delta \tau \right) \left[ \hat{\varsigma}\left( \delta \tau \right) -\hat{\varsigma}\left( 0 \right) \right] 
\nonumber \\ 
=1-\hat{\varsigma}^{\dagger}\left( \delta \tau \right) \frac{\partial \hat{\varsigma}\left( \delta \tau \right)}{\partial \tau}\delta \tau \approx \exp \left( -\hat{\varsigma}^{\dagger}\frac{\partial \hat{\varsigma}}{\partial \tau}\delta \tau \right) ,
\label{12}
\end{gather}
we can rewrite Eq. \eqref{11} as
\begin{gather}
\left< \boldsymbol{s}_j\left( \tau \right) \mid \boldsymbol{s}_k\left( \tau \right) \right> =\int{\exp \left( i\mathcal{S} \right)}\mathcal{D} ^2\left( \frac{\boldsymbol{s}\left( \tau \right)}{2\pi} \right) 
\nonumber \\ 
\Rightarrow \mathcal{S} =i\int_0^{\tau}{\hat{\varsigma}^{\dagger}\frac{\partial \hat{\varsigma}}{\partial \tau}}d\tau 
\label{13}
\end{gather}
with the spin action $\mathcal{S} $ and measure $\mathcal{D} $ of the space $\boldsymbol{s}$, and Eq. \eqref{13} represents ferromagnetic Abelian phases. 
For ferrimagnetism, we expand the spin $\boldsymbol{s}_i$ of each lattice point to the sum of Néel vectors and net spin vectors: $\boldsymbol{s}_j=\left( -1 \right) ^j\boldsymbol{n}+\boldsymbol{s}$.
Thus, the above Abelian Berry phases become non-Abelian, inducing the following motion equation:
\begin{gather}
2\delta \mathcal{S} \left( \hat{\varsigma} \right) /\delta \hat{\varsigma}=\boldsymbol{n}\cdot \left( \partial _{\tau}\boldsymbol{n}\times \boldsymbol{s} \right) +\boldsymbol{n}\cdot \left( \partial _{\tau}\boldsymbol{n}\times \partial _{\boldsymbol{r}}\boldsymbol{n} \right) .
\label{14}
\end{gather}
Using the transformation relation \cite{tveten2016intrinsic} between $\boldsymbol{n}$ and $\boldsymbol{s}$ to eliminate $\boldsymbol{s}$ and then expressing $\boldsymbol{n}$ by spinors, we can obtain the Berry phase of Eq. \eqref{1}.
The first and second terms in Eq. \eqref{14} are dynamic and topological terms, respectively.
This topological term $\mathcal{S} _{\mathrm{top}}=\frac{1}{4\pi}\int{\boldsymbol{n}}\cdot \left( \partial _{\tau}\boldsymbol{n}\times \partial _{\boldsymbol{r}}\boldsymbol{n} \right) d\boldsymbol{r}d \tau $ has some similar effects with DMIs and is also the common origin of the topological spin Hall effect \cite{akosa2018theory} and topological magneto-optical effect \cite{feng2020topological}.
By variational operations for non-Abelian phases and considering \cite{chen2019landaulifshitzbloch} spin-flip and spin-phonon scatterings, we obtain the ferrimagnetic Landau-Lifshitz-Bloch equation $\rho _s\partial _{\tau}\boldsymbol{n}+\rho _n\boldsymbol{n}\times \partial _{\tau}^{2}\boldsymbol{n}=- \boldsymbol{n}\times \boldsymbol{h}_{\mathrm{eff}}+\alpha _{\mathrm{eff}}\times \partial _{\tau}\boldsymbol{n}-\tilde{\alpha}\left( \boldsymbol{n}\cdot \boldsymbol{h}_{\mathrm{eff}} \right) \boldsymbol{n}$, where non-Abelian Berry phases generate spin nutation and precession, while spin scatterings induce transverse and longitudinal damping.
We also derive this equation from the usual spin-up and spin-down double-lattice-coupling method in \textcolor{blue}{Supplementary materials}.
\\

\noindent\textbf{{Supplementary materials}}\\
A more detailed derivation of Eqs. \eqref{1} and \eqref{7} can be found in the supplementary information.\\

\nocite{*}

\bibliography{cite}

\begin{thebibliography}{225}%
\makeatletter
\providecommand \@ifxundefined [1]{%
 \@ifx{#1\undefined}
}%
\providecommand \@ifnum [1]{%
 \ifnum #1\expandafter \@firstoftwo
 \else \expandafter \@secondoftwo
 \fi
}%
\providecommand \@ifx [1]{%
 \ifx #1\expandafter \@firstoftwo
 \else \expandafter \@secondoftwo
 \fi
}%
\providecommand \natexlab [1]{#1}%
\providecommand \enquote  [1]{``#1''}%
\providecommand \bibnamefont  [1]{#1}%
\providecommand \bibfnamefont [1]{#1}%
\providecommand \citenamefont [1]{#1}%
\providecommand \href@noop [0]{\@secondoftwo}%
\providecommand \href [0]{\begingroup \@sanitize@url \@href}%
\providecommand \@href[1]{\@@startlink{#1}\@@href}%
\providecommand \@@href[1]{\endgroup#1\@@endlink}%
\providecommand \@sanitize@url [0]{\catcode `\\12\catcode `\$12\catcode
  `\&12\catcode `\#12\catcode `\^12\catcode `\_12\catcode `\%12\relax}%
\providecommand \@@startlink[1]{}%
\providecommand \@@endlink[0]{}%
\providecommand \url  [0]{\begingroup\@sanitize@url \@url }%
\providecommand \@url [1]{\endgroup\@href {#1}{\urlprefix }}%
\providecommand \urlprefix  [0]{URL }%
\providecommand \Eprint [0]{\href }%
\providecommand \doibase [0]{https://doi.org/}%
\providecommand \selectlanguage [0]{\@gobble}%
\providecommand \bibinfo  [0]{\@secondoftwo}%
\providecommand \bibfield  [0]{\@secondoftwo}%
\providecommand \translation [1]{[#1]}%
\providecommand \BibitemOpen [0]{}%
\providecommand \bibitemStop [0]{}%
\providecommand \bibitemNoStop [0]{.\EOS\space}%
\providecommand \EOS [0]{\spacefactor3000\relax}%
\providecommand \BibitemShut  [1]{\csname bibitem#1\endcsname}%
\let\auto@bib@innerbib\@empty
\bibitem [{\citenamefont {Parkin}\ \emph {et~al.}(2008)\citenamefont {Parkin},
  \citenamefont {Hayashi},\ and\ \citenamefont {Thomas}}]{parkin2008Magnetic}%
  \BibitemOpen
  \bibfield  {author} {\bibinfo {author} {\bibfnamefont {S.~S.~P.}\
  \bibnamefont {Parkin}}, \bibinfo {author} {\bibfnamefont {M.}~\bibnamefont
  {Hayashi}},\ and\ \bibinfo {author} {\bibfnamefont {L.}~\bibnamefont
  {Thomas}},\ }\bibinfo {title} {Magnetic {{Domain-Wall Racetrack Memory}}},\
  \href {https://www.science.org/doi/10.1126/science.1145799} {\bibfield
  {journal} {\bibinfo  {journal} {Science}\ }\textbf {\bibinfo {volume}
  {320}},\ \bibinfo {pages} {190} (\bibinfo {year} {2008})}\BibitemShut
  {NoStop}%
\bibitem [{\citenamefont {Luo}\ \emph {et~al.}(2020)\citenamefont {Luo},
  \citenamefont {Hrabec}, \citenamefont {Dao}, \citenamefont {Sala},
  \citenamefont {Finizio}, \citenamefont {Feng}, \citenamefont {Mayr},
  \citenamefont {Raabe}, \citenamefont {Gambardella},\ and\ \citenamefont
  {Heyderman}}]{luo2020Currentdriven}%
  \BibitemOpen
  \bibfield  {author} {\bibinfo {author} {\bibfnamefont {Z.}~\bibnamefont
  {Luo}}, \bibinfo {author} {\bibfnamefont {A.}~\bibnamefont {Hrabec}},
  \bibinfo {author} {\bibfnamefont {T.~P.}\ \bibnamefont {Dao}}, \bibinfo
  {author} {\bibfnamefont {G.}~\bibnamefont {Sala}}, \bibinfo {author}
  {\bibfnamefont {S.}~\bibnamefont {Finizio}}, \bibinfo {author} {\bibfnamefont
  {J.}~\bibnamefont {Feng}}, \bibinfo {author} {\bibfnamefont {S.}~\bibnamefont
  {Mayr}}, \bibinfo {author} {\bibfnamefont {J.}~\bibnamefont {Raabe}},
  \bibinfo {author} {\bibfnamefont {P.}~\bibnamefont {Gambardella}},\ and\
  \bibinfo {author} {\bibfnamefont {L.~J.}\ \bibnamefont {Heyderman}},\
  }\bibinfo {title} {Current-Driven Magnetic Domain-Wall Logic},\ \href
  {http://www.nature.com/articles/s41586-020-2061-y} {\bibfield  {journal}
  {\bibinfo  {journal} {Nature}\ }\textbf {\bibinfo {volume} {579}},\ \bibinfo
  {pages} {214} (\bibinfo {year} {2020})}\BibitemShut {NoStop}%
\bibitem [{\citenamefont {Yokouchi}\ \emph {et~al.}(2020)\citenamefont
  {Yokouchi}, \citenamefont {Kagawa}, \citenamefont {Hirschberger},
  \citenamefont {Otani}, \citenamefont {Nagaosa},\ and\ \citenamefont
  {Tokura}}]{yokouchi2020Emergent}%
  \BibitemOpen
  \bibfield  {author} {\bibinfo {author} {\bibfnamefont {T.}~\bibnamefont
  {Yokouchi}}, \bibinfo {author} {\bibfnamefont {F.}~\bibnamefont {Kagawa}},
  \bibinfo {author} {\bibfnamefont {M.}~\bibnamefont {Hirschberger}}, \bibinfo
  {author} {\bibfnamefont {Y.}~\bibnamefont {Otani}}, \bibinfo {author}
  {\bibfnamefont {N.}~\bibnamefont {Nagaosa}},\ and\ \bibinfo {author}
  {\bibfnamefont {Y.}~\bibnamefont {Tokura}},\ }\bibinfo {title} {Emergent
  Electromagnetic Induction in a Helical-Spin Magnet},\ \href
  {http://www.nature.com/articles/s41586-020-2775-x} {\bibfield  {journal}
  {\bibinfo  {journal} {Nature}\ }\textbf {\bibinfo {volume} {586}},\ \bibinfo
  {pages} {232} (\bibinfo {year} {2020})}\BibitemShut {NoStop}%
\bibitem [{\citenamefont {Kitaori}\ \emph {et~al.}(2021)\citenamefont
  {Kitaori}, \citenamefont {Kanazawa}, \citenamefont {Yokouchi}, \citenamefont
  {Kagawa}, \citenamefont {Nagaosa},\ and\ \citenamefont
  {Tokura}}]{kitaori2021emergenta}%
  \BibitemOpen
  \bibfield  {author} {\bibinfo {author} {\bibfnamefont {A.}~\bibnamefont
  {Kitaori}}, \bibinfo {author} {\bibfnamefont {N.}~\bibnamefont {Kanazawa}},
  \bibinfo {author} {\bibfnamefont {T.}~\bibnamefont {Yokouchi}}, \bibinfo
  {author} {\bibfnamefont {F.}~\bibnamefont {Kagawa}}, \bibinfo {author}
  {\bibfnamefont {N.}~\bibnamefont {Nagaosa}},\ and\ \bibinfo {author}
  {\bibfnamefont {Y.}~\bibnamefont {Tokura}},\ }\bibinfo {title} {Emergent
  Electromagnetic Induction beyond Room Temperature},\ \href
  {http://www.pnas.org/lookup/doi/10.1073/pnas.2105422118} {\bibfield
  {journal} {\bibinfo  {journal} {Proc. Natl. Acad. Sci. U.S.A.}\ }\textbf
  {\bibinfo {volume} {118}},\ \bibinfo {pages} {e2105422118} (\bibinfo {year}
  {2021})}\BibitemShut {NoStop}%
\bibitem [{\citenamefont {Xiang}\ \emph {et~al.}(2020)\citenamefont {Xiang},
  \citenamefont {Hu}, \citenamefont {Lyu}, \citenamefont {Zhu}, \citenamefont
  {Ma}, \citenamefont {Chen}, \citenamefont {Steglich}, \citenamefont {Chen},
  \citenamefont {Sun}, \citenamefont {Xiang}, \citenamefont {Hu}, \citenamefont
  {Lyu}, \citenamefont {Zhu}, \citenamefont {Ma}, \citenamefont {Chen},
  \citenamefont {Steglich}, \citenamefont {Chen},\ and\ \citenamefont
  {Sun}}]{xiang2020large}%
  \BibitemOpen
  \bibfield  {author} {\bibinfo {author} {\bibfnamefont {J.}~\bibnamefont
  {Xiang}}, \bibinfo {author} {\bibfnamefont {S.}~\bibnamefont {Hu}}, \bibinfo
  {author} {\bibfnamefont {M.}~\bibnamefont {Lyu}}, \bibinfo {author}
  {\bibfnamefont {W.}~\bibnamefont {Zhu}}, \bibinfo {author} {\bibfnamefont
  {C.}~\bibnamefont {Ma}}, \bibinfo {author} {\bibfnamefont {Z.}~\bibnamefont
  {Chen}}, \bibinfo {author} {\bibfnamefont {F.}~\bibnamefont {Steglich}},
  \bibinfo {author} {\bibfnamefont {G.}~\bibnamefont {Chen}}, \bibinfo {author}
  {\bibfnamefont {P.}~\bibnamefont {Sun}}, \bibinfo {author} {\bibfnamefont
  {J.}~\bibnamefont {Xiang}}, \bibinfo {author} {\bibfnamefont
  {S.}~\bibnamefont {Hu}}, \bibinfo {author} {\bibfnamefont {M.}~\bibnamefont
  {Lyu}}, \bibinfo {author} {\bibfnamefont {W.}~\bibnamefont {Zhu}}, \bibinfo
  {author} {\bibfnamefont {C.}~\bibnamefont {Ma}}, \bibinfo {author}
  {\bibfnamefont {Z.}~\bibnamefont {Chen}}, \bibinfo {author} {\bibfnamefont
  {F.}~\bibnamefont {Steglich}}, \bibinfo {author} {\bibfnamefont
  {G.}~\bibnamefont {Chen}},\ and\ \bibinfo {author} {\bibfnamefont
  {P.}~\bibnamefont {Sun}},\ }\bibinfo {title} {Large Transverse Thermoelectric
  Figure of Merit in a Topological {{Dirac}} Semimetal},\ \href
  {http://link.springer.com/content/pdf/10.1007/s11433-019-1445-4.pdf}
  {\bibfield  {journal} {\bibinfo  {journal} {Sci. China-Phys. Mech. Astron.}\
  }\textbf {\bibinfo {volume} {63}},\ \bibinfo {pages} {237011} (\bibinfo
  {year} {2020})}\BibitemShut {NoStop}%
\bibitem [{\citenamefont {Roychowdhury}\ \emph {et~al.}(2022)\citenamefont
  {Roychowdhury}, \citenamefont {Ochs}, \citenamefont {Guin}, \citenamefont
  {Samanta}, \citenamefont {Noky}, \citenamefont {Shekhar}, \citenamefont
  {Vergniory}, \citenamefont {Goldberger},\ and\ \citenamefont
  {Felser}}]{roychowdhury2022large}%
  \BibitemOpen
  \bibfield  {author} {\bibinfo {author} {\bibfnamefont {S.}~\bibnamefont
  {Roychowdhury}}, \bibinfo {author} {\bibfnamefont {A.~M.}\ \bibnamefont
  {Ochs}}, \bibinfo {author} {\bibfnamefont {S.~N.}\ \bibnamefont {Guin}},
  \bibinfo {author} {\bibfnamefont {K.}~\bibnamefont {Samanta}}, \bibinfo
  {author} {\bibfnamefont {J.}~\bibnamefont {Noky}}, \bibinfo {author}
  {\bibfnamefont {C.}~\bibnamefont {Shekhar}}, \bibinfo {author} {\bibfnamefont
  {M.~G.}\ \bibnamefont {Vergniory}}, \bibinfo {author} {\bibfnamefont {J.~E.}\
  \bibnamefont {Goldberger}},\ and\ \bibinfo {author} {\bibfnamefont
  {C.}~\bibnamefont {Felser}},\ }\bibinfo {title} {Large {{Room Temperature
  Anomalous Transverse Thermoelectric Effect}} in {{Kagome Antiferromagnet
  YMn}} {\textsubscript{6}} {{Sn}} {\textsubscript{6}}},\ \href
  {https://onlinelibrary.wiley.com/doi/10.1002/adma.202201350} {\bibfield
  {journal} {\bibinfo  {journal} {Adv. Mater.}\ ,\ \bibinfo {pages} {2201350}}
  (\bibinfo {year} {2022})}\BibitemShut {NoStop}%
\bibitem [{\citenamefont {You}\ \emph {et~al.}(2022)\citenamefont {You},
  \citenamefont {Lam}, \citenamefont {Wan}, \citenamefont {Wan}, \citenamefont
  {Zhu}, \citenamefont {Han}, \citenamefont {Bai}, \citenamefont {Zhou},
  \citenamefont {Qiao}, \citenamefont {Chen}, \citenamefont {Pan},
  \citenamefont {Liu},\ and\ \citenamefont {Song}}]{you2022anomalous}%
  \BibitemOpen
  \bibfield  {author} {\bibinfo {author} {\bibfnamefont {Y.}~\bibnamefont
  {You}}, \bibinfo {author} {\bibfnamefont {H.}~\bibnamefont {Lam}}, \bibinfo
  {author} {\bibfnamefont {C.}~\bibnamefont {Wan}}, \bibinfo {author}
  {\bibfnamefont {C.}~\bibnamefont {Wan}}, \bibinfo {author} {\bibfnamefont
  {W.}~\bibnamefont {Zhu}}, \bibinfo {author} {\bibfnamefont {L.}~\bibnamefont
  {Han}}, \bibinfo {author} {\bibfnamefont {H.}~\bibnamefont {Bai}}, \bibinfo
  {author} {\bibfnamefont {Y.}~\bibnamefont {Zhou}}, \bibinfo {author}
  {\bibfnamefont {L.}~\bibnamefont {Qiao}}, \bibinfo {author} {\bibfnamefont
  {T.}~\bibnamefont {Chen}}, \bibinfo {author} {\bibfnamefont {F.}~\bibnamefont
  {Pan}}, \bibinfo {author} {\bibfnamefont {J.}~\bibnamefont {Liu}},\ and\
  \bibinfo {author} {\bibfnamefont {C.}~\bibnamefont {Song}},\ }\bibinfo
  {title} {Anomalous {{Nernst Effect}} in an {{Antiperovskite
  Antiferromagnet}}},\ \href
  {https://link.aps.org/doi/10.1103/PhysRevApplied.18.024007} {\bibfield
  {journal} {\bibinfo  {journal} {Phys. Rev. Applied}\ }\textbf {\bibinfo
  {volume} {18}},\ \bibinfo {pages} {024007} (\bibinfo {year}
  {2022})}\BibitemShut {NoStop}%
\bibitem [{\citenamefont {Xu}\ \emph {et~al.}(2022)\citenamefont {Xu},
  \citenamefont {He}, \citenamefont {Zhou}, \citenamefont {Qu}, \citenamefont
  {Huang},\ and\ \citenamefont {Chien}}]{xu2022observation}%
  \BibitemOpen
  \bibfield  {author} {\bibinfo {author} {\bibfnamefont {J.}~\bibnamefont
  {Xu}}, \bibinfo {author} {\bibfnamefont {J.}~\bibnamefont {He}}, \bibinfo
  {author} {\bibfnamefont {J.-S.}\ \bibnamefont {Zhou}}, \bibinfo {author}
  {\bibfnamefont {D.}~\bibnamefont {Qu}}, \bibinfo {author} {\bibfnamefont
  {S.-Y.}\ \bibnamefont {Huang}},\ and\ \bibinfo {author} {\bibfnamefont
  {C.~L.}\ \bibnamefont {Chien}},\ }\bibinfo {title} {Observation of {{Vector
  Spin Seebeck Effect}} in a {{Noncollinear Antiferromagnet}}},\ \href
  {https://link.aps.org/doi/10.1103/PhysRevLett.129.117202} {\bibfield
  {journal} {\bibinfo  {journal} {Phys. Rev. Lett.}\ }\textbf {\bibinfo
  {volume} {129}},\ \bibinfo {pages} {117202} (\bibinfo {year}
  {2022})}\BibitemShut {NoStop}%
\bibitem [{\citenamefont {Lin}\ \emph {et~al.}(2022)\citenamefont {Lin},
  \citenamefont {He}, \citenamefont {Ma}, \citenamefont {Matzelle},
  \citenamefont {Xu}, \citenamefont {Freeland}, \citenamefont {Choi},
  \citenamefont {Haskel}, \citenamefont {Barbiellini}, \citenamefont {Bansil},
  \citenamefont {Fiete}, \citenamefont {Zhou},\ and\ \citenamefont
  {Chien}}]{lin2022evidence}%
  \BibitemOpen
  \bibfield  {author} {\bibinfo {author} {\bibfnamefont {W.}~\bibnamefont
  {Lin}}, \bibinfo {author} {\bibfnamefont {J.}~\bibnamefont {He}}, \bibinfo
  {author} {\bibfnamefont {B.}~\bibnamefont {Ma}}, \bibinfo {author}
  {\bibfnamefont {M.}~\bibnamefont {Matzelle}}, \bibinfo {author}
  {\bibfnamefont {J.}~\bibnamefont {Xu}}, \bibinfo {author} {\bibfnamefont
  {J.}~\bibnamefont {Freeland}}, \bibinfo {author} {\bibfnamefont
  {Y.}~\bibnamefont {Choi}}, \bibinfo {author} {\bibfnamefont {D.}~\bibnamefont
  {Haskel}}, \bibinfo {author} {\bibfnamefont {B.}~\bibnamefont {Barbiellini}},
  \bibinfo {author} {\bibfnamefont {A.}~\bibnamefont {Bansil}}, \bibinfo
  {author} {\bibfnamefont {G.~A.}\ \bibnamefont {Fiete}}, \bibinfo {author}
  {\bibfnamefont {J.}~\bibnamefont {Zhou}},\ and\ \bibinfo {author}
  {\bibfnamefont {C.~L.}\ \bibnamefont {Chien}},\ }\bibinfo {title} {Evidence
  for Spin Swapping in an Antiferromagnet},\ \href
  {https://www.nature.com/articles/s41567-022-01608-w} {\bibfield  {journal}
  {\bibinfo  {journal} {Nat. Phys.}\ }\textbf {\bibinfo {volume} {18}},\
  \bibinfo {pages} {800} (\bibinfo {year} {2022})}\BibitemShut {NoStop}%
\bibitem [{\citenamefont {Chen}\ \emph
  {et~al.}(2021{\natexlab{a}})\citenamefont {Chen}, \citenamefont {Tomita},
  \citenamefont {Minami}, \citenamefont {Fu}, \citenamefont {Koretsune},
  \citenamefont {Kitatani}, \citenamefont {Muhammad}, \citenamefont
  {{Nishio-Hamane}}, \citenamefont {Ishii}, \citenamefont {Ishii},
  \citenamefont {Arita},\ and\ \citenamefont {Nakatsuji}}]{chen2021anomalous}%
  \BibitemOpen
  \bibfield  {author} {\bibinfo {author} {\bibfnamefont {T.}~\bibnamefont
  {Chen}}, \bibinfo {author} {\bibfnamefont {T.}~\bibnamefont {Tomita}},
  \bibinfo {author} {\bibfnamefont {S.}~\bibnamefont {Minami}}, \bibinfo
  {author} {\bibfnamefont {M.}~\bibnamefont {Fu}}, \bibinfo {author}
  {\bibfnamefont {T.}~\bibnamefont {Koretsune}}, \bibinfo {author}
  {\bibfnamefont {M.}~\bibnamefont {Kitatani}}, \bibinfo {author}
  {\bibfnamefont {I.}~\bibnamefont {Muhammad}}, \bibinfo {author}
  {\bibfnamefont {D.}~\bibnamefont {{Nishio-Hamane}}}, \bibinfo {author}
  {\bibfnamefont {R.}~\bibnamefont {Ishii}}, \bibinfo {author} {\bibfnamefont
  {F.}~\bibnamefont {Ishii}}, \bibinfo {author} {\bibfnamefont
  {R.}~\bibnamefont {Arita}},\ and\ \bibinfo {author} {\bibfnamefont
  {S.}~\bibnamefont {Nakatsuji}},\ }\bibinfo {title} {Anomalous Transport Due
  to {{Weyl}} Fermions in the Chiral Antiferromagnets {{Mn3X}},
  {{X}}\,=\,{{Sn}}, {{Ge}}},\ \href
  {http://www.nature.com/articles/s41467-020-20838-1} {\bibfield  {journal}
  {\bibinfo  {journal} {Nat. Commun.}\ }\textbf {\bibinfo {volume} {12}},\
  \bibinfo {pages} {572} (\bibinfo {year} {2021}{\natexlab{a}})}\BibitemShut
  {NoStop}%
\bibitem [{\citenamefont {Ikhlas}\ \emph {et~al.}(2017)\citenamefont {Ikhlas},
  \citenamefont {Tomita}, \citenamefont {Koretsune}, \citenamefont {Suzuki},
  \citenamefont {{Nishio-Hamane}}, \citenamefont {Arita}, \citenamefont
  {Otani},\ and\ \citenamefont {Nakatsuji}}]{ikhlas2017large}%
  \BibitemOpen
  \bibfield  {author} {\bibinfo {author} {\bibfnamefont {M.}~\bibnamefont
  {Ikhlas}}, \bibinfo {author} {\bibfnamefont {T.}~\bibnamefont {Tomita}},
  \bibinfo {author} {\bibfnamefont {T.}~\bibnamefont {Koretsune}}, \bibinfo
  {author} {\bibfnamefont {M.-T.}\ \bibnamefont {Suzuki}}, \bibinfo {author}
  {\bibfnamefont {D.}~\bibnamefont {{Nishio-Hamane}}}, \bibinfo {author}
  {\bibfnamefont {R.}~\bibnamefont {Arita}}, \bibinfo {author} {\bibfnamefont
  {Y.}~\bibnamefont {Otani}},\ and\ \bibinfo {author} {\bibfnamefont
  {S.}~\bibnamefont {Nakatsuji}},\ }\bibinfo {title} {Large Anomalous
  {{Nernst}} Effect at Room Temperature in a Chiral Antiferromagnet},\ \href
  {http://www.nature.com/articles/nphys4181} {\bibfield  {journal} {\bibinfo
  {journal} {Nat. Phys.}\ }\textbf {\bibinfo {volume} {13}},\ \bibinfo {pages}
  {1085} (\bibinfo {year} {2017})}\BibitemShut {NoStop}%
\bibitem [{\citenamefont {Uchida}\ \emph {et~al.}(2021)\citenamefont {Uchida},
  \citenamefont {Zhou},\ and\ \citenamefont {Sakuraba}}]{uchida2021transverse}%
  \BibitemOpen
  \bibfield  {author} {\bibinfo {author} {\bibfnamefont {K.-i.}\ \bibnamefont
  {Uchida}}, \bibinfo {author} {\bibfnamefont {W.}~\bibnamefont {Zhou}},\ and\
  \bibinfo {author} {\bibfnamefont {Y.}~\bibnamefont {Sakuraba}},\ }\bibinfo
  {title} {Transverse Thermoelectric Generation Using Magnetic Materials},\
  \href {https://aip.scitation.org/doi/pdf/10.1063/5.0046877} {\bibfield
  {journal} {\bibinfo  {journal} {Appl. Phys. Lett.}\ }\textbf {\bibinfo
  {volume} {118}},\ \bibinfo {pages} {140504} (\bibinfo {year}
  {2021})}\BibitemShut {NoStop}%
\bibitem [{\citenamefont {Kuroyama}\ \emph {et~al.}(2022)\citenamefont
  {Kuroyama}, \citenamefont {Matsuo}, \citenamefont {Muramoto}, \citenamefont
  {Yabunaka}, \citenamefont {Valentin}, \citenamefont {Ludwig}, \citenamefont
  {Wieck}, \citenamefont {Tokura},\ and\ \citenamefont
  {Tarucha}}]{kuroyama2022realtime}%
  \BibitemOpen
  \bibfield  {author} {\bibinfo {author} {\bibfnamefont {K.}~\bibnamefont
  {Kuroyama}}, \bibinfo {author} {\bibfnamefont {S.}~\bibnamefont {Matsuo}},
  \bibinfo {author} {\bibfnamefont {J.}~\bibnamefont {Muramoto}}, \bibinfo
  {author} {\bibfnamefont {S.}~\bibnamefont {Yabunaka}}, \bibinfo {author}
  {\bibfnamefont {S.~R.}\ \bibnamefont {Valentin}}, \bibinfo {author}
  {\bibfnamefont {A.}~\bibnamefont {Ludwig}}, \bibinfo {author} {\bibfnamefont
  {A.~D.}\ \bibnamefont {Wieck}}, \bibinfo {author} {\bibfnamefont
  {Y.}~\bibnamefont {Tokura}},\ and\ \bibinfo {author} {\bibfnamefont
  {S.}~\bibnamefont {Tarucha}},\ }\bibinfo {title} {Real-{{Time Observation}}
  of {{Charge-Spin Cooperative Dynamics Driven}} by a {{Nonequilibrium Phonon
  Environment}}},\ \href
  {https://link.aps.org/accepted/10.1103/PhysRevLett.129.095901} {\bibfield
  {journal} {\bibinfo  {journal} {Phys. Rev. Lett.}\ }\textbf {\bibinfo
  {volume} {129}},\ \bibinfo {pages} {6} (\bibinfo {year} {2022})}\BibitemShut
  {NoStop}%
\bibitem [{\citenamefont {Krzysteczko}\ \emph {et~al.}(2015)\citenamefont
  {Krzysteczko}, \citenamefont {Hu}, \citenamefont {Liebing}, \citenamefont
  {Sievers},\ and\ \citenamefont {Schumacher}}]{krzysteczko2015domain}%
  \BibitemOpen
  \bibfield  {author} {\bibinfo {author} {\bibfnamefont {P.}~\bibnamefont
  {Krzysteczko}}, \bibinfo {author} {\bibfnamefont {X.}~\bibnamefont {Hu}},
  \bibinfo {author} {\bibfnamefont {N.}~\bibnamefont {Liebing}}, \bibinfo
  {author} {\bibfnamefont {S.}~\bibnamefont {Sievers}},\ and\ \bibinfo {author}
  {\bibfnamefont {H.~W.}\ \bibnamefont {Schumacher}},\ }\bibinfo {title}
  {Domain Wall Magneto-{{Seebeck}} Effect},\ \href
  {https://link.aps.org/doi/10.1103/PhysRevB.92.140405} {\bibfield  {journal}
  {\bibinfo  {journal} {Phys. Rev. B}\ }\textbf {\bibinfo {volume} {92}},\
  \bibinfo {pages} {140405} (\bibinfo {year} {2015})}\BibitemShut {NoStop}%
\bibitem [{\citenamefont {Niemann}\ \emph {et~al.}(2016)\citenamefont
  {Niemann}, \citenamefont {B{\"o}hnert}, \citenamefont {Michel}, \citenamefont
  {B{\"a}{\ss}ler}, \citenamefont {Gotsmann}, \citenamefont {Neur{\'o}hr},
  \citenamefont {T{\'o}th}, \citenamefont {P{\'e}ter}, \citenamefont {Bakonyi},
  \citenamefont {Vega}, \citenamefont {Prida}, \citenamefont {Gooth},\ and\
  \citenamefont {Nielsch}}]{niemann2016thermoelectric}%
  \BibitemOpen
  \bibfield  {author} {\bibinfo {author} {\bibfnamefont {A.~C.}\ \bibnamefont
  {Niemann}}, \bibinfo {author} {\bibfnamefont {T.}~\bibnamefont
  {B{\"o}hnert}}, \bibinfo {author} {\bibfnamefont {A.-K.}\ \bibnamefont
  {Michel}}, \bibinfo {author} {\bibfnamefont {S.}~\bibnamefont
  {B{\"a}{\ss}ler}}, \bibinfo {author} {\bibfnamefont {B.}~\bibnamefont
  {Gotsmann}}, \bibinfo {author} {\bibfnamefont {K.}~\bibnamefont
  {Neur{\'o}hr}}, \bibinfo {author} {\bibfnamefont {B.}~\bibnamefont
  {T{\'o}th}}, \bibinfo {author} {\bibfnamefont {L.}~\bibnamefont {P{\'e}ter}},
  \bibinfo {author} {\bibfnamefont {I.}~\bibnamefont {Bakonyi}}, \bibinfo
  {author} {\bibfnamefont {V.}~\bibnamefont {Vega}}, \bibinfo {author}
  {\bibfnamefont {V.~M.}\ \bibnamefont {Prida}}, \bibinfo {author}
  {\bibfnamefont {J.}~\bibnamefont {Gooth}},\ and\ \bibinfo {author}
  {\bibfnamefont {K.}~\bibnamefont {Nielsch}},\ }\bibinfo {title}
  {Thermoelectric {{Power Factor Enhancement}} by {{Spin-Polarized Currents-A
  Nanowire Case Study}}},\ \href
  {https://onlinelibrary.wiley.com/doi/10.1002/aelm.201600058} {\bibfield
  {journal} {\bibinfo  {journal} {Adv. Electron. Mater.}\ }\textbf {\bibinfo
  {volume} {2}},\ \bibinfo {pages} {1600058} (\bibinfo {year}
  {2016})}\BibitemShut {NoStop}%
\bibitem [{\citenamefont {Lee}\ \emph {et~al.}(2022)\citenamefont {Lee},
  \citenamefont {Nakata}, \citenamefont {Tchernyshyov},\ and\ \citenamefont
  {Kim}}]{lee2022magnon}%
  \BibitemOpen
  \bibfield  {author} {\bibinfo {author} {\bibfnamefont {S.}~\bibnamefont
  {Lee}}, \bibinfo {author} {\bibfnamefont {K.}~\bibnamefont {Nakata}},
  \bibinfo {author} {\bibfnamefont {O.}~\bibnamefont {Tchernyshyov}},\ and\
  \bibinfo {author} {\bibfnamefont {S.~K.}\ \bibnamefont {Kim}},\ }\bibinfo
  {title} {Magnon Dynamics in a {{Skyrmion-textured}} Domain Wall of
  Antiferromagnets},\ \href {http://arxiv.org/abs/2211.00030} {\bibfield
  {journal} {\bibinfo  {journal} {arXiv:2211.00030}\ } (\bibinfo {year}
  {2022})}\BibitemShut {NoStop}%
\bibitem [{\citenamefont {Keimer}\ and\ \citenamefont
  {Moore}(2017)}]{keimer2017physics}%
  \BibitemOpen
  \bibfield  {author} {\bibinfo {author} {\bibfnamefont {B.}~\bibnamefont
  {Keimer}}\ and\ \bibinfo {author} {\bibfnamefont {J.~E.}\ \bibnamefont
  {Moore}},\ }\bibinfo {title} {The Physics of Quantum Materials},\ \href
  {http://www.nature.com/articles/nphys4302} {\bibfield  {journal} {\bibinfo
  {journal} {Nat. Phys.}\ }\textbf {\bibinfo {volume} {13}},\ \bibinfo {pages}
  {1045} (\bibinfo {year} {2017})}\BibitemShut {NoStop}%
\bibitem [{\citenamefont {Tokura}(2022)}]{tokura2022quantum}%
  \BibitemOpen
  \bibfield  {author} {\bibinfo {author} {\bibfnamefont {Y.}~\bibnamefont
  {Tokura}},\ }\bibinfo {title} {Quantum Materials at the Crossroads of Strong
  Correlation and Topology},\ \href
  {https://www.nature.com/articles/s41563-022-01339-6} {\bibfield  {journal}
  {\bibinfo  {journal} {Nat. Mater.}\ }\textbf {\bibinfo {volume} {21}},\
  \bibinfo {pages} {971} (\bibinfo {year} {2022})}\BibitemShut {NoStop}%
\bibitem [{\citenamefont {Uchida}\ \emph {et~al.}(2008)\citenamefont {Uchida},
  \citenamefont {Takahashi}, \citenamefont {Harii}, \citenamefont {Ieda},
  \citenamefont {Koshibae}, \citenamefont {Ando}, \citenamefont {Maekawa},\
  and\ \citenamefont {Saitoh}}]{uchida2008observation}%
  \BibitemOpen
  \bibfield  {author} {\bibinfo {author} {\bibfnamefont {K.}~\bibnamefont
  {Uchida}}, \bibinfo {author} {\bibfnamefont {S.}~\bibnamefont {Takahashi}},
  \bibinfo {author} {\bibfnamefont {K.}~\bibnamefont {Harii}}, \bibinfo
  {author} {\bibfnamefont {J.}~\bibnamefont {Ieda}}, \bibinfo {author}
  {\bibfnamefont {W.}~\bibnamefont {Koshibae}}, \bibinfo {author}
  {\bibfnamefont {K.}~\bibnamefont {Ando}}, \bibinfo {author} {\bibfnamefont
  {S.}~\bibnamefont {Maekawa}},\ and\ \bibinfo {author} {\bibfnamefont
  {E.}~\bibnamefont {Saitoh}},\ }\bibinfo {title} {Observation of the Spin
  {{Seebeck}} Effect},\ \href {http://www.nature.com/articles/nature07321}
  {\bibfield  {journal} {\bibinfo  {journal} {Nature}\ }\textbf {\bibinfo
  {volume} {455}},\ \bibinfo {pages} {778} (\bibinfo {year}
  {2008})}\BibitemShut {NoStop}%
\bibitem [{\citenamefont {Cheng}\ \emph {et~al.}(2016)\citenamefont {Cheng},
  \citenamefont {Okamoto},\ and\ \citenamefont {Xiao}}]{cheng2016Spin}%
  \BibitemOpen
  \bibfield  {author} {\bibinfo {author} {\bibfnamefont {R.}~\bibnamefont
  {Cheng}}, \bibinfo {author} {\bibfnamefont {S.}~\bibnamefont {Okamoto}},\
  and\ \bibinfo {author} {\bibfnamefont {D.}~\bibnamefont {Xiao}},\ }\bibinfo
  {title} {Spin {{Nernst Effect}} of {{Magnons}} in {{Collinear
  Antiferromagnets}}},\ \href
  {https://link.aps.org/doi/10.1103/PhysRevLett.117.217202} {\bibfield
  {journal} {\bibinfo  {journal} {Phys. Rev. Lett.}\ }\textbf {\bibinfo
  {volume} {117}},\ \bibinfo {pages} {217202} (\bibinfo {year}
  {2016})}\BibitemShut {NoStop}%
\bibitem [{\citenamefont {Raimondo}\ \emph {et~al.}(2022)\citenamefont
  {Raimondo}, \citenamefont {Saugar}, \citenamefont {Barker}, \citenamefont
  {Rodrigues}, \citenamefont {Giordano}, \citenamefont {Carpentieri},
  \citenamefont {Jiang}, \citenamefont {{Chubykalo-Fesenko}}, \citenamefont
  {Tomasello},\ and\ \citenamefont
  {Finocchio}}]{raimondo2022temperaturegradientdriven}%
  \BibitemOpen
  \bibfield  {author} {\bibinfo {author} {\bibfnamefont {E.}~\bibnamefont
  {Raimondo}}, \bibinfo {author} {\bibfnamefont {E.}~\bibnamefont {Saugar}},
  \bibinfo {author} {\bibfnamefont {J.}~\bibnamefont {Barker}}, \bibinfo
  {author} {\bibfnamefont {D.}~\bibnamefont {Rodrigues}}, \bibinfo {author}
  {\bibfnamefont {A.}~\bibnamefont {Giordano}}, \bibinfo {author}
  {\bibfnamefont {M.}~\bibnamefont {Carpentieri}}, \bibinfo {author}
  {\bibfnamefont {W.}~\bibnamefont {Jiang}}, \bibinfo {author} {\bibfnamefont
  {O.}~\bibnamefont {{Chubykalo-Fesenko}}}, \bibinfo {author} {\bibfnamefont
  {R.}~\bibnamefont {Tomasello}},\ and\ \bibinfo {author} {\bibfnamefont
  {G.}~\bibnamefont {Finocchio}},\ }\bibinfo {title}
  {Temperature-{{Gradient-Driven Magnetic Skyrmion Motion}}},\ \href
  {https://link.aps.org/doi/10.1103/PhysRevApplied.18.024062} {\bibfield
  {journal} {\bibinfo  {journal} {Phys. Rev. Applied}\ }\textbf {\bibinfo
  {volume} {18}},\ \bibinfo {pages} {024062} (\bibinfo {year}
  {2022})}\BibitemShut {NoStop}%
\bibitem [{\citenamefont {Slonczewski}(2010)}]{slonczewski2010Initiation}%
  \BibitemOpen
  \bibfield  {author} {\bibinfo {author} {\bibfnamefont {J.~C.}\ \bibnamefont
  {Slonczewski}},\ }\bibinfo {title} {Initiation of Spin-Transfer Torque by
  Thermal Transport from Magnons},\ \href
  {https://link.aps.org/doi/10.1103/PhysRevB.82.054403} {\bibfield  {journal}
  {\bibinfo  {journal} {Phys. Rev. B}\ }\textbf {\bibinfo {volume} {82}},\
  \bibinfo {pages} {054403} (\bibinfo {year} {2010})}\BibitemShut {NoStop}%
\bibitem [{\citenamefont {Matsuo}\ \emph {et~al.}(2018)\citenamefont {Matsuo},
  \citenamefont {Ohnuma}, \citenamefont {Kato},\ and\ \citenamefont
  {Maekawa}}]{matsuo2018spin}%
  \BibitemOpen
  \bibfield  {author} {\bibinfo {author} {\bibfnamefont {M.}~\bibnamefont
  {Matsuo}}, \bibinfo {author} {\bibfnamefont {Y.}~\bibnamefont {Ohnuma}},
  \bibinfo {author} {\bibfnamefont {T.}~\bibnamefont {Kato}},\ and\ \bibinfo
  {author} {\bibfnamefont {S.}~\bibnamefont {Maekawa}},\ }\bibinfo {title}
  {Spin {{Current Noise}} of the {{Spin Seebeck Effect}} and {{Spin
  Pumping}}},\ \href {https://link.aps.org/doi/10.1103/PhysRevLett.120.037201}
  {\bibfield  {journal} {\bibinfo  {journal} {Phys. Rev. Lett.}\ }\textbf
  {\bibinfo {volume} {120}},\ \bibinfo {pages} {037201} (\bibinfo {year}
  {2018})}\BibitemShut {NoStop}%
\bibitem [{\citenamefont {Choi}\ \emph {et~al.}(2015)\citenamefont {Choi},
  \citenamefont {Moon}, \citenamefont {Min}, \citenamefont {Lee},\ and\
  \citenamefont {Cahill}}]{choi2015Thermal}%
  \BibitemOpen
  \bibfield  {author} {\bibinfo {author} {\bibfnamefont {G.-M.}\ \bibnamefont
  {Choi}}, \bibinfo {author} {\bibfnamefont {C.-H.}\ \bibnamefont {Moon}},
  \bibinfo {author} {\bibfnamefont {B.-C.}\ \bibnamefont {Min}}, \bibinfo
  {author} {\bibfnamefont {K.-J.}\ \bibnamefont {Lee}},\ and\ \bibinfo {author}
  {\bibfnamefont {D.~G.}\ \bibnamefont {Cahill}},\ }\bibinfo {title} {Thermal
  Spin-Transfer Torque Driven by the Spin-Dependent {{Seebeck}} Effect in
  Metallic Spin-Valves},\ \href {http://www.nature.com/articles/nphys3355}
  {\bibfield  {journal} {\bibinfo  {journal} {Nat. Phys.}\ }\textbf {\bibinfo
  {volume} {11}},\ \bibinfo {pages} {576} (\bibinfo {year} {2015})}\BibitemShut
  {NoStop}%
\bibitem [{\citenamefont {Bauer}\ \emph {et~al.}(2012)\citenamefont {Bauer},
  \citenamefont {Saitoh},\ and\ \citenamefont {{van Wees}}}]{bauer2012Spin}%
  \BibitemOpen
  \bibfield  {author} {\bibinfo {author} {\bibfnamefont {G.~E.~W.}\
  \bibnamefont {Bauer}}, \bibinfo {author} {\bibfnamefont {E.}~\bibnamefont
  {Saitoh}},\ and\ \bibinfo {author} {\bibfnamefont {B.~J.}\ \bibnamefont {{van
  Wees}}},\ }\bibinfo {title} {Spin Caloritronics},\ \href
  {http://www.nature.com/articles/nmat3301} {\bibfield  {journal} {\bibinfo
  {journal} {Nat. Mater.}\ }\textbf {\bibinfo {volume} {11}},\ \bibinfo {pages}
  {391} (\bibinfo {year} {2012})}\BibitemShut {NoStop}%
\bibitem [{\citenamefont {Wang}\ \emph {et~al.}(2020)\citenamefont {Wang},
  \citenamefont {Guo}, \citenamefont {Zhou}, \citenamefont {Zhao},
  \citenamefont {Xu}, \citenamefont {Tomasello}, \citenamefont {Bai},
  \citenamefont {Dong}, \citenamefont {Je}, \citenamefont {Chao}, \citenamefont
  {Han}, \citenamefont {Lee}, \citenamefont {Lee}, \citenamefont {Yao},
  \citenamefont {Han}, \citenamefont {Song}, \citenamefont {Wu}, \citenamefont
  {Carpentieri}, \citenamefont {Finocchio}, \citenamefont {Im}, \citenamefont
  {Lin},\ and\ \citenamefont {Jiang}}]{wang2020thermal}%
  \BibitemOpen
  \bibfield  {author} {\bibinfo {author} {\bibfnamefont {Z.}~\bibnamefont
  {Wang}}, \bibinfo {author} {\bibfnamefont {M.}~\bibnamefont {Guo}}, \bibinfo
  {author} {\bibfnamefont {H.-A.}\ \bibnamefont {Zhou}}, \bibinfo {author}
  {\bibfnamefont {L.}~\bibnamefont {Zhao}}, \bibinfo {author} {\bibfnamefont
  {T.}~\bibnamefont {Xu}}, \bibinfo {author} {\bibfnamefont {R.}~\bibnamefont
  {Tomasello}}, \bibinfo {author} {\bibfnamefont {H.}~\bibnamefont {Bai}},
  \bibinfo {author} {\bibfnamefont {Y.}~\bibnamefont {Dong}}, \bibinfo {author}
  {\bibfnamefont {S.-G.}\ \bibnamefont {Je}}, \bibinfo {author} {\bibfnamefont
  {W.}~\bibnamefont {Chao}}, \bibinfo {author} {\bibfnamefont {H.-S.}\
  \bibnamefont {Han}}, \bibinfo {author} {\bibfnamefont {S.}~\bibnamefont
  {Lee}}, \bibinfo {author} {\bibfnamefont {K.-S.}\ \bibnamefont {Lee}},
  \bibinfo {author} {\bibfnamefont {Y.}~\bibnamefont {Yao}}, \bibinfo {author}
  {\bibfnamefont {W.}~\bibnamefont {Han}}, \bibinfo {author} {\bibfnamefont
  {C.}~\bibnamefont {Song}}, \bibinfo {author} {\bibfnamefont {H.}~\bibnamefont
  {Wu}}, \bibinfo {author} {\bibfnamefont {M.}~\bibnamefont {Carpentieri}},
  \bibinfo {author} {\bibfnamefont {G.}~\bibnamefont {Finocchio}}, \bibinfo
  {author} {\bibfnamefont {M.-Y.}\ \bibnamefont {Im}}, \bibinfo {author}
  {\bibfnamefont {S.-Z.}\ \bibnamefont {Lin}},\ and\ \bibinfo {author}
  {\bibfnamefont {W.}~\bibnamefont {Jiang}},\ }\bibinfo {title} {Thermal
  Generation, Manipulation and Thermoelectric Detection of Skyrmions},\ \href
  {http://www.nature.com/articles/s41928-020-00489-2} {\bibfield  {journal}
  {\bibinfo  {journal} {Nat. Electron.}\ }\textbf {\bibinfo {volume} {3}},\
  \bibinfo {pages} {672} (\bibinfo {year} {2020})}\BibitemShut {NoStop}%
\bibitem [{\citenamefont {{Mazo-Zuluaga}}\ \emph {et~al.}(2016)\citenamefont
  {{Mazo-Zuluaga}}, \citenamefont {Vel{\'a}squez}, \citenamefont {Altbir},\
  and\ \citenamefont {{Mej{\'i}a-L{\'o}pez}}}]{mazo-zuluaga2016controlling}%
  \BibitemOpen
  \bibfield  {author} {\bibinfo {author} {\bibfnamefont {J.}~\bibnamefont
  {{Mazo-Zuluaga}}}, \bibinfo {author} {\bibfnamefont {E.~A.}\ \bibnamefont
  {Vel{\'a}squez}}, \bibinfo {author} {\bibfnamefont {D.}~\bibnamefont
  {Altbir}},\ and\ \bibinfo {author} {\bibfnamefont {J.}~\bibnamefont
  {{Mej{\'i}a-L{\'o}pez}}},\ }\bibinfo {title} {Controlling Domain Wall
  Nucleation and Propagation with Temperature Gradients},\ \href
  {http://aip.scitation.org/doi/10.1063/1.4963181} {\bibfield  {journal}
  {\bibinfo  {journal} {Appl. Phys. Lett.}\ }\textbf {\bibinfo {volume}
  {109}},\ \bibinfo {pages} {122408} (\bibinfo {year} {2016})}\BibitemShut
  {NoStop}%
\bibitem [{\citenamefont {Yang}\ \emph {et~al.}(2009)\citenamefont {Yang},
  \citenamefont {Beach}, \citenamefont {Knutson}, \citenamefont {Xiao},
  \citenamefont {Niu}, \citenamefont {Tsoi},\ and\ \citenamefont
  {Erskine}}]{yang2009universal}%
  \BibitemOpen
  \bibfield  {author} {\bibinfo {author} {\bibfnamefont {S.~A.}\ \bibnamefont
  {Yang}}, \bibinfo {author} {\bibfnamefont {G.~S.~D.}\ \bibnamefont {Beach}},
  \bibinfo {author} {\bibfnamefont {C.}~\bibnamefont {Knutson}}, \bibinfo
  {author} {\bibfnamefont {D.}~\bibnamefont {Xiao}}, \bibinfo {author}
  {\bibfnamefont {Q.}~\bibnamefont {Niu}}, \bibinfo {author} {\bibfnamefont
  {M.}~\bibnamefont {Tsoi}},\ and\ \bibinfo {author} {\bibfnamefont {J.~L.}\
  \bibnamefont {Erskine}},\ }\bibinfo {title} {Universal {{Electromotive Force
  Induced}} by {{Domain Wall Motion}}},\ \href
  {https://link.aps.org/doi/10.1103/PhysRevLett.102.067201} {\bibfield
  {journal} {\bibinfo  {journal} {Phys. Rev. Lett.}\ }\textbf {\bibinfo
  {volume} {102}},\ \bibinfo {pages} {067201} (\bibinfo {year}
  {2009})}\BibitemShut {NoStop}%
\bibitem [{\citenamefont {Kovalev}\ and\ \citenamefont
  {Tserkovnyak}(2012)}]{kovalev2012thermomagnonic}%
  \BibitemOpen
  \bibfield  {author} {\bibinfo {author} {\bibfnamefont {A.~A.}\ \bibnamefont
  {Kovalev}}\ and\ \bibinfo {author} {\bibfnamefont {Y.}~\bibnamefont
  {Tserkovnyak}},\ }\bibinfo {title} {Thermomagnonic Spin Transfer and
  {{Peltier}} Effects in Insulating Magnets},\ \href
  {https://iopscience.iop.org/article/10.1209/0295-5075/97/67002/pdf}
  {\bibfield  {journal} {\bibinfo  {journal} {Europhys. Lett.}\ }\textbf
  {\bibinfo {volume} {97}},\ \bibinfo {pages} {67002} (\bibinfo {year}
  {2012})}\BibitemShut {NoStop}%
\bibitem [{\citenamefont {Kovalev}\ and\ \citenamefont
  {Tserkovnyak}(2010)}]{kovalev2010magnetocaloritronic}%
  \BibitemOpen
  \bibfield  {author} {\bibinfo {author} {\bibfnamefont {A.~A.}\ \bibnamefont
  {Kovalev}}\ and\ \bibinfo {author} {\bibfnamefont {Y.}~\bibnamefont
  {Tserkovnyak}},\ }\bibinfo {title} {Magnetocaloritronic Nanomachines},\ \href
  {https://linkinghub.elsevier.com/retrieve/pii/S0038109809007005} {\bibfield
  {journal} {\bibinfo  {journal} {Solid State Commun.}\ }\textbf {\bibinfo
  {volume} {150}},\ \bibinfo {pages} {500} (\bibinfo {year}
  {2010})}\BibitemShut {NoStop}%
\bibitem [{\citenamefont {Kovalev}\ and\ \citenamefont
  {Tserkovnyak}(2009)}]{kovalev2009Thermoelectric}%
  \BibitemOpen
  \bibfield  {author} {\bibinfo {author} {\bibfnamefont {A.~A.}\ \bibnamefont
  {Kovalev}}\ and\ \bibinfo {author} {\bibfnamefont {Y.}~\bibnamefont
  {Tserkovnyak}},\ }\bibinfo {title} {Thermoelectric Spin Transfer in Textured
  Magnets},\ \href {https://link.aps.org/doi/10.1103/PhysRevB.80.100408}
  {\bibfield  {journal} {\bibinfo  {journal} {Phys. Rev. B}\ }\textbf {\bibinfo
  {volume} {80}},\ \bibinfo {pages} {100408} (\bibinfo {year}
  {2009})}\BibitemShut {NoStop}%
\bibitem [{\citenamefont {Shen}\ \emph {et~al.}(2021)\citenamefont {Shen},
  \citenamefont {Xia}, \citenamefont {Ezawa}, \citenamefont {Tretiakov},
  \citenamefont {Zhao},\ and\ \citenamefont {Zhou}}]{shen2021Signal}%
  \BibitemOpen
  \bibfield  {author} {\bibinfo {author} {\bibfnamefont {L.}~\bibnamefont
  {Shen}}, \bibinfo {author} {\bibfnamefont {J.}~\bibnamefont {Xia}}, \bibinfo
  {author} {\bibfnamefont {M.}~\bibnamefont {Ezawa}}, \bibinfo {author}
  {\bibfnamefont {O.~A.}\ \bibnamefont {Tretiakov}}, \bibinfo {author}
  {\bibfnamefont {G.}~\bibnamefont {Zhao}},\ and\ \bibinfo {author}
  {\bibfnamefont {Y.}~\bibnamefont {Zhou}},\ }\bibinfo {title} {Signal
  Detection Based on the Chaotic Motion of an Antiferromagnetic Domain Wall},\
  \href {http://aip.scitation.org/doi/10.1063/5.0034997} {\bibfield  {journal}
  {\bibinfo  {journal} {Appl. Phys. Lett.}\ }\textbf {\bibinfo {volume}
  {118}},\ \bibinfo {pages} {012402} (\bibinfo {year} {2021})}\BibitemShut
  {NoStop}%
\bibitem [{\citenamefont {Z{\'a}zvorka}\ \emph {et~al.}(2019)\citenamefont
  {Z{\'a}zvorka}, \citenamefont {Jakobs}, \citenamefont {Heinze}, \citenamefont
  {Keil}, \citenamefont {Kromin}, \citenamefont {Jaiswal}, \citenamefont
  {Litzius}, \citenamefont {Jakob}, \citenamefont {Virnau}, \citenamefont
  {Pinna}, \citenamefont {{Everschor-Sitte}}, \citenamefont {R{\'o}zsa},
  \citenamefont {Donges}, \citenamefont {Nowak},\ and\ \citenamefont
  {Kl{\"a}ui}}]{zazvorka2019Thermal}%
  \BibitemOpen
  \bibfield  {author} {\bibinfo {author} {\bibfnamefont {J.}~\bibnamefont
  {Z{\'a}zvorka}}, \bibinfo {author} {\bibfnamefont {F.}~\bibnamefont
  {Jakobs}}, \bibinfo {author} {\bibfnamefont {D.}~\bibnamefont {Heinze}},
  \bibinfo {author} {\bibfnamefont {N.}~\bibnamefont {Keil}}, \bibinfo {author}
  {\bibfnamefont {S.}~\bibnamefont {Kromin}}, \bibinfo {author} {\bibfnamefont
  {S.}~\bibnamefont {Jaiswal}}, \bibinfo {author} {\bibfnamefont
  {K.}~\bibnamefont {Litzius}}, \bibinfo {author} {\bibfnamefont
  {G.}~\bibnamefont {Jakob}}, \bibinfo {author} {\bibfnamefont
  {P.}~\bibnamefont {Virnau}}, \bibinfo {author} {\bibfnamefont
  {D.}~\bibnamefont {Pinna}}, \bibinfo {author} {\bibfnamefont
  {K.}~\bibnamefont {{Everschor-Sitte}}}, \bibinfo {author} {\bibfnamefont
  {L.}~\bibnamefont {R{\'o}zsa}}, \bibinfo {author} {\bibfnamefont
  {A.}~\bibnamefont {Donges}}, \bibinfo {author} {\bibfnamefont
  {U.}~\bibnamefont {Nowak}},\ and\ \bibinfo {author} {\bibfnamefont
  {M.}~\bibnamefont {Kl{\"a}ui}},\ }\bibinfo {title} {Thermal Skyrmion
  Diffusion Used in a Reshuffler Device},\ \href
  {http://www.nature.com/articles/s41565-019-0436-8} {\bibfield  {journal}
  {\bibinfo  {journal} {Nat. Nanotechnol.}\ }\textbf {\bibinfo {volume} {14}},\
  \bibinfo {pages} {658} (\bibinfo {year} {2019})}\BibitemShut {NoStop}%
\bibitem [{\citenamefont {Fern{\'a}ndez~Scarioni}\ \emph
  {et~al.}(2021)\citenamefont {Fern{\'a}ndez~Scarioni}, \citenamefont {Barton},
  \citenamefont {{Corte-Le{\'o}n}}, \citenamefont {Sievers}, \citenamefont
  {Hu}, \citenamefont {Ajejas}, \citenamefont {Legrand}, \citenamefont
  {Reyren}, \citenamefont {Cros}, \citenamefont {Kazakova},\ and\ \citenamefont
  {Schumacher}}]{fernandezscarioni2021thermoelectric}%
  \BibitemOpen
  \bibfield  {author} {\bibinfo {author} {\bibfnamefont {A.}~\bibnamefont
  {Fern{\'a}ndez~Scarioni}}, \bibinfo {author} {\bibfnamefont {C.}~\bibnamefont
  {Barton}}, \bibinfo {author} {\bibfnamefont {H.}~\bibnamefont
  {{Corte-Le{\'o}n}}}, \bibinfo {author} {\bibfnamefont {S.}~\bibnamefont
  {Sievers}}, \bibinfo {author} {\bibfnamefont {X.}~\bibnamefont {Hu}},
  \bibinfo {author} {\bibfnamefont {F.}~\bibnamefont {Ajejas}}, \bibinfo
  {author} {\bibfnamefont {W.}~\bibnamefont {Legrand}}, \bibinfo {author}
  {\bibfnamefont {N.}~\bibnamefont {Reyren}}, \bibinfo {author} {\bibfnamefont
  {V.}~\bibnamefont {Cros}}, \bibinfo {author} {\bibfnamefont {O.}~\bibnamefont
  {Kazakova}},\ and\ \bibinfo {author} {\bibfnamefont {H.~W.}\ \bibnamefont
  {Schumacher}},\ }\bibinfo {title} {Thermoelectric {{Signature}} of
  {{Individual Skyrmions}}},\ \href
  {https://link.aps.org/doi/10.1103/PhysRevLett.126.077202} {\bibfield
  {journal} {\bibinfo  {journal} {Phys. Rev. Lett.}\ }\textbf {\bibinfo
  {volume} {126}},\ \bibinfo {pages} {077202} (\bibinfo {year}
  {2021})}\BibitemShut {NoStop}%
\bibitem [{\citenamefont {Bartell}\ \emph {et~al.}(2015)\citenamefont
  {Bartell}, \citenamefont {Ngai}, \citenamefont {Leng},\ and\ \citenamefont
  {Fuchs}}]{bartell2015Tabletop}%
  \BibitemOpen
  \bibfield  {author} {\bibinfo {author} {\bibfnamefont {J.~M.}\ \bibnamefont
  {Bartell}}, \bibinfo {author} {\bibfnamefont {D.~H.}\ \bibnamefont {Ngai}},
  \bibinfo {author} {\bibfnamefont {Z.}~\bibnamefont {Leng}},\ and\ \bibinfo
  {author} {\bibfnamefont {G.~D.}\ \bibnamefont {Fuchs}},\ }\bibinfo {title}
  {Towards a Table-Top Microscope for Nanoscale Magnetic Imaging Using
  Picosecond Thermal Gradients},\ \href
  {http://www.nature.com/articles/ncomms9460} {\bibfield  {journal} {\bibinfo
  {journal} {Nat. Commun.}\ }\textbf {\bibinfo {volume} {6}},\ \bibinfo {pages}
  {8460} (\bibinfo {year} {2015})}\BibitemShut {NoStop}%
\bibitem [{\citenamefont {Katsura}\ \emph {et~al.}(2010)\citenamefont
  {Katsura}, \citenamefont {Nagaosa},\ and\ \citenamefont
  {Lee}}]{katsura2010Theory}%
  \BibitemOpen
  \bibfield  {author} {\bibinfo {author} {\bibfnamefont {H.}~\bibnamefont
  {Katsura}}, \bibinfo {author} {\bibfnamefont {N.}~\bibnamefont {Nagaosa}},\
  and\ \bibinfo {author} {\bibfnamefont {P.~A.}\ \bibnamefont {Lee}},\
  }\bibinfo {title} {Theory of the {{Thermal Hall Effect}} in {{Quantum
  Magnets}}},\ \href {https://link.aps.org/doi/10.1103/PhysRevLett.104.066403}
  {\bibfield  {journal} {\bibinfo  {journal} {Phys. Rev. Lett.}\ }\textbf
  {\bibinfo {volume} {104}},\ \bibinfo {pages} {066403} (\bibinfo {year}
  {2010})}\BibitemShut {NoStop}%
\bibitem [{\citenamefont {Chernyshev}(2015)}]{chernyshev2015strong}%
  \BibitemOpen
  \bibfield  {author} {\bibinfo {author} {\bibfnamefont {A.~L.}\ \bibnamefont
  {Chernyshev}},\ }\bibinfo {title} {Strong Quantum Effects in an Almost
  Classical Antiferromagnet on a Kagome Lattice},\ \href
  {https://link.aps.org/doi/10.1103/PhysRevB.92.094409} {\bibfield  {journal}
  {\bibinfo  {journal} {Phys. Rev. B}\ }\textbf {\bibinfo {volume} {92}},\
  \bibinfo {pages} {094409} (\bibinfo {year} {2015})}\BibitemShut {NoStop}%
\bibitem [{\citenamefont {Yu}\ \emph {et~al.}(2019)\citenamefont {Yu},
  \citenamefont {Bang}, \citenamefont {Mishra}, \citenamefont {Ramaswamy},
  \citenamefont {Oh}, \citenamefont {Park}, \citenamefont {Jeong},
  \citenamefont {Van~Thach}, \citenamefont {Lee}, \citenamefont {Go},
  \citenamefont {Lee}, \citenamefont {Wang}, \citenamefont {Shi}, \citenamefont
  {Qiu}, \citenamefont {Awano}, \citenamefont {Lee},\ and\ \citenamefont
  {Yang}}]{yu2019Long}%
  \BibitemOpen
  \bibfield  {author} {\bibinfo {author} {\bibfnamefont {J.}~\bibnamefont
  {Yu}}, \bibinfo {author} {\bibfnamefont {D.}~\bibnamefont {Bang}}, \bibinfo
  {author} {\bibfnamefont {R.}~\bibnamefont {Mishra}}, \bibinfo {author}
  {\bibfnamefont {R.}~\bibnamefont {Ramaswamy}}, \bibinfo {author}
  {\bibfnamefont {J.~H.}\ \bibnamefont {Oh}}, \bibinfo {author} {\bibfnamefont
  {H.-J.}\ \bibnamefont {Park}}, \bibinfo {author} {\bibfnamefont
  {Y.}~\bibnamefont {Jeong}}, \bibinfo {author} {\bibfnamefont
  {P.}~\bibnamefont {Van~Thach}}, \bibinfo {author} {\bibfnamefont {D.-K.}\
  \bibnamefont {Lee}}, \bibinfo {author} {\bibfnamefont {G.}~\bibnamefont
  {Go}}, \bibinfo {author} {\bibfnamefont {S.-W.}\ \bibnamefont {Lee}},
  \bibinfo {author} {\bibfnamefont {Y.}~\bibnamefont {Wang}}, \bibinfo {author}
  {\bibfnamefont {S.}~\bibnamefont {Shi}}, \bibinfo {author} {\bibfnamefont
  {X.}~\bibnamefont {Qiu}}, \bibinfo {author} {\bibfnamefont {H.}~\bibnamefont
  {Awano}}, \bibinfo {author} {\bibfnamefont {K.-J.}\ \bibnamefont {Lee}},\
  and\ \bibinfo {author} {\bibfnamefont {H.}~\bibnamefont {Yang}},\ }\bibinfo
  {title} {Long Spin Coherence Length and Bulk-like Spin\textendash Orbit
  Torque in Ferrimagnetic Multilayers},\ \href
  {http://www.nature.com/articles/s41563-018-0236-9} {\bibfield  {journal}
  {\bibinfo  {journal} {Nat. Mater.}\ }\textbf {\bibinfo {volume} {18}},\
  \bibinfo {pages} {29} (\bibinfo {year} {2019})}\BibitemShut {NoStop}%
\bibitem [{\citenamefont {Kim}\ \emph {et~al.}(2019{\natexlab{a}})\citenamefont
  {Kim}, \citenamefont {Okuno}, \citenamefont {Kim}, \citenamefont {Oh},
  \citenamefont {Nishimura}, \citenamefont {Hirata}, \citenamefont {Futakawa},
  \citenamefont {Yoshikawa}, \citenamefont {Tsukamoto}, \citenamefont
  {Tserkovnyak}, \citenamefont {Shiota}, \citenamefont {Moriyama},
  \citenamefont {Kim}, \citenamefont {Lee},\ and\ \citenamefont
  {Ono}}]{kim2019Low}%
  \BibitemOpen
  \bibfield  {author} {\bibinfo {author} {\bibfnamefont {D.-H.}\ \bibnamefont
  {Kim}}, \bibinfo {author} {\bibfnamefont {T.}~\bibnamefont {Okuno}}, \bibinfo
  {author} {\bibfnamefont {S.~K.}\ \bibnamefont {Kim}}, \bibinfo {author}
  {\bibfnamefont {S.-H.}\ \bibnamefont {Oh}}, \bibinfo {author} {\bibfnamefont
  {T.}~\bibnamefont {Nishimura}}, \bibinfo {author} {\bibfnamefont
  {Y.}~\bibnamefont {Hirata}}, \bibinfo {author} {\bibfnamefont
  {Y.}~\bibnamefont {Futakawa}}, \bibinfo {author} {\bibfnamefont
  {H.}~\bibnamefont {Yoshikawa}}, \bibinfo {author} {\bibfnamefont
  {A.}~\bibnamefont {Tsukamoto}}, \bibinfo {author} {\bibfnamefont
  {Y.}~\bibnamefont {Tserkovnyak}}, \bibinfo {author} {\bibfnamefont
  {Y.}~\bibnamefont {Shiota}}, \bibinfo {author} {\bibfnamefont
  {T.}~\bibnamefont {Moriyama}}, \bibinfo {author} {\bibfnamefont {K.-J.}\
  \bibnamefont {Kim}}, \bibinfo {author} {\bibfnamefont {K.-J.}\ \bibnamefont
  {Lee}},\ and\ \bibinfo {author} {\bibfnamefont {T.}~\bibnamefont {Ono}},\
  }\bibinfo {title} {Low {{Magnetic Damping}} of {{Ferrimagnetic GdFeCo
  Alloys}}},\ \href {https://link.aps.org/doi/10.1103/PhysRevLett.122.127203}
  {\bibfield  {journal} {\bibinfo  {journal} {Phys. Rev. Lett.}\ }\textbf
  {\bibinfo {volume} {122}},\ \bibinfo {pages} {127203} (\bibinfo {year}
  {2019}{\natexlab{a}})}\BibitemShut {NoStop}%
\bibitem [{\citenamefont {Okuno}\ \emph {et~al.}(2019)\citenamefont {Okuno},
  \citenamefont {Kim}, \citenamefont {Oh}, \citenamefont {Kim}, \citenamefont
  {Hirata}, \citenamefont {Nishimura}, \citenamefont {Ham}, \citenamefont
  {Futakawa}, \citenamefont {Yoshikawa}, \citenamefont {Tsukamoto},
  \citenamefont {Tserkovnyak}, \citenamefont {Shiota}, \citenamefont
  {Moriyama}, \citenamefont {Kim}, \citenamefont {Lee},\ and\ \citenamefont
  {Ono}}]{okuno2019Spintransfer}%
  \BibitemOpen
  \bibfield  {author} {\bibinfo {author} {\bibfnamefont {T.}~\bibnamefont
  {Okuno}}, \bibinfo {author} {\bibfnamefont {D.-H.}\ \bibnamefont {Kim}},
  \bibinfo {author} {\bibfnamefont {S.-H.}\ \bibnamefont {Oh}}, \bibinfo
  {author} {\bibfnamefont {S.~K.}\ \bibnamefont {Kim}}, \bibinfo {author}
  {\bibfnamefont {Y.}~\bibnamefont {Hirata}}, \bibinfo {author} {\bibfnamefont
  {T.}~\bibnamefont {Nishimura}}, \bibinfo {author} {\bibfnamefont {W.~S.}\
  \bibnamefont {Ham}}, \bibinfo {author} {\bibfnamefont {Y.}~\bibnamefont
  {Futakawa}}, \bibinfo {author} {\bibfnamefont {H.}~\bibnamefont {Yoshikawa}},
  \bibinfo {author} {\bibfnamefont {A.}~\bibnamefont {Tsukamoto}}, \bibinfo
  {author} {\bibfnamefont {Y.}~\bibnamefont {Tserkovnyak}}, \bibinfo {author}
  {\bibfnamefont {Y.}~\bibnamefont {Shiota}}, \bibinfo {author} {\bibfnamefont
  {T.}~\bibnamefont {Moriyama}}, \bibinfo {author} {\bibfnamefont {K.-J.}\
  \bibnamefont {Kim}}, \bibinfo {author} {\bibfnamefont {K.-J.}\ \bibnamefont
  {Lee}},\ and\ \bibinfo {author} {\bibfnamefont {T.}~\bibnamefont {Ono}},\
  }\bibinfo {title} {Spin-Transfer Torques for Domain Wall Motion in
  Antiferromagnetically Coupled Ferrimagnets},\ \href
  {http://www.nature.com/articles/s41928-019-0303-5} {\bibfield  {journal}
  {\bibinfo  {journal} {Nat. Electron.}\ }\textbf {\bibinfo {volume} {2}},\
  \bibinfo {pages} {389} (\bibinfo {year} {2019})}\BibitemShut {NoStop}%
\bibitem [{\citenamefont {Siddiqui}\ \emph {et~al.}(2018)\citenamefont
  {Siddiqui}, \citenamefont {Han}, \citenamefont {Finley}, \citenamefont
  {Ross},\ and\ \citenamefont {Liu}}]{siddiqui2018CurrentInduced}%
  \BibitemOpen
  \bibfield  {author} {\bibinfo {author} {\bibfnamefont {S.~A.}\ \bibnamefont
  {Siddiqui}}, \bibinfo {author} {\bibfnamefont {J.}~\bibnamefont {Han}},
  \bibinfo {author} {\bibfnamefont {J.~T.}\ \bibnamefont {Finley}}, \bibinfo
  {author} {\bibfnamefont {C.~A.}\ \bibnamefont {Ross}},\ and\ \bibinfo
  {author} {\bibfnamefont {L.}~\bibnamefont {Liu}},\ }\bibinfo {title}
  {Current-{{Induced Domain Wall Motion}} in a {{Compensated Ferrimagnet}}},\
  \href {https://link.aps.org/doi/10.1103/PhysRevLett.121.057701} {\bibfield
  {journal} {\bibinfo  {journal} {Phys. Rev. Lett.}\ }\textbf {\bibinfo
  {volume} {121}},\ \bibinfo {pages} {057701} (\bibinfo {year}
  {2018})}\BibitemShut {NoStop}%
\bibitem [{\citenamefont {S{\"u}rgers}\ \emph {et~al.}(2014)\citenamefont
  {S{\"u}rgers}, \citenamefont {Fischer}, \citenamefont {Winkel},\ and\
  \citenamefont {v.~L{\"o}hneysen}}]{surgers2014large}%
  \BibitemOpen
  \bibfield  {author} {\bibinfo {author} {\bibfnamefont {C.}~\bibnamefont
  {S{\"u}rgers}}, \bibinfo {author} {\bibfnamefont {G.}~\bibnamefont
  {Fischer}}, \bibinfo {author} {\bibfnamefont {P.}~\bibnamefont {Winkel}},\
  and\ \bibinfo {author} {\bibfnamefont {H.}~\bibnamefont {v.~L{\"o}hneysen}},\
  }\bibinfo {title} {Large Topological {{Hall}} Effect in the Non-Collinear
  Phase of an Antiferromagnet},\ \href
  {http://www.nature.com/articles/ncomms4400.pdf} {\bibfield  {journal}
  {\bibinfo  {journal} {Nat. Commun.}\ }\textbf {\bibinfo {volume} {5}},\
  \bibinfo {pages} {8} (\bibinfo {year} {2014})}\BibitemShut {NoStop}%
\bibitem [{\citenamefont {Park}\ \emph {et~al.}(2020)\citenamefont {Park},
  \citenamefont {Jeong}, \citenamefont {Oh}, \citenamefont {Go}, \citenamefont
  {Oh}, \citenamefont {Kim}, \citenamefont {Lee},\ and\ \citenamefont
  {Lee}}]{park2020numerical}%
  \BibitemOpen
  \bibfield  {author} {\bibinfo {author} {\bibfnamefont {H.-J.}\ \bibnamefont
  {Park}}, \bibinfo {author} {\bibfnamefont {Y.}~\bibnamefont {Jeong}},
  \bibinfo {author} {\bibfnamefont {S.-H.}\ \bibnamefont {Oh}}, \bibinfo
  {author} {\bibfnamefont {G.}~\bibnamefont {Go}}, \bibinfo {author}
  {\bibfnamefont {J.~H.}\ \bibnamefont {Oh}}, \bibinfo {author} {\bibfnamefont
  {K.-W.}\ \bibnamefont {Kim}}, \bibinfo {author} {\bibfnamefont {H.-W.}\
  \bibnamefont {Lee}},\ and\ \bibinfo {author} {\bibfnamefont {K.-J.}\
  \bibnamefont {Lee}},\ }\bibinfo {title} {Numerical Computation of
  Spin-Transfer Torques for Antiferromagnetic Domain Walls},\ \href
  {https://link.aps.org/doi/10.1103/PhysRevB.101.144431} {\bibfield  {journal}
  {\bibinfo  {journal} {Phys. Rev. B}\ }\textbf {\bibinfo {volume} {101}},\
  \bibinfo {pages} {144431} (\bibinfo {year} {2020})}\BibitemShut {NoStop}%
\bibitem [{\citenamefont {Okuno}(2020)}]{okuno2020Magnetic}%
  \BibitemOpen
  \bibfield  {author} {\bibinfo {author} {\bibfnamefont {T.}~\bibnamefont
  {Okuno}},\ }\bibinfo {title} {Magnetic {{Dynamics}} in
  {{Antiferromagnetically-Coupled Ferrimagnets}}: {{The Role}} of {{Angular
  Momentum}}},\ \href {http://link.springer.com/10.1007/978-981-15-9176-1}
  {\bibfield  {journal} {\bibinfo  {journal} {Springer Theses, Singapore}\ }
  (\bibinfo {year} {2020})}\BibitemShut {NoStop}%
\bibitem [{\citenamefont {Yu}\ \emph {et~al.}(2018)\citenamefont {Yu},
  \citenamefont {Lan},\ and\ \citenamefont
  {Xiao}}]{yu2018polarizationselective}%
  \BibitemOpen
  \bibfield  {author} {\bibinfo {author} {\bibfnamefont {W.}~\bibnamefont
  {Yu}}, \bibinfo {author} {\bibfnamefont {J.}~\bibnamefont {Lan}},\ and\
  \bibinfo {author} {\bibfnamefont {J.}~\bibnamefont {Xiao}},\ }\bibinfo
  {title} {Polarization-Selective Spin Wave Driven Domain-Wall Motion in
  Antiferromagnets},\ \href
  {https://link.aps.org/doi/10.1103/PhysRevB.98.144422} {\bibfield  {journal}
  {\bibinfo  {journal} {Phys. Rev. B}\ }\textbf {\bibinfo {volume} {98}},\
  \bibinfo {pages} {144422} (\bibinfo {year} {2018})}\BibitemShut {NoStop}%
\bibitem [{\citenamefont {Dasgupta}\ and\ \citenamefont
  {Zou}(2021)}]{dasgupta2021zeeman}%
  \BibitemOpen
  \bibfield  {author} {\bibinfo {author} {\bibfnamefont {S.}~\bibnamefont
  {Dasgupta}}\ and\ \bibinfo {author} {\bibfnamefont {J.}~\bibnamefont {Zou}},\
  }\bibinfo {title} {Zeeman Term for the {{N\'eel}} Vector in a Two Sublattice
  Antiferromagnet},\ \href
  {https://link.aps.org/article/10.1103/PhysRevB.104.064415} {\bibfield
  {journal} {\bibinfo  {journal} {Phys. Rev. B}\ }\textbf {\bibinfo {volume}
  {104}},\ \bibinfo {pages} {064415} (\bibinfo {year} {2021})}\BibitemShut
  {NoStop}%
\bibitem [{\citenamefont {Kim}\ \emph {et~al.}(2017)\citenamefont {Kim},
  \citenamefont {Kim}, \citenamefont {Hirata}, \citenamefont {Oh},
  \citenamefont {Tono}, \citenamefont {Kim}, \citenamefont {Okuno},
  \citenamefont {Ham}, \citenamefont {Kim}, \citenamefont {Go}, \citenamefont
  {Tserkovnyak}, \citenamefont {Tsukamoto}, \citenamefont {Moriyama},
  \citenamefont {Lee},\ and\ \citenamefont {Ono}}]{kim2017fast}%
  \BibitemOpen
  \bibfield  {author} {\bibinfo {author} {\bibfnamefont {K.-J.}\ \bibnamefont
  {Kim}}, \bibinfo {author} {\bibfnamefont {S.~K.}\ \bibnamefont {Kim}},
  \bibinfo {author} {\bibfnamefont {Y.}~\bibnamefont {Hirata}}, \bibinfo
  {author} {\bibfnamefont {S.-H.}\ \bibnamefont {Oh}}, \bibinfo {author}
  {\bibfnamefont {T.}~\bibnamefont {Tono}}, \bibinfo {author} {\bibfnamefont
  {D.-H.}\ \bibnamefont {Kim}}, \bibinfo {author} {\bibfnamefont
  {T.}~\bibnamefont {Okuno}}, \bibinfo {author} {\bibfnamefont {W.~S.}\
  \bibnamefont {Ham}}, \bibinfo {author} {\bibfnamefont {S.}~\bibnamefont
  {Kim}}, \bibinfo {author} {\bibfnamefont {G.}~\bibnamefont {Go}}, \bibinfo
  {author} {\bibfnamefont {Y.}~\bibnamefont {Tserkovnyak}}, \bibinfo {author}
  {\bibfnamefont {A.}~\bibnamefont {Tsukamoto}}, \bibinfo {author}
  {\bibfnamefont {T.}~\bibnamefont {Moriyama}}, \bibinfo {author}
  {\bibfnamefont {K.-J.}\ \bibnamefont {Lee}},\ and\ \bibinfo {author}
  {\bibfnamefont {T.}~\bibnamefont {Ono}},\ }\bibinfo {title} {Fast Domain Wall
  Motion in the Vicinity of the Angular Momentum Compensation Temperature of
  Ferrimagnets},\ \href {http://www.nature.com/articles/nmat4990} {\bibfield
  {journal} {\bibinfo  {journal} {Nat. Mater.}\ }\textbf {\bibinfo {volume}
  {16}},\ \bibinfo {pages} {1187} (\bibinfo {year} {2017})}\BibitemShut
  {NoStop}%
\bibitem [{\citenamefont {Caretta}\ \emph {et~al.}(2020)\citenamefont
  {Caretta}, \citenamefont {Oh}, \citenamefont {Fakhrul}, \citenamefont {Lee},
  \citenamefont {Lee}, \citenamefont {Kim}, \citenamefont {Ross}, \citenamefont
  {Lee},\ and\ \citenamefont {Beach}}]{caretta2020Relativistic}%
  \BibitemOpen
  \bibfield  {author} {\bibinfo {author} {\bibfnamefont {L.}~\bibnamefont
  {Caretta}}, \bibinfo {author} {\bibfnamefont {S.-H.}\ \bibnamefont {Oh}},
  \bibinfo {author} {\bibfnamefont {T.}~\bibnamefont {Fakhrul}}, \bibinfo
  {author} {\bibfnamefont {D.-K.}\ \bibnamefont {Lee}}, \bibinfo {author}
  {\bibfnamefont {B.~H.}\ \bibnamefont {Lee}}, \bibinfo {author} {\bibfnamefont
  {S.~K.}\ \bibnamefont {Kim}}, \bibinfo {author} {\bibfnamefont {C.~A.}\
  \bibnamefont {Ross}}, \bibinfo {author} {\bibfnamefont {K.-J.}\ \bibnamefont
  {Lee}},\ and\ \bibinfo {author} {\bibfnamefont {G.~S.~D.}\ \bibnamefont
  {Beach}},\ }\bibinfo {title} {Relativistic Kinematics of a Magnetic
  Soliton},\ \href {https://www.science.org/doi/10.1126/science.aba5555}
  {\bibfield  {journal} {\bibinfo  {journal} {Science}\ }\textbf {\bibinfo
  {volume} {370}},\ \bibinfo {pages} {1438} (\bibinfo {year}
  {2020})}\BibitemShut {NoStop}%
\bibitem [{\citenamefont {Cheng}\ and\ \citenamefont
  {Niu}(2012)}]{cheng2012Electron}%
  \BibitemOpen
  \bibfield  {author} {\bibinfo {author} {\bibfnamefont {R.}~\bibnamefont
  {Cheng}}\ and\ \bibinfo {author} {\bibfnamefont {Q.}~\bibnamefont {Niu}},\
  }\bibinfo {title} {Electron Dynamics in Slowly Varying Antiferromagnetic
  Texture},\ \href {https://link.aps.org/doi/10.1103/PhysRevB.86.245118}
  {\bibfield  {journal} {\bibinfo  {journal} {Phys. Rev. B}\ }\textbf {\bibinfo
  {volume} {86}},\ \bibinfo {pages} {245118} (\bibinfo {year}
  {2012})}\BibitemShut {NoStop}%
\bibitem [{\citenamefont {Mondal}\ and\ \citenamefont
  {Kamra}(2021)}]{mondal2021spin}%
  \BibitemOpen
  \bibfield  {author} {\bibinfo {author} {\bibfnamefont {R.}~\bibnamefont
  {Mondal}}\ and\ \bibinfo {author} {\bibfnamefont {A.}~\bibnamefont {Kamra}},\
  }\bibinfo {title} {Spin Pumping at Terahertz Nutation Resonances},\ \href
  {https://link.aps.org/article/10.1103/PhysRevB.104.214426} {\bibfield
  {journal} {\bibinfo  {journal} {Phys. Rev. B}\ }\textbf {\bibinfo {volume}
  {104}},\ \bibinfo {pages} {214426} (\bibinfo {year} {2021})}\BibitemShut
  {NoStop}%
\bibitem [{\citenamefont {Pacchioni}(2020)}]{pacchioni2020heat}%
  \BibitemOpen
  \bibfield  {author} {\bibinfo {author} {\bibfnamefont {G.}~\bibnamefont
  {Pacchioni}},\ }\bibinfo {title} {The Heat Is On},\ \href
  {http://www.nature.com/articles/s41578-020-00266-9} {\bibfield  {journal}
  {\bibinfo  {journal} {Nat. Rev. Mater.}\ }\textbf {\bibinfo {volume} {5}},\
  \bibinfo {pages} {868} (\bibinfo {year} {2020})}\BibitemShut {NoStop}%
\bibitem [{\citenamefont {Yu}\ \emph {et~al.}(2021)\citenamefont {Yu},
  \citenamefont {Kagawa}, \citenamefont {Seki}, \citenamefont {Kubota},
  \citenamefont {Masell}, \citenamefont {Yasin}, \citenamefont {Nakajima},
  \citenamefont {Nakamura}, \citenamefont {Kawasaki}, \citenamefont {Nagaosa},\
  and\ \citenamefont {Tokura}}]{yu2021realspace}%
  \BibitemOpen
  \bibfield  {author} {\bibinfo {author} {\bibfnamefont {X.}~\bibnamefont
  {Yu}}, \bibinfo {author} {\bibfnamefont {F.}~\bibnamefont {Kagawa}}, \bibinfo
  {author} {\bibfnamefont {S.}~\bibnamefont {Seki}}, \bibinfo {author}
  {\bibfnamefont {M.}~\bibnamefont {Kubota}}, \bibinfo {author} {\bibfnamefont
  {J.}~\bibnamefont {Masell}}, \bibinfo {author} {\bibfnamefont {F.~S.}\
  \bibnamefont {Yasin}}, \bibinfo {author} {\bibfnamefont {K.}~\bibnamefont
  {Nakajima}}, \bibinfo {author} {\bibfnamefont {M.}~\bibnamefont {Nakamura}},
  \bibinfo {author} {\bibfnamefont {M.}~\bibnamefont {Kawasaki}}, \bibinfo
  {author} {\bibfnamefont {N.}~\bibnamefont {Nagaosa}},\ and\ \bibinfo {author}
  {\bibfnamefont {Y.}~\bibnamefont {Tokura}},\ }\bibinfo {title} {Real-Space
  Observations of 60-Nm Skyrmion Dynamics in an Insulating Magnet under Low
  Heat Flow},\ \href {https://www.nature.com/articles/s41467-021-25291-2}
  {\bibfield  {journal} {\bibinfo  {journal} {Nat. Commun.}\ }\textbf {\bibinfo
  {volume} {12}},\ \bibinfo {pages} {5079} (\bibinfo {year}
  {2021})}\BibitemShut {NoStop}%
\bibitem [{\citenamefont {Jiang}\ \emph {et~al.}(2013)\citenamefont {Jiang},
  \citenamefont {Upadhyaya}, \citenamefont {Fan}, \citenamefont {Zhao},
  \citenamefont {Wang}, \citenamefont {Chang}, \citenamefont {Lang},
  \citenamefont {Wong}, \citenamefont {Lewis}, \citenamefont {Lin},
  \citenamefont {Tang}, \citenamefont {Cherepov}, \citenamefont {Zhou},
  \citenamefont {Tserkovnyak}, \citenamefont {Schwartz},\ and\ \citenamefont
  {Wang}}]{jiang2013direct}%
  \BibitemOpen
  \bibfield  {author} {\bibinfo {author} {\bibfnamefont {W.}~\bibnamefont
  {Jiang}}, \bibinfo {author} {\bibfnamefont {P.}~\bibnamefont {Upadhyaya}},
  \bibinfo {author} {\bibfnamefont {Y.}~\bibnamefont {Fan}}, \bibinfo {author}
  {\bibfnamefont {J.}~\bibnamefont {Zhao}}, \bibinfo {author} {\bibfnamefont
  {M.}~\bibnamefont {Wang}}, \bibinfo {author} {\bibfnamefont {L.-T.}\
  \bibnamefont {Chang}}, \bibinfo {author} {\bibfnamefont {M.}~\bibnamefont
  {Lang}}, \bibinfo {author} {\bibfnamefont {K.~L.}\ \bibnamefont {Wong}},
  \bibinfo {author} {\bibfnamefont {M.}~\bibnamefont {Lewis}}, \bibinfo
  {author} {\bibfnamefont {Y.-T.}\ \bibnamefont {Lin}}, \bibinfo {author}
  {\bibfnamefont {J.}~\bibnamefont {Tang}}, \bibinfo {author} {\bibfnamefont
  {S.}~\bibnamefont {Cherepov}}, \bibinfo {author} {\bibfnamefont
  {X.}~\bibnamefont {Zhou}}, \bibinfo {author} {\bibfnamefont {Y.}~\bibnamefont
  {Tserkovnyak}}, \bibinfo {author} {\bibfnamefont {R.~N.}\ \bibnamefont
  {Schwartz}},\ and\ \bibinfo {author} {\bibfnamefont {K.~L.}\ \bibnamefont
  {Wang}},\ }\bibinfo {title} {Direct {{Imaging}} of {{Thermally Driven Domain
  Wall Motion}} in {{Magnetic Insulators}}},\ \href
  {https://link.aps.org/doi/10.1103/PhysRevLett.110.177202} {\bibfield
  {journal} {\bibinfo  {journal} {Phys. Rev. Lett.}\ }\textbf {\bibinfo
  {volume} {110}},\ \bibinfo {pages} {177202} (\bibinfo {year}
  {2013})}\BibitemShut {NoStop}%
\bibitem [{\citenamefont {Tolley}\ \emph {et~al.}(2015)\citenamefont {Tolley},
  \citenamefont {Liu}, \citenamefont {Xu}, \citenamefont {Le~Gall},
  \citenamefont {Gottwald}, \citenamefont {Hauet}, \citenamefont {Hehn},
  \citenamefont {Montaigne}, \citenamefont {Fullerton},\ and\ \citenamefont
  {Mangin}}]{tolley2015generation}%
  \BibitemOpen
  \bibfield  {author} {\bibinfo {author} {\bibfnamefont {R.}~\bibnamefont
  {Tolley}}, \bibinfo {author} {\bibfnamefont {T.}~\bibnamefont {Liu}},
  \bibinfo {author} {\bibfnamefont {Y.}~\bibnamefont {Xu}}, \bibinfo {author}
  {\bibfnamefont {S.}~\bibnamefont {Le~Gall}}, \bibinfo {author} {\bibfnamefont
  {M.}~\bibnamefont {Gottwald}}, \bibinfo {author} {\bibfnamefont
  {T.}~\bibnamefont {Hauet}}, \bibinfo {author} {\bibfnamefont
  {M.}~\bibnamefont {Hehn}}, \bibinfo {author} {\bibfnamefont {F.}~\bibnamefont
  {Montaigne}}, \bibinfo {author} {\bibfnamefont {E.~E.}\ \bibnamefont
  {Fullerton}},\ and\ \bibinfo {author} {\bibfnamefont {S.}~\bibnamefont
  {Mangin}},\ }\bibinfo {title} {Generation and Manipulation of Domain Walls
  Using a Thermal Gradient in a Ferrimagnetic {{TbCo}} Wire},\ \href
  {http://aip.scitation.org/doi/10.1063/1.4922603} {\bibfield  {journal}
  {\bibinfo  {journal} {Appl. Phys. Lett.}\ }\textbf {\bibinfo {volume}
  {106}},\ \bibinfo {pages} {242403} (\bibinfo {year} {2015})}\BibitemShut
  {NoStop}%
\bibitem [{\citenamefont {Torrejon}\ \emph {et~al.}(2012)\citenamefont
  {Torrejon}, \citenamefont {Malinowski}, \citenamefont {Pelloux},
  \citenamefont {Weil}, \citenamefont {Thiaville}, \citenamefont {Curiale},
  \citenamefont {Lacour}, \citenamefont {Montaigne},\ and\ \citenamefont
  {Hehn}}]{torrejon2012unidirectional}%
  \BibitemOpen
  \bibfield  {author} {\bibinfo {author} {\bibfnamefont {J.}~\bibnamefont
  {Torrejon}}, \bibinfo {author} {\bibfnamefont {G.}~\bibnamefont
  {Malinowski}}, \bibinfo {author} {\bibfnamefont {M.}~\bibnamefont {Pelloux}},
  \bibinfo {author} {\bibfnamefont {R.}~\bibnamefont {Weil}}, \bibinfo {author}
  {\bibfnamefont {A.}~\bibnamefont {Thiaville}}, \bibinfo {author}
  {\bibfnamefont {J.}~\bibnamefont {Curiale}}, \bibinfo {author} {\bibfnamefont
  {D.}~\bibnamefont {Lacour}}, \bibinfo {author} {\bibfnamefont
  {F.}~\bibnamefont {Montaigne}},\ and\ \bibinfo {author} {\bibfnamefont
  {M.}~\bibnamefont {Hehn}},\ }\bibinfo {title} {Unidirectional {{Thermal
  Effects}} in {{Current-Induced Domain Wall Motion}}},\ \href
  {https://link.aps.org/doi/10.1103/PhysRevLett.109.106601} {\bibfield
  {journal} {\bibinfo  {journal} {Phys. Rev. Lett.}\ }\textbf {\bibinfo
  {volume} {109}},\ \bibinfo {pages} {106601} (\bibinfo {year}
  {2012})}\BibitemShut {NoStop}%
\bibitem [{\citenamefont {Qin}\ \emph {et~al.}(2022)\citenamefont {Qin},
  \citenamefont {Zhang}, \citenamefont {Zhang}, \citenamefont {Pei},
  \citenamefont {Yang}, \citenamefont {Xu}, \citenamefont {Zhou}, \citenamefont
  {Wu}, \citenamefont {Du},\ and\ \citenamefont {Che}}]{qin2022dynamics}%
  \BibitemOpen
  \bibfield  {author} {\bibinfo {author} {\bibfnamefont {G.}~\bibnamefont
  {Qin}}, \bibinfo {author} {\bibfnamefont {X.}~\bibnamefont {Zhang}}, \bibinfo
  {author} {\bibfnamefont {R.}~\bibnamefont {Zhang}}, \bibinfo {author}
  {\bibfnamefont {K.}~\bibnamefont {Pei}}, \bibinfo {author} {\bibfnamefont
  {C.}~\bibnamefont {Yang}}, \bibinfo {author} {\bibfnamefont {C.}~\bibnamefont
  {Xu}}, \bibinfo {author} {\bibfnamefont {Y.}~\bibnamefont {Zhou}}, \bibinfo
  {author} {\bibfnamefont {Y.}~\bibnamefont {Wu}}, \bibinfo {author}
  {\bibfnamefont {H.}~\bibnamefont {Du}},\ and\ \bibinfo {author}
  {\bibfnamefont {R.}~\bibnamefont {Che}},\ }\bibinfo {title} {Dynamics of
  Magnetic Skyrmions Driven by a Temperature Gradient in a Chiral Magnet
  {{FeGe}}},\ \href {https://link.aps.org/article/10.1103/PhysRevB.106.024415}
  {\bibfield  {journal} {\bibinfo  {journal} {Phys. Rev. B}\ }\textbf {\bibinfo
  {volume} {106}},\ \bibinfo {pages} {024415} (\bibinfo {year}
  {2022})}\BibitemShut {NoStop}%
\bibitem [{\citenamefont {Shokr}\ \emph {et~al.}(2019)\citenamefont {Shokr},
  \citenamefont {Sandig}, \citenamefont {Erkovan}, \citenamefont {Zhang},
  \citenamefont {Bernien}, \citenamefont {{\"U}nal}, \citenamefont {Kronast},
  \citenamefont {Parlak}, \citenamefont {Vogel},\ and\ \citenamefont
  {Kuch}}]{shokr2019steeringa}%
  \BibitemOpen
  \bibfield  {author} {\bibinfo {author} {\bibfnamefont {Y.~A.}\ \bibnamefont
  {Shokr}}, \bibinfo {author} {\bibfnamefont {O.}~\bibnamefont {Sandig}},
  \bibinfo {author} {\bibfnamefont {M.}~\bibnamefont {Erkovan}}, \bibinfo
  {author} {\bibfnamefont {B.}~\bibnamefont {Zhang}}, \bibinfo {author}
  {\bibfnamefont {M.}~\bibnamefont {Bernien}}, \bibinfo {author} {\bibfnamefont
  {A.~A.}\ \bibnamefont {{\"U}nal}}, \bibinfo {author} {\bibfnamefont
  {F.}~\bibnamefont {Kronast}}, \bibinfo {author} {\bibfnamefont
  {U.}~\bibnamefont {Parlak}}, \bibinfo {author} {\bibfnamefont
  {J.}~\bibnamefont {Vogel}},\ and\ \bibinfo {author} {\bibfnamefont
  {W.}~\bibnamefont {Kuch}},\ }\bibinfo {title} {Steering of Magnetic Domain
  Walls by Single Ultrashort Laser Pulses},\ \href
  {https://link.aps.org/doi/10.1103/PhysRevB.99.214404} {\bibfield  {journal}
  {\bibinfo  {journal} {Phys. Rev. B}\ }\textbf {\bibinfo {volume} {99}},\
  \bibinfo {pages} {214404} (\bibinfo {year} {2019})}\BibitemShut {NoStop}%
\bibitem [{\citenamefont {Jen}\ and\ \citenamefont
  {Berger}(1986)}]{jen1986thermal}%
  \BibitemOpen
  \bibfield  {author} {\bibinfo {author} {\bibfnamefont {S.~U.}\ \bibnamefont
  {Jen}}\ and\ \bibinfo {author} {\bibfnamefont {L.}~\bibnamefont {Berger}},\
  }\bibinfo {title} {Thermal Domain Drag Effect in Amorphous Ferromagnetic
  Materials. {{II}}. {{Experiments}}},\ \href
  {http://aip.scitation.org/doi/pdf/10.1063/1.336518} {\bibfield  {journal}
  {\bibinfo  {journal} {J. Appl. Phys.}\ }\textbf {\bibinfo {volume} {59}},\
  \bibinfo {pages} {1285} (\bibinfo {year} {1986})}\BibitemShut {NoStop}%
\bibitem [{\citenamefont {Yan}\ \emph {et~al.}(2011)\citenamefont {Yan},
  \citenamefont {Wang},\ and\ \citenamefont {Wang}}]{yan2011allmagnonic}%
  \BibitemOpen
  \bibfield  {author} {\bibinfo {author} {\bibfnamefont {P.}~\bibnamefont
  {Yan}}, \bibinfo {author} {\bibfnamefont {X.~S.}\ \bibnamefont {Wang}},\ and\
  \bibinfo {author} {\bibfnamefont {X.~R.}\ \bibnamefont {Wang}},\ }\bibinfo
  {title} {All-{{Magnonic Spin-Transfer Torque}} and {{Domain Wall
  Propagation}}},\ \href
  {https://link.aps.org/doi/10.1103/PhysRevLett.107.177207} {\bibfield
  {journal} {\bibinfo  {journal} {Phys. Rev. Lett.}\ }\textbf {\bibinfo
  {volume} {107}},\ \bibinfo {pages} {177207} (\bibinfo {year}
  {2011})}\BibitemShut {NoStop}%
\bibitem [{\citenamefont {Schlickeiser}\ \emph {et~al.}(2014)\citenamefont
  {Schlickeiser}, \citenamefont {Ritzmann}, \citenamefont {Hinzke},\ and\
  \citenamefont {Nowak}}]{schlickeiser2014role}%
  \BibitemOpen
  \bibfield  {author} {\bibinfo {author} {\bibfnamefont {F.}~\bibnamefont
  {Schlickeiser}}, \bibinfo {author} {\bibfnamefont {U.}~\bibnamefont
  {Ritzmann}}, \bibinfo {author} {\bibfnamefont {D.}~\bibnamefont {Hinzke}},\
  and\ \bibinfo {author} {\bibfnamefont {U.}~\bibnamefont {Nowak}},\ }\bibinfo
  {title} {Role of {{Entropy}} in {{Domain Wall Motion}} in {{Thermal
  Gradients}}},\ \href
  {https://link.aps.org/doi/10.1103/PhysRevLett.113.097201} {\bibfield
  {journal} {\bibinfo  {journal} {Phys. Rev. Lett.}\ }\textbf {\bibinfo
  {volume} {113}},\ \bibinfo {pages} {097201} (\bibinfo {year}
  {2014})}\BibitemShut {NoStop}%
\bibitem [{\citenamefont {Wang}\ \emph {et~al.}(2016)\citenamefont {Wang},
  \citenamefont {Chotorlishvili}, \citenamefont {Guo}, \citenamefont {Sukhov},
  \citenamefont {Dugaev}, \citenamefont {Barna{\'s}},\ and\ \citenamefont
  {Berakdar}}]{wang2016thermally}%
  \BibitemOpen
  \bibfield  {author} {\bibinfo {author} {\bibfnamefont {X.-G.}\ \bibnamefont
  {Wang}}, \bibinfo {author} {\bibfnamefont {L.}~\bibnamefont
  {Chotorlishvili}}, \bibinfo {author} {\bibfnamefont {G.-H.}\ \bibnamefont
  {Guo}}, \bibinfo {author} {\bibfnamefont {A.}~\bibnamefont {Sukhov}},
  \bibinfo {author} {\bibfnamefont {V.}~\bibnamefont {Dugaev}}, \bibinfo
  {author} {\bibfnamefont {J.}~\bibnamefont {Barna{\'s}}},\ and\ \bibinfo
  {author} {\bibfnamefont {J.}~\bibnamefont {Berakdar}},\ }\bibinfo {title}
  {Thermally Induced Magnonic Spin Current, Thermomagnonic Torques, and
  Domain-Wall Dynamics in the Presence of {{Dzyaloshinskii-Moriya}}
  Interaction},\ \href {https://link.aps.org/doi/10.1103/PhysRevB.94.104410}
  {\bibfield  {journal} {\bibinfo  {journal} {Phys. Rev. B}\ }\textbf {\bibinfo
  {volume} {94}},\ \bibinfo {pages} {104410} (\bibinfo {year}
  {2016})}\BibitemShut {NoStop}%
\bibitem [{\citenamefont {Sukhov}\ \emph {et~al.}(2016)\citenamefont {Sukhov},
  \citenamefont {Chotorlishvili}, \citenamefont {Ernst}, \citenamefont
  {Zubizarreta}, \citenamefont {Ostanin}, \citenamefont {Mertig}, \citenamefont
  {Gross},\ and\ \citenamefont {Berakdar}}]{sukhov2016swift}%
  \BibitemOpen
  \bibfield  {author} {\bibinfo {author} {\bibfnamefont {A.}~\bibnamefont
  {Sukhov}}, \bibinfo {author} {\bibfnamefont {L.}~\bibnamefont
  {Chotorlishvili}}, \bibinfo {author} {\bibfnamefont {A.}~\bibnamefont
  {Ernst}}, \bibinfo {author} {\bibfnamefont {X.}~\bibnamefont {Zubizarreta}},
  \bibinfo {author} {\bibfnamefont {S.}~\bibnamefont {Ostanin}}, \bibinfo
  {author} {\bibfnamefont {I.}~\bibnamefont {Mertig}}, \bibinfo {author}
  {\bibfnamefont {E.~K.~U.}\ \bibnamefont {Gross}},\ and\ \bibinfo {author}
  {\bibfnamefont {J.}~\bibnamefont {Berakdar}},\ }\bibinfo {title} {Swift
  Thermal Steering of Domain Walls in Ferromagnetic {{MnBi}} Stripes},\ \href
  {http://www.nature.com/articles/srep24411} {\bibfield  {journal} {\bibinfo
  {journal} {Sci. Rep.}\ }\textbf {\bibinfo {volume} {6}},\ \bibinfo {pages}
  {24411} (\bibinfo {year} {2016})}\BibitemShut {NoStop}%
\bibitem [{\citenamefont {Donges}\ \emph {et~al.}(2020)\citenamefont {Donges},
  \citenamefont {Grimm}, \citenamefont {Jakobs}, \citenamefont {Selzer},
  \citenamefont {Ritzmann}, \citenamefont {Atxitia},\ and\ \citenamefont
  {Nowak}}]{donges2020unveiling}%
  \BibitemOpen
  \bibfield  {author} {\bibinfo {author} {\bibfnamefont {A.}~\bibnamefont
  {Donges}}, \bibinfo {author} {\bibfnamefont {N.}~\bibnamefont {Grimm}},
  \bibinfo {author} {\bibfnamefont {F.}~\bibnamefont {Jakobs}}, \bibinfo
  {author} {\bibfnamefont {S.}~\bibnamefont {Selzer}}, \bibinfo {author}
  {\bibfnamefont {U.}~\bibnamefont {Ritzmann}}, \bibinfo {author}
  {\bibfnamefont {U.}~\bibnamefont {Atxitia}},\ and\ \bibinfo {author}
  {\bibfnamefont {U.}~\bibnamefont {Nowak}},\ }\bibinfo {title} {Unveiling
  Domain Wall Dynamics of Ferrimagnets in Thermal Magnon Currents:
  {{Competition}} of Angular Momentum Transfer and Entropic Torque},\ \href
  {https://link.aps.org/doi/10.1103/PhysRevResearch.2.013293} {\bibfield
  {journal} {\bibinfo  {journal} {Phys. Rev. Research}\ }\textbf {\bibinfo
  {volume} {2}},\ \bibinfo {pages} {013293} (\bibinfo {year}
  {2020})}\BibitemShut {NoStop}%
\bibitem [{\citenamefont {Hinzke}\ and\ \citenamefont
  {Nowak}(2011)}]{hinzke2011domain}%
  \BibitemOpen
  \bibfield  {author} {\bibinfo {author} {\bibfnamefont {D.}~\bibnamefont
  {Hinzke}}\ and\ \bibinfo {author} {\bibfnamefont {U.}~\bibnamefont {Nowak}},\
  }\bibinfo {title} {Domain {{Wall Motion}} by the {{Magnonic Spin Seebeck
  Effect}}},\ \href {https://link.aps.org/doi/10.1103/PhysRevLett.107.027205}
  {\bibfield  {journal} {\bibinfo  {journal} {Phys. Rev. Lett.}\ }\textbf
  {\bibinfo {volume} {107}},\ \bibinfo {pages} {027205} (\bibinfo {year}
  {2011})}\BibitemShut {NoStop}%
\bibitem [{\citenamefont {Kim}\ and\ \citenamefont
  {Tserkovnyak}(2015)}]{kim2015landaulifshitz}%
  \BibitemOpen
  \bibfield  {author} {\bibinfo {author} {\bibfnamefont {S.~K.}\ \bibnamefont
  {Kim}}\ and\ \bibinfo {author} {\bibfnamefont {Y.}~\bibnamefont
  {Tserkovnyak}},\ }\bibinfo {title} {Landau-{{Lifshitz}} Theory of
  Thermomagnonic Torque},\ \href
  {https://link.aps.org/doi/10.1103/PhysRevB.92.020410} {\bibfield  {journal}
  {\bibinfo  {journal} {Phys. Rev. B}\ }\textbf {\bibinfo {volume} {92}},\
  \bibinfo {pages} {020410} (\bibinfo {year} {2015})}\BibitemShut {NoStop}%
\bibitem [{\citenamefont {Selzer}\ \emph {et~al.}(2016)\citenamefont {Selzer},
  \citenamefont {Atxitia}, \citenamefont {Ritzmann}, \citenamefont {Hinzke},\
  and\ \citenamefont {Nowak}}]{selzer2016inertiafree}%
  \BibitemOpen
  \bibfield  {author} {\bibinfo {author} {\bibfnamefont {S.}~\bibnamefont
  {Selzer}}, \bibinfo {author} {\bibfnamefont {U.}~\bibnamefont {Atxitia}},
  \bibinfo {author} {\bibfnamefont {U.}~\bibnamefont {Ritzmann}}, \bibinfo
  {author} {\bibfnamefont {D.}~\bibnamefont {Hinzke}},\ and\ \bibinfo {author}
  {\bibfnamefont {U.}~\bibnamefont {Nowak}},\ }\bibinfo {title} {Inertia-{{Free
  Thermally Driven Domain-Wall Motion}} in {{Antiferromagnets}}},\ \href
  {https://link.aps.org/doi/10.1103/PhysRevLett.117.107201} {\bibfield
  {journal} {\bibinfo  {journal} {Phys. Rev. Lett.}\ }\textbf {\bibinfo
  {volume} {117}},\ \bibinfo {pages} {107201} (\bibinfo {year}
  {2016})}\BibitemShut {NoStop}%
\bibitem [{\citenamefont {Wang}\ and\ \citenamefont
  {Wang}(2014)}]{wang2014thermodynamic}%
  \BibitemOpen
  \bibfield  {author} {\bibinfo {author} {\bibfnamefont {X.~S.}\ \bibnamefont
  {Wang}}\ and\ \bibinfo {author} {\bibfnamefont {X.~R.}\ \bibnamefont
  {Wang}},\ }\bibinfo {title} {Thermodynamic Theory for Thermal-Gradient-Driven
  Domain-Wall Motion},\ \href
  {https://link.aps.org/doi/10.1103/PhysRevB.90.014414} {\bibfield  {journal}
  {\bibinfo  {journal} {Phys. Rev. B}\ }\textbf {\bibinfo {volume} {90}},\
  \bibinfo {pages} {014414} (\bibinfo {year} {2014})}\BibitemShut {NoStop}%
\bibitem [{\citenamefont {Yan}\ \emph {et~al.}(2015)\citenamefont {Yan},
  \citenamefont {Cao},\ and\ \citenamefont {Sinova}}]{yan2015thermodynamic}%
  \BibitemOpen
  \bibfield  {author} {\bibinfo {author} {\bibfnamefont {P.}~\bibnamefont
  {Yan}}, \bibinfo {author} {\bibfnamefont {Y.}~\bibnamefont {Cao}},\ and\
  \bibinfo {author} {\bibfnamefont {J.}~\bibnamefont {Sinova}},\ }\bibinfo
  {title} {Thermodynamic Magnon Recoil for Domain Wall Motion},\ \href
  {https://link.aps.org/doi/10.1103/PhysRevB.92.100408} {\bibfield  {journal}
  {\bibinfo  {journal} {Phys. Rev. B}\ }\textbf {\bibinfo {volume} {92}},\
  \bibinfo {pages} {100408} (\bibinfo {year} {2015})}\BibitemShut {NoStop}%
\bibitem [{\citenamefont {Wang}\ \emph {et~al.}(2021)\citenamefont {Wang},
  \citenamefont {Shimada}, \citenamefont {Wang}, \citenamefont {Kitamura},\
  and\ \citenamefont {Hirakata}}]{wang2021rectilinear}%
  \BibitemOpen
  \bibfield  {author} {\bibinfo {author} {\bibfnamefont {Y.}~\bibnamefont
  {Wang}}, \bibinfo {author} {\bibfnamefont {T.}~\bibnamefont {Shimada}},
  \bibinfo {author} {\bibfnamefont {J.}~\bibnamefont {Wang}}, \bibinfo {author}
  {\bibfnamefont {T.}~\bibnamefont {Kitamura}},\ and\ \bibinfo {author}
  {\bibfnamefont {H.}~\bibnamefont {Hirakata}},\ }\bibinfo {title} {The
  Rectilinear Motion of the Individual Asymmetrical Skyrmion Driven by
  Temperature Gradients},\ \href
  {https://linkinghub.elsevier.com/retrieve/pii/S135964542100762X} {\bibfield
  {journal} {\bibinfo  {journal} {Acta Mater.}\ ,\ \bibinfo {pages} {117383}}
  (\bibinfo {year} {2021})}\BibitemShut {NoStop}%
\bibitem [{\citenamefont {Gorshkov}\ \emph {et~al.}(2022)\citenamefont
  {Gorshkov}, \citenamefont {Gorev}, \citenamefont {Sapozhnikov},\ and\
  \citenamefont {Udalov}}]{gorshkov2022dmigradientdriven}%
  \BibitemOpen
  \bibfield  {author} {\bibinfo {author} {\bibfnamefont {I.~O.}\ \bibnamefont
  {Gorshkov}}, \bibinfo {author} {\bibfnamefont {R.~V.}\ \bibnamefont {Gorev}},
  \bibinfo {author} {\bibfnamefont {M.~V.}\ \bibnamefont {Sapozhnikov}},\ and\
  \bibinfo {author} {\bibfnamefont {O.~G.}\ \bibnamefont {Udalov}},\ }\bibinfo
  {title} {{{DMI-Gradient-Driven Skyrmion Motion}}},\ \href
  {https://pubs.acs.org/doi/pdf/10.1021/acsaelm.2c00404} {\bibfield  {journal}
  {\bibinfo  {journal} {ACS Appl. Electron. Mater.}\ ,\ \bibinfo {pages}
  {acsaelm.2c00404}} (\bibinfo {year} {2022})}\BibitemShut {NoStop}%
\bibitem [{\citenamefont {Yershov}\ \emph {et~al.}(2020)\citenamefont
  {Yershov}, \citenamefont {Kravchuk}, \citenamefont {Sheka}, \citenamefont
  {{van den Brink}},\ and\ \citenamefont {Saxena}}]{yershov2020domain}%
  \BibitemOpen
  \bibfield  {author} {\bibinfo {author} {\bibfnamefont {K.~V.}\ \bibnamefont
  {Yershov}}, \bibinfo {author} {\bibfnamefont {V.~P.}\ \bibnamefont
  {Kravchuk}}, \bibinfo {author} {\bibfnamefont {D.~D.}\ \bibnamefont {Sheka}},
  \bibinfo {author} {\bibfnamefont {J.}~\bibnamefont {{van den Brink}}},\ and\
  \bibinfo {author} {\bibfnamefont {A.}~\bibnamefont {Saxena}},\ }\bibinfo
  {title} {Domain Wall Diode Based on Functionally Graded
  {{Dzyaloshinskii}}\textendash{{Moriya}} Interaction},\ \href
  {http://aip.scitation.org/doi/10.1063/5.0010107} {\bibfield  {journal}
  {\bibinfo  {journal} {Appl. Phys. Lett.}\ }\textbf {\bibinfo {volume}
  {116}},\ \bibinfo {pages} {222406} (\bibinfo {year} {2020})}\BibitemShut
  {NoStop}%
\bibitem [{\citenamefont {Shen}\ \emph
  {et~al.}(2018{\natexlab{a}})\citenamefont {Shen}, \citenamefont {Xia},
  \citenamefont {Zhao}, \citenamefont {Zhang}, \citenamefont {Ezawa},
  \citenamefont {Tretiakov}, \citenamefont {Liu},\ and\ \citenamefont
  {Zhou}}]{shen2018dynamics}%
  \BibitemOpen
  \bibfield  {author} {\bibinfo {author} {\bibfnamefont {L.}~\bibnamefont
  {Shen}}, \bibinfo {author} {\bibfnamefont {J.}~\bibnamefont {Xia}}, \bibinfo
  {author} {\bibfnamefont {G.}~\bibnamefont {Zhao}}, \bibinfo {author}
  {\bibfnamefont {X.}~\bibnamefont {Zhang}}, \bibinfo {author} {\bibfnamefont
  {M.}~\bibnamefont {Ezawa}}, \bibinfo {author} {\bibfnamefont {O.~A.}\
  \bibnamefont {Tretiakov}}, \bibinfo {author} {\bibfnamefont {X.}~\bibnamefont
  {Liu}},\ and\ \bibinfo {author} {\bibfnamefont {Y.}~\bibnamefont {Zhou}},\
  }\bibinfo {title} {Dynamics of the Antiferromagnetic Skyrmion Induced by a
  Magnetic Anisotropy Gradient},\ \href
  {https://link.aps.org/doi/10.1103/PhysRevB.98.134448} {\bibfield  {journal}
  {\bibinfo  {journal} {Phys. Rev. B}\ }\textbf {\bibinfo {volume} {98}},\
  \bibinfo {pages} {134448} (\bibinfo {year} {2018}{\natexlab{a}})}\BibitemShut
  {NoStop}%
\bibitem [{\citenamefont {Tomasello}\ \emph {et~al.}(2018)\citenamefont
  {Tomasello}, \citenamefont {Komineas}, \citenamefont {Siracusano},
  \citenamefont {Carpentieri},\ and\ \citenamefont
  {Finocchio}}]{tomasello2018chiral}%
  \BibitemOpen
  \bibfield  {author} {\bibinfo {author} {\bibfnamefont {R.}~\bibnamefont
  {Tomasello}}, \bibinfo {author} {\bibfnamefont {S.}~\bibnamefont {Komineas}},
  \bibinfo {author} {\bibfnamefont {G.}~\bibnamefont {Siracusano}}, \bibinfo
  {author} {\bibfnamefont {M.}~\bibnamefont {Carpentieri}},\ and\ \bibinfo
  {author} {\bibfnamefont {G.}~\bibnamefont {Finocchio}},\ }\bibinfo {title}
  {Chiral Skyrmions in an Anisotropy Gradient},\ \href
  {https://link.aps.org/article/10.1103/PhysRevB.98.024421} {\bibfield
  {journal} {\bibinfo  {journal} {Phys. Rev. B}\ }\textbf {\bibinfo {volume}
  {98}},\ \bibinfo {pages} {024421} (\bibinfo {year} {2018})}\BibitemShut
  {NoStop}%
\bibitem [{\citenamefont {Moretti}\ \emph {et~al.}(2017)\citenamefont
  {Moretti}, \citenamefont {Raposo}, \citenamefont {Martinez},\ and\
  \citenamefont {{Lopez-Diaz}}}]{moretti2017domain}%
  \BibitemOpen
  \bibfield  {author} {\bibinfo {author} {\bibfnamefont {S.}~\bibnamefont
  {Moretti}}, \bibinfo {author} {\bibfnamefont {V.}~\bibnamefont {Raposo}},
  \bibinfo {author} {\bibfnamefont {E.}~\bibnamefont {Martinez}},\ and\
  \bibinfo {author} {\bibfnamefont {L.}~\bibnamefont {{Lopez-Diaz}}},\
  }\bibinfo {title} {Domain Wall Motion by Localized Temperature Gradients},\
  \href {https://link.aps.org/doi/10.1103/PhysRevB.95.064419} {\bibfield
  {journal} {\bibinfo  {journal} {Phys. Rev. B}\ }\textbf {\bibinfo {volume}
  {95}},\ \bibinfo {pages} {064419} (\bibinfo {year} {2017})}\BibitemShut
  {NoStop}%
\bibitem [{\citenamefont {Kim}\ \emph {et~al.}(2019{\natexlab{b}})\citenamefont
  {Kim}, \citenamefont {Nakata}, \citenamefont {Loss},\ and\ \citenamefont
  {Tserkovnyak}}]{kim2019tunable}%
  \BibitemOpen
  \bibfield  {author} {\bibinfo {author} {\bibfnamefont {S.~K.}\ \bibnamefont
  {Kim}}, \bibinfo {author} {\bibfnamefont {K.}~\bibnamefont {Nakata}},
  \bibinfo {author} {\bibfnamefont {D.}~\bibnamefont {Loss}},\ and\ \bibinfo
  {author} {\bibfnamefont {Y.}~\bibnamefont {Tserkovnyak}},\ }\bibinfo {title}
  {Tunable {{Magnonic Thermal Hall Effect}} in {{Skyrmion Crystal Phases}} of
  {{Ferrimagnets}}},\ \href
  {https://link.aps.org/doi/10.1103/PhysRevLett.122.057204} {\bibfield
  {journal} {\bibinfo  {journal} {Phys. Rev. Lett.}\ }\textbf {\bibinfo
  {volume} {122}},\ \bibinfo {pages} {057204} (\bibinfo {year}
  {2019}{\natexlab{b}})}\BibitemShut {NoStop}%
\bibitem [{\citenamefont {Kovalev}(2014)}]{kovalev2014skyrmionic}%
  \BibitemOpen
  \bibfield  {author} {\bibinfo {author} {\bibfnamefont {A.~A.}\ \bibnamefont
  {Kovalev}},\ }\bibinfo {title} {Skyrmionic Spin {{Seebeck}} Effect via
  Dissipative Thermomagnonic Torques},\ \href
  {https://link.aps.org/doi/10.1103/PhysRevB.89.241101} {\bibfield  {journal}
  {\bibinfo  {journal} {Phys. Rev. B}\ }\textbf {\bibinfo {volume} {89}},\
  \bibinfo {pages} {241101} (\bibinfo {year} {2014})}\BibitemShut {NoStop}%
\bibitem [{\citenamefont {Bayer}\ \emph {et~al.}(2005)\citenamefont {Bayer},
  \citenamefont {Schultheiss}, \citenamefont {Hillebrands},\ and\ \citenamefont
  {Stamps}}]{bayer2005phase}%
  \BibitemOpen
  \bibfield  {author} {\bibinfo {author} {\bibfnamefont {C.}~\bibnamefont
  {Bayer}}, \bibinfo {author} {\bibfnamefont {H.}~\bibnamefont {Schultheiss}},
  \bibinfo {author} {\bibfnamefont {B.}~\bibnamefont {Hillebrands}},\ and\
  \bibinfo {author} {\bibfnamefont {R.}~\bibnamefont {Stamps}},\ }\bibinfo
  {title} {Phase Shift of Spin Waves Traveling through a 180/Spl Deg/
  {{Bloch-domain}} Wall},\ \href {http://ieeexplore.ieee.org/document/1519218/}
  {\bibfield  {journal} {\bibinfo  {journal} {IEEE Trans. Magn.}\ }\textbf
  {\bibinfo {volume} {41}},\ \bibinfo {pages} {3094} (\bibinfo {year}
  {2005})}\BibitemShut {NoStop}%
\bibitem [{\citenamefont {Hertel}\ \emph {et~al.}(2004)\citenamefont {Hertel},
  \citenamefont {Wulfhekel},\ and\ \citenamefont
  {Kirschner}}]{hertel2004domainwall}%
  \BibitemOpen
  \bibfield  {author} {\bibinfo {author} {\bibfnamefont {R.}~\bibnamefont
  {Hertel}}, \bibinfo {author} {\bibfnamefont {W.}~\bibnamefont {Wulfhekel}},\
  and\ \bibinfo {author} {\bibfnamefont {J.}~\bibnamefont {Kirschner}},\
  }\bibinfo {title} {Domain-{{Wall Induced Phase Shifts}} in {{Spin Waves}}},\
  \href {https://link.aps.org/doi/10.1103/PhysRevLett.93.257202} {\bibfield
  {journal} {\bibinfo  {journal} {Phys. Rev. Lett.}\ }\textbf {\bibinfo
  {volume} {93}},\ \bibinfo {pages} {257202} (\bibinfo {year}
  {2004})}\BibitemShut {NoStop}%
\bibitem [{\citenamefont {Zhao}\ \emph {et~al.}(2020)\citenamefont {Zhao},
  \citenamefont {He}, \citenamefont {Cai},\ and\ \citenamefont
  {Li}}]{zhao2020spin}%
  \BibitemOpen
  \bibfield  {author} {\bibinfo {author} {\bibfnamefont {Z.-X.}\ \bibnamefont
  {Zhao}}, \bibinfo {author} {\bibfnamefont {P.-B.}\ \bibnamefont {He}},
  \bibinfo {author} {\bibfnamefont {M.-Q.}\ \bibnamefont {Cai}},\ and\ \bibinfo
  {author} {\bibfnamefont {Z.-D.}\ \bibnamefont {Li}},\ }\bibinfo {title} {Spin
  Waves and Transverse Domain Walls Driven by Spin Waves: {{Role}} of
  Damping},\ \href
  {https://iopscience.iop.org/article/10.1088/1674-1056/ab90e5} {\bibfield
  {journal} {\bibinfo  {journal} {Chin. Phys. B}\ }\textbf {\bibinfo {volume}
  {29}},\ \bibinfo {pages} {077502} (\bibinfo {year} {2020})}\BibitemShut
  {NoStop}%
\bibitem [{\citenamefont {Kiriushcheva}\ and\ \citenamefont
  {Kuzmin}(1998)}]{kiriushcheva1998scattering}%
  \BibitemOpen
  \bibfield  {author} {\bibinfo {author} {\bibfnamefont {N.}~\bibnamefont
  {Kiriushcheva}}\ and\ \bibinfo {author} {\bibfnamefont {S.}~\bibnamefont
  {Kuzmin}},\ }\bibinfo {title} {Scattering of a {{Gaussian}} Wave Packet by a
  Reflectionless Potential},\ \href
  {http://aapt.scitation.org/doi/10.1119/1.18985} {\bibfield  {journal}
  {\bibinfo  {journal} {Am. J. Phys.}\ }\textbf {\bibinfo {volume} {66}},\
  \bibinfo {pages} {867} (\bibinfo {year} {1998})}\BibitemShut {NoStop}%
\bibitem [{\citenamefont {Kim}\ \emph {et~al.}(2012)\citenamefont {Kim},
  \citenamefont {St{\"a}rk}, \citenamefont {Kl{\"a}ui}, \citenamefont {Yoon},
  \citenamefont {You}, \citenamefont {{Lopez-Diaz}},\ and\ \citenamefont
  {Martinez}}]{kim2012interaction}%
  \BibitemOpen
  \bibfield  {author} {\bibinfo {author} {\bibfnamefont {J.-S.}\ \bibnamefont
  {Kim}}, \bibinfo {author} {\bibfnamefont {M.}~\bibnamefont {St{\"a}rk}},
  \bibinfo {author} {\bibfnamefont {M.}~\bibnamefont {Kl{\"a}ui}}, \bibinfo
  {author} {\bibfnamefont {J.}~\bibnamefont {Yoon}}, \bibinfo {author}
  {\bibfnamefont {C.-Y.}\ \bibnamefont {You}}, \bibinfo {author} {\bibfnamefont
  {L.}~\bibnamefont {{Lopez-Diaz}}},\ and\ \bibinfo {author} {\bibfnamefont
  {E.}~\bibnamefont {Martinez}},\ }\bibinfo {title} {Interaction between
  Propagating Spin Waves and Domain Walls on a Ferromagnetic Nanowire},\ \href
  {https://link.aps.org/doi/10.1103/PhysRevB.85.174428} {\bibfield  {journal}
  {\bibinfo  {journal} {Phys. Rev. B}\ }\textbf {\bibinfo {volume} {85}},\
  \bibinfo {pages} {174428} (\bibinfo {year} {2012})}\BibitemShut {NoStop}%
\bibitem [{\citenamefont {Ross}\ \emph {et~al.}(2020)\citenamefont {Ross},
  \citenamefont {Lebrun}, \citenamefont {Gomonay}, \citenamefont {Grave},
  \citenamefont {Kay}, \citenamefont {Baldrati}, \citenamefont {Becker},
  \citenamefont {Qaiumzadeh}, \citenamefont {Ulloa}, \citenamefont {Jakob},
  \citenamefont {Kronast}, \citenamefont {Sinova}, \citenamefont {Duine},
  \citenamefont {Brataas}, \citenamefont {Rothschild},\ and\ \citenamefont
  {Kl{\"a}ui}}]{ross2020propagation}%
  \BibitemOpen
  \bibfield  {author} {\bibinfo {author} {\bibfnamefont {A.}~\bibnamefont
  {Ross}}, \bibinfo {author} {\bibfnamefont {R.}~\bibnamefont {Lebrun}},
  \bibinfo {author} {\bibfnamefont {O.}~\bibnamefont {Gomonay}}, \bibinfo
  {author} {\bibfnamefont {D.~A.}\ \bibnamefont {Grave}}, \bibinfo {author}
  {\bibfnamefont {A.}~\bibnamefont {Kay}}, \bibinfo {author} {\bibfnamefont
  {L.}~\bibnamefont {Baldrati}}, \bibinfo {author} {\bibfnamefont
  {S.}~\bibnamefont {Becker}}, \bibinfo {author} {\bibfnamefont
  {A.}~\bibnamefont {Qaiumzadeh}}, \bibinfo {author} {\bibfnamefont
  {C.}~\bibnamefont {Ulloa}}, \bibinfo {author} {\bibfnamefont
  {G.}~\bibnamefont {Jakob}}, \bibinfo {author} {\bibfnamefont
  {F.}~\bibnamefont {Kronast}}, \bibinfo {author} {\bibfnamefont
  {J.}~\bibnamefont {Sinova}}, \bibinfo {author} {\bibfnamefont
  {R.}~\bibnamefont {Duine}}, \bibinfo {author} {\bibfnamefont
  {A.}~\bibnamefont {Brataas}}, \bibinfo {author} {\bibfnamefont
  {A.}~\bibnamefont {Rothschild}},\ and\ \bibinfo {author} {\bibfnamefont
  {M.}~\bibnamefont {Kl{\"a}ui}},\ }\bibinfo {title} {Propagation {{Length}} of
  {{Antiferromagnetic Magnons Governed}} by {{Domain Configurations}}},\ \href
  {https://pubs.acs.org/doi/10.1021/acs.nanolett.9b03837} {\bibfield  {journal}
  {\bibinfo  {journal} {Nano Lett.}\ }\textbf {\bibinfo {volume} {20}},\
  \bibinfo {pages} {306} (\bibinfo {year} {2020})}\BibitemShut {NoStop}%
\bibitem [{\citenamefont {Qaiumzadeh}\ \emph {et~al.}(2018)\citenamefont
  {Qaiumzadeh}, \citenamefont {Kristiansen},\ and\ \citenamefont
  {Brataas}}]{qaiumzadeh2018controlling}%
  \BibitemOpen
  \bibfield  {author} {\bibinfo {author} {\bibfnamefont {A.}~\bibnamefont
  {Qaiumzadeh}}, \bibinfo {author} {\bibfnamefont {L.~A.}\ \bibnamefont
  {Kristiansen}},\ and\ \bibinfo {author} {\bibfnamefont {A.}~\bibnamefont
  {Brataas}},\ }\bibinfo {title} {Controlling Chiral Domain Walls in
  Antiferromagnets Using Spin-Wave Helicity},\ \href
  {https://link.aps.org/doi/10.1103/PhysRevB.97.020402} {\bibfield  {journal}
  {\bibinfo  {journal} {Phys. Rev. B}\ }\textbf {\bibinfo {volume} {97}},\
  \bibinfo {pages} {020402} (\bibinfo {year} {2018})}\BibitemShut {NoStop}%
\bibitem [{\citenamefont {H{\"a}m{\"a}l{\"a}inen}\ \emph
  {et~al.}(2018)\citenamefont {H{\"a}m{\"a}l{\"a}inen}, \citenamefont {Madami},
  \citenamefont {Qin}, \citenamefont {Gubbiotti},\ and\ \citenamefont {{van
  Dijken}}}]{hamalainen2018control}%
  \BibitemOpen
  \bibfield  {author} {\bibinfo {author} {\bibfnamefont {S.~J.}\ \bibnamefont
  {H{\"a}m{\"a}l{\"a}inen}}, \bibinfo {author} {\bibfnamefont {M.}~\bibnamefont
  {Madami}}, \bibinfo {author} {\bibfnamefont {H.}~\bibnamefont {Qin}},
  \bibinfo {author} {\bibfnamefont {G.}~\bibnamefont {Gubbiotti}},\ and\
  \bibinfo {author} {\bibfnamefont {S.}~\bibnamefont {{van Dijken}}},\
  }\bibinfo {title} {Control of Spin-Wave Transmission by a Programmable Domain
  Wall},\ \href {http://www.nature.com/articles/s41467-018-07372-x} {\bibfield
  {journal} {\bibinfo  {journal} {Nat. Commun.}\ }\textbf {\bibinfo {volume}
  {9}},\ \bibinfo {pages} {4853} (\bibinfo {year} {2018})}\BibitemShut
  {NoStop}%
\bibitem [{\citenamefont {Chang}\ \emph {et~al.}(2018)\citenamefont {Chang},
  \citenamefont {Liu}, \citenamefont {Kao}, \citenamefont {Tsai}, \citenamefont
  {Liang},\ and\ \citenamefont {Lee}}]{chang2018ferromagnetic}%
  \BibitemOpen
  \bibfield  {author} {\bibinfo {author} {\bibfnamefont {L.-J.}\ \bibnamefont
  {Chang}}, \bibinfo {author} {\bibfnamefont {Y.-F.}\ \bibnamefont {Liu}},
  \bibinfo {author} {\bibfnamefont {M.-Y.}\ \bibnamefont {Kao}}, \bibinfo
  {author} {\bibfnamefont {L.-Z.}\ \bibnamefont {Tsai}}, \bibinfo {author}
  {\bibfnamefont {J.-Z.}\ \bibnamefont {Liang}},\ and\ \bibinfo {author}
  {\bibfnamefont {S.-F.}\ \bibnamefont {Lee}},\ }\bibinfo {title}
  {Ferromagnetic Domain Walls as Spin Wave Filters and the Interplay between
  Domain Walls and Spin Waves},\ \href
  {http://www.nature.com/articles/s41598-018-22272-2} {\bibfield  {journal}
  {\bibinfo  {journal} {Sci. Rep.}\ }\textbf {\bibinfo {volume} {8}},\ \bibinfo
  {pages} {3910} (\bibinfo {year} {2018})}\BibitemShut {NoStop}%
\bibitem [{\citenamefont {Lan}\ \emph {et~al.}(2017)\citenamefont {Lan},
  \citenamefont {Yu},\ and\ \citenamefont {Xiao}}]{lan2017antiferromagnetic}%
  \BibitemOpen
  \bibfield  {author} {\bibinfo {author} {\bibfnamefont {J.}~\bibnamefont
  {Lan}}, \bibinfo {author} {\bibfnamefont {W.}~\bibnamefont {Yu}},\ and\
  \bibinfo {author} {\bibfnamefont {J.}~\bibnamefont {Xiao}},\ }\bibinfo
  {title} {Antiferromagnetic Domain Wall as Spin Wave Polarizer and Retarder},\
  \href {http://www.nature.com/articles/s41467-017-00265-5} {\bibfield
  {journal} {\bibinfo  {journal} {Nat. Commun.}\ }\textbf {\bibinfo {volume}
  {8}},\ \bibinfo {pages} {178} (\bibinfo {year} {2017})}\BibitemShut {NoStop}%
\bibitem [{\citenamefont {Borys}\ \emph {et~al.}(2016)\citenamefont {Borys},
  \citenamefont {{Garcia-Sanchez}}, \citenamefont {Kim},\ and\ \citenamefont
  {Stamps}}]{borys2016spinwave}%
  \BibitemOpen
  \bibfield  {author} {\bibinfo {author} {\bibfnamefont {P.}~\bibnamefont
  {Borys}}, \bibinfo {author} {\bibfnamefont {F.}~\bibnamefont
  {{Garcia-Sanchez}}}, \bibinfo {author} {\bibfnamefont {J.-V.}\ \bibnamefont
  {Kim}},\ and\ \bibinfo {author} {\bibfnamefont {R.~L.}\ \bibnamefont
  {Stamps}},\ }\bibinfo {title} {Spin-{{Wave Eigenmodes}} of {{Dzyaloshinskii
  Domain Walls}}},\ \href
  {https://onlinelibrary.wiley.com/doi/10.1002/aelm.201500202} {\bibfield
  {journal} {\bibinfo  {journal} {Adv. Electron. Mater.}\ }\textbf {\bibinfo
  {volume} {2}},\ \bibinfo {pages} {1500202} (\bibinfo {year}
  {2016})}\BibitemShut {NoStop}%
\bibitem [{\citenamefont {Wang}\ \emph {et~al.}(2015)\citenamefont {Wang},
  \citenamefont {Albert}, \citenamefont {Beg}, \citenamefont {Bisotti},
  \citenamefont {Chernyshenko}, \citenamefont {{Cort{\'e}s-Ortu{\~n}o}},
  \citenamefont {Hawke},\ and\ \citenamefont {Fangohr}}]{wang2015magnondriven}%
  \BibitemOpen
  \bibfield  {author} {\bibinfo {author} {\bibfnamefont {W.}~\bibnamefont
  {Wang}}, \bibinfo {author} {\bibfnamefont {M.}~\bibnamefont {Albert}},
  \bibinfo {author} {\bibfnamefont {M.}~\bibnamefont {Beg}}, \bibinfo {author}
  {\bibfnamefont {M.-A.}\ \bibnamefont {Bisotti}}, \bibinfo {author}
  {\bibfnamefont {D.}~\bibnamefont {Chernyshenko}}, \bibinfo {author}
  {\bibfnamefont {D.}~\bibnamefont {{Cort{\'e}s-Ortu{\~n}o}}}, \bibinfo
  {author} {\bibfnamefont {I.}~\bibnamefont {Hawke}},\ and\ \bibinfo {author}
  {\bibfnamefont {H.}~\bibnamefont {Fangohr}},\ }\bibinfo {title}
  {Magnon-{{Driven Domain-Wall Motion}} with the {{Dzyaloshinskii-Moriya
  Interaction}}},\ \href
  {https://link.aps.org/doi/10.1103/PhysRevLett.114.087203} {\bibfield
  {journal} {\bibinfo  {journal} {Phys. Rev. Lett.}\ }\textbf {\bibinfo
  {volume} {114}},\ \bibinfo {pages} {087203} (\bibinfo {year}
  {2015})}\BibitemShut {NoStop}%
\bibitem [{\citenamefont {Schroeter}\ and\ \citenamefont
  {Garst}(2015)}]{schroeter2015scattering}%
  \BibitemOpen
  \bibfield  {author} {\bibinfo {author} {\bibfnamefont {S.}~\bibnamefont
  {Schroeter}}\ and\ \bibinfo {author} {\bibfnamefont {M.}~\bibnamefont
  {Garst}},\ }\bibinfo {title} {Scattering of High-Energy Magnons off a
  Magnetic Skyrmion},\ \href {http://aip.scitation.org/doi/10.1063/1.4932356}
  {\bibfield  {journal} {\bibinfo  {journal} {Low Temp. Phys.}\ }\textbf
  {\bibinfo {volume} {41}},\ \bibinfo {pages} {817} (\bibinfo {year}
  {2015})}\BibitemShut {NoStop}%
\bibitem [{\citenamefont {Pirro}\ \emph {et~al.}(2015)\citenamefont {Pirro},
  \citenamefont {Koyama}, \citenamefont {Br{\"a}cher}, \citenamefont
  {Sebastian}, \citenamefont {Leven},\ and\ \citenamefont
  {Hillebrands}}]{pirro2015experimental}%
  \BibitemOpen
  \bibfield  {author} {\bibinfo {author} {\bibfnamefont {P.}~\bibnamefont
  {Pirro}}, \bibinfo {author} {\bibfnamefont {T.}~\bibnamefont {Koyama}},
  \bibinfo {author} {\bibfnamefont {T.}~\bibnamefont {Br{\"a}cher}}, \bibinfo
  {author} {\bibfnamefont {T.}~\bibnamefont {Sebastian}}, \bibinfo {author}
  {\bibfnamefont {B.}~\bibnamefont {Leven}},\ and\ \bibinfo {author}
  {\bibfnamefont {B.}~\bibnamefont {Hillebrands}},\ }\bibinfo {title}
  {Experimental Observation of the Interaction of Propagating Spin Waves with
  {{N\'eel}} Domain Walls in a {{Landau}} Domain Structure},\ \href
  {http://aip.scitation.org/doi/10.1063/1.4922396} {\bibfield  {journal}
  {\bibinfo  {journal} {Appl. Phys. Lett.}\ }\textbf {\bibinfo {volume}
  {106}},\ \bibinfo {pages} {232405} (\bibinfo {year} {2015})}\BibitemShut
  {NoStop}%
\bibitem [{\citenamefont {Iwasaki}\ \emph {et~al.}(2014)\citenamefont
  {Iwasaki}, \citenamefont {Beekman},\ and\ \citenamefont
  {Nagaosa}}]{iwasaki2014theory}%
  \BibitemOpen
  \bibfield  {author} {\bibinfo {author} {\bibfnamefont {J.}~\bibnamefont
  {Iwasaki}}, \bibinfo {author} {\bibfnamefont {A.~J.}\ \bibnamefont
  {Beekman}},\ and\ \bibinfo {author} {\bibfnamefont {N.}~\bibnamefont
  {Nagaosa}},\ }\bibinfo {title} {Theory of Magnon-Skyrmion Scattering in
  Chiral Magnets},\ \href {https://link.aps.org/doi/10.1103/PhysRevB.89.064412}
  {\bibfield  {journal} {\bibinfo  {journal} {Phys. Rev. B}\ }\textbf {\bibinfo
  {volume} {89}},\ \bibinfo {pages} {064412} (\bibinfo {year}
  {2014})}\BibitemShut {NoStop}%
\bibitem [{\citenamefont {Hata}\ \emph {et~al.}(2014)\citenamefont {Hata},
  \citenamefont {Taniguchi}, \citenamefont {Lee}, \citenamefont {Moriyama},\
  and\ \citenamefont {Ono}}]{hata2014spinwaveinduced}%
  \BibitemOpen
  \bibfield  {author} {\bibinfo {author} {\bibfnamefont {H.}~\bibnamefont
  {Hata}}, \bibinfo {author} {\bibfnamefont {T.}~\bibnamefont {Taniguchi}},
  \bibinfo {author} {\bibfnamefont {H.-W.}\ \bibnamefont {Lee}}, \bibinfo
  {author} {\bibfnamefont {T.}~\bibnamefont {Moriyama}},\ and\ \bibinfo
  {author} {\bibfnamefont {T.}~\bibnamefont {Ono}},\ }\bibinfo {title}
  {Spin-Wave-Induced Domain Wall Motion in Perpendicularly Magnetized System},\
  \href {https://iopscience.iop.org/article/10.7567/APEX.7.033001} {\bibfield
  {journal} {\bibinfo  {journal} {Appl. Phys. Express}\ }\textbf {\bibinfo
  {volume} {7}},\ \bibinfo {pages} {033001} (\bibinfo {year}
  {2014})}\BibitemShut {NoStop}%
\bibitem [{\citenamefont {Bogdan}\ and\ \citenamefont
  {Charkina}(2014)}]{bogdan2014spin}%
  \BibitemOpen
  \bibfield  {author} {\bibinfo {author} {\bibfnamefont {M.~M.}\ \bibnamefont
  {Bogdan}}\ and\ \bibinfo {author} {\bibfnamefont {O.~V.}\ \bibnamefont
  {Charkina}},\ }\bibinfo {title} {Spin Waves in Easy-Axis Antiferromagnets
  with Precessing Domain Walls},\ \href
  {http://aip.scitation.org/doi/10.1063/1.4862465} {\bibfield  {journal}
  {\bibinfo  {journal} {Low Temp. Phys.}\ }\textbf {\bibinfo {volume} {40}},\
  \bibinfo {pages} {84} (\bibinfo {year} {2014})}\BibitemShut {NoStop}%
\bibitem [{\citenamefont {Wang}\ \emph
  {et~al.}(2013{\natexlab{a}})\citenamefont {Wang}, \citenamefont {Guo},
  \citenamefont {Zhang}, \citenamefont {Nie},\ and\ \citenamefont
  {Xia}}]{wang2013spinwave}%
  \BibitemOpen
  \bibfield  {author} {\bibinfo {author} {\bibfnamefont {X.-G.}\ \bibnamefont
  {Wang}}, \bibinfo {author} {\bibfnamefont {G.-H.}\ \bibnamefont {Guo}},
  \bibinfo {author} {\bibfnamefont {G.-F.}\ \bibnamefont {Zhang}}, \bibinfo
  {author} {\bibfnamefont {Y.-Z.}\ \bibnamefont {Nie}},\ and\ \bibinfo {author}
  {\bibfnamefont {Q.-L.}\ \bibnamefont {Xia}},\ }\bibinfo {title} {Spin-Wave
  Resonance Reflection and Spin-Wave Induced Domain Wall Displacement},\ \href
  {http://aip.scitation.org/doi/10.1063/1.4808298} {\bibfield  {journal}
  {\bibinfo  {journal} {J. Appl. Phys.}\ }\textbf {\bibinfo {volume} {113}},\
  \bibinfo {pages} {213904} (\bibinfo {year} {2013}{\natexlab{a}})}\BibitemShut
  {NoStop}%
\bibitem [{\citenamefont {Wang}\ \emph
  {et~al.}(2013{\natexlab{b}})\citenamefont {Wang}, \citenamefont {Guo},
  \citenamefont {Zhang}, \citenamefont {Nie},\ and\ \citenamefont
  {Xia}}]{wang2013analytical}%
  \BibitemOpen
  \bibfield  {author} {\bibinfo {author} {\bibfnamefont {X.-g.}\ \bibnamefont
  {Wang}}, \bibinfo {author} {\bibfnamefont {G.-h.}\ \bibnamefont {Guo}},
  \bibinfo {author} {\bibfnamefont {G.-f.}\ \bibnamefont {Zhang}}, \bibinfo
  {author} {\bibfnamefont {Y.-z.}\ \bibnamefont {Nie}},\ and\ \bibinfo {author}
  {\bibfnamefont {Q.-l.}\ \bibnamefont {Xia}},\ }\bibinfo {title} {An
  Analytical Approach to the Interaction of a Propagating Spin Wave and a
  {{Bloch}} Wall},\ \href {http://aip.scitation.org/doi/10.1063/1.4799285}
  {\bibfield  {journal} {\bibinfo  {journal} {Appl. Phys. Lett.}\ }\textbf
  {\bibinfo {volume} {102}},\ \bibinfo {pages} {132401} (\bibinfo {year}
  {2013}{\natexlab{b}})}\BibitemShut {NoStop}%
\bibitem [{\citenamefont {Wang}\ \emph
  {et~al.}(2013{\natexlab{c}})\citenamefont {Wang}, \citenamefont {Wang},\ and\
  \citenamefont {Guo}}]{wang2013magnonic}%
  \BibitemOpen
  \bibfield  {author} {\bibinfo {author} {\bibfnamefont {D.}~\bibnamefont
  {Wang}}, \bibinfo {author} {\bibfnamefont {X.-g.}\ \bibnamefont {Wang}},\
  and\ \bibinfo {author} {\bibfnamefont {G.-h.}\ \bibnamefont {Guo}},\
  }\bibinfo {title} {Magnonic Momentum Transfer Force on Domain Walls Confined
  in Space},\ \href
  {https://iopscience.iop.org/article/10.1209/0295-5075/101/27007} {\bibfield
  {journal} {\bibinfo  {journal} {Europhys. Lett.}\ }\textbf {\bibinfo {volume}
  {101}},\ \bibinfo {pages} {27007} (\bibinfo {year}
  {2013}{\natexlab{c}})}\BibitemShut {NoStop}%
\bibitem [{\citenamefont {Wang}\ \emph {et~al.}(2012)\citenamefont {Wang},
  \citenamefont {Guo}, \citenamefont {Nie}, \citenamefont {Zhang},\ and\
  \citenamefont {Li}}]{wang2012domain}%
  \BibitemOpen
  \bibfield  {author} {\bibinfo {author} {\bibfnamefont {X.-g.}\ \bibnamefont
  {Wang}}, \bibinfo {author} {\bibfnamefont {G.-h.}\ \bibnamefont {Guo}},
  \bibinfo {author} {\bibfnamefont {Y.-z.}\ \bibnamefont {Nie}}, \bibinfo
  {author} {\bibfnamefont {G.-f.}\ \bibnamefont {Zhang}},\ and\ \bibinfo
  {author} {\bibfnamefont {Z.-x.}\ \bibnamefont {Li}},\ }\bibinfo {title}
  {Domain Wall Motion Induced by the Magnonic Spin Current},\ \href
  {https://link.aps.org/doi/10.1103/PhysRevB.86.054445} {\bibfield  {journal}
  {\bibinfo  {journal} {Phys. Rev. B}\ }\textbf {\bibinfo {volume} {86}},\
  \bibinfo {pages} {054445} (\bibinfo {year} {2012})}\BibitemShut {NoStop}%
\bibitem [{\citenamefont {Macke}\ and\ \citenamefont
  {Goll}(2010)}]{macke2010transmission}%
  \BibitemOpen
  \bibfield  {author} {\bibinfo {author} {\bibfnamefont {S.}~\bibnamefont
  {Macke}}\ and\ \bibinfo {author} {\bibfnamefont {D.}~\bibnamefont {Goll}},\
  }\bibinfo {title} {Transmission and Reflection of Spin Waves in the Presence
  of {{N\'eel}} Walls},\ \href
  {https://iopscience.iop.org/article/10.1088/1742-6596/200/4/042015}
  {\bibfield  {journal} {\bibinfo  {journal} {J. Phys.: Conf. Ser.}\ }\textbf
  {\bibinfo {volume} {200}},\ \bibinfo {pages} {042015} (\bibinfo {year}
  {2010})}\BibitemShut {NoStop}%
\bibitem [{\citenamefont {Yuan}\ \emph {et~al.}(2006)\citenamefont {Yuan},
  \citenamefont {De~Raedt},\ and\ \citenamefont {Miyashita}}]{yuan2006quantum}%
  \BibitemOpen
  \bibfield  {author} {\bibinfo {author} {\bibfnamefont {S.}~\bibnamefont
  {Yuan}}, \bibinfo {author} {\bibfnamefont {H.}~\bibnamefont {De~Raedt}},\
  and\ \bibinfo {author} {\bibfnamefont {S.}~\bibnamefont {Miyashita}},\
  }\bibinfo {title} {Quantum {{Dynamics}} of {{Spin Wave Propagation}} through
  {{Domain Walls}}},\ \href
  {https://journals.jps.jp/doi/10.1143/JPSJ.75.084703} {\bibfield  {journal}
  {\bibinfo  {journal} {J. Phys. Soc. Jpn.}\ }\textbf {\bibinfo {volume}
  {75}},\ \bibinfo {pages} {084703} (\bibinfo {year} {2006})}\BibitemShut
  {NoStop}%
\bibitem [{\citenamefont {Liu}(1979)}]{liu1979spin}%
  \BibitemOpen
  \bibfield  {author} {\bibinfo {author} {\bibfnamefont {S.~H.}\ \bibnamefont
  {Liu}},\ }\bibinfo {title} {Spin Waves in Static Non-Periodic Magnetic
  Structures},\ \href
  {https://api.elsevier.com/content/article/PII:030488537990088X?httpAccept=text/xml}
  {\bibfield  {journal} {\bibinfo  {journal} {J. Magn. Magn. Mater.}\ }\textbf
  {\bibinfo {volume} {12}},\ \bibinfo {pages} {262} (\bibinfo {year}
  {1979})}\BibitemShut {NoStop}%
\bibitem [{\citenamefont {Mochizuki}\ \emph {et~al.}(2014)\citenamefont
  {Mochizuki}, \citenamefont {Yu}, \citenamefont {Seki}, \citenamefont
  {Kanazawa}, \citenamefont {Koshibae}, \citenamefont {Zang}, \citenamefont
  {Mostovoy}, \citenamefont {Tokura},\ and\ \citenamefont
  {Nagaosa}}]{mochizuki2014thermally}%
  \BibitemOpen
  \bibfield  {author} {\bibinfo {author} {\bibfnamefont {M.}~\bibnamefont
  {Mochizuki}}, \bibinfo {author} {\bibfnamefont {X.~Z.}\ \bibnamefont {Yu}},
  \bibinfo {author} {\bibfnamefont {S.}~\bibnamefont {Seki}}, \bibinfo {author}
  {\bibfnamefont {N.}~\bibnamefont {Kanazawa}}, \bibinfo {author}
  {\bibfnamefont {W.}~\bibnamefont {Koshibae}}, \bibinfo {author}
  {\bibfnamefont {J.}~\bibnamefont {Zang}}, \bibinfo {author} {\bibfnamefont
  {M.}~\bibnamefont {Mostovoy}}, \bibinfo {author} {\bibfnamefont
  {Y.}~\bibnamefont {Tokura}},\ and\ \bibinfo {author} {\bibfnamefont
  {N.}~\bibnamefont {Nagaosa}},\ }\bibinfo {title} {Thermally Driven Ratchet
  Motion of a Skyrmion Microcrystal and Topological Magnon {{Hall}} Effect},\
  \href {http://www.nature.com/articles/nmat3862.pdf} {\bibfield  {journal}
  {\bibinfo  {journal} {Nat. Mater.}\ }\textbf {\bibinfo {volume} {13}},\
  \bibinfo {pages} {241} (\bibinfo {year} {2014})}\BibitemShut {NoStop}%
\bibitem [{\citenamefont {Shen}\ \emph
  {et~al.}(2020{\natexlab{a}})\citenamefont {Shen}, \citenamefont
  {Tserkovnyak},\ and\ \citenamefont {Kim}}]{shen2020driving}%
  \BibitemOpen
  \bibfield  {author} {\bibinfo {author} {\bibfnamefont {P.}~\bibnamefont
  {Shen}}, \bibinfo {author} {\bibfnamefont {Y.}~\bibnamefont {Tserkovnyak}},\
  and\ \bibinfo {author} {\bibfnamefont {S.~K.}\ \bibnamefont {Kim}},\
  }\bibinfo {title} {Driving a Magnetized Domain Wall in an Antiferromagnet by
  Magnons},\ \href {http://aip.scitation.org/doi/am-pdf/10.1063/5.0006038}
  {\bibfield  {journal} {\bibinfo  {journal} {J. Appl. Phys.}\ }\textbf
  {\bibinfo {volume} {127}},\ \bibinfo {pages} {223905} (\bibinfo {year}
  {2020}{\natexlab{a}})}\BibitemShut {NoStop}%
\bibitem [{\citenamefont {Sch{\"u}tte}\ and\ \citenamefont
  {Garst}(2014)}]{schutte2014magnonskyrmion}%
  \BibitemOpen
  \bibfield  {author} {\bibinfo {author} {\bibfnamefont {C.}~\bibnamefont
  {Sch{\"u}tte}}\ and\ \bibinfo {author} {\bibfnamefont {M.}~\bibnamefont
  {Garst}},\ }\bibinfo {title} {Magnon-Skyrmion Scattering in Chiral Magnets},\
  \href {https://link.aps.org/doi/10.1103/PhysRevB.90.094423} {\bibfield
  {journal} {\bibinfo  {journal} {Phys. Rev. B}\ }\textbf {\bibinfo {volume}
  {90}},\ \bibinfo {pages} {094423} (\bibinfo {year} {2014})}\BibitemShut
  {NoStop}%
\bibitem [{\citenamefont {Laliena}\ \emph {et~al.}(2022)\citenamefont
  {Laliena}, \citenamefont {Athanasopoulos},\ and\ \citenamefont
  {Campo}}]{laliena2022scattering}%
  \BibitemOpen
  \bibfield  {author} {\bibinfo {author} {\bibfnamefont {V.}~\bibnamefont
  {Laliena}}, \bibinfo {author} {\bibfnamefont {A.}~\bibnamefont
  {Athanasopoulos}},\ and\ \bibinfo {author} {\bibfnamefont {J.}~\bibnamefont
  {Campo}},\ }\bibinfo {title} {Scattering of Spin Waves by a {{Bloch}} Domain
  Wall: Effect of the Dipolar Interaction},\ \href
  {http://arxiv.org/abs/2203.11140} {\bibfield  {journal} {\bibinfo  {journal}
  {arXiv:2203.11140}\ } (\bibinfo {year} {2022})}\BibitemShut {NoStop}%
\bibitem [{\citenamefont {Sterk}\ \emph {et~al.}(2021)\citenamefont {Sterk},
  \citenamefont {Yuan}, \citenamefont {R{\"u}ckriegel}, \citenamefont
  {Rameshti},\ and\ \citenamefont {Duine}}]{sterk2021green}%
  \BibitemOpen
  \bibfield  {author} {\bibinfo {author} {\bibfnamefont {W.~P.}\ \bibnamefont
  {Sterk}}, \bibinfo {author} {\bibfnamefont {H.~Y.}\ \bibnamefont {Yuan}},
  \bibinfo {author} {\bibfnamefont {A.}~\bibnamefont {R{\"u}ckriegel}},
  \bibinfo {author} {\bibfnamefont {B.~Z.}\ \bibnamefont {Rameshti}},\ and\
  \bibinfo {author} {\bibfnamefont {R.~A.}\ \bibnamefont {Duine}},\ }\bibinfo
  {title} {Green's Function Formalism for Nonlocal Elliptical Magnon
  Transport},\ \href {https://link.aps.org/article/10.1103/PhysRevB.104.174404}
  {\bibfield  {journal} {\bibinfo  {journal} {Phys. Rev. B}\ }\textbf {\bibinfo
  {volume} {104}},\ \bibinfo {pages} {174404} (\bibinfo {year}
  {2021})}\BibitemShut {NoStop}%
\bibitem [{\citenamefont {Faridi}\ \emph
  {et~al.}(2022{\natexlab{a}})\citenamefont {Faridi}, \citenamefont {Kim},\
  and\ \citenamefont {Vignale}}]{faridi2022atomicscale}%
  \BibitemOpen
  \bibfield  {author} {\bibinfo {author} {\bibfnamefont {E.}~\bibnamefont
  {Faridi}}, \bibinfo {author} {\bibfnamefont {S.~K.}\ \bibnamefont {Kim}},\
  and\ \bibinfo {author} {\bibfnamefont {G.}~\bibnamefont {Vignale}},\
  }\bibinfo {title} {Atomic-{{Scale Spin-Wave Polarizer Based}} on a {{Sharp
  Antiferromagnetic Domain Wall}}},\ \href {http://arxiv.org/abs/2203.01453}
  {\bibfield  {journal} {\bibinfo  {journal} {arXiv:2203.01453}\ } (\bibinfo
  {year} {2022}{\natexlab{a}})}\BibitemShut {NoStop}%
\bibitem [{\citenamefont {Daniels}\ \emph {et~al.}(2019)\citenamefont
  {Daniels}, \citenamefont {Yu}, \citenamefont {Cheng}, \citenamefont {Xiao},\
  and\ \citenamefont {Xiao}}]{daniels2019topological}%
  \BibitemOpen
  \bibfield  {author} {\bibinfo {author} {\bibfnamefont {M.~W.}\ \bibnamefont
  {Daniels}}, \bibinfo {author} {\bibfnamefont {W.}~\bibnamefont {Yu}},
  \bibinfo {author} {\bibfnamefont {R.}~\bibnamefont {Cheng}}, \bibinfo
  {author} {\bibfnamefont {J.}~\bibnamefont {Xiao}},\ and\ \bibinfo {author}
  {\bibfnamefont {D.}~\bibnamefont {Xiao}},\ }\bibinfo {title} {Topological
  Spin {{Hall}} Effects and Tunable Skyrmion {{Hall}} Effects in Uniaxial
  Antiferromagnetic Insulators},\ \href
  {https://link.aps.org/accepted/10.1103/PhysRevB.99.224433} {\bibfield
  {journal} {\bibinfo  {journal} {Phys. Rev. B}\ }\textbf {\bibinfo {volume}
  {99}},\ \bibinfo {pages} {224433} (\bibinfo {year} {2019})}\BibitemShut
  {NoStop}%
\bibitem [{\citenamefont {Zhang}\ \emph {et~al.}(2018)\citenamefont {Zhang},
  \citenamefont {Tian}, \citenamefont {Deng}, \citenamefont {Jiang},\ and\
  \citenamefont {Deng}}]{zhang2018spinwavedriven}%
  \BibitemOpen
  \bibfield  {author} {\bibinfo {author} {\bibfnamefont {G.}~\bibnamefont
  {Zhang}}, \bibinfo {author} {\bibfnamefont {Y.}~\bibnamefont {Tian}},
  \bibinfo {author} {\bibfnamefont {Y.}~\bibnamefont {Deng}}, \bibinfo {author}
  {\bibfnamefont {D.}~\bibnamefont {Jiang}},\ and\ \bibinfo {author}
  {\bibfnamefont {S.}~\bibnamefont {Deng}},\ }\bibinfo {title}
  {Spin-{{Wave-Driven Skyrmion Motion}} in {{Magnetic Nanostrip}}},\ \href
  {https://www.hindawi.com/journals/jnt/2018/2602913/} {\bibfield  {journal}
  {\bibinfo  {journal} {J. Nanotechnol.}\ }\textbf {\bibinfo {volume} {2018}},\
  \bibinfo {pages} {1} (\bibinfo {year} {2018})}\BibitemShut {NoStop}%
\bibitem [{\citenamefont {Shen}\ \emph
  {et~al.}(2018{\natexlab{b}})\citenamefont {Shen}, \citenamefont {Zhang},
  \citenamefont {{Ou-Yang}}, \citenamefont {Yang},\ and\ \citenamefont
  {You}}]{shen2018motion}%
  \BibitemOpen
  \bibfield  {author} {\bibinfo {author} {\bibfnamefont {M.}~\bibnamefont
  {Shen}}, \bibinfo {author} {\bibfnamefont {Y.}~\bibnamefont {Zhang}},
  \bibinfo {author} {\bibfnamefont {J.}~\bibnamefont {{Ou-Yang}}}, \bibinfo
  {author} {\bibfnamefont {X.}~\bibnamefont {Yang}},\ and\ \bibinfo {author}
  {\bibfnamefont {L.}~\bibnamefont {You}},\ }\bibinfo {title} {Motion of a
  Skyrmionium Driven by Spin Wave},\ \href
  {http://aip.scitation.org/doi/10.1063/1.5010605} {\bibfield  {journal}
  {\bibinfo  {journal} {Appl. Phys. Lett.}\ }\textbf {\bibinfo {volume}
  {112}},\ \bibinfo {pages} {062403} (\bibinfo {year}
  {2018}{\natexlab{b}})}\BibitemShut {NoStop}%
\bibitem [{\citenamefont {Kim}\ and\ \citenamefont
  {Kim}(2022)}]{kim2022interaction}%
  \BibitemOpen
  \bibfield  {author} {\bibinfo {author} {\bibfnamefont {W.}~\bibnamefont
  {Kim}}\ and\ \bibinfo {author} {\bibfnamefont {S.~K.}\ \bibnamefont {Kim}},\
  }\bibinfo {title} {Interaction of Gapless Spin Waves and a Domain Wall in an
  Easy-Cone Ferromagnet},\ \href {http://arxiv.org/abs/2211.03331} {\bibfield
  {journal} {\bibinfo  {journal} {arXiv:2211.03331}\ } (\bibinfo {year}
  {2022})}\BibitemShut {NoStop}%
\bibitem [{\citenamefont {Dadoenkova}\ \emph {et~al.}(2019)\citenamefont
  {Dadoenkova}, \citenamefont {Dadoenkova}, \citenamefont {Lyubchanskii},
  \citenamefont {Krawczyk},\ and\ \citenamefont
  {Guslienko}}]{dadoenkova2019inelastic}%
  \BibitemOpen
  \bibfield  {author} {\bibinfo {author} {\bibfnamefont {N.~N.}\ \bibnamefont
  {Dadoenkova}}, \bibinfo {author} {\bibfnamefont {Y.~S.}\ \bibnamefont
  {Dadoenkova}}, \bibinfo {author} {\bibfnamefont {I.~L.}\ \bibnamefont
  {Lyubchanskii}}, \bibinfo {author} {\bibfnamefont {M.}~\bibnamefont
  {Krawczyk}},\ and\ \bibinfo {author} {\bibfnamefont {K.~Y.}\ \bibnamefont
  {Guslienko}},\ }\bibinfo {title} {Inelastic {{Spin}}-{{Wave Scattering}} by
  {{Bloch Domain Wall Flexure Oscillations}}},\ \href
  {https://api.wiley.com/onlinelibrary/tdm/v1/articles/10.1002%2Fpssr.201800589}
  {\bibfield  {journal} {\bibinfo  {journal} {Phys. Status Solidi RRL}\
  }\textbf {\bibinfo {volume} {13}},\ \bibinfo {pages} {1800589} (\bibinfo
  {year} {2019})}\BibitemShut {NoStop}%
\bibitem [{\citenamefont {Janutka}(2013)}]{janutka2013resonance}%
  \BibitemOpen
  \bibfield  {author} {\bibinfo {author} {\bibfnamefont {A.}~\bibnamefont
  {Janutka}},\ }\bibinfo {title} {Resonance of {{Spin Waves}} and {{Domain-Wall
  Excitations}} in {{Ferromagnetic Stripes}}},\ \href
  {http://xplorestaging.ieee.org/ielx7/5165412/6423904/06519419.pdf?arnumber=6519419}
  {\bibfield  {journal} {\bibinfo  {journal} {IEEE Magn. Lett.}\ }\textbf
  {\bibinfo {volume} {4}},\ \bibinfo {pages} {4000104} (\bibinfo {year}
  {2013})}\BibitemShut {NoStop}%
\bibitem [{\citenamefont {Seo}\ \emph {et~al.}(2011)\citenamefont {Seo},
  \citenamefont {Lee}, \citenamefont {Kohno},\ and\ \citenamefont
  {Lee}}]{seo2011magnetic}%
  \BibitemOpen
  \bibfield  {author} {\bibinfo {author} {\bibfnamefont {S.-M.}\ \bibnamefont
  {Seo}}, \bibinfo {author} {\bibfnamefont {H.-W.}\ \bibnamefont {Lee}},
  \bibinfo {author} {\bibfnamefont {H.}~\bibnamefont {Kohno}},\ and\ \bibinfo
  {author} {\bibfnamefont {K.-J.}\ \bibnamefont {Lee}},\ }\bibinfo {title}
  {Magnetic Vortex Wall Motion Driven by Spin Waves},\ \href
  {http://aip.scitation.org/doi/pdf/10.1063/1.3541651} {\bibfield  {journal}
  {\bibinfo  {journal} {Appl. Phys. Lett.}\ }\textbf {\bibinfo {volume} {98}},\
  \bibinfo {pages} {012514} (\bibinfo {year} {2011})}\BibitemShut {NoStop}%
\bibitem [{\citenamefont {Han}\ \emph {et~al.}(2009)\citenamefont {Han},
  \citenamefont {Kim}, \citenamefont {Lee}, \citenamefont {Hermsdoerfer},
  \citenamefont {Schultheiss}, \citenamefont {Leven},\ and\ \citenamefont
  {Hillebrands}}]{han2009magnetic}%
  \BibitemOpen
  \bibfield  {author} {\bibinfo {author} {\bibfnamefont {D.-S.}\ \bibnamefont
  {Han}}, \bibinfo {author} {\bibfnamefont {S.-K.}\ \bibnamefont {Kim}},
  \bibinfo {author} {\bibfnamefont {J.-Y.}\ \bibnamefont {Lee}}, \bibinfo
  {author} {\bibfnamefont {S.~J.}\ \bibnamefont {Hermsdoerfer}}, \bibinfo
  {author} {\bibfnamefont {H.}~\bibnamefont {Schultheiss}}, \bibinfo {author}
  {\bibfnamefont {B.}~\bibnamefont {Leven}},\ and\ \bibinfo {author}
  {\bibfnamefont {B.}~\bibnamefont {Hillebrands}},\ }\bibinfo {title} {Magnetic
  Domain-Wall Motion by Propagating Spin Waves},\ \href
  {http://aip.scitation.org/doi/pdf/10.1063/1.3098409} {\bibfield  {journal}
  {\bibinfo  {journal} {Appl. Phys. Lett.}\ }\textbf {\bibinfo {volume} {94}},\
  \bibinfo {pages} {112502} (\bibinfo {year} {2009})}\BibitemShut {NoStop}%
\bibitem [{\citenamefont {Jin}\ \emph {et~al.}(2022)\citenamefont {Jin},
  \citenamefont {Li}, \citenamefont {Zhang}, \citenamefont {Wang},
  \citenamefont {Wang}, \citenamefont {Lian}, \citenamefont {Gong},\ and\
  \citenamefont {Shi}}]{jin2022spinwave}%
  \BibitemOpen
  \bibfield  {author} {\bibinfo {author} {\bibfnamefont {C.}~\bibnamefont
  {Jin}}, \bibinfo {author} {\bibfnamefont {S.}~\bibnamefont {Li}}, \bibinfo
  {author} {\bibfnamefont {H.}~\bibnamefont {Zhang}}, \bibinfo {author}
  {\bibfnamefont {R.}~\bibnamefont {Wang}}, \bibinfo {author} {\bibfnamefont
  {J.}~\bibnamefont {Wang}}, \bibinfo {author} {\bibfnamefont {R.}~\bibnamefont
  {Lian}}, \bibinfo {author} {\bibfnamefont {P.}~\bibnamefont {Gong}},\ and\
  \bibinfo {author} {\bibfnamefont {X.}~\bibnamefont {Shi}},\ }\bibinfo {title}
  {Spin-Wave Modes of Elliptical Skyrmions in Magnetic Nanodots},\ \href
  {https://iopscience.iop.org/article/10.1088/1367-2630/ac5df9} {\bibfield
  {journal} {\bibinfo  {journal} {New J. Phys.}\ }\textbf {\bibinfo {volume}
  {24}},\ \bibinfo {pages} {043005} (\bibinfo {year} {2022})}\BibitemShut
  {NoStop}%
\bibitem [{\citenamefont {Gruszecki}\ and\ \citenamefont
  {Kisielewski}(2022)}]{gruszecki2022influence}%
  \BibitemOpen
  \bibfield  {author} {\bibinfo {author} {\bibfnamefont {P.}~\bibnamefont
  {Gruszecki}}\ and\ \bibinfo {author} {\bibfnamefont {J.}~\bibnamefont
  {Kisielewski}},\ }\bibinfo {title} {Influence of
  {{Dzyaloshinskii}}\textendash{{Moriya}} Interaction and Perpendicular
  Anisotropy on Spin Waves Propagation in Stripe Domain Patterns and Spin
  Spirals},\ \href {https://www.researchsquare.com/article/rs-2190167/v1}
  {\bibfield  {journal} {\bibinfo  {journal} {Research Square}\ } (\bibinfo
  {year} {2022})}\BibitemShut {NoStop}%
\bibitem [{\citenamefont {Hu}\ \emph {et~al.}(2022)\citenamefont {Hu},
  \citenamefont {Shao}, \citenamefont {{Lopez-Dominguez}},\ and\ \citenamefont
  {Amiri}}]{hu2022micromagnetic}%
  \BibitemOpen
  \bibfield  {author} {\bibinfo {author} {\bibfnamefont {Z.}~\bibnamefont
  {Hu}}, \bibinfo {author} {\bibfnamefont {Y.}~\bibnamefont {Shao}}, \bibinfo
  {author} {\bibfnamefont {V.}~\bibnamefont {{Lopez-Dominguez}}},\ and\
  \bibinfo {author} {\bibfnamefont {P.~K.}\ \bibnamefont {Amiri}},\ }\bibinfo
  {title} {Micromagnetic {{Investigation}} of a {{Voltage-Controlled Skyrmionic
  Magnon Switch}}},\ \href
  {https://link.aps.org/article/10.1103/PhysRevApplied.17.044055} {\bibfield
  {journal} {\bibinfo  {journal} {Phys. Rev. Applied}\ }\textbf {\bibinfo
  {volume} {17}},\ \bibinfo {pages} {9} (\bibinfo {year} {2022})}\BibitemShut
  {NoStop}%
\bibitem [{\citenamefont {Lan}\ and\ \citenamefont {Xiao}(2021)}]{lan2021skew}%
  \BibitemOpen
  \bibfield  {author} {\bibinfo {author} {\bibfnamefont {J.}~\bibnamefont
  {Lan}}\ and\ \bibinfo {author} {\bibfnamefont {J.}~\bibnamefont {Xiao}},\
  }\bibinfo {title} {Skew Scattering and Side Jump of Spin Wave across Magnetic
  Texture},\ \href {https://link.aps.org/article/10.1103/PhysRevB.103.054428}
  {\bibfield  {journal} {\bibinfo  {journal} {Phys. Rev. B}\ }\textbf {\bibinfo
  {volume} {103}},\ \bibinfo {pages} {054428} (\bibinfo {year}
  {2021})}\BibitemShut {NoStop}%
\bibitem [{\citenamefont {Liu}\ \emph {et~al.}(2022)\citenamefont {Liu},
  \citenamefont {Liu}, \citenamefont {Jin}, \citenamefont {Hou}, \citenamefont
  {Chen}, \citenamefont {Fan}, \citenamefont {Zeng}, \citenamefont {Lu},
  \citenamefont {Gao}, \citenamefont {Qin},\ and\ \citenamefont
  {Liu}}]{liu2022spinwavedriven}%
  \BibitemOpen
  \bibfield  {author} {\bibinfo {author} {\bibfnamefont {Y.}~\bibnamefont
  {Liu}}, \bibinfo {author} {\bibfnamefont {T.~T.}\ \bibnamefont {Liu}},
  \bibinfo {author} {\bibfnamefont {Z.}~\bibnamefont {Jin}}, \bibinfo {author}
  {\bibfnamefont {Z.~P.}\ \bibnamefont {Hou}}, \bibinfo {author} {\bibfnamefont
  {D.~Y.}\ \bibnamefont {Chen}}, \bibinfo {author} {\bibfnamefont
  {Z.}~\bibnamefont {Fan}}, \bibinfo {author} {\bibfnamefont {M.}~\bibnamefont
  {Zeng}}, \bibinfo {author} {\bibfnamefont {X.~B.}\ \bibnamefont {Lu}},
  \bibinfo {author} {\bibfnamefont {X.~S.}\ \bibnamefont {Gao}}, \bibinfo
  {author} {\bibfnamefont {M.~H.}\ \bibnamefont {Qin}},\ and\ \bibinfo {author}
  {\bibfnamefont {J.-M.}\ \bibnamefont {Liu}},\ }\bibinfo {title}
  {Spin-Wave-Driven Skyrmion Dynamics in Ferrimagnets: {{Effect}} of Net
  Angular Momentum},\ \href
  {https://link.aps.org/doi/10.1103/PhysRevB.106.064424} {\bibfield  {journal}
  {\bibinfo  {journal} {Phys. Rev. B}\ }\textbf {\bibinfo {volume} {106}}
  (\bibinfo {year} {2022})}\BibitemShut {NoStop}%
\bibitem [{\citenamefont {Wang}\ \emph
  {et~al.}(2022{\natexlab{a}})\citenamefont {Wang}, \citenamefont {Bao},
  \citenamefont {Cao},\ and\ \citenamefont {Yan}}]{wang2022allmagnonic}%
  \BibitemOpen
  \bibfield  {author} {\bibinfo {author} {\bibfnamefont {Z.}~\bibnamefont
  {Wang}}, \bibinfo {author} {\bibfnamefont {W.}~\bibnamefont {Bao}}, \bibinfo
  {author} {\bibfnamefont {Y.}~\bibnamefont {Cao}},\ and\ \bibinfo {author}
  {\bibfnamefont {P.}~\bibnamefont {Yan}},\ }\bibinfo {title} {All-Magnonic
  {{Stern-Gerlach}} Effect in Antiferromagnets},\ \href
  {https://aip.scitation.org/doi/pdf/10.1063/5.0096968} {\bibfield  {journal}
  {\bibinfo  {journal} {Appl. Phys. Lett.}\ }\textbf {\bibinfo {volume}
  {120}},\ \bibinfo {pages} {242403} (\bibinfo {year}
  {2022}{\natexlab{a}})}\BibitemShut {NoStop}%
\bibitem [{\citenamefont {Zhang}\ \emph {et~al.}(2019)\citenamefont {Zhang},
  \citenamefont {Zhang}, \citenamefont {Okamoto},\ and\ \citenamefont
  {Xiao}}]{zhang2019thermala}%
  \BibitemOpen
  \bibfield  {author} {\bibinfo {author} {\bibfnamefont {X.}~\bibnamefont
  {Zhang}}, \bibinfo {author} {\bibfnamefont {Y.}~\bibnamefont {Zhang}},
  \bibinfo {author} {\bibfnamefont {S.}~\bibnamefont {Okamoto}},\ and\ \bibinfo
  {author} {\bibfnamefont {D.}~\bibnamefont {Xiao}},\ }\bibinfo {title}
  {Thermal {{Hall Effect Induced}} by {{Magnon-Phonon Interactions}}},\ \href
  {https://link.aps.org/doi/10.1103/PhysRevLett.123.167202} {\bibfield
  {journal} {\bibinfo  {journal} {Phys. Rev. Lett.}\ }\textbf {\bibinfo
  {volume} {123}},\ \bibinfo {pages} {6} (\bibinfo {year} {2019})}\BibitemShut
  {NoStop}%
\bibitem [{\citenamefont {Yu}\ \emph {et~al.}(2020)\citenamefont {Yu},
  \citenamefont {Lan},\ and\ \citenamefont {Xiao}}]{yu2020magnetic}%
  \BibitemOpen
  \bibfield  {author} {\bibinfo {author} {\bibfnamefont {W.}~\bibnamefont
  {Yu}}, \bibinfo {author} {\bibfnamefont {J.}~\bibnamefont {Lan}},\ and\
  \bibinfo {author} {\bibfnamefont {J.}~\bibnamefont {Xiao}},\ }\bibinfo
  {title} {Magnetic {{Logic Gate Based}} on {{Polarized Spin Waves}}},\ \href
  {https://link.aps.org/doi/10.1103/PhysRevApplied.13.024055} {\bibfield
  {journal} {\bibinfo  {journal} {Phys. Rev. Applied}\ }\textbf {\bibinfo
  {volume} {13}},\ \bibinfo {pages} {024055} (\bibinfo {year}
  {2020})}\BibitemShut {NoStop}%
\bibitem [{\citenamefont {Ye}\ and\ \citenamefont
  {Lan}(2021)}]{ye2021magnetically}%
  \BibitemOpen
  \bibfield  {author} {\bibinfo {author} {\bibfnamefont {F.}~\bibnamefont
  {Ye}}\ and\ \bibinfo {author} {\bibfnamefont {J.}~\bibnamefont {Lan}},\
  }\bibinfo {title} {Magnetically Switchable Spin-Wave Retarder with
  90\textdegree{} Antiferromagnetic Domain Wall},\ \href
  {https://link.aps.org/article/10.1103/PhysRevB.104.L180401} {\bibfield
  {journal} {\bibinfo  {journal} {Phys. Rev. B}\ }\textbf {\bibinfo {volume}
  {104}},\ \bibinfo {pages} {6} (\bibinfo {year} {2021})}\BibitemShut {NoStop}%
\bibitem [{\citenamefont {Yu}\ \emph {et~al.}(2016)\citenamefont {Yu},
  \citenamefont {Lan}, \citenamefont {Wu},\ and\ \citenamefont
  {Xiao}}]{yu2016magnetic}%
  \BibitemOpen
  \bibfield  {author} {\bibinfo {author} {\bibfnamefont {W.}~\bibnamefont
  {Yu}}, \bibinfo {author} {\bibfnamefont {J.}~\bibnamefont {Lan}}, \bibinfo
  {author} {\bibfnamefont {R.}~\bibnamefont {Wu}},\ and\ \bibinfo {author}
  {\bibfnamefont {J.}~\bibnamefont {Xiao}},\ }\bibinfo {title} {Magnetic
  {{Snell}}'s Law and Spin-Wave Fiber with {{Dzyaloshinskii-Moriya}}
  Interaction},\ \href {https://link.aps.org/doi/10.1103/PhysRevB.94.140410}
  {\bibfield  {journal} {\bibinfo  {journal} {Phys. Rev. B}\ }\textbf {\bibinfo
  {volume} {94}},\ \bibinfo {pages} {140410} (\bibinfo {year}
  {2016})}\BibitemShut {NoStop}%
\bibitem [{\citenamefont {Moon}\ \emph {et~al.}(2013)\citenamefont {Moon},
  \citenamefont {Sun~Chun}, \citenamefont {Kim},\ and\ \citenamefont
  {Hwang}}]{moon2013control}%
  \BibitemOpen
  \bibfield  {author} {\bibinfo {author} {\bibfnamefont {K.-W.}\ \bibnamefont
  {Moon}}, \bibinfo {author} {\bibfnamefont {B.}~\bibnamefont {Sun~Chun}},
  \bibinfo {author} {\bibfnamefont {W.}~\bibnamefont {Kim}},\ and\ \bibinfo
  {author} {\bibfnamefont {C.}~\bibnamefont {Hwang}},\ }\bibinfo {title}
  {Control of Domain Wall Motion by Interference of Spin Wave},\ \href
  {http://aip.scitation.org/doi/10.1063/1.4822314} {\bibfield  {journal}
  {\bibinfo  {journal} {J. Appl. Phys.}\ }\textbf {\bibinfo {volume} {114}},\
  \bibinfo {pages} {123908} (\bibinfo {year} {2013})}\BibitemShut {NoStop}%
\bibitem [{\citenamefont {Laliena}\ and\ \citenamefont
  {Campo}(2022)}]{laliena2022magnonic}%
  \BibitemOpen
  \bibfield  {author} {\bibinfo {author} {\bibfnamefont {V.}~\bibnamefont
  {Laliena}}\ and\ \bibinfo {author} {\bibfnamefont {J.}~\bibnamefont
  {Campo}},\ }\bibinfo {title} {Magnonic {{Goos-H\"anchen Effect Induced}} by
  {{1D Solitons}}},\ \href
  {https://onlinelibrary.wiley.com/doi/pdf/10.1002/aelm.202100782} {\bibfield
  {journal} {\bibinfo  {journal} {Adv. Electron. Mater.}\ }\textbf {\bibinfo
  {volume} {8}},\ \bibinfo {pages} {2100782} (\bibinfo {year}
  {2022})}\BibitemShut {NoStop}%
\bibitem [{\citenamefont {Wang}\ \emph {et~al.}(2019)\citenamefont {Wang},
  \citenamefont {Cao},\ and\ \citenamefont {Yan}}]{wang2019gooshanchen}%
  \BibitemOpen
  \bibfield  {author} {\bibinfo {author} {\bibfnamefont {Z.}~\bibnamefont
  {Wang}}, \bibinfo {author} {\bibfnamefont {Y.}~\bibnamefont {Cao}},\ and\
  \bibinfo {author} {\bibfnamefont {P.}~\bibnamefont {Yan}},\ }\bibinfo {title}
  {Goos-{{H\"anchen}} Effect of Spin Waves at Heterochiral Interfaces},\ \href
  {https://link.aps.org/article/10.1103/PhysRevB.100.064421} {\bibfield
  {journal} {\bibinfo  {journal} {Phys. Rev. B}\ }\textbf {\bibinfo {volume}
  {100}},\ \bibinfo {pages} {6} (\bibinfo {year} {2019})}\BibitemShut {NoStop}%
\bibitem [{\citenamefont {Bao}\ \emph {et~al.}(2020)\citenamefont {Bao},
  \citenamefont {Wang}, \citenamefont {Cao},\ and\ \citenamefont
  {Yan}}]{bao2020offaxial}%
  \BibitemOpen
  \bibfield  {author} {\bibinfo {author} {\bibfnamefont {W.}~\bibnamefont
  {Bao}}, \bibinfo {author} {\bibfnamefont {Z.}~\bibnamefont {Wang}}, \bibinfo
  {author} {\bibfnamefont {Y.}~\bibnamefont {Cao}},\ and\ \bibinfo {author}
  {\bibfnamefont {P.}~\bibnamefont {Yan}},\ }\bibinfo {title} {Off-Axial
  Focusing of a Spin-Wave Lens in the Presence of {{Dzyaloshinskii-Moriya}}
  Interaction},\ \href
  {https://link.aps.org/article/10.1103/PhysRevB.102.014423} {\bibfield
  {journal} {\bibinfo  {journal} {Phys. Rev. B}\ }\textbf {\bibinfo {volume}
  {102}},\ \bibinfo {pages} {014423} (\bibinfo {year} {2020})}\BibitemShut
  {NoStop}%
\bibitem [{\citenamefont {Schlickeiser}\ \emph {et~al.}(2012)\citenamefont
  {Schlickeiser}, \citenamefont {Atxitia}, \citenamefont {Wienholdt},
  \citenamefont {Hinzke}, \citenamefont {{Chubykalo-Fesenko}},\ and\
  \citenamefont {Nowak}}]{schlickeiser2012temperature}%
  \BibitemOpen
  \bibfield  {author} {\bibinfo {author} {\bibfnamefont {F.}~\bibnamefont
  {Schlickeiser}}, \bibinfo {author} {\bibfnamefont {U.}~\bibnamefont
  {Atxitia}}, \bibinfo {author} {\bibfnamefont {S.}~\bibnamefont {Wienholdt}},
  \bibinfo {author} {\bibfnamefont {D.}~\bibnamefont {Hinzke}}, \bibinfo
  {author} {\bibfnamefont {O.}~\bibnamefont {{Chubykalo-Fesenko}}},\ and\
  \bibinfo {author} {\bibfnamefont {U.}~\bibnamefont {Nowak}},\ }\bibinfo
  {title} {Temperature Dependence of the Frequencies and Effective Damping
  Parameters of Ferrimagnetic Resonance},\ \href
  {https://link.aps.org/doi/10.1103/PhysRevB.86.214416} {\bibfield  {journal}
  {\bibinfo  {journal} {Phys. Rev. B}\ }\textbf {\bibinfo {volume} {86}},\
  \bibinfo {pages} {214416} (\bibinfo {year} {2012})}\BibitemShut {NoStop}%
\bibitem [{\citenamefont {Stanciu}\ \emph {et~al.}(2006)\citenamefont
  {Stanciu}, \citenamefont {Kimel}, \citenamefont {Hansteen}, \citenamefont
  {Tsukamoto}, \citenamefont {Itoh}, \citenamefont {Kirilyuk},\ and\
  \citenamefont {Rasing}}]{stanciu2006ultrafast}%
  \BibitemOpen
  \bibfield  {author} {\bibinfo {author} {\bibfnamefont {C.~D.}\ \bibnamefont
  {Stanciu}}, \bibinfo {author} {\bibfnamefont {A.~V.}\ \bibnamefont {Kimel}},
  \bibinfo {author} {\bibfnamefont {F.}~\bibnamefont {Hansteen}}, \bibinfo
  {author} {\bibfnamefont {A.}~\bibnamefont {Tsukamoto}}, \bibinfo {author}
  {\bibfnamefont {A.}~\bibnamefont {Itoh}}, \bibinfo {author} {\bibfnamefont
  {A.}~\bibnamefont {Kirilyuk}},\ and\ \bibinfo {author} {\bibfnamefont
  {{\relax Th}.}~\bibnamefont {Rasing}},\ }\bibinfo {title} {Ultrafast Spin
  Dynamics across Compensation Points in Ferrimagnetic {{GdFeCo}} : {{The}}
  Role of Angular Momentum Compensation},\ \href
  {https://link.aps.org/doi/10.1103/PhysRevB.73.220402} {\bibfield  {journal}
  {\bibinfo  {journal} {Phys. Rev. B}\ }\textbf {\bibinfo {volume} {73}},\
  \bibinfo {pages} {220402} (\bibinfo {year} {2006})}\BibitemShut {NoStop}%
\bibitem [{\citenamefont {Binder}\ \emph {et~al.}(2006)\citenamefont {Binder},
  \citenamefont {Weber}, \citenamefont {Mosendz}, \citenamefont {Woltersdorf},
  \citenamefont {Izquierdo}, \citenamefont {Neudecker}, \citenamefont {Dahn},
  \citenamefont {Hatchard}, \citenamefont {Thiele}, \citenamefont {Back},\ and\
  \citenamefont {Scheinfein}}]{binder2006magnetization}%
  \BibitemOpen
  \bibfield  {author} {\bibinfo {author} {\bibfnamefont {M.}~\bibnamefont
  {Binder}}, \bibinfo {author} {\bibfnamefont {A.}~\bibnamefont {Weber}},
  \bibinfo {author} {\bibfnamefont {O.}~\bibnamefont {Mosendz}}, \bibinfo
  {author} {\bibfnamefont {G.}~\bibnamefont {Woltersdorf}}, \bibinfo {author}
  {\bibfnamefont {M.}~\bibnamefont {Izquierdo}}, \bibinfo {author}
  {\bibfnamefont {I.}~\bibnamefont {Neudecker}}, \bibinfo {author}
  {\bibfnamefont {J.~R.}\ \bibnamefont {Dahn}}, \bibinfo {author}
  {\bibfnamefont {T.~D.}\ \bibnamefont {Hatchard}}, \bibinfo {author}
  {\bibfnamefont {J.-U.}\ \bibnamefont {Thiele}}, \bibinfo {author}
  {\bibfnamefont {C.~H.}\ \bibnamefont {Back}},\ and\ \bibinfo {author}
  {\bibfnamefont {M.~R.}\ \bibnamefont {Scheinfein}},\ }\bibinfo {title}
  {Magnetization Dynamics of the Ferrimagnet {{CoGd}} near the Compensation of
  Magnetization and Angular Momentum},\ \href
  {https://link.aps.org/doi/10.1103/PhysRevB.74.134404} {\bibfield  {journal}
  {\bibinfo  {journal} {Phys. Rev. B}\ }\textbf {\bibinfo {volume} {74}},\
  \bibinfo {pages} {134404} (\bibinfo {year} {2006})}\BibitemShut {NoStop}%
\bibitem [{\citenamefont {Caretta}\ \emph {et~al.}(2018)\citenamefont
  {Caretta}, \citenamefont {Mann}, \citenamefont {B{\"u}ttner}, \citenamefont
  {Ueda}, \citenamefont {Pfau}, \citenamefont {G{\"u}nther}, \citenamefont
  {Hessing}, \citenamefont {Churikova}, \citenamefont {Klose}, \citenamefont
  {Schneider}, \citenamefont {Engel}, \citenamefont {Marcus}, \citenamefont
  {Bono}, \citenamefont {Bagschik}, \citenamefont {Eisebitt},\ and\
  \citenamefont {Beach}}]{caretta2018fast}%
  \BibitemOpen
  \bibfield  {author} {\bibinfo {author} {\bibfnamefont {L.}~\bibnamefont
  {Caretta}}, \bibinfo {author} {\bibfnamefont {M.}~\bibnamefont {Mann}},
  \bibinfo {author} {\bibfnamefont {F.}~\bibnamefont {B{\"u}ttner}}, \bibinfo
  {author} {\bibfnamefont {K.}~\bibnamefont {Ueda}}, \bibinfo {author}
  {\bibfnamefont {B.}~\bibnamefont {Pfau}}, \bibinfo {author} {\bibfnamefont
  {C.~M.}\ \bibnamefont {G{\"u}nther}}, \bibinfo {author} {\bibfnamefont
  {P.}~\bibnamefont {Hessing}}, \bibinfo {author} {\bibfnamefont
  {A.}~\bibnamefont {Churikova}}, \bibinfo {author} {\bibfnamefont
  {C.}~\bibnamefont {Klose}}, \bibinfo {author} {\bibfnamefont
  {M.}~\bibnamefont {Schneider}}, \bibinfo {author} {\bibfnamefont
  {D.}~\bibnamefont {Engel}}, \bibinfo {author} {\bibfnamefont
  {C.}~\bibnamefont {Marcus}}, \bibinfo {author} {\bibfnamefont
  {D.}~\bibnamefont {Bono}}, \bibinfo {author} {\bibfnamefont {K.}~\bibnamefont
  {Bagschik}}, \bibinfo {author} {\bibfnamefont {S.}~\bibnamefont {Eisebitt}},\
  and\ \bibinfo {author} {\bibfnamefont {G.~S.~D.}\ \bibnamefont {Beach}},\
  }\bibinfo {title} {Fast Current-Driven Domain Walls and Small Skyrmions in a
  Compensated Ferrimagnet},\ \href
  {http://www.nature.com/articles/s41565-018-0255-3} {\bibfield  {journal}
  {\bibinfo  {journal} {Nat. Nanotechnol.}\ }\textbf {\bibinfo {volume} {13}},\
  \bibinfo {pages} {1154} (\bibinfo {year} {2018})}\BibitemShut {NoStop}%
\bibitem [{\citenamefont {Kim}\ \emph {et~al.}(2014)\citenamefont {Kim},
  \citenamefont {Tserkovnyak},\ and\ \citenamefont
  {Tchernyshyov}}]{kim2014propulsion}%
  \BibitemOpen
  \bibfield  {author} {\bibinfo {author} {\bibfnamefont {S.~K.}\ \bibnamefont
  {Kim}}, \bibinfo {author} {\bibfnamefont {Y.}~\bibnamefont {Tserkovnyak}},\
  and\ \bibinfo {author} {\bibfnamefont {O.}~\bibnamefont {Tchernyshyov}},\
  }\bibinfo {title} {Propulsion of a Domain Wall in an Antiferromagnet by
  Magnons},\ \href {https://link.aps.org/doi/10.1103/PhysRevB.90.104406}
  {\bibfield  {journal} {\bibinfo  {journal} {Phys. Rev. B}\ }\textbf {\bibinfo
  {volume} {90}},\ \bibinfo {pages} {104406} (\bibinfo {year}
  {2014})}\BibitemShut {NoStop}%
\bibitem [{\citenamefont {Tatara}\ and\ \citenamefont
  {Kohno}(2004)}]{tatara2004theory}%
  \BibitemOpen
  \bibfield  {author} {\bibinfo {author} {\bibfnamefont {G.}~\bibnamefont
  {Tatara}}\ and\ \bibinfo {author} {\bibfnamefont {H.}~\bibnamefont {Kohno}},\
  }\bibinfo {title} {Theory of {{Current-Driven Domain Wall Motion}}: {{Spin
  Transfer}} versus {{Momentum Transfer}}},\ \href
  {https://link.aps.org/doi/10.1103/PhysRevLett.92.086601} {\bibfield
  {journal} {\bibinfo  {journal} {Phys. Rev. Lett.}\ }\textbf {\bibinfo
  {volume} {92}},\ \bibinfo {pages} {086601} (\bibinfo {year}
  {2004})}\BibitemShut {NoStop}%
\bibitem [{\citenamefont {Gong}\ \emph {et~al.}(2022)\citenamefont {Gong},
  \citenamefont {Zhou},\ and\ \citenamefont {Zhao}}]{gong2022dynamics}%
  \BibitemOpen
  \bibfield  {author} {\bibinfo {author} {\bibfnamefont {C.}~\bibnamefont
  {Gong}}, \bibinfo {author} {\bibfnamefont {Y.}~\bibnamefont {Zhou}},\ and\
  \bibinfo {author} {\bibfnamefont {G.}~\bibnamefont {Zhao}},\ }\bibinfo
  {title} {Dynamics of Magnetic Skyrmions under Temperature Gradients},\ \href
  {https://aip.scitation.org/doi/10.1063/5.0080778} {\bibfield  {journal}
  {\bibinfo  {journal} {Appl. Phys. Lett.}\ }\textbf {\bibinfo {volume}
  {120}},\ \bibinfo {pages} {052402} (\bibinfo {year} {2022})}\BibitemShut
  {NoStop}%
\bibitem [{\citenamefont {Abbout}\ \emph {et~al.}(2018)\citenamefont {Abbout},
  \citenamefont {Weston}, \citenamefont {Waintal},\ and\ \citenamefont
  {Manchon}}]{abbout2018cooperative}%
  \BibitemOpen
  \bibfield  {author} {\bibinfo {author} {\bibfnamefont {A.}~\bibnamefont
  {Abbout}}, \bibinfo {author} {\bibfnamefont {J.}~\bibnamefont {Weston}},
  \bibinfo {author} {\bibfnamefont {X.}~\bibnamefont {Waintal}},\ and\ \bibinfo
  {author} {\bibfnamefont {A.}~\bibnamefont {Manchon}},\ }\bibinfo {title}
  {Cooperative {{Charge Pumping}} and {{Enhanced Skyrmion Mobility}}},\ \href
  {https://link.aps.org/doi/10.1103/PhysRevLett.121.257203} {\bibfield
  {journal} {\bibinfo  {journal} {Phys. Rev. Lett.}\ }\textbf {\bibinfo
  {volume} {121}},\ \bibinfo {pages} {257203} (\bibinfo {year}
  {2018})}\BibitemShut {NoStop}%
\bibitem [{\citenamefont {Tatara}(2015)}]{tatara2015thermala}%
  \BibitemOpen
  \bibfield  {author} {\bibinfo {author} {\bibfnamefont {G.}~\bibnamefont
  {Tatara}},\ }\bibinfo {title} {Thermal Vector Potential Theory of
  Magnon-Driven Magnetization Dynamics},\ \href
  {https://link.aps.org/doi/10.1103/PhysRevB.92.064405} {\bibfield  {journal}
  {\bibinfo  {journal} {Phys. Rev. B}\ }\textbf {\bibinfo {volume} {92}},\
  \bibinfo {pages} {064405} (\bibinfo {year} {2015})}\BibitemShut {NoStop}%
\bibitem [{\citenamefont {Shen}\ \emph
  {et~al.}(2020{\natexlab{b}})\citenamefont {Shen}, \citenamefont {Xia},
  \citenamefont {Zhang}, \citenamefont {Ezawa}, \citenamefont {Tretiakov},
  \citenamefont {Liu}, \citenamefont {Zhao},\ and\ \citenamefont
  {Zhou}}]{shen2020currentinduced}%
  \BibitemOpen
  \bibfield  {author} {\bibinfo {author} {\bibfnamefont {L.}~\bibnamefont
  {Shen}}, \bibinfo {author} {\bibfnamefont {J.}~\bibnamefont {Xia}}, \bibinfo
  {author} {\bibfnamefont {X.}~\bibnamefont {Zhang}}, \bibinfo {author}
  {\bibfnamefont {M.}~\bibnamefont {Ezawa}}, \bibinfo {author} {\bibfnamefont
  {O.~A.}\ \bibnamefont {Tretiakov}}, \bibinfo {author} {\bibfnamefont
  {X.}~\bibnamefont {Liu}}, \bibinfo {author} {\bibfnamefont {G.}~\bibnamefont
  {Zhao}},\ and\ \bibinfo {author} {\bibfnamefont {Y.}~\bibnamefont {Zhou}},\
  }\bibinfo {title} {Current-{{Induced Dynamics}} and {{Chaos}} of
  {{Antiferromagnetic Bimerons}}},\ \href
  {https://link.aps.org/doi/10.1103/PhysRevLett.124.037202} {\bibfield
  {journal} {\bibinfo  {journal} {Phys. Rev. Lett.}\ }\textbf {\bibinfo
  {volume} {124}},\ \bibinfo {pages} {037202} (\bibinfo {year}
  {2020}{\natexlab{b}})}\BibitemShut {NoStop}%
\bibitem [{\citenamefont {Woo}\ and\ \citenamefont
  {Beach}(2017)}]{woo2017control}%
  \BibitemOpen
  \bibfield  {author} {\bibinfo {author} {\bibfnamefont {S.}~\bibnamefont
  {Woo}}\ and\ \bibinfo {author} {\bibfnamefont {G.~S.~D.}\ \bibnamefont
  {Beach}},\ }\bibinfo {title} {Control of Propagating Spin-Wave Attenuation by
  the Spin-{{Hall}} Effect},\ \href
  {http://aip.scitation.org/doi/pdf/10.1063/1.4999828} {\bibfield  {journal}
  {\bibinfo  {journal} {J. Appl. Phys.}\ }\textbf {\bibinfo {volume} {122}},\
  \bibinfo {pages} {093901} (\bibinfo {year} {2017})}\BibitemShut {NoStop}%
\bibitem [{\citenamefont {Kim}\ \emph {et~al.}(2015)\citenamefont {Kim},
  \citenamefont {Tchernyshyov},\ and\ \citenamefont
  {Tserkovnyak}}]{kim2015thermophoresis}%
  \BibitemOpen
  \bibfield  {author} {\bibinfo {author} {\bibfnamefont {S.~K.}\ \bibnamefont
  {Kim}}, \bibinfo {author} {\bibfnamefont {O.}~\bibnamefont {Tchernyshyov}},\
  and\ \bibinfo {author} {\bibfnamefont {Y.}~\bibnamefont {Tserkovnyak}},\
  }\bibinfo {title} {Thermophoresis of an Antiferromagnetic Soliton},\ \href
  {https://link.aps.org/doi/10.1103/PhysRevB.92.020402} {\bibfield  {journal}
  {\bibinfo  {journal} {Phys. Rev. B}\ }\textbf {\bibinfo {volume} {92}},\
  \bibinfo {pages} {020402} (\bibinfo {year} {2015})}\BibitemShut {NoStop}%
\bibitem [{\citenamefont {Lifshitzs}\ and\ \citenamefont
  {Landau}(1982)}]{lifshitzs1982course}%
  \BibitemOpen
  \bibfield  {author} {\bibinfo {author} {\bibfnamefont {E.~M.}\ \bibnamefont
  {Lifshitzs}}\ and\ \bibinfo {author} {\bibfnamefont {L.~D.}\ \bibnamefont
  {Landau}},\ }\bibinfo {title} {Course of {{Theoretical Physics}}: {{Vol}}. 4,
  {{Quantum Electrodynamics}}, 2rd Edition},\ \href
  {https://www.amazon.com/Quantum-Electrodynamics-Course-Theoretical-Physics-ebook/dp/B00DBELAJC}
  {\bibfield  {journal} {\bibinfo  {journal} {Butterworth-Heinemann}\ }
  (\bibinfo {year} {1982})}\BibitemShut {NoStop}%
\bibitem [{\citenamefont {Wen}(2010)}]{wen2010quantum}%
  \BibitemOpen
  \bibfield  {author} {\bibinfo {author} {\bibfnamefont {X.-G.}\ \bibnamefont
  {Wen}},\ }\href@noop {} {\emph {\bibinfo {title} {Quantum Field Theory of
  Many-Body Systems: From the Origin of Sound to an Origin of Light and
  Electrons}}},\ Oxford Graduate Texts\ (\bibinfo  {publisher} {{Oxford
  University Press}},\ \bibinfo {address} {{Oxford}},\ \bibinfo {year}
  {2010})\BibitemShut {NoStop}%
\bibitem [{\citenamefont {Tchernyshyov}(2015)}]{tchernyshyov2015conserved}%
  \BibitemOpen
  \bibfield  {author} {\bibinfo {author} {\bibfnamefont {O.}~\bibnamefont
  {Tchernyshyov}},\ }\bibinfo {title} {Conserved Momenta of a Ferromagnetic
  Soliton},\ \href
  {https://linkinghub.elsevier.com/retrieve/pii/S0003491615003395} {\bibfield
  {journal} {\bibinfo  {journal} {Ann. Phys.}\ }\textbf {\bibinfo {volume}
  {363}},\ \bibinfo {pages} {98} (\bibinfo {year} {2015})}\BibitemShut
  {NoStop}%
\bibitem [{\citenamefont {Nieves}\ \emph {et~al.}(2016)\citenamefont {Nieves},
  \citenamefont {Serantes},\ and\ \citenamefont
  {{Chubykalo-Fesenko}}}]{nieves2016selfconsistent}%
  \BibitemOpen
  \bibfield  {author} {\bibinfo {author} {\bibfnamefont {P.}~\bibnamefont
  {Nieves}}, \bibinfo {author} {\bibfnamefont {D.}~\bibnamefont {Serantes}},\
  and\ \bibinfo {author} {\bibfnamefont {O.}~\bibnamefont
  {{Chubykalo-Fesenko}}},\ }\bibinfo {title} {Self-Consistent Description of
  Spin-Phonon Dynamics in Ferromagnets},\ \href
  {https://link.aps.org/doi/10.1103/PhysRevB.94.014409} {\bibfield  {journal}
  {\bibinfo  {journal} {Phys. Rev. B}\ }\textbf {\bibinfo {volume} {94}},\
  \bibinfo {pages} {014409} (\bibinfo {year} {2016})}\BibitemShut {NoStop}%
\bibitem [{\citenamefont {Kim}\ \emph {et~al.}(2018)\citenamefont {Kim},
  \citenamefont {Tchernyshyov}, \citenamefont {Galitski},\ and\ \citenamefont
  {Tserkovnyak}}]{kim2018magnoninduced}%
  \BibitemOpen
  \bibfield  {author} {\bibinfo {author} {\bibfnamefont {S.~K.}\ \bibnamefont
  {Kim}}, \bibinfo {author} {\bibfnamefont {O.}~\bibnamefont {Tchernyshyov}},
  \bibinfo {author} {\bibfnamefont {V.}~\bibnamefont {Galitski}},\ and\
  \bibinfo {author} {\bibfnamefont {Y.}~\bibnamefont {Tserkovnyak}},\ }\bibinfo
  {title} {Magnon-Induced Non-{{Markovian}} Friction of a Domain Wall in a
  Ferromagnet},\ \href {https://link.aps.org/doi/10.1103/PhysRevB.97.174433}
  {\bibfield  {journal} {\bibinfo  {journal} {Phys. Rev. B}\ }\textbf {\bibinfo
  {volume} {97}},\ \bibinfo {pages} {174433} (\bibinfo {year}
  {2018})}\BibitemShut {NoStop}%
\bibitem [{\citenamefont {Akosa}\ \emph {et~al.}(2016)\citenamefont {Akosa},
  \citenamefont {Miron}, \citenamefont {Gaudin},\ and\ \citenamefont
  {Manchon}}]{akosa2016phenomenology}%
  \BibitemOpen
  \bibfield  {author} {\bibinfo {author} {\bibfnamefont {C.~A.}\ \bibnamefont
  {Akosa}}, \bibinfo {author} {\bibfnamefont {I.~M.}\ \bibnamefont {Miron}},
  \bibinfo {author} {\bibfnamefont {G.}~\bibnamefont {Gaudin}},\ and\ \bibinfo
  {author} {\bibfnamefont {A.}~\bibnamefont {Manchon}},\ }\bibinfo {title}
  {Phenomenology of Chiral Damping in Noncentrosymmetric Magnets},\ \href
  {https://link.aps.org/doi/10.1103/PhysRevB.93.214429} {\bibfield  {journal}
  {\bibinfo  {journal} {Phys. Rev. B}\ }\textbf {\bibinfo {volume} {93}},\
  \bibinfo {pages} {5} (\bibinfo {year} {2016})}\BibitemShut {NoStop}%
\bibitem [{\citenamefont {Akosa}\ \emph {et~al.}(2017)\citenamefont {Akosa},
  \citenamefont {Ndiaye},\ and\ \citenamefont {Manchon}}]{akosa2017intrinsic}%
  \BibitemOpen
  \bibfield  {author} {\bibinfo {author} {\bibfnamefont {C.~A.}\ \bibnamefont
  {Akosa}}, \bibinfo {author} {\bibfnamefont {P.~B.}\ \bibnamefont {Ndiaye}},\
  and\ \bibinfo {author} {\bibfnamefont {A.}~\bibnamefont {Manchon}},\
  }\bibinfo {title} {Intrinsic Nonadiabatic Topological Torque in Magnetic
  Skyrmions and Vortices},\ \href
  {https://link.aps.org/doi/10.1103/PhysRevB.95.054434} {\bibfield  {journal}
  {\bibinfo  {journal} {Phys. Rev. B}\ }\textbf {\bibinfo {volume} {95}},\
  \bibinfo {pages} {054434} (\bibinfo {year} {2017})}\BibitemShut {NoStop}%
\bibitem [{\citenamefont {Hals}\ \emph {et~al.}(2009)\citenamefont {Hals},
  \citenamefont {Nguyen},\ and\ \citenamefont {Brataas}}]{hals2009intrinsica}%
  \BibitemOpen
  \bibfield  {author} {\bibinfo {author} {\bibfnamefont {K.~M.~D.}\
  \bibnamefont {Hals}}, \bibinfo {author} {\bibfnamefont {A.~K.}\ \bibnamefont
  {Nguyen}},\ and\ \bibinfo {author} {\bibfnamefont {A.}~\bibnamefont
  {Brataas}},\ }\bibinfo {title} {Intrinsic {{Coupling}} between {{Current}}
  and {{Domain Wall Motion}} in ({{Ga}},{{Mn}}){{As}}},\ \href
  {https://link.aps.org/doi/10.1103/PhysRevLett.102.256601} {\bibfield
  {journal} {\bibinfo  {journal} {Phys. Rev. Lett.}\ }\textbf {\bibinfo
  {volume} {102}},\ \bibinfo {pages} {256601} (\bibinfo {year}
  {2009})}\BibitemShut {NoStop}%
\bibitem [{\citenamefont {Wei{\ss}enhofer}\ \emph {et~al.}(2021)\citenamefont
  {Wei{\ss}enhofer}, \citenamefont {R{\'o}zsa},\ and\ \citenamefont
  {Nowak}}]{weissenhofer2021skyrmion}%
  \BibitemOpen
  \bibfield  {author} {\bibinfo {author} {\bibfnamefont {M.}~\bibnamefont
  {Wei{\ss}enhofer}}, \bibinfo {author} {\bibfnamefont {L.}~\bibnamefont
  {R{\'o}zsa}},\ and\ \bibinfo {author} {\bibfnamefont {U.}~\bibnamefont
  {Nowak}},\ }\bibinfo {title} {Skyrmion {{Dynamics}} at {{Finite
  Temperatures}}: {{Beyond Thiele}}'s {{Equation}}},\ \href
  {https://link.aps.org/doi/10.1103/PhysRevLett.127.047203} {\bibfield
  {journal} {\bibinfo  {journal} {Phys. Rev. Lett.}\ }\textbf {\bibinfo
  {volume} {127}},\ \bibinfo {pages} {047203} (\bibinfo {year}
  {2021})}\BibitemShut {NoStop}%
\bibitem [{\citenamefont {Kim}\ \emph {et~al.}(2013{\natexlab{a}})\citenamefont
  {Kim}, \citenamefont {Hiramatsu}, \citenamefont {Koyama}, \citenamefont
  {Ueda}, \citenamefont {Yoshimura}, \citenamefont {Chiba}, \citenamefont
  {Kobayashi}, \citenamefont {Nakatani}, \citenamefont {Fukami}, \citenamefont
  {Yamanouchi}, \citenamefont {Ohno}, \citenamefont {Kohno}, \citenamefont
  {Tatara},\ and\ \citenamefont {Ono}}]{kim2013twobarrier}%
  \BibitemOpen
  \bibfield  {author} {\bibinfo {author} {\bibfnamefont {K.-J.}\ \bibnamefont
  {Kim}}, \bibinfo {author} {\bibfnamefont {R.}~\bibnamefont {Hiramatsu}},
  \bibinfo {author} {\bibfnamefont {T.}~\bibnamefont {Koyama}}, \bibinfo
  {author} {\bibfnamefont {K.}~\bibnamefont {Ueda}}, \bibinfo {author}
  {\bibfnamefont {Y.}~\bibnamefont {Yoshimura}}, \bibinfo {author}
  {\bibfnamefont {D.}~\bibnamefont {Chiba}}, \bibinfo {author} {\bibfnamefont
  {K.}~\bibnamefont {Kobayashi}}, \bibinfo {author} {\bibfnamefont
  {Y.}~\bibnamefont {Nakatani}}, \bibinfo {author} {\bibfnamefont
  {S.}~\bibnamefont {Fukami}}, \bibinfo {author} {\bibfnamefont
  {M.}~\bibnamefont {Yamanouchi}}, \bibinfo {author} {\bibfnamefont
  {H.}~\bibnamefont {Ohno}}, \bibinfo {author} {\bibfnamefont {H.}~\bibnamefont
  {Kohno}}, \bibinfo {author} {\bibfnamefont {G.}~\bibnamefont {Tatara}},\ and\
  \bibinfo {author} {\bibfnamefont {T.}~\bibnamefont {Ono}},\ }\bibinfo {title}
  {Two-Barrier Stability That Allows Low-Power Operation in Current-Induced
  Domain-Wall Motion},\ \href {http://www.nature.com/articles/ncomms3011.pdf}
  {\bibfield  {journal} {\bibinfo  {journal} {Nat. Commun.}\ }\textbf {\bibinfo
  {volume} {4}},\ \bibinfo {pages} {2011} (\bibinfo {year}
  {2013}{\natexlab{a}})}\BibitemShut {NoStop}%
\bibitem [{\citenamefont {Kohno}\ \emph {et~al.}(2006)\citenamefont {Kohno},
  \citenamefont {Tatara},\ and\ \citenamefont
  {Shibata}}]{kohno2006microscopic}%
  \BibitemOpen
  \bibfield  {author} {\bibinfo {author} {\bibfnamefont {H.}~\bibnamefont
  {Kohno}}, \bibinfo {author} {\bibfnamefont {G.}~\bibnamefont {Tatara}},\ and\
  \bibinfo {author} {\bibfnamefont {J.}~\bibnamefont {Shibata}},\ }\bibinfo
  {title} {Microscopic {{Calculation}} of {{Spin Torques}} in {{Disordered
  Ferromagnets}}},\ \href {https://journals.jps.jp/doi/10.1143/JPSJ.75.113706}
  {\bibfield  {journal} {\bibinfo  {journal} {J. Phys. Soc. Jpn.}\ }\textbf
  {\bibinfo {volume} {75}},\ \bibinfo {pages} {113706} (\bibinfo {year}
  {2006})}\BibitemShut {NoStop}%
\bibitem [{\citenamefont {Litzius}\ \emph {et~al.}(2017)\citenamefont
  {Litzius}, \citenamefont {Lemesh}, \citenamefont {Kr{\"u}ger}, \citenamefont
  {Bassirian}, \citenamefont {Caretta}, \citenamefont {Richter}, \citenamefont
  {B{\"u}ttner}, \citenamefont {Sato}, \citenamefont {Tretiakov}, \citenamefont
  {F{\"o}rster}, \citenamefont {Reeve}, \citenamefont {Weigand}, \citenamefont
  {Bykova}, \citenamefont {Stoll}, \citenamefont {Sch{\"u}tz}, \citenamefont
  {Beach},\ and\ \citenamefont {Kl{\"a}ui}}]{litzius2017skyrmion}%
  \BibitemOpen
  \bibfield  {author} {\bibinfo {author} {\bibfnamefont {K.}~\bibnamefont
  {Litzius}}, \bibinfo {author} {\bibfnamefont {I.}~\bibnamefont {Lemesh}},
  \bibinfo {author} {\bibfnamefont {B.}~\bibnamefont {Kr{\"u}ger}}, \bibinfo
  {author} {\bibfnamefont {P.}~\bibnamefont {Bassirian}}, \bibinfo {author}
  {\bibfnamefont {L.}~\bibnamefont {Caretta}}, \bibinfo {author} {\bibfnamefont
  {K.}~\bibnamefont {Richter}}, \bibinfo {author} {\bibfnamefont
  {F.}~\bibnamefont {B{\"u}ttner}}, \bibinfo {author} {\bibfnamefont
  {K.}~\bibnamefont {Sato}}, \bibinfo {author} {\bibfnamefont {O.~A.}\
  \bibnamefont {Tretiakov}}, \bibinfo {author} {\bibfnamefont {J.}~\bibnamefont
  {F{\"o}rster}}, \bibinfo {author} {\bibfnamefont {R.~M.}\ \bibnamefont
  {Reeve}}, \bibinfo {author} {\bibfnamefont {M.}~\bibnamefont {Weigand}},
  \bibinfo {author} {\bibfnamefont {I.}~\bibnamefont {Bykova}}, \bibinfo
  {author} {\bibfnamefont {H.}~\bibnamefont {Stoll}}, \bibinfo {author}
  {\bibfnamefont {G.}~\bibnamefont {Sch{\"u}tz}}, \bibinfo {author}
  {\bibfnamefont {G.~S.~D.}\ \bibnamefont {Beach}},\ and\ \bibinfo {author}
  {\bibfnamefont {M.}~\bibnamefont {Kl{\"a}ui}},\ }\bibinfo {title} {Skyrmion
  {{Hall}} Effect Revealed by Direct Time-Resolved {{X-ray}} Microscopy},\
  \href {http://www.nature.com/articles/nphys4000} {\bibfield  {journal}
  {\bibinfo  {journal} {Nat. Phys.}\ }\textbf {\bibinfo {volume} {13}},\
  \bibinfo {pages} {170} (\bibinfo {year} {2017})}\BibitemShut {NoStop}%
\bibitem [{\citenamefont {Mougin}\ \emph {et~al.}(2007)\citenamefont {Mougin},
  \citenamefont {Cormier}, \citenamefont {Adam}, \citenamefont {Metaxas},\ and\
  \citenamefont {Ferr{\'e}}}]{mougin2007domain}%
  \BibitemOpen
  \bibfield  {author} {\bibinfo {author} {\bibfnamefont {A.}~\bibnamefont
  {Mougin}}, \bibinfo {author} {\bibfnamefont {M.}~\bibnamefont {Cormier}},
  \bibinfo {author} {\bibfnamefont {J.~P.}\ \bibnamefont {Adam}}, \bibinfo
  {author} {\bibfnamefont {P.~J.}\ \bibnamefont {Metaxas}},\ and\ \bibinfo
  {author} {\bibfnamefont {J.}~\bibnamefont {Ferr{\'e}}},\ }\bibinfo {title}
  {Domain Wall Mobility, Stability and {{Walker}} Breakdown in Magnetic
  Nanowires},\ \href
  {https://iopscience.iop.org/article/10.1209/0295-5075/78/57007} {\bibfield
  {journal} {\bibinfo  {journal} {Europhys. Lett.}\ }\textbf {\bibinfo {volume}
  {78}},\ \bibinfo {pages} {57007} (\bibinfo {year} {2007})}\BibitemShut
  {NoStop}%
\bibitem [{\citenamefont {Kohno}\ and\ \citenamefont
  {Shibata}(2007)}]{kohno2007gauge}%
  \BibitemOpen
  \bibfield  {author} {\bibinfo {author} {\bibfnamefont {H.}~\bibnamefont
  {Kohno}}\ and\ \bibinfo {author} {\bibfnamefont {J.}~\bibnamefont
  {Shibata}},\ }\bibinfo {title} {Gauge {{Field Formulation}} of {{Adiabatic
  Spin Torques}}},\ \href {https://journals.jps.jp/doi/10.1143/JPSJ.76.063710}
  {\bibfield  {journal} {\bibinfo  {journal} {J. Phys. Soc. Jpn.}\ }\textbf
  {\bibinfo {volume} {76}},\ \bibinfo {pages} {063710} (\bibinfo {year}
  {2007})}\BibitemShut {NoStop}%
\bibitem [{\citenamefont {Gruber}\ \emph {et~al.}(2022)\citenamefont {Gruber},
  \citenamefont {Z{\'a}zvorka}, \citenamefont {Brems}, \citenamefont
  {Rodrigues}, \citenamefont {Dohi}, \citenamefont {Kerber}, \citenamefont
  {Seng}, \citenamefont {Vafaee}, \citenamefont {{Everschor-Sitte}},
  \citenamefont {Virnau},\ and\ \citenamefont
  {Kl{\"a}ui}}]{gruber2022skyrmion}%
  \BibitemOpen
  \bibfield  {author} {\bibinfo {author} {\bibfnamefont {R.}~\bibnamefont
  {Gruber}}, \bibinfo {author} {\bibfnamefont {J.}~\bibnamefont
  {Z{\'a}zvorka}}, \bibinfo {author} {\bibfnamefont {M.~A.}\ \bibnamefont
  {Brems}}, \bibinfo {author} {\bibfnamefont {D.~R.}\ \bibnamefont
  {Rodrigues}}, \bibinfo {author} {\bibfnamefont {T.}~\bibnamefont {Dohi}},
  \bibinfo {author} {\bibfnamefont {N.}~\bibnamefont {Kerber}}, \bibinfo
  {author} {\bibfnamefont {B.}~\bibnamefont {Seng}}, \bibinfo {author}
  {\bibfnamefont {M.}~\bibnamefont {Vafaee}}, \bibinfo {author} {\bibfnamefont
  {K.}~\bibnamefont {{Everschor-Sitte}}}, \bibinfo {author} {\bibfnamefont
  {P.}~\bibnamefont {Virnau}},\ and\ \bibinfo {author} {\bibfnamefont
  {M.}~\bibnamefont {Kl{\"a}ui}},\ }\bibinfo {title} {Skyrmion Pinning
  Energetics in Thin Film Systems},\ \href
  {https://www.nature.com/articles/s41467-022-30743-4.pdf} {\bibfield
  {journal} {\bibinfo  {journal} {Nat. Commun.}\ }\textbf {\bibinfo {volume}
  {13}},\ \bibinfo {pages} {3144} (\bibinfo {year} {2022})}\BibitemShut
  {NoStop}%
\bibitem [{\citenamefont {Berges}\ \emph {et~al.}(2022)\citenamefont {Berges},
  \citenamefont {Haltz}, \citenamefont {Panigrahy}, \citenamefont {Mallick},
  \citenamefont {Weil}, \citenamefont {Rohart}, \citenamefont {Mougin},\ and\
  \citenamefont {Sampaio}}]{berges2022sizedependent}%
  \BibitemOpen
  \bibfield  {author} {\bibinfo {author} {\bibfnamefont {L.}~\bibnamefont
  {Berges}}, \bibinfo {author} {\bibfnamefont {E.}~\bibnamefont {Haltz}},
  \bibinfo {author} {\bibfnamefont {S.}~\bibnamefont {Panigrahy}}, \bibinfo
  {author} {\bibfnamefont {S.}~\bibnamefont {Mallick}}, \bibinfo {author}
  {\bibfnamefont {R.}~\bibnamefont {Weil}}, \bibinfo {author} {\bibfnamefont
  {S.}~\bibnamefont {Rohart}}, \bibinfo {author} {\bibfnamefont
  {A.}~\bibnamefont {Mougin}},\ and\ \bibinfo {author} {\bibfnamefont
  {J.}~\bibnamefont {Sampaio}},\ }\bibinfo {title} {Size-Dependent Mobility of
  Skyrmions beyond Pinning in Ferrimagnetic {{GdCo}} Thin Films},\ \href
  {https://link.aps.org/doi/10.1103/PhysRevB.106.144408} {\bibfield  {journal}
  {\bibinfo  {journal} {Phys. Rev. B}\ }\textbf {\bibinfo {volume} {106}},\
  \bibinfo {pages} {144408} (\bibinfo {year} {2022})}\BibitemShut {NoStop}%
\bibitem [{\citenamefont {Buijnsters}\ \emph {et~al.}(2014)\citenamefont
  {Buijnsters}, \citenamefont {Fasolino},\ and\ \citenamefont
  {Katsnelson}}]{buijnsters2014motion}%
  \BibitemOpen
  \bibfield  {author} {\bibinfo {author} {\bibfnamefont {F.~J.}\ \bibnamefont
  {Buijnsters}}, \bibinfo {author} {\bibfnamefont {A.}~\bibnamefont
  {Fasolino}},\ and\ \bibinfo {author} {\bibfnamefont {M.~I.}\ \bibnamefont
  {Katsnelson}},\ }\bibinfo {title} {Motion of {{Domain Walls}} and the
  {{Dynamics}} of {{Kinks}} in the {{Magnetic Peierls Potential}}},\ \href
  {http://link.aps.org/article/10.1103/PhysRevLett.113.217202} {\bibfield
  {journal} {\bibinfo  {journal} {Phys. Rev. Lett.}\ }\textbf {\bibinfo
  {volume} {113}},\ \bibinfo {pages} {217202} (\bibinfo {year}
  {2014})}\BibitemShut {NoStop}%
\bibitem [{\citenamefont {Ju{\'e}}\ \emph {et~al.}(2016)\citenamefont
  {Ju{\'e}}, \citenamefont {Safeer}, \citenamefont {Drouard}, \citenamefont
  {Lopez}, \citenamefont {Balint}, \citenamefont {{Buda-Prejbeanu}},
  \citenamefont {Boulle}, \citenamefont {Auffret}, \citenamefont {Schuhl},
  \citenamefont {Manchon}, \citenamefont {Miron},\ and\ \citenamefont
  {Gaudin}}]{jue2016chirala}%
  \BibitemOpen
  \bibfield  {author} {\bibinfo {author} {\bibfnamefont {E.}~\bibnamefont
  {Ju{\'e}}}, \bibinfo {author} {\bibfnamefont {C.~K.}\ \bibnamefont {Safeer}},
  \bibinfo {author} {\bibfnamefont {M.}~\bibnamefont {Drouard}}, \bibinfo
  {author} {\bibfnamefont {A.}~\bibnamefont {Lopez}}, \bibinfo {author}
  {\bibfnamefont {P.}~\bibnamefont {Balint}}, \bibinfo {author} {\bibfnamefont
  {L.}~\bibnamefont {{Buda-Prejbeanu}}}, \bibinfo {author} {\bibfnamefont
  {O.}~\bibnamefont {Boulle}}, \bibinfo {author} {\bibfnamefont
  {S.}~\bibnamefont {Auffret}}, \bibinfo {author} {\bibfnamefont
  {A.}~\bibnamefont {Schuhl}}, \bibinfo {author} {\bibfnamefont
  {A.}~\bibnamefont {Manchon}}, \bibinfo {author} {\bibfnamefont {I.~M.}\
  \bibnamefont {Miron}},\ and\ \bibinfo {author} {\bibfnamefont
  {G.}~\bibnamefont {Gaudin}},\ }\bibinfo {title} {Chiral Damping of Magnetic
  Domain Walls},\ \href {https://www.nature.com/articles/nmat4518} {\bibfield
  {journal} {\bibinfo  {journal} {Nat. Mater.}\ }\textbf {\bibinfo {volume}
  {15}},\ \bibinfo {pages} {272} (\bibinfo {year} {2016})}\BibitemShut
  {NoStop}%
\bibitem [{\citenamefont {Landau}\ and\ \citenamefont
  {Lifshits}(1981)}]{landau1981course}%
  \BibitemOpen
  \bibfield  {author} {\bibinfo {author} {\bibfnamefont {L.~D.}\ \bibnamefont
  {Landau}}\ and\ \bibinfo {author} {\bibfnamefont {E.~M.}\ \bibnamefont
  {Lifshits}},\ }\bibinfo {title} {Course of {{Theoretical Physics}}: {{Vol}}.
  3, {{Quantum Mechanics}}: {{Non-Relativistic Theory}}, 3rd Edition},\ \href
  {https://www.amazon.com/gp/product/0080291406/ref=x_gr_w_bb_glide_sout?ie=UTF8&tag=x_gr_w_bb_glide_sout-20&linkCode=as2&camp=1789&creative=9325&creativeASIN=0080291406&SubscriptionId=1MGPYB6YW3HWK55XCGG2}
  {\bibfield  {journal} {\bibinfo  {journal} {Butterworth-Heinemann}\ }
  (\bibinfo {year} {1981})}\BibitemShut {NoStop}%
\bibitem [{\citenamefont {Su}\ \emph {et~al.}(1994)\citenamefont {Su},
  \citenamefont {Chen},\ and\ \citenamefont {Ge}}]{su1994temperature}%
  \BibitemOpen
  \bibfield  {author} {\bibinfo {author} {\bibfnamefont {G.}~\bibnamefont
  {Su}}, \bibinfo {author} {\bibfnamefont {B.}~\bibnamefont {Chen}},\ and\
  \bibinfo {author} {\bibfnamefont {M.}~\bibnamefont {Ge}},\ }\bibinfo {title}
  {{Temperature Dependence of the Coercivity for Some Intermetallic Magnetic
  Particles at Low Temperatures}},\ \href
  {https://api.wiley.com/onlinelibrary/tdm/v1/articles/10.1002%2Fpssb.2221810132}
  {\bibfield  {journal} {\bibinfo  {journal} {Phys. Status Solidi B}\ }\textbf
  {\bibinfo {volume} {181}},\ \bibinfo {pages} {K33} (\bibinfo {year}
  {1994})}\BibitemShut {NoStop}%
\bibitem [{\citenamefont {{Derras-Chouk}}\ \emph {et~al.}(2022)\citenamefont
  {{Derras-Chouk}}, \citenamefont {Chudnovsky},\ and\ \citenamefont
  {Garanin}}]{derras-chouk2022dynamics}%
  \BibitemOpen
  \bibfield  {author} {\bibinfo {author} {\bibfnamefont {A.}~\bibnamefont
  {{Derras-Chouk}}}, \bibinfo {author} {\bibfnamefont {E.~M.}\ \bibnamefont
  {Chudnovsky}},\ and\ \bibinfo {author} {\bibfnamefont {D.~A.}\ \bibnamefont
  {Garanin}},\ }\bibinfo {title} {Dynamics of the Collapse of a Ferromagnetic
  Skyrmion in a Centrosymmetric Lattice},\ \href
  {https://link.aps.org/article/10.1103/PhysRevB.105.134432} {\bibfield
  {journal} {\bibinfo  {journal} {Phys. Rev. B}\ }\textbf {\bibinfo {volume}
  {105}},\ \bibinfo {pages} {134432} (\bibinfo {year} {2022})}\BibitemShut
  {NoStop}%
\bibitem [{\citenamefont {Dasgupta}\ and\ \citenamefont
  {Tchernyshyov}(2018)}]{dasgupta2018energymomentum}%
  \BibitemOpen
  \bibfield  {author} {\bibinfo {author} {\bibfnamefont {S.}~\bibnamefont
  {Dasgupta}}\ and\ \bibinfo {author} {\bibfnamefont {O.}~\bibnamefont
  {Tchernyshyov}},\ }\bibinfo {title} {Energy-Momentum Tensor of a
  Ferromagnet},\ \href
  {https://link.aps.org/accepted/10.1103/PhysRevB.98.224401} {\bibfield
  {journal} {\bibinfo  {journal} {Phys. Rev. B}\ }\textbf {\bibinfo {volume}
  {98}},\ \bibinfo {pages} {224401} (\bibinfo {year} {2018})}\BibitemShut
  {NoStop}%
\bibitem [{\citenamefont {Thiaville}\ \emph {et~al.}(2005)\citenamefont
  {Thiaville}, \citenamefont {Nakatani}, \citenamefont {Miltat},\ and\
  \citenamefont {Suzuki}}]{thiaville2005micromagnetic}%
  \BibitemOpen
  \bibfield  {author} {\bibinfo {author} {\bibfnamefont {A.}~\bibnamefont
  {Thiaville}}, \bibinfo {author} {\bibfnamefont {Y.}~\bibnamefont {Nakatani}},
  \bibinfo {author} {\bibfnamefont {J.}~\bibnamefont {Miltat}},\ and\ \bibinfo
  {author} {\bibfnamefont {Y.}~\bibnamefont {Suzuki}},\ }\bibinfo {title}
  {Micromagnetic Understanding of Current-Driven Domain Wall Motion in
  Patterned Nanowires},\ \href
  {https://iopscience.iop.org/article/10.1209/epl/i2004-10452-6} {\bibfield
  {journal} {\bibinfo  {journal} {Europhys. Lett.}\ }\textbf {\bibinfo {volume}
  {69}},\ \bibinfo {pages} {990} (\bibinfo {year} {2005})}\BibitemShut
  {NoStop}%
\bibitem [{\citenamefont {Jeudy}\ \emph {et~al.}(2016)\citenamefont {Jeudy},
  \citenamefont {Mougin}, \citenamefont {Bustingorry}, \citenamefont
  {Savero~Torres}, \citenamefont {Gorchon}, \citenamefont {Kolton},
  \citenamefont {Lema{\^i}tre},\ and\ \citenamefont
  {Jamet}}]{jeudy2016universal}%
  \BibitemOpen
  \bibfield  {author} {\bibinfo {author} {\bibfnamefont {V.}~\bibnamefont
  {Jeudy}}, \bibinfo {author} {\bibfnamefont {A.}~\bibnamefont {Mougin}},
  \bibinfo {author} {\bibfnamefont {S.}~\bibnamefont {Bustingorry}}, \bibinfo
  {author} {\bibfnamefont {W.}~\bibnamefont {Savero~Torres}}, \bibinfo {author}
  {\bibfnamefont {J.}~\bibnamefont {Gorchon}}, \bibinfo {author} {\bibfnamefont
  {A.~B.}\ \bibnamefont {Kolton}}, \bibinfo {author} {\bibfnamefont
  {A.}~\bibnamefont {Lema{\^i}tre}},\ and\ \bibinfo {author} {\bibfnamefont
  {J.-P.}\ \bibnamefont {Jamet}},\ }\bibinfo {title} {Universal {{Pinning
  Energy Barrier}} for {{Driven Domain Walls}} in {{Thin Ferromagnetic
  Films}}},\ \href {https://link.aps.org/doi/10.1103/PhysRevLett.117.057201}
  {\bibfield  {journal} {\bibinfo  {journal} {Phys. Rev. Lett.}\ }\textbf
  {\bibinfo {volume} {117}},\ \bibinfo {pages} {057201} (\bibinfo {year}
  {2016})}\BibitemShut {NoStop}%
\bibitem [{\citenamefont {Eltschka}\ \emph {et~al.}(2010)\citenamefont
  {Eltschka}, \citenamefont {W{\"o}tzel}, \citenamefont {Rhensius},
  \citenamefont {Krzyk}, \citenamefont {Nowak}, \citenamefont {Kl{\"a}ui},
  \citenamefont {Kasama}, \citenamefont {{Dunin-Borkowski}}, \citenamefont
  {Heyderman}, \citenamefont {{van Driel}},\ and\ \citenamefont
  {Duine}}]{eltschka2010nonadiabatic}%
  \BibitemOpen
  \bibfield  {author} {\bibinfo {author} {\bibfnamefont {M.}~\bibnamefont
  {Eltschka}}, \bibinfo {author} {\bibfnamefont {M.}~\bibnamefont
  {W{\"o}tzel}}, \bibinfo {author} {\bibfnamefont {J.}~\bibnamefont
  {Rhensius}}, \bibinfo {author} {\bibfnamefont {S.}~\bibnamefont {Krzyk}},
  \bibinfo {author} {\bibfnamefont {U.}~\bibnamefont {Nowak}}, \bibinfo
  {author} {\bibfnamefont {M.}~\bibnamefont {Kl{\"a}ui}}, \bibinfo {author}
  {\bibfnamefont {T.}~\bibnamefont {Kasama}}, \bibinfo {author} {\bibfnamefont
  {R.~E.}\ \bibnamefont {{Dunin-Borkowski}}}, \bibinfo {author} {\bibfnamefont
  {L.~J.}\ \bibnamefont {Heyderman}}, \bibinfo {author} {\bibfnamefont {H.~J.}\
  \bibnamefont {{van Driel}}},\ and\ \bibinfo {author} {\bibfnamefont {R.~A.}\
  \bibnamefont {Duine}},\ }\bibinfo {title} {Nonadiabatic {{Spin Torque
  Investigated Using Thermally Activated Magnetic Domain Wall Dynamics}}},\
  \href {http://link.aps.org/article/10.1103/PhysRevLett.105.056601} {\bibfield
   {journal} {\bibinfo  {journal} {Phys. Rev. Lett.}\ }\textbf {\bibinfo
  {volume} {105}},\ \bibinfo {pages} {056601} (\bibinfo {year}
  {2010})}\BibitemShut {NoStop}%
\bibitem [{\citenamefont {Kim}\ \emph {et~al.}(2013{\natexlab{b}})\citenamefont
  {Kim}, \citenamefont {Moon}, \citenamefont {Yoo}, \citenamefont {Min},
  \citenamefont {Shin},\ and\ \citenamefont {Choe}}]{kim2013method}%
  \BibitemOpen
  \bibfield  {author} {\bibinfo {author} {\bibfnamefont {D.-H.}\ \bibnamefont
  {Kim}}, \bibinfo {author} {\bibfnamefont {K.-W.}\ \bibnamefont {Moon}},
  \bibinfo {author} {\bibfnamefont {S.-C.}\ \bibnamefont {Yoo}}, \bibinfo
  {author} {\bibfnamefont {B.-C.}\ \bibnamefont {Min}}, \bibinfo {author}
  {\bibfnamefont {K.-H.}\ \bibnamefont {Shin}},\ and\ \bibinfo {author}
  {\bibfnamefont {S.-B.}\ \bibnamefont {Choe}},\ }\bibinfo {title} {A
  {{Method}} for {{Compensating}} the {{Joule-Heating Effects}} in
  {{Current-Induced Domain Wall Motion}}},\ \href
  {http://xplorestaging.ieee.org/ielx7/20/6558881/06559072.pdf?arnumber=6559072}
  {\bibfield  {journal} {\bibinfo  {journal} {IEEE Trans. Magn.}\ }\textbf
  {\bibinfo {volume} {49}},\ \bibinfo {pages} {3207} (\bibinfo {year}
  {2013}{\natexlab{b}})}\BibitemShut {NoStop}%
\bibitem [{\citenamefont {Litzius}\ \emph {et~al.}(2020)\citenamefont
  {Litzius}, \citenamefont {Leliaert}, \citenamefont {Bassirian}, \citenamefont
  {Rodrigues}, \citenamefont {Kromin}, \citenamefont {Lemesh}, \citenamefont
  {Zazvorka}, \citenamefont {Lee}, \citenamefont {Mulkers}, \citenamefont
  {Kerber}, \citenamefont {Heinze}, \citenamefont {Keil}, \citenamefont
  {Reeve}, \citenamefont {Weigand}, \citenamefont {Van~Waeyenberge},
  \citenamefont {Sch{\"u}tz}, \citenamefont {{Everschor-Sitte}}, \citenamefont
  {Beach},\ and\ \citenamefont {Kl{\"a}ui}}]{litzius2020role}%
  \BibitemOpen
  \bibfield  {author} {\bibinfo {author} {\bibfnamefont {K.}~\bibnamefont
  {Litzius}}, \bibinfo {author} {\bibfnamefont {J.}~\bibnamefont {Leliaert}},
  \bibinfo {author} {\bibfnamefont {P.}~\bibnamefont {Bassirian}}, \bibinfo
  {author} {\bibfnamefont {D.}~\bibnamefont {Rodrigues}}, \bibinfo {author}
  {\bibfnamefont {S.}~\bibnamefont {Kromin}}, \bibinfo {author} {\bibfnamefont
  {I.}~\bibnamefont {Lemesh}}, \bibinfo {author} {\bibfnamefont
  {J.}~\bibnamefont {Zazvorka}}, \bibinfo {author} {\bibfnamefont {K.-J.}\
  \bibnamefont {Lee}}, \bibinfo {author} {\bibfnamefont {J.}~\bibnamefont
  {Mulkers}}, \bibinfo {author} {\bibfnamefont {N.}~\bibnamefont {Kerber}},
  \bibinfo {author} {\bibfnamefont {D.}~\bibnamefont {Heinze}}, \bibinfo
  {author} {\bibfnamefont {N.}~\bibnamefont {Keil}}, \bibinfo {author}
  {\bibfnamefont {R.~M.}\ \bibnamefont {Reeve}}, \bibinfo {author}
  {\bibfnamefont {M.}~\bibnamefont {Weigand}}, \bibinfo {author} {\bibfnamefont
  {B.}~\bibnamefont {Van~Waeyenberge}}, \bibinfo {author} {\bibfnamefont
  {G.}~\bibnamefont {Sch{\"u}tz}}, \bibinfo {author} {\bibfnamefont
  {K.}~\bibnamefont {{Everschor-Sitte}}}, \bibinfo {author} {\bibfnamefont
  {G.~S.~D.}\ \bibnamefont {Beach}},\ and\ \bibinfo {author} {\bibfnamefont
  {M.}~\bibnamefont {Kl{\"a}ui}},\ }\bibinfo {title} {The Role of Temperature
  and Drive Current in Skyrmion Dynamics},\ \href
  {http://www.nature.com/articles/s41928-019-0359-2} {\bibfield  {journal}
  {\bibinfo  {journal} {Nat. Electron.}\ }\textbf {\bibinfo {volume} {3}},\
  \bibinfo {pages} {30} (\bibinfo {year} {2020})}\BibitemShut {NoStop}%
\bibitem [{\citenamefont {Freimuth}\ \emph {et~al.}(2015)\citenamefont
  {Freimuth}, \citenamefont {Bl{\"u}gel},\ and\ \citenamefont
  {Mokrousov}}]{freimuth2015direct}%
  \BibitemOpen
  \bibfield  {author} {\bibinfo {author} {\bibfnamefont {F.}~\bibnamefont
  {Freimuth}}, \bibinfo {author} {\bibfnamefont {S.}~\bibnamefont
  {Bl{\"u}gel}},\ and\ \bibinfo {author} {\bibfnamefont {Y.}~\bibnamefont
  {Mokrousov}},\ }\bibinfo {title} {Direct and Inverse Spin-Orbit Torques},\
  \href {http://link.aps.org/article/10.1103/PhysRevB.92.064415} {\bibfield
  {journal} {\bibinfo  {journal} {Phys. Rev. B}\ }\textbf {\bibinfo {volume}
  {92}},\ \bibinfo {pages} {19} (\bibinfo {year} {2015})}\BibitemShut {NoStop}%
\bibitem [{\citenamefont {Zhang}\ \emph {et~al.}(2020)\citenamefont {Zhang},
  \citenamefont {Wang}, \citenamefont {Jin}, \citenamefont {Zeng},
  \citenamefont {Xia}, \citenamefont {Wang},\ and\ \citenamefont
  {Liu}}]{zhang2020spin}%
  \BibitemOpen
  \bibfield  {author} {\bibinfo {author} {\bibfnamefont {C.}~\bibnamefont
  {Zhang}}, \bibinfo {author} {\bibfnamefont {J.}~\bibnamefont {Wang}},
  \bibinfo {author} {\bibfnamefont {C.}~\bibnamefont {Jin}}, \bibinfo {author}
  {\bibfnamefont {Z.}~\bibnamefont {Zeng}}, \bibinfo {author} {\bibfnamefont
  {H.}~\bibnamefont {Xia}}, \bibinfo {author} {\bibfnamefont {J.}~\bibnamefont
  {Wang}},\ and\ \bibinfo {author} {\bibfnamefont {Q.}~\bibnamefont {Liu}},\
  }\bibinfo {title} {Spin Current Pumped by Confined Breathing Skyrmion},\
  \href {https://iopscience.iop.org/article/10.1088/1367-2630/ab83d6}
  {\bibfield  {journal} {\bibinfo  {journal} {New J. Phys.}\ }\textbf {\bibinfo
  {volume} {22}},\ \bibinfo {pages} {053029} (\bibinfo {year}
  {2020})}\BibitemShut {NoStop}%
\bibitem [{\citenamefont {Reiss}\ and\ \citenamefont
  {Brouwer}(2022)}]{reiss2022finitefrequency}%
  \BibitemOpen
  \bibfield  {author} {\bibinfo {author} {\bibfnamefont {D.~A.}\ \bibnamefont
  {Reiss}}\ and\ \bibinfo {author} {\bibfnamefont {P.~W.}\ \bibnamefont
  {Brouwer}},\ }\bibinfo {title} {Finite-Frequency Spin Conductance of the
  Interface between a Ferro- or Ferrimagnetic Insulator and a Normal Metal},\
  \href {https://link.aps.org/doi/10.1103/PhysRevB.106.144423} {\bibfield
  {journal} {\bibinfo  {journal} {Phys. Rev. B}\ }\textbf {\bibinfo {volume}
  {106}},\ \bibinfo {pages} {144423} (\bibinfo {year} {2022})}\BibitemShut
  {NoStop}%
\bibitem [{\citenamefont {Bauer}\ \emph {et~al.}(2010)\citenamefont {Bauer},
  \citenamefont {Bretzel}, \citenamefont {Brataas},\ and\ \citenamefont
  {Tserkovnyak}}]{bauer2010nanoscale}%
  \BibitemOpen
  \bibfield  {author} {\bibinfo {author} {\bibfnamefont {G.~E.~W.}\
  \bibnamefont {Bauer}}, \bibinfo {author} {\bibfnamefont {S.}~\bibnamefont
  {Bretzel}}, \bibinfo {author} {\bibfnamefont {A.}~\bibnamefont {Brataas}},\
  and\ \bibinfo {author} {\bibfnamefont {Y.}~\bibnamefont {Tserkovnyak}},\
  }\bibinfo {title} {Nanoscale Magnetic Heat Pumps and Engines},\ \href
  {http://link.aps.org/article/10.1103/PhysRevB.81.024427} {\bibfield
  {journal} {\bibinfo  {journal} {Phys. Rev. B}\ }\textbf {\bibinfo {volume}
  {81}},\ \bibinfo {pages} {11} (\bibinfo {year} {2010})}\BibitemShut {NoStop}%
\bibitem [{\citenamefont {Wong}\ \emph {et~al.}(2012)\citenamefont {Wong},
  \citenamefont {{van Driel}}, \citenamefont {Kittinaradorn}, \citenamefont
  {Stoof},\ and\ \citenamefont {Duine}}]{wong2012spin}%
  \BibitemOpen
  \bibfield  {author} {\bibinfo {author} {\bibfnamefont {C.~H.}\ \bibnamefont
  {Wong}}, \bibinfo {author} {\bibfnamefont {H.~J.}\ \bibnamefont {{van
  Driel}}}, \bibinfo {author} {\bibfnamefont {R.}~\bibnamefont
  {Kittinaradorn}}, \bibinfo {author} {\bibfnamefont {H.~T.~C.}\ \bibnamefont
  {Stoof}},\ and\ \bibinfo {author} {\bibfnamefont {R.~A.}\ \bibnamefont
  {Duine}},\ }\bibinfo {title} {Spin {{Caloritronics}} in {{Noncondensed Bose
  Gases}}},\ \href {https://link.aps.org/doi/10.1103/PhysRevLett.108.075301}
  {\bibfield  {journal} {\bibinfo  {journal} {Phys. Rev. Lett.}\ }\textbf
  {\bibinfo {volume} {108}},\ \bibinfo {pages} {075301} (\bibinfo {year}
  {2012})}\BibitemShut {NoStop}%
\bibitem [{\citenamefont {Huang}\ \emph {et~al.}(2015)\citenamefont {Huang},
  \citenamefont {Lai}, \citenamefont {Hou},\ and\ \citenamefont
  {Wei}}]{huang2015influence}%
  \BibitemOpen
  \bibfield  {author} {\bibinfo {author} {\bibfnamefont {H.-T.}\ \bibnamefont
  {Huang}}, \bibinfo {author} {\bibfnamefont {M.-F.}\ \bibnamefont {Lai}},
  \bibinfo {author} {\bibfnamefont {Y.-F.}\ \bibnamefont {Hou}},\ and\ \bibinfo
  {author} {\bibfnamefont {Z.-H.}\ \bibnamefont {Wei}},\ }\bibinfo {title}
  {Influence of {{Magnetic Domain Walls}} and {{Magnetic Field}} on the
  {{Thermal Conductivity}} of {{Magnetic Nanowires}}},\ \href
  {https://pubs.acs.org/doi/10.1021/nl502577y} {\bibfield  {journal} {\bibinfo
  {journal} {Nano Lett.}\ }\textbf {\bibinfo {volume} {15}},\ \bibinfo {pages}
  {2773} (\bibinfo {year} {2015})}\BibitemShut {NoStop}%
\bibitem [{\citenamefont {Boulle}\ \emph {et~al.}(2014)\citenamefont {Boulle},
  \citenamefont {{Buda-Prejbeanu}}, \citenamefont {Ju{\'e}}, \citenamefont
  {Miron},\ and\ \citenamefont {Gaudin}}]{boulle2014current}%
  \BibitemOpen
  \bibfield  {author} {\bibinfo {author} {\bibfnamefont {O.}~\bibnamefont
  {Boulle}}, \bibinfo {author} {\bibfnamefont {L.~D.}\ \bibnamefont
  {{Buda-Prejbeanu}}}, \bibinfo {author} {\bibfnamefont {E.}~\bibnamefont
  {Ju{\'e}}}, \bibinfo {author} {\bibfnamefont {I.~M.}\ \bibnamefont {Miron}},\
  and\ \bibinfo {author} {\bibfnamefont {G.}~\bibnamefont {Gaudin}},\ }\bibinfo
  {title} {Current Induced Domain Wall Dynamics in the Presence of Spin Orbit
  Torques},\ \href {http://aip.scitation.org/doi/10.1063/1.4860946} {\bibfield
  {journal} {\bibinfo  {journal} {J. Appl. Phys.}\ }\textbf {\bibinfo {volume}
  {115}},\ \bibinfo {pages} {17D502} (\bibinfo {year} {2014})}\BibitemShut
  {NoStop}%
\bibitem [{\citenamefont {Saitoh}\ \emph {et~al.}(2004)\citenamefont {Saitoh},
  \citenamefont {Miyajima}, \citenamefont {Yamaoka},\ and\ \citenamefont
  {Tatara}}]{saitoh2004currentinduced}%
  \BibitemOpen
  \bibfield  {author} {\bibinfo {author} {\bibfnamefont {E.}~\bibnamefont
  {Saitoh}}, \bibinfo {author} {\bibfnamefont {H.}~\bibnamefont {Miyajima}},
  \bibinfo {author} {\bibfnamefont {T.}~\bibnamefont {Yamaoka}},\ and\ \bibinfo
  {author} {\bibfnamefont {G.}~\bibnamefont {Tatara}},\ }\bibinfo {title}
  {Current-Induced Resonance and Mass Determination of a Single Magnetic Domain
  Wall},\ \href {http://www.nature.com/articles/nature03009} {\bibfield
  {journal} {\bibinfo  {journal} {Nature}\ }\textbf {\bibinfo {volume} {432}},\
  \bibinfo {pages} {203} (\bibinfo {year} {2004})}\BibitemShut {NoStop}%
\bibitem [{\citenamefont {Tatara}\ and\ \citenamefont
  {Fukuyama}(1997)}]{tatara1997resistivity}%
  \BibitemOpen
  \bibfield  {author} {\bibinfo {author} {\bibfnamefont {G.}~\bibnamefont
  {Tatara}}\ and\ \bibinfo {author} {\bibfnamefont {H.}~\bibnamefont
  {Fukuyama}},\ }\bibinfo {title} {Resistivity Due to a {{Domain Wall}} in
  {{Ferromagnetic Metal}}},\ \href
  {http://link.aps.org/article/10.1103/PhysRevLett.78.3773} {\bibfield
  {journal} {\bibinfo  {journal} {Phys. Rev. Lett.}\ }\textbf {\bibinfo
  {volume} {78}},\ \bibinfo {pages} {3773} (\bibinfo {year}
  {1997})}\BibitemShut {NoStop}%
\bibitem [{\citenamefont {Tatara}(2000)}]{tatara2000domain}%
  \BibitemOpen
  \bibfield  {author} {\bibinfo {author} {\bibfnamefont {G.}~\bibnamefont
  {Tatara}},\ }\bibinfo {title} {Domain {{Wall Resistance Based}} on
  {{Landauer}}'s {{Formula}}},\ \href
  {https://journals.jps.jp/doi/pdf/10.1143/JPSJ.69.2969} {\bibfield  {journal}
  {\bibinfo  {journal} {J. Phys. Soc. Jpn.}\ }\textbf {\bibinfo {volume}
  {69}},\ \bibinfo {pages} {2969} (\bibinfo {year} {2000})}\BibitemShut
  {NoStop}%
\bibitem [{\citenamefont {TATARA}(2001)}]{tatara2001domain}%
  \BibitemOpen
  \bibfield  {author} {\bibinfo {author} {\bibfnamefont {{\relax
  GEN}.}~\bibnamefont {TATARA}},\ }\bibinfo {title} {{{DOMAIN WALL RESISTIVITY
  BASED ON A LINEAR RESPONSE THEORY}}},\ \href
  {https://www.worldscientific.com/doi/pdf/10.1142/S0217979201002540}
  {\bibfield  {journal} {\bibinfo  {journal} {Int. J. Mod. Phys. B}\ }\textbf
  {\bibinfo {volume} {15}},\ \bibinfo {pages} {321} (\bibinfo {year}
  {2001})}\BibitemShut {NoStop}%
\bibitem [{\citenamefont {Lucassen}\ \emph {et~al.}(2011)\citenamefont
  {Lucassen}, \citenamefont {Wong}, \citenamefont {Duine},\ and\ \citenamefont
  {Tserkovnyak}}]{lucassen2011spintransfer}%
  \BibitemOpen
  \bibfield  {author} {\bibinfo {author} {\bibfnamefont {M.~E.}\ \bibnamefont
  {Lucassen}}, \bibinfo {author} {\bibfnamefont {C.~H.}\ \bibnamefont {Wong}},
  \bibinfo {author} {\bibfnamefont {R.~A.}\ \bibnamefont {Duine}},\ and\
  \bibinfo {author} {\bibfnamefont {Y.}~\bibnamefont {Tserkovnyak}},\ }\bibinfo
  {title} {Spin-Transfer Mechanism for Magnon-Drag Thermopower},\ \href
  {http://aip.scitation.org/doi/10.1063/1.3672207} {\bibfield  {journal}
  {\bibinfo  {journal} {Appl. Phys. Lett.}\ }\textbf {\bibinfo {volume} {99}},\
  \bibinfo {pages} {262506} (\bibinfo {year} {2011})}\BibitemShut {NoStop}%
\bibitem [{\citenamefont {Tian}\ \emph {et~al.}(2022)\citenamefont {Tian},
  \citenamefont {Zhang}, \citenamefont {Xiao}, \citenamefont {Gamage},
  \citenamefont {Tian}, \citenamefont {Yue}, \citenamefont {Hadjiev},
  \citenamefont {Bao}, \citenamefont {Ren}, \citenamefont {Liang},\ and\
  \citenamefont {Zhao}}]{tian2022ultraweak}%
  \BibitemOpen
  \bibfield  {author} {\bibinfo {author} {\bibfnamefont {Z.~Y.}\ \bibnamefont
  {Tian}}, \bibinfo {author} {\bibfnamefont {Q.~Y.}\ \bibnamefont {Zhang}},
  \bibinfo {author} {\bibfnamefont {Y.~W.}\ \bibnamefont {Xiao}}, \bibinfo
  {author} {\bibfnamefont {G.~A.}\ \bibnamefont {Gamage}}, \bibinfo {author}
  {\bibfnamefont {F.}~\bibnamefont {Tian}}, \bibinfo {author} {\bibfnamefont
  {S.}~\bibnamefont {Yue}}, \bibinfo {author} {\bibfnamefont {V.~G.}\
  \bibnamefont {Hadjiev}}, \bibinfo {author} {\bibfnamefont {J.}~\bibnamefont
  {Bao}}, \bibinfo {author} {\bibfnamefont {Z.}~\bibnamefont {Ren}}, \bibinfo
  {author} {\bibfnamefont {E.}~\bibnamefont {Liang}},\ and\ \bibinfo {author}
  {\bibfnamefont {J.}~\bibnamefont {Zhao}},\ }\bibinfo {title} {Ultraweak
  Electron-Phonon Coupling Strength in Cubic Boron Arsenide Unveiled by
  Ultrafast Dynamics},\ \href
  {https://link.aps.org/article/10.1103/PhysRevB.105.174306} {\bibfield
  {journal} {\bibinfo  {journal} {Phys. Rev. B}\ }\textbf {\bibinfo {volume}
  {105}},\ \bibinfo {pages} {11} (\bibinfo {year} {2022})}\BibitemShut
  {NoStop}%
\bibitem [{\citenamefont {Brechet}\ \emph {et~al.}(2013)\citenamefont
  {Brechet}, \citenamefont {Vetro}, \citenamefont {Papa}, \citenamefont
  {Barnes},\ and\ \citenamefont {Ansermet}}]{brechet2013evidence}%
  \BibitemOpen
  \bibfield  {author} {\bibinfo {author} {\bibfnamefont {S.~D.}\ \bibnamefont
  {Brechet}}, \bibinfo {author} {\bibfnamefont {F.~A.}\ \bibnamefont {Vetro}},
  \bibinfo {author} {\bibfnamefont {E.}~\bibnamefont {Papa}}, \bibinfo {author}
  {\bibfnamefont {S.~E.}\ \bibnamefont {Barnes}},\ and\ \bibinfo {author}
  {\bibfnamefont {J.-P.}\ \bibnamefont {Ansermet}},\ }\bibinfo {title}
  {Evidence for a {{Magnetic Seebeck Effect}}},\ \href
  {https://link.aps.org/doi/10.1103/PhysRevLett.111.087205} {\bibfield
  {journal} {\bibinfo  {journal} {Phys. Rev. Lett.}\ }\textbf {\bibinfo
  {volume} {111}},\ \bibinfo {pages} {087205} (\bibinfo {year}
  {2013})}\BibitemShut {NoStop}%
\bibitem [{\citenamefont {Nikitin}\ \emph {et~al.}(2022)\citenamefont
  {Nikitin}, \citenamefont {F{\aa}k}, \citenamefont {Kr{\"a}mer}, \citenamefont
  {Fennell}, \citenamefont {Normand}, \citenamefont {L{\"a}uchli},\ and\
  \citenamefont {R{\"u}egg}}]{nikitin2022thermal}%
  \BibitemOpen
  \bibfield  {author} {\bibinfo {author} {\bibfnamefont {S.~E.}\ \bibnamefont
  {Nikitin}}, \bibinfo {author} {\bibfnamefont {B.}~\bibnamefont {F{\aa}k}},
  \bibinfo {author} {\bibfnamefont {K.~W.}\ \bibnamefont {Kr{\"a}mer}},
  \bibinfo {author} {\bibfnamefont {T.}~\bibnamefont {Fennell}}, \bibinfo
  {author} {\bibfnamefont {B.}~\bibnamefont {Normand}}, \bibinfo {author}
  {\bibfnamefont {A.~M.}\ \bibnamefont {L{\"a}uchli}},\ and\ \bibinfo {author}
  {\bibfnamefont {{\relax Ch}.}~\bibnamefont {R{\"u}egg}},\ }\bibinfo {title}
  {Thermal {{Evolution}} of {{Dirac Magnons}} in the {{Honeycomb Ferromagnet}}
  {{CrBr}}{\textsubscript{3}}},\ \href
  {https://link.aps.org/doi/10.1103/PhysRevLett.129.127201} {\bibfield
  {journal} {\bibinfo  {journal} {Phys. Rev. Lett.}\ }\textbf {\bibinfo
  {volume} {129}},\ \bibinfo {pages} {127201} (\bibinfo {year}
  {2022})}\BibitemShut {NoStop}%
\bibitem [{\citenamefont {Cai}\ \emph {et~al.}(2020)\citenamefont {Cai},
  \citenamefont {Zhu}, \citenamefont {Lee}, \citenamefont {Mishra},
  \citenamefont {Ren}, \citenamefont {Pollard}, \citenamefont {He},
  \citenamefont {Liang}, \citenamefont {Teo},\ and\ \citenamefont
  {Yang}}]{cai2020ultrafast}%
  \BibitemOpen
  \bibfield  {author} {\bibinfo {author} {\bibfnamefont {K.}~\bibnamefont
  {Cai}}, \bibinfo {author} {\bibfnamefont {Z.}~\bibnamefont {Zhu}}, \bibinfo
  {author} {\bibfnamefont {J.~M.}\ \bibnamefont {Lee}}, \bibinfo {author}
  {\bibfnamefont {R.}~\bibnamefont {Mishra}}, \bibinfo {author} {\bibfnamefont
  {L.}~\bibnamefont {Ren}}, \bibinfo {author} {\bibfnamefont {S.~D.}\
  \bibnamefont {Pollard}}, \bibinfo {author} {\bibfnamefont {P.}~\bibnamefont
  {He}}, \bibinfo {author} {\bibfnamefont {G.}~\bibnamefont {Liang}}, \bibinfo
  {author} {\bibfnamefont {K.~L.}\ \bibnamefont {Teo}},\ and\ \bibinfo {author}
  {\bibfnamefont {H.}~\bibnamefont {Yang}},\ }\bibinfo {title} {Ultrafast and
  Energy-Efficient Spin\textendash Orbit Torque Switching in Compensated
  Ferrimagnets},\ \href {https://www.nature.com/articles/s41928-019-0345-8}
  {\bibfield  {journal} {\bibinfo  {journal} {Nat. Electron.}\ }\textbf
  {\bibinfo {volume} {3}},\ \bibinfo {pages} {37} (\bibinfo {year}
  {2020})}\BibitemShut {NoStop}%
\bibitem [{\citenamefont {Khoshlahni}\ \emph {et~al.}(2019)\citenamefont
  {Khoshlahni}, \citenamefont {Qaiumzadeh}, \citenamefont {Bergman},\ and\
  \citenamefont {Brataas}}]{khoshlahni2019ultrafasta}%
  \BibitemOpen
  \bibfield  {author} {\bibinfo {author} {\bibfnamefont {R.}~\bibnamefont
  {Khoshlahni}}, \bibinfo {author} {\bibfnamefont {A.}~\bibnamefont
  {Qaiumzadeh}}, \bibinfo {author} {\bibfnamefont {A.}~\bibnamefont
  {Bergman}},\ and\ \bibinfo {author} {\bibfnamefont {A.}~\bibnamefont
  {Brataas}},\ }\bibinfo {title} {Ultrafast Generation and Dynamics of Isolated
  Skyrmions in Antiferromagnetic Insulators},\ \href
  {https://link.aps.org/article/10.1103/PhysRevB.99.054423} {\bibfield
  {journal} {\bibinfo  {journal} {Phys. Rev. B}\ }\textbf {\bibinfo {volume}
  {99}},\ \bibinfo {pages} {054423} (\bibinfo {year} {2019})}\BibitemShut
  {NoStop}%
\bibitem [{\citenamefont {Lin}\ \emph {et~al.}(2014)\citenamefont {Lin},
  \citenamefont {Batista}, \citenamefont {Reichhardt},\ and\ \citenamefont
  {Saxena}}]{lin2014aca}%
  \BibitemOpen
  \bibfield  {author} {\bibinfo {author} {\bibfnamefont {S.-Z.}\ \bibnamefont
  {Lin}}, \bibinfo {author} {\bibfnamefont {C.~D.}\ \bibnamefont {Batista}},
  \bibinfo {author} {\bibfnamefont {C.}~\bibnamefont {Reichhardt}},\ and\
  \bibinfo {author} {\bibfnamefont {A.}~\bibnamefont {Saxena}},\ }\bibinfo
  {title} {Ac {{Current Generation}} in {{Chiral Magnetic Insulators}} and
  {{Skyrmion Motion}} Induced by the {{Spin Seebeck Effect}}},\ \href
  {https://link.aps.org/doi/10.1103/PhysRevLett.112.187203} {\bibfield
  {journal} {\bibinfo  {journal} {Phys. Rev. Lett.}\ }\textbf {\bibinfo
  {volume} {112}},\ \bibinfo {pages} {187203} (\bibinfo {year}
  {2014})}\BibitemShut {NoStop}%
\bibitem [{\citenamefont {Wang}\ \emph
  {et~al.}(2022{\natexlab{b}})\citenamefont {Wang}, \citenamefont {Yang},
  \citenamefont {Wang},\ and\ \citenamefont {Su}}]{wang2022domainwalla}%
  \BibitemOpen
  \bibfield  {author} {\bibinfo {author} {\bibfnamefont {Y.-R.}\ \bibnamefont
  {Wang}}, \bibinfo {author} {\bibfnamefont {C.}~\bibnamefont {Yang}}, \bibinfo
  {author} {\bibfnamefont {Z.-C.}\ \bibnamefont {Wang}},\ and\ \bibinfo
  {author} {\bibfnamefont {G.}~\bibnamefont {Su}},\ }\bibinfo {title}
  {Domain-Wall Dynamics Driven by Thermal and Electrical Spin-Transfer
  Torque},\ \href {https://link.aps.org/doi/10.1103/PhysRevB.106.054432}
  {\bibfield  {journal} {\bibinfo  {journal} {Phys. Rev. B}\ }\textbf {\bibinfo
  {volume} {106}},\ \bibinfo {pages} {054432} (\bibinfo {year}
  {2022}{\natexlab{b}})}\BibitemShut {NoStop}%
\bibitem [{\citenamefont {Faridi}\ \emph
  {et~al.}(2022{\natexlab{b}})\citenamefont {Faridi}, \citenamefont {Kim},\
  and\ \citenamefont {Vignale}}]{faridi2022atomicscalea}%
  \BibitemOpen
  \bibfield  {author} {\bibinfo {author} {\bibfnamefont {E.}~\bibnamefont
  {Faridi}}, \bibinfo {author} {\bibfnamefont {S.~K.}\ \bibnamefont {Kim}},\
  and\ \bibinfo {author} {\bibfnamefont {G.}~\bibnamefont {Vignale}},\
  }\bibinfo {title} {Atomic-Scale Spin-Wave Polarizer Based on a Sharp
  Antiferromagnetic Domain Wall},\ \href
  {https://link.aps.org/doi/10.1103/PhysRevB.106.094411} {\bibfield  {journal}
  {\bibinfo  {journal} {Phys. Rev. B}\ }\textbf {\bibinfo {volume} {106}},\
  \bibinfo {pages} {094411} (\bibinfo {year} {2022}{\natexlab{b}})}\BibitemShut
  {NoStop}%
\bibitem [{\citenamefont {Thiaville}\ \emph {et~al.}(2012)\citenamefont
  {Thiaville}, \citenamefont {Rohart}, \citenamefont {Ju{\'e}}, \citenamefont
  {Cros},\ and\ \citenamefont {Fert}}]{thiaville2012dynamics}%
  \BibitemOpen
  \bibfield  {author} {\bibinfo {author} {\bibfnamefont {A.}~\bibnamefont
  {Thiaville}}, \bibinfo {author} {\bibfnamefont {S.}~\bibnamefont {Rohart}},
  \bibinfo {author} {\bibfnamefont {{\'E}.}~\bibnamefont {Ju{\'e}}}, \bibinfo
  {author} {\bibfnamefont {V.}~\bibnamefont {Cros}},\ and\ \bibinfo {author}
  {\bibfnamefont {A.}~\bibnamefont {Fert}},\ }\bibinfo {title} {Dynamics of
  {{Dzyaloshinskii}} Domain Walls in Ultrathin Magnetic Films},\ \href
  {https://iopscience.iop.org/article/10.1209/0295-5075/100/57002} {\bibfield
  {journal} {\bibinfo  {journal} {Europhys. Lett.}\ }\textbf {\bibinfo {volume}
  {100}},\ \bibinfo {pages} {57002} (\bibinfo {year} {2012})}\BibitemShut
  {NoStop}%
\bibitem [{\citenamefont {Tretiakov}\ and\ \citenamefont
  {Abanov}(2010)}]{tretiakov2010current}%
  \BibitemOpen
  \bibfield  {author} {\bibinfo {author} {\bibfnamefont {O.~A.}\ \bibnamefont
  {Tretiakov}}\ and\ \bibinfo {author} {\bibfnamefont {{\relax
  Ar}.}~\bibnamefont {Abanov}},\ }\bibinfo {title} {Current {{Driven
  Magnetization Dynamics}} in {{Ferromagnetic Nanowires}} with a
  {{Dzyaloshinskii-Moriya Interaction}}},\ \href
  {https://link.aps.org/doi/10.1103/PhysRevLett.105.157201} {\bibfield
  {journal} {\bibinfo  {journal} {Phys. Rev. Lett.}\ }\textbf {\bibinfo
  {volume} {105}},\ \bibinfo {pages} {157201} (\bibinfo {year}
  {2010})}\BibitemShut {NoStop}%
\bibitem [{\citenamefont {Chen}\ \emph
  {et~al.}(2021{\natexlab{b}})\citenamefont {Chen}, \citenamefont {Hu},\ and\
  \citenamefont {Yu}}]{chen2021chiral}%
  \BibitemOpen
  \bibfield  {author} {\bibinfo {author} {\bibfnamefont {J.}~\bibnamefont
  {Chen}}, \bibinfo {author} {\bibfnamefont {J.}~\bibnamefont {Hu}},\ and\
  \bibinfo {author} {\bibfnamefont {H.}~\bibnamefont {Yu}},\ }\bibinfo {title}
  {Chiral {{Emission}} of {{Exchange Spin Waves}} by {{Magnetic Skyrmions}}},\
  \href {https://pubs.acs.org/doi/10.1021/acsnano.0c07805} {\bibfield
  {journal} {\bibinfo  {journal} {ACS Nano}\ }\textbf {\bibinfo {volume}
  {15}},\ \bibinfo {pages} {4372} (\bibinfo {year}
  {2021}{\natexlab{b}})}\BibitemShut {NoStop}%
\bibitem [{\citenamefont {Liang}\ \emph {et~al.}(2019)\citenamefont {Liang},
  \citenamefont {He}, \citenamefont {Cai},\ and\ \citenamefont
  {Li}}]{liang2019motion}%
  \BibitemOpen
  \bibfield  {author} {\bibinfo {author} {\bibfnamefont {Y.-C.}\ \bibnamefont
  {Liang}}, \bibinfo {author} {\bibfnamefont {P.-B.}\ \bibnamefont {He}},
  \bibinfo {author} {\bibfnamefont {M.-Q.}\ \bibnamefont {Cai}},\ and\ \bibinfo
  {author} {\bibfnamefont {Z.-D.}\ \bibnamefont {Li}},\ }\bibinfo {title}
  {Motion and Stability of Chiral Domain Walls Driven by Non-Gradient Spin
  Torques: {{Antiferromagnets}} and Ferromagnets Compared},\ \href
  {https://linkinghub.elsevier.com/retrieve/pii/S0304885318337545} {\bibfield
  {journal} {\bibinfo  {journal} {J. Magn. Magn. Mater.}\ }\textbf {\bibinfo
  {volume} {479}},\ \bibinfo {pages} {291} (\bibinfo {year}
  {2019})}\BibitemShut {NoStop}%
\bibitem [{\citenamefont {Han}\ \emph {et~al.}(2019)\citenamefont {Han},
  \citenamefont {Zhang}, \citenamefont {Hou}, \citenamefont {Siddiqui},\ and\
  \citenamefont {Liu}}]{han2019mutuala}%
  \BibitemOpen
  \bibfield  {author} {\bibinfo {author} {\bibfnamefont {J.}~\bibnamefont
  {Han}}, \bibinfo {author} {\bibfnamefont {P.}~\bibnamefont {Zhang}}, \bibinfo
  {author} {\bibfnamefont {J.~T.}\ \bibnamefont {Hou}}, \bibinfo {author}
  {\bibfnamefont {S.~A.}\ \bibnamefont {Siddiqui}},\ and\ \bibinfo {author}
  {\bibfnamefont {L.}~\bibnamefont {Liu}},\ }\bibinfo {title} {Mutual Control
  of Coherent Spin Waves and Magnetic Domain Walls in a Magnonic Device},\
  \href {https://syndication.highwire.org/content/doi/10.1126/science.aau2610}
  {\bibfield  {journal} {\bibinfo  {journal} {Science}\ }\textbf {\bibinfo
  {volume} {366}},\ \bibinfo {pages} {1121} (\bibinfo {year}
  {2019})}\BibitemShut {NoStop}%
\bibitem [{\citenamefont {Kong}\ \emph {et~al.}(2021)\citenamefont {Kong},
  \citenamefont {Chen}, \citenamefont {Wang}, \citenamefont {Song},\ and\
  \citenamefont {Du}}]{kong2021dynamics}%
  \BibitemOpen
  \bibfield  {author} {\bibinfo {author} {\bibfnamefont {L.}~\bibnamefont
  {Kong}}, \bibinfo {author} {\bibfnamefont {X.}~\bibnamefont {Chen}}, \bibinfo
  {author} {\bibfnamefont {W.}~\bibnamefont {Wang}}, \bibinfo {author}
  {\bibfnamefont {D.}~\bibnamefont {Song}},\ and\ \bibinfo {author}
  {\bibfnamefont {H.}~\bibnamefont {Du}},\ }\bibinfo {title} {Dynamics of
  Interstitial Skyrmions in the Presence of Temperature Gradients},\ \href
  {https://link.aps.org/doi/10.1103/PhysRevB.104.214407} {\bibfield  {journal}
  {\bibinfo  {journal} {Phys. Rev. B}\ }\textbf {\bibinfo {volume} {104}},\
  \bibinfo {pages} {214407} (\bibinfo {year} {2021})}\BibitemShut {NoStop}%
\bibitem [{\citenamefont {Knapman}\ \emph {et~al.}(2021)\citenamefont
  {Knapman}, \citenamefont {Rodrigues}, \citenamefont {Masell},\ and\
  \citenamefont {{Everschor-Sitte}}}]{knapman2021currentinduced}%
  \BibitemOpen
  \bibfield  {author} {\bibinfo {author} {\bibfnamefont {R.}~\bibnamefont
  {Knapman}}, \bibinfo {author} {\bibfnamefont {D.~R.}\ \bibnamefont
  {Rodrigues}}, \bibinfo {author} {\bibfnamefont {J.}~\bibnamefont {Masell}},\
  and\ \bibinfo {author} {\bibfnamefont {K.}~\bibnamefont
  {{Everschor-Sitte}}},\ }\bibinfo {title} {Current-Induced
  {{H-shaped-skyrmion}} Creation and Their Dynamics in the Helical Phase},\
  \href {https://iopscience.iop.org/article/10.1088/1361-6463/ac0e5a}
  {\bibfield  {journal} {\bibinfo  {journal} {J. Phys. D: Appl. Phys.}\
  }\textbf {\bibinfo {volume} {54}},\ \bibinfo {pages} {404003} (\bibinfo
  {year} {2021})}\BibitemShut {NoStop}%
\bibitem [{\citenamefont {Boulle}\ \emph {et~al.}(2013)\citenamefont {Boulle},
  \citenamefont {Rohart}, \citenamefont {{Buda-Prejbeanu}}, \citenamefont
  {Ju{\'e}}, \citenamefont {Miron}, \citenamefont {Pizzini}, \citenamefont
  {Vogel}, \citenamefont {Gaudin},\ and\ \citenamefont
  {Thiaville}}]{boulle2013domain}%
  \BibitemOpen
  \bibfield  {author} {\bibinfo {author} {\bibfnamefont {O.}~\bibnamefont
  {Boulle}}, \bibinfo {author} {\bibfnamefont {S.}~\bibnamefont {Rohart}},
  \bibinfo {author} {\bibfnamefont {L.~D.}\ \bibnamefont {{Buda-Prejbeanu}}},
  \bibinfo {author} {\bibfnamefont {E.}~\bibnamefont {Ju{\'e}}}, \bibinfo
  {author} {\bibfnamefont {I.~M.}\ \bibnamefont {Miron}}, \bibinfo {author}
  {\bibfnamefont {S.}~\bibnamefont {Pizzini}}, \bibinfo {author} {\bibfnamefont
  {J.}~\bibnamefont {Vogel}}, \bibinfo {author} {\bibfnamefont
  {G.}~\bibnamefont {Gaudin}},\ and\ \bibinfo {author} {\bibfnamefont
  {A.}~\bibnamefont {Thiaville}},\ }\bibinfo {title} {Domain {{Wall Tilting}}
  in the {{Presence}} of the {{Dzyaloshinskii-Moriya Interaction}} in
  {{Out-of-Plane Magnetized Magnetic Nanotracks}}},\ \href
  {http://link.aps.org/article/10.1103/PhysRevLett.111.217203} {\bibfield
  {journal} {\bibinfo  {journal} {Phys. Rev. Lett.}\ }\textbf {\bibinfo
  {volume} {111}},\ \bibinfo {pages} {5} (\bibinfo {year} {2013})}\BibitemShut
  {NoStop}%
\bibitem [{\citenamefont {Dohi}\ \emph {et~al.}(2022)\citenamefont {Dohi},
  \citenamefont {Wei{\ss}enhofer}, \citenamefont {Kerber}, \citenamefont
  {Kammerbauer}, \citenamefont {Ge}, \citenamefont {Raab}, \citenamefont
  {Z{\`a}zvorka}, \citenamefont {Syskaki}, \citenamefont {Shahee},\ and\
  \citenamefont {Ruhwedel}}]{dohi2022enhanced}%
  \BibitemOpen
  \bibfield  {author} {\bibinfo {author} {\bibfnamefont {T.}~\bibnamefont
  {Dohi}}, \bibinfo {author} {\bibfnamefont {M.}~\bibnamefont
  {Wei{\ss}enhofer}}, \bibinfo {author} {\bibfnamefont {N.}~\bibnamefont
  {Kerber}}, \bibinfo {author} {\bibfnamefont {F.}~\bibnamefont {Kammerbauer}},
  \bibinfo {author} {\bibfnamefont {Y.}~\bibnamefont {Ge}}, \bibinfo {author}
  {\bibfnamefont {K.}~\bibnamefont {Raab}}, \bibinfo {author} {\bibfnamefont
  {J.}~\bibnamefont {Z{\`a}zvorka}}, \bibinfo {author} {\bibfnamefont {M.-A.}\
  \bibnamefont {Syskaki}}, \bibinfo {author} {\bibfnamefont {A.}~\bibnamefont
  {Shahee}},\ and\ \bibinfo {author} {\bibfnamefont {M.}~\bibnamefont
  {Ruhwedel}},\ }\bibinfo {title} {Enhanced Thermally-Activated Skyrmion
  Diffusion in Synthetic Antiferromagnetic Systems with Tunable Effective
  Topological Charge},\ \href {http://arxiv.org/abs/2206.00791} {\bibfield
  {journal} {\bibinfo  {journal} {arXiv:2206.00791}\ } (\bibinfo {year}
  {2022})}\BibitemShut {NoStop}%
\bibitem [{\citenamefont {Jen}\ and\ \citenamefont
  {Berger}(1982)}]{jen1982dragging}%
  \BibitemOpen
  \bibfield  {author} {\bibinfo {author} {\bibfnamefont {S.}~\bibnamefont
  {Jen}}\ and\ \bibinfo {author} {\bibfnamefont {L.}~\bibnamefont {Berger}},\
  }\bibinfo {title} {Dragging of Stripe Domains by a Temperature Gradient in
  {{Metglas}} 2826 {{MB}} (Invited)},\ \href
  {http://aip.scitation.org/doi/10.1063/1.330843} {\bibfield  {journal}
  {\bibinfo  {journal} {J. Appl. Phys.}\ }\textbf {\bibinfo {volume} {53}},\
  \bibinfo {pages} {2298} (\bibinfo {year} {1982})}\BibitemShut {NoStop}%
\bibitem [{\citenamefont {Tanabe}\ \emph {et~al.}(2012)\citenamefont {Tanabe},
  \citenamefont {Chiba}, \citenamefont {Ohe}, \citenamefont {Kasai},
  \citenamefont {Kohno}, \citenamefont {Barnes}, \citenamefont {Maekawa},
  \citenamefont {Kobayashi},\ and\ \citenamefont
  {Ono}}]{tanabe2012spinmotivea}%
  \BibitemOpen
  \bibfield  {author} {\bibinfo {author} {\bibfnamefont {K.}~\bibnamefont
  {Tanabe}}, \bibinfo {author} {\bibfnamefont {D.}~\bibnamefont {Chiba}},
  \bibinfo {author} {\bibfnamefont {J.}~\bibnamefont {Ohe}}, \bibinfo {author}
  {\bibfnamefont {S.}~\bibnamefont {Kasai}}, \bibinfo {author} {\bibfnamefont
  {H.}~\bibnamefont {Kohno}}, \bibinfo {author} {\bibfnamefont {S.~E.}\
  \bibnamefont {Barnes}}, \bibinfo {author} {\bibfnamefont {S.}~\bibnamefont
  {Maekawa}}, \bibinfo {author} {\bibfnamefont {K.}~\bibnamefont {Kobayashi}},\
  and\ \bibinfo {author} {\bibfnamefont {T.}~\bibnamefont {Ono}},\ }\bibinfo
  {title} {Spin-Motive Force Due to a Gyrating Magnetic Vortex},\ \href
  {http://www.nature.com/articles/ncomms1824} {\bibfield  {journal} {\bibinfo
  {journal} {Nat. Commun.}\ }\textbf {\bibinfo {volume} {3}},\ \bibinfo {pages}
  {845} (\bibinfo {year} {2012})}\BibitemShut {NoStop}%
\bibitem [{\citenamefont {Agrawal}\ \emph {et~al.}(2013)\citenamefont
  {Agrawal}, \citenamefont {Vasyuchka}, \citenamefont {Serga}, \citenamefont
  {Karenowska}, \citenamefont {Melkov},\ and\ \citenamefont
  {Hillebrands}}]{agrawal2013direct}%
  \BibitemOpen
  \bibfield  {author} {\bibinfo {author} {\bibfnamefont {M.}~\bibnamefont
  {Agrawal}}, \bibinfo {author} {\bibfnamefont {V.~I.}\ \bibnamefont
  {Vasyuchka}}, \bibinfo {author} {\bibfnamefont {A.~A.}\ \bibnamefont
  {Serga}}, \bibinfo {author} {\bibfnamefont {A.~D.}\ \bibnamefont
  {Karenowska}}, \bibinfo {author} {\bibfnamefont {G.~A.}\ \bibnamefont
  {Melkov}},\ and\ \bibinfo {author} {\bibfnamefont {B.}~\bibnamefont
  {Hillebrands}},\ }\bibinfo {title} {Direct {{Measurement}} of {{Magnon
  Temperature}}: {{New Insight}} into {{Magnon-Phonon Coupling}} in {{Magnetic
  Insulators}}},\ \href
  {http://link.aps.org/article/10.1103/PhysRevLett.111.107204} {\bibfield
  {journal} {\bibinfo  {journal} {Phys. Rev. Lett.}\ }\textbf {\bibinfo
  {volume} {111}},\ \bibinfo {pages} {107204} (\bibinfo {year}
  {2013})}\BibitemShut {NoStop}%
\bibitem [{\citenamefont {An}\ \emph {et~al.}(2013)\citenamefont {An},
  \citenamefont {Vasyuchka}, \citenamefont {Uchida}, \citenamefont {Chumak},
  \citenamefont {Yamaguchi}, \citenamefont {Harii}, \citenamefont {Ohe},
  \citenamefont {Jungfleisch}, \citenamefont {Kajiwara}, \citenamefont
  {Adachi}, \citenamefont {Hillebrands}, \citenamefont {Maekawa},\ and\
  \citenamefont {Saitoh}}]{an2013unidirectional}%
  \BibitemOpen
  \bibfield  {author} {\bibinfo {author} {\bibfnamefont {T.}~\bibnamefont
  {An}}, \bibinfo {author} {\bibfnamefont {V.~I.}\ \bibnamefont {Vasyuchka}},
  \bibinfo {author} {\bibfnamefont {K.}~\bibnamefont {Uchida}}, \bibinfo
  {author} {\bibfnamefont {A.~V.}\ \bibnamefont {Chumak}}, \bibinfo {author}
  {\bibfnamefont {K.}~\bibnamefont {Yamaguchi}}, \bibinfo {author}
  {\bibfnamefont {K.}~\bibnamefont {Harii}}, \bibinfo {author} {\bibfnamefont
  {J.}~\bibnamefont {Ohe}}, \bibinfo {author} {\bibfnamefont {M.~B.}\
  \bibnamefont {Jungfleisch}}, \bibinfo {author} {\bibfnamefont
  {Y.}~\bibnamefont {Kajiwara}}, \bibinfo {author} {\bibfnamefont
  {H.}~\bibnamefont {Adachi}}, \bibinfo {author} {\bibfnamefont
  {B.}~\bibnamefont {Hillebrands}}, \bibinfo {author} {\bibfnamefont
  {S.}~\bibnamefont {Maekawa}},\ and\ \bibinfo {author} {\bibfnamefont
  {E.}~\bibnamefont {Saitoh}},\ }\bibinfo {title} {Unidirectional Spin-Wave
  Heat Conveyer},\ \href {https://www.nature.com/articles/nmat3628} {\bibfield
  {journal} {\bibinfo  {journal} {Nat. Mater.}\ }\textbf {\bibinfo {volume}
  {12}},\ \bibinfo {pages} {549} (\bibinfo {year} {2013})}\BibitemShut
  {NoStop}%
\bibitem [{\citenamefont {Yan}\ and\ \citenamefont
  {Bauer}(2013)}]{yan2013magnon}%
  \BibitemOpen
  \bibfield  {author} {\bibinfo {author} {\bibfnamefont {P.}~\bibnamefont
  {Yan}}\ and\ \bibinfo {author} {\bibfnamefont {G.~E.~W.}\ \bibnamefont
  {Bauer}},\ }\bibinfo {title} {Magnon {{Mediated Domain Wall Heat
  Conductance}} in {{Ferromagnetic Wires}}},\ \href
  {http://ieeexplore.ieee.org/document/6559056/} {\bibfield  {journal}
  {\bibinfo  {journal} {IEEE Trans. Magn.}\ }\textbf {\bibinfo {volume} {49}},\
  \bibinfo {pages} {3109} (\bibinfo {year} {2013})}\BibitemShut {NoStop}%
\bibitem [{\citenamefont {Lee}\ \emph {et~al.}(2017)\citenamefont {Lee},
  \citenamefont {Hippalgaonkar}, \citenamefont {Yang}, \citenamefont {Hong},
  \citenamefont {Ko}, \citenamefont {Suh}, \citenamefont {Liu}, \citenamefont
  {Wang}, \citenamefont {Urban}, \citenamefont {Zhang}, \citenamefont {Dames},
  \citenamefont {Hartnoll}, \citenamefont {Delaire},\ and\ \citenamefont
  {Wu}}]{lee2017anomalously}%
  \BibitemOpen
  \bibfield  {author} {\bibinfo {author} {\bibfnamefont {S.}~\bibnamefont
  {Lee}}, \bibinfo {author} {\bibfnamefont {K.}~\bibnamefont {Hippalgaonkar}},
  \bibinfo {author} {\bibfnamefont {F.}~\bibnamefont {Yang}}, \bibinfo {author}
  {\bibfnamefont {J.}~\bibnamefont {Hong}}, \bibinfo {author} {\bibfnamefont
  {C.}~\bibnamefont {Ko}}, \bibinfo {author} {\bibfnamefont {J.}~\bibnamefont
  {Suh}}, \bibinfo {author} {\bibfnamefont {K.}~\bibnamefont {Liu}}, \bibinfo
  {author} {\bibfnamefont {K.}~\bibnamefont {Wang}}, \bibinfo {author}
  {\bibfnamefont {J.~J.}\ \bibnamefont {Urban}}, \bibinfo {author}
  {\bibfnamefont {X.}~\bibnamefont {Zhang}}, \bibinfo {author} {\bibfnamefont
  {C.}~\bibnamefont {Dames}}, \bibinfo {author} {\bibfnamefont {S.~A.}\
  \bibnamefont {Hartnoll}}, \bibinfo {author} {\bibfnamefont {O.}~\bibnamefont
  {Delaire}},\ and\ \bibinfo {author} {\bibfnamefont {J.}~\bibnamefont {Wu}},\
  }\bibinfo {title} {Anomalously Low Electronic Thermal Conductivity in
  Metallic Vanadium Dioxide},\ \href
  {https://syndication.highwire.org/content/doi/10.1126/science.aag0410}
  {\bibfield  {journal} {\bibinfo  {journal} {Science}\ }\textbf {\bibinfo
  {volume} {355}},\ \bibinfo {pages} {371} (\bibinfo {year}
  {2017})}\BibitemShut {NoStop}%
\bibitem [{\citenamefont {Qiu}\ and\ \citenamefont
  {Shen}(2022)}]{qiu2022tunable}%
  \BibitemOpen
  \bibfield  {author} {\bibinfo {author} {\bibfnamefont {L.}~\bibnamefont
  {Qiu}}\ and\ \bibinfo {author} {\bibfnamefont {K.}~\bibnamefont {Shen}},\
  }\bibinfo {title} {Tunable Spin-Wave Nonreciprocity in Synthetic
  Antiferromagnetic Domain Walls},\ \href
  {https://link.aps.org/doi/10.1103/PhysRevB.105.094436} {\bibfield  {journal}
  {\bibinfo  {journal} {Phys. Rev. B}\ }\textbf {\bibinfo {volume} {105}},\
  \bibinfo {pages} {094436} (\bibinfo {year} {2022})}\BibitemShut {NoStop}%
\bibitem [{\citenamefont {Flebus}\ \emph {et~al.}(2016)\citenamefont {Flebus},
  \citenamefont {Bender}, \citenamefont {Tserkovnyak},\ and\ \citenamefont
  {Duine}}]{flebus2016twofluid}%
  \BibitemOpen
  \bibfield  {author} {\bibinfo {author} {\bibfnamefont {B.}~\bibnamefont
  {Flebus}}, \bibinfo {author} {\bibfnamefont {S.~A.}\ \bibnamefont {Bender}},
  \bibinfo {author} {\bibfnamefont {Y.}~\bibnamefont {Tserkovnyak}},\ and\
  \bibinfo {author} {\bibfnamefont {R.~A.}\ \bibnamefont {Duine}},\ }\bibinfo
  {title} {Two-{{Fluid Theory}} for {{Spin Superfluidity}} in {{Magnetic
  Insulators}}},\ \href
  {https://link.aps.org/doi/10.1103/PhysRevLett.116.117201} {\bibfield
  {journal} {\bibinfo  {journal} {Phys. Rev. Lett.}\ }\textbf {\bibinfo
  {volume} {116}},\ \bibinfo {pages} {117201} (\bibinfo {year}
  {2016})}\BibitemShut {NoStop}%
\bibitem [{\citenamefont {Boona}\ \emph {et~al.}(2016)\citenamefont {Boona},
  \citenamefont {Watzman},\ and\ \citenamefont {Heremans}}]{boona2016research}%
  \BibitemOpen
  \bibfield  {author} {\bibinfo {author} {\bibfnamefont {S.~R.}\ \bibnamefont
  {Boona}}, \bibinfo {author} {\bibfnamefont {S.~J.}\ \bibnamefont {Watzman}},\
  and\ \bibinfo {author} {\bibfnamefont {J.~P.}\ \bibnamefont {Heremans}},\
  }\bibinfo {title} {Research {{Update}}: {{Utilizing}} Magnetization Dynamics
  in Solid-State Thermal Energy Conversion},\ \href
  {http://aip.scitation.org/doi/pdf/10.1063/1.4955027} {\bibfield  {journal}
  {\bibinfo  {journal} {APL Mater.}\ }\textbf {\bibinfo {volume} {4}},\
  \bibinfo {pages} {104502} (\bibinfo {year} {2016})}\BibitemShut {NoStop}%
\bibitem [{\citenamefont {Yamaguchi}\ \emph {et~al.}(2019)\citenamefont
  {Yamaguchi}, \citenamefont {Kohno},\ and\ \citenamefont
  {Duine}}]{yamaguchi2019microscopic}%
  \BibitemOpen
  \bibfield  {author} {\bibinfo {author} {\bibfnamefont {T.}~\bibnamefont
  {Yamaguchi}}, \bibinfo {author} {\bibfnamefont {H.}~\bibnamefont {Kohno}},\
  and\ \bibinfo {author} {\bibfnamefont {R.~A.}\ \bibnamefont {Duine}},\
  }\bibinfo {title} {Microscopic Theory of Magnon-Drag Electron Flow in
  Ferromagnetic Metals},\ \href
  {https://link.aps.org/article/10.1103/PhysRevB.99.094425} {\bibfield
  {journal} {\bibinfo  {journal} {Phys. Rev. B}\ }\textbf {\bibinfo {volume}
  {99}},\ \bibinfo {pages} {094425} (\bibinfo {year} {2019})}\BibitemShut
  {NoStop}%
\bibitem [{\citenamefont {Miura}\ and\ \citenamefont
  {Sakuma}(2012)}]{miura2012microscopic}%
  \BibitemOpen
  \bibfield  {author} {\bibinfo {author} {\bibfnamefont {D.}~\bibnamefont
  {Miura}}\ and\ \bibinfo {author} {\bibfnamefont {A.}~\bibnamefont {Sakuma}},\
  }\bibinfo {title} {Microscopic {{Theory}} of {{Magnon-Drag Thermoelectric
  Transport}} in {{Ferromagnetic Metals}}},\ \href
  {http://journals.jps.jp/doi/pdf/10.1143/JPSJ.81.113602} {\bibfield  {journal}
  {\bibinfo  {journal} {J. Phys. Soc. Jpn.}\ }\textbf {\bibinfo {volume}
  {81}},\ \bibinfo {pages} {113602} (\bibinfo {year} {2012})}\BibitemShut
  {NoStop}%
\bibitem [{\citenamefont {LIU}\ and\ \citenamefont
  {ZHANG}(2019)}]{liu2019topologically}%
  \BibitemOpen
  \bibfield  {author} {\bibinfo {author} {\bibfnamefont {E.}~\bibnamefont
  {LIU}}\ and\ \bibinfo {author} {\bibfnamefont {S.}~\bibnamefont {ZHANG}},\
  }\bibinfo {title} {Topologically Enhanced Zero-Field Transverse
  {{Nernstthermoelectric}} Effect in Magnetic Topological Semimetals
  ({{inChinese}})},\ \href
  {http://engine.scichina.com/doi/10.1360/SSPMA-2019-0367} {\bibfield
  {journal} {\bibinfo  {journal} {Sci. China-Phys. Mech. Astron.}\ }\textbf
  {\bibinfo {volume} {49}},\ \bibinfo {pages} {127001} (\bibinfo {year}
  {2019})}\BibitemShut {NoStop}%
\bibitem [{\citenamefont {Kasahara}\ \emph {et~al.}(2018)\citenamefont
  {Kasahara}, \citenamefont {Ohnishi}, \citenamefont {Mizukami}, \citenamefont
  {Tanaka}, \citenamefont {Ma}, \citenamefont {Sugii}, \citenamefont {Kurita},
  \citenamefont {Tanaka}, \citenamefont {Nasu}, \citenamefont {Motome},
  \citenamefont {Shibauchi},\ and\ \citenamefont
  {Matsuda}}]{kasahara2018majorana}%
  \BibitemOpen
  \bibfield  {author} {\bibinfo {author} {\bibfnamefont {Y.}~\bibnamefont
  {Kasahara}}, \bibinfo {author} {\bibfnamefont {T.}~\bibnamefont {Ohnishi}},
  \bibinfo {author} {\bibfnamefont {Y.}~\bibnamefont {Mizukami}}, \bibinfo
  {author} {\bibfnamefont {O.}~\bibnamefont {Tanaka}}, \bibinfo {author}
  {\bibfnamefont {S.}~\bibnamefont {Ma}}, \bibinfo {author} {\bibfnamefont
  {K.}~\bibnamefont {Sugii}}, \bibinfo {author} {\bibfnamefont
  {N.}~\bibnamefont {Kurita}}, \bibinfo {author} {\bibfnamefont
  {H.}~\bibnamefont {Tanaka}}, \bibinfo {author} {\bibfnamefont
  {J.}~\bibnamefont {Nasu}}, \bibinfo {author} {\bibfnamefont {Y.}~\bibnamefont
  {Motome}}, \bibinfo {author} {\bibfnamefont {T.}~\bibnamefont {Shibauchi}},\
  and\ \bibinfo {author} {\bibfnamefont {Y.}~\bibnamefont {Matsuda}},\
  }\bibinfo {title} {Majorana Quantization and Half-Integer Thermal Quantum
  {{Hall}} Effect in a {{Kitaev}} Spin Liquid},\ \href
  {http://www.nature.com/articles/s41586-018-0274-0} {\bibfield  {journal}
  {\bibinfo  {journal} {Nature}\ }\textbf {\bibinfo {volume} {559}},\ \bibinfo
  {pages} {227} (\bibinfo {year} {2018})}\BibitemShut {NoStop}%
\bibitem [{\citenamefont {Wei}\ \emph {et~al.}(2016)\citenamefont {Wei},
  \citenamefont {Yang}, \citenamefont {Guo}, \citenamefont {Wang},
  \citenamefont {Wu}, \citenamefont {Xu}, \citenamefont {Zhao}, \citenamefont
  {Zhang}, \citenamefont {Zhang}, \citenamefont {Dresselhaus},\ and\
  \citenamefont {Yang}}]{wei2016minimum}%
  \BibitemOpen
  \bibfield  {author} {\bibinfo {author} {\bibfnamefont {P.}~\bibnamefont
  {Wei}}, \bibinfo {author} {\bibfnamefont {J.}~\bibnamefont {Yang}}, \bibinfo
  {author} {\bibfnamefont {L.}~\bibnamefont {Guo}}, \bibinfo {author}
  {\bibfnamefont {S.}~\bibnamefont {Wang}}, \bibinfo {author} {\bibfnamefont
  {L.}~\bibnamefont {Wu}}, \bibinfo {author} {\bibfnamefont {X.}~\bibnamefont
  {Xu}}, \bibinfo {author} {\bibfnamefont {W.}~\bibnamefont {Zhao}}, \bibinfo
  {author} {\bibfnamefont {Q.}~\bibnamefont {Zhang}}, \bibinfo {author}
  {\bibfnamefont {W.}~\bibnamefont {Zhang}}, \bibinfo {author} {\bibfnamefont
  {M.~S.}\ \bibnamefont {Dresselhaus}},\ and\ \bibinfo {author} {\bibfnamefont
  {J.}~\bibnamefont {Yang}},\ }\bibinfo {title} {Minimum {{Thermal
  Conductivity}} in {{Weak Topological Insulators}} with {{Bismuth-Based Stack
  Structure}}},\ \href
  {https://api.wiley.com/onlinelibrary/tdm/v1/articles/10.1002%2Fadfm.201600718}
  {\bibfield  {journal} {\bibinfo  {journal} {Adv. Funct. Mater.}\ }\textbf
  {\bibinfo {volume} {26}},\ \bibinfo {pages} {5360} (\bibinfo {year}
  {2016})}\BibitemShut {NoStop}%
\bibitem [{\citenamefont {Liang}\ \emph {et~al.}(2016)\citenamefont {Liang},
  \citenamefont {Cheng}, \citenamefont {Zhang}, \citenamefont {Liu},\ and\
  \citenamefont {Zhang}}]{liang2016maximizing}%
  \BibitemOpen
  \bibfield  {author} {\bibinfo {author} {\bibfnamefont {J.}~\bibnamefont
  {Liang}}, \bibinfo {author} {\bibfnamefont {L.}~\bibnamefont {Cheng}},
  \bibinfo {author} {\bibfnamefont {J.}~\bibnamefont {Zhang}}, \bibinfo
  {author} {\bibfnamefont {H.}~\bibnamefont {Liu}},\ and\ \bibinfo {author}
  {\bibfnamefont {Z.}~\bibnamefont {Zhang}},\ }\bibinfo {title} {Maximizing the
  Thermoelectric Performance of Topological Insulator
  {{Bi}}{\textsubscript{2}}{{Te}}{\textsubscript{3}}films in the Few-Quintuple
  Layer Regime},\ \href
  {http://pubs.rsc.org/en/content/articlepdf/2016/NR/C6NR00724D} {\bibfield
  {journal} {\bibinfo  {journal} {Nanoscale}\ }\textbf {\bibinfo {volume}
  {8}},\ \bibinfo {pages} {8855} (\bibinfo {year} {2016})}\BibitemShut
  {NoStop}%
\bibitem [{\citenamefont {Xu}\ \emph {et~al.}(2014)\citenamefont {Xu},
  \citenamefont {Gan},\ and\ \citenamefont {Zhang}}]{xu2014enhanced}%
  \BibitemOpen
  \bibfield  {author} {\bibinfo {author} {\bibfnamefont {Y.}~\bibnamefont
  {Xu}}, \bibinfo {author} {\bibfnamefont {Z.}~\bibnamefont {Gan}},\ and\
  \bibinfo {author} {\bibfnamefont {S.-C.}\ \bibnamefont {Zhang}},\ }\bibinfo
  {title} {Enhanced {{Thermoelectric Performance}} and {{Anomalous Seebeck
  Effects}} in {{Topological Insulators}}},\ \href
  {http://link.aps.org/article/10.1103/PhysRevLett.112.226801} {\bibfield
  {journal} {\bibinfo  {journal} {Phys. Rev. Lett.}\ }\textbf {\bibinfo
  {volume} {112}},\ \bibinfo {pages} {226801} (\bibinfo {year}
  {2014})}\BibitemShut {NoStop}%
\bibitem [{\citenamefont {Zahid}\ and\ \citenamefont
  {Lake}(2010)}]{zahid2010thermoelectric}%
  \BibitemOpen
  \bibfield  {author} {\bibinfo {author} {\bibfnamefont {F.}~\bibnamefont
  {Zahid}}\ and\ \bibinfo {author} {\bibfnamefont {R.}~\bibnamefont {Lake}},\
  }\bibinfo {title} {Thermoelectric Properties of {{Bi2Te3}} Atomic Quintuple
  Thin Films},\ \href {http://aip.scitation.org/doi/pdf/10.1063/1.3518078}
  {\bibfield  {journal} {\bibinfo  {journal} {Appl. Phys. Lett.}\ }\textbf
  {\bibinfo {volume} {97}},\ \bibinfo {pages} {212102} (\bibinfo {year}
  {2010})}\BibitemShut {NoStop}%
\bibitem [{\citenamefont {Goyal}\ \emph {et~al.}(2010)\citenamefont {Goyal},
  \citenamefont {Teweldebrhan},\ and\ \citenamefont
  {Balandin}}]{goyal2010mechanicallyexfoliated}%
  \BibitemOpen
  \bibfield  {author} {\bibinfo {author} {\bibfnamefont {V.}~\bibnamefont
  {Goyal}}, \bibinfo {author} {\bibfnamefont {D.}~\bibnamefont
  {Teweldebrhan}},\ and\ \bibinfo {author} {\bibfnamefont {A.~A.}\ \bibnamefont
  {Balandin}},\ }\bibinfo {title} {Mechanically-Exfoliated Stacks of Thin Films
  of {{Bi2Te3}} Topological Insulators with Enhanced Thermoelectric
  Performance},\ \href {http://aip.scitation.org/doi/pdf/10.1063/1.3494529}
  {\bibfield  {journal} {\bibinfo  {journal} {Appl. Phys. Lett.}\ }\textbf
  {\bibinfo {volume} {97}},\ \bibinfo {pages} {133117} (\bibinfo {year}
  {2010})}\BibitemShut {NoStop}%
\bibitem [{\citenamefont {Ye}\ \emph {et~al.}(1999)\citenamefont {Ye},
  \citenamefont {Kim}, \citenamefont {Millis}, \citenamefont {Shraiman},
  \citenamefont {Majumdar},\ and\ \citenamefont {Te{\v
  s}anovi{\'c}}}]{ye1999berrya}%
  \BibitemOpen
  \bibfield  {author} {\bibinfo {author} {\bibfnamefont {J.}~\bibnamefont
  {Ye}}, \bibinfo {author} {\bibfnamefont {Y.~B.}\ \bibnamefont {Kim}},
  \bibinfo {author} {\bibfnamefont {A.~J.}\ \bibnamefont {Millis}}, \bibinfo
  {author} {\bibfnamefont {B.~I.}\ \bibnamefont {Shraiman}}, \bibinfo {author}
  {\bibfnamefont {P.}~\bibnamefont {Majumdar}},\ and\ \bibinfo {author}
  {\bibfnamefont {Z.}~\bibnamefont {Te{\v s}anovi{\'c}}},\ }\bibinfo {title}
  {Berry {{Phase Theory}} of the {{Anomalous Hall Effect}}: {{Application}} to
  {{Colossal Magnetoresistance Manganites}}},\ \href
  {https://link.aps.org/doi/10.1103/PhysRevLett.83.3737} {\bibfield  {journal}
  {\bibinfo  {journal} {Phys. Rev. Lett.}\ }\textbf {\bibinfo {volume} {83}},\
  \bibinfo {pages} {3737} (\bibinfo {year} {1999})}\BibitemShut {NoStop}%
\bibitem [{\citenamefont {Hamamoto}\ \emph {et~al.}(2015)\citenamefont
  {Hamamoto}, \citenamefont {Ezawa},\ and\ \citenamefont
  {Nagaosa}}]{hamamoto2015quantized}%
  \BibitemOpen
  \bibfield  {author} {\bibinfo {author} {\bibfnamefont {K.}~\bibnamefont
  {Hamamoto}}, \bibinfo {author} {\bibfnamefont {M.}~\bibnamefont {Ezawa}},\
  and\ \bibinfo {author} {\bibfnamefont {N.}~\bibnamefont {Nagaosa}},\
  }\bibinfo {title} {Quantized Topological {{Hall}} Effect in Skyrmion
  Crystal},\ \href {https://link.aps.org/doi/10.1103/PhysRevB.92.115417}
  {\bibfield  {journal} {\bibinfo  {journal} {Phys. Rev. B}\ }\textbf {\bibinfo
  {volume} {92}},\ \bibinfo {pages} {115417} (\bibinfo {year}
  {2015})}\BibitemShut {NoStop}%
\bibitem [{\citenamefont {Wang}\ \emph {et~al.}(2003)\citenamefont {Wang},
  \citenamefont {Rogado}, \citenamefont {Cava},\ and\ \citenamefont
  {Ong}}]{wang2003spin}%
  \BibitemOpen
  \bibfield  {author} {\bibinfo {author} {\bibfnamefont {Y.}~\bibnamefont
  {Wang}}, \bibinfo {author} {\bibfnamefont {N.~S.}\ \bibnamefont {Rogado}},
  \bibinfo {author} {\bibfnamefont {R.~J.}\ \bibnamefont {Cava}},\ and\
  \bibinfo {author} {\bibfnamefont {N.~P.}\ \bibnamefont {Ong}},\ }\bibinfo
  {title} {Spin Entropy as the Likely Source of Enhanced Thermopower in
  {{NaxCo2O4}}},\ \href {http://www.nature.com/articles/nature01639.pdf}
  {\bibfield  {journal} {\bibinfo  {journal} {Nature}\ }\textbf {\bibinfo
  {volume} {423}},\ \bibinfo {pages} {425} (\bibinfo {year}
  {2003})}\BibitemShut {NoStop}%
\bibitem [{\citenamefont {Flipse}\ \emph {et~al.}(2014)\citenamefont {Flipse},
  \citenamefont {Dejene}, \citenamefont {Wagenaar}, \citenamefont {Bauer},
  \citenamefont {Youssef},\ and\ \citenamefont {{van
  Wees}}}]{flipse2014observationa}%
  \BibitemOpen
  \bibfield  {author} {\bibinfo {author} {\bibfnamefont {J.}~\bibnamefont
  {Flipse}}, \bibinfo {author} {\bibfnamefont {F.~K.}\ \bibnamefont {Dejene}},
  \bibinfo {author} {\bibfnamefont {D.}~\bibnamefont {Wagenaar}}, \bibinfo
  {author} {\bibfnamefont {G.~E.~W.}\ \bibnamefont {Bauer}}, \bibinfo {author}
  {\bibfnamefont {J.~B.}\ \bibnamefont {Youssef}},\ and\ \bibinfo {author}
  {\bibfnamefont {B.~J.}\ \bibnamefont {{van Wees}}},\ }\bibinfo {title}
  {Observation of the {{Spin Peltier Effect}} for {{Magnetic Insulators}}},\
  \href {http://link.aps.org/article/10.1103/PhysRevLett.113.027601} {\bibfield
   {journal} {\bibinfo  {journal} {Phys. Rev. Lett.}\ }\textbf {\bibinfo
  {volume} {113}},\ \bibinfo {pages} {027601} (\bibinfo {year}
  {2014})}\BibitemShut {NoStop}%
\bibitem [{\citenamefont {Yang}\ \emph {et~al.}(2010)\citenamefont {Yang},
  \citenamefont {Beach}, \citenamefont {Knutson}, \citenamefont {Xiao},
  \citenamefont {Zhang}, \citenamefont {Tsoi}, \citenamefont {Niu},
  \citenamefont {MacDonald},\ and\ \citenamefont
  {Erskine}}]{yang2010topological}%
  \BibitemOpen
  \bibfield  {author} {\bibinfo {author} {\bibfnamefont {S.~A.}\ \bibnamefont
  {Yang}}, \bibinfo {author} {\bibfnamefont {G.~S.~D.}\ \bibnamefont {Beach}},
  \bibinfo {author} {\bibfnamefont {C.}~\bibnamefont {Knutson}}, \bibinfo
  {author} {\bibfnamefont {D.}~\bibnamefont {Xiao}}, \bibinfo {author}
  {\bibfnamefont {Z.}~\bibnamefont {Zhang}}, \bibinfo {author} {\bibfnamefont
  {M.}~\bibnamefont {Tsoi}}, \bibinfo {author} {\bibfnamefont {Q.}~\bibnamefont
  {Niu}}, \bibinfo {author} {\bibfnamefont {A.~H.}\ \bibnamefont {MacDonald}},\
  and\ \bibinfo {author} {\bibfnamefont {J.~L.}\ \bibnamefont {Erskine}},\
  }\bibinfo {title} {Topological Electromotive Force from Domain-Wall Dynamics
  in a Ferromagnet},\ \href
  {https://link.aps.org/doi/10.1103/PhysRevB.82.054410} {\bibfield  {journal}
  {\bibinfo  {journal} {Phys. Rev. B}\ }\textbf {\bibinfo {volume} {82}},\
  \bibinfo {pages} {054410} (\bibinfo {year} {2010})}\BibitemShut {NoStop}%
\bibitem [{\citenamefont {Yamane}\ and\ \citenamefont
  {Ieda}(2019)}]{yamane2019skyrmiongenerateda}%
  \BibitemOpen
  \bibfield  {author} {\bibinfo {author} {\bibfnamefont {Y.}~\bibnamefont
  {Yamane}}\ and\ \bibinfo {author} {\bibfnamefont {J.}~\bibnamefont {Ieda}},\
  }\bibinfo {title} {Skyrmion-Generated Spinmotive Forces in Inversion Broken
  Ferromagnets},\ \href
  {https://linkinghub.elsevier.com/retrieve/pii/S0304885319320335} {\bibfield
  {journal} {\bibinfo  {journal} {J. Magn. Magn. Mater.}\ }\textbf {\bibinfo
  {volume} {491}},\ \bibinfo {pages} {165550} (\bibinfo {year}
  {2019})}\BibitemShut {NoStop}%
\bibitem [{\citenamefont {{\v S}mejkal}\ \emph {et~al.}(2018)\citenamefont {{\v
  S}mejkal}, \citenamefont {Mokrousov}, \citenamefont {Yan},\ and\
  \citenamefont {MacDonald}}]{smejkal2018topological}%
  \BibitemOpen
  \bibfield  {author} {\bibinfo {author} {\bibfnamefont {L.}~\bibnamefont {{\v
  S}mejkal}}, \bibinfo {author} {\bibfnamefont {Y.}~\bibnamefont {Mokrousov}},
  \bibinfo {author} {\bibfnamefont {B.}~\bibnamefont {Yan}},\ and\ \bibinfo
  {author} {\bibfnamefont {A.~H.}\ \bibnamefont {MacDonald}},\ }\bibinfo
  {title} {Topological Antiferromagnetic Spintronics},\ \href
  {http://www.nature.com/articles/s41567-018-0064-5} {\bibfield  {journal}
  {\bibinfo  {journal} {Nat. Phys.}\ }\textbf {\bibinfo {volume} {14}},\
  \bibinfo {pages} {242} (\bibinfo {year} {2018})}\BibitemShut {NoStop}%
\bibitem [{\citenamefont {Tveten}\ \emph {et~al.}(2016)\citenamefont {Tveten},
  \citenamefont {M{\"u}ller}, \citenamefont {Linder},\ and\ \citenamefont
  {Brataas}}]{tveten2016intrinsic}%
  \BibitemOpen
  \bibfield  {author} {\bibinfo {author} {\bibfnamefont {E.~G.}\ \bibnamefont
  {Tveten}}, \bibinfo {author} {\bibfnamefont {T.}~\bibnamefont {M{\"u}ller}},
  \bibinfo {author} {\bibfnamefont {J.}~\bibnamefont {Linder}},\ and\ \bibinfo
  {author} {\bibfnamefont {A.}~\bibnamefont {Brataas}},\ }\bibinfo {title}
  {Intrinsic Magnetization of Antiferromagnetic Textures},\ \href
  {https://link.aps.org/doi/10.1103/PhysRevB.93.104408} {\bibfield  {journal}
  {\bibinfo  {journal} {Phys. Rev. B}\ }\textbf {\bibinfo {volume} {93}},\
  \bibinfo {pages} {104408} (\bibinfo {year} {2016})}\BibitemShut {NoStop}%
\bibitem [{\citenamefont {Akosa}\ \emph {et~al.}(2018)\citenamefont {Akosa},
  \citenamefont {Tretiakov}, \citenamefont {Tatara},\ and\ \citenamefont
  {Manchon}}]{akosa2018theory}%
  \BibitemOpen
  \bibfield  {author} {\bibinfo {author} {\bibfnamefont {C.~A.}\ \bibnamefont
  {Akosa}}, \bibinfo {author} {\bibfnamefont {O.~A.}\ \bibnamefont
  {Tretiakov}}, \bibinfo {author} {\bibfnamefont {G.}~\bibnamefont {Tatara}},\
  and\ \bibinfo {author} {\bibfnamefont {A.}~\bibnamefont {Manchon}},\
  }\bibinfo {title} {Theory of the {{Topological Spin Hall Effect}} in
  {{Antiferromagnetic Skyrmions}}: {{Impact}} on {{Current-Induced Motion}}},\
  \href {https://link.aps.org/doi/10.1103/PhysRevLett.121.097204} {\bibfield
  {journal} {\bibinfo  {journal} {Phys. Rev. Lett.}\ }\textbf {\bibinfo
  {volume} {121}},\ \bibinfo {pages} {097204} (\bibinfo {year}
  {2018})}\BibitemShut {NoStop}%
\bibitem [{\citenamefont {Feng}\ \emph {et~al.}(2020)\citenamefont {Feng},
  \citenamefont {Hanke}, \citenamefont {Zhou}, \citenamefont {Guo},
  \citenamefont {Bl{\"u}gel}, \citenamefont {Mokrousov},\ and\ \citenamefont
  {Yao}}]{feng2020topological}%
  \BibitemOpen
  \bibfield  {author} {\bibinfo {author} {\bibfnamefont {W.}~\bibnamefont
  {Feng}}, \bibinfo {author} {\bibfnamefont {J.-P.}\ \bibnamefont {Hanke}},
  \bibinfo {author} {\bibfnamefont {X.}~\bibnamefont {Zhou}}, \bibinfo {author}
  {\bibfnamefont {G.-Y.}\ \bibnamefont {Guo}}, \bibinfo {author} {\bibfnamefont
  {S.}~\bibnamefont {Bl{\"u}gel}}, \bibinfo {author} {\bibfnamefont
  {Y.}~\bibnamefont {Mokrousov}},\ and\ \bibinfo {author} {\bibfnamefont
  {Y.}~\bibnamefont {Yao}},\ }\bibinfo {title} {Topological Magneto-Optical
  Effects and Their Quantization in Noncoplanar Antiferromagnets},\ \href
  {http://www.nature.com/articles/s41467-019-13968-8.pdf} {\bibfield  {journal}
  {\bibinfo  {journal} {Nat. Commun.}\ }\textbf {\bibinfo {volume} {11}},\
  \bibinfo {pages} {118} (\bibinfo {year} {2020})}\BibitemShut {NoStop}%
\bibitem [{\citenamefont {Chen}\ \emph {et~al.}(2019)\citenamefont {Chen},
  \citenamefont {Yan}, \citenamefont {Qin},\ and\ \citenamefont
  {Liu}}]{chen2019landaulifshitzbloch}%
  \BibitemOpen
  \bibfield  {author} {\bibinfo {author} {\bibfnamefont {Z.~Y.}\ \bibnamefont
  {Chen}}, \bibinfo {author} {\bibfnamefont {Z.~R.}\ \bibnamefont {Yan}},
  \bibinfo {author} {\bibfnamefont {M.~H.}\ \bibnamefont {Qin}},\ and\ \bibinfo
  {author} {\bibfnamefont {J.-M.}\ \bibnamefont {Liu}},\ }\bibinfo {title}
  {Landau-{{Lifshitz-Bloch}} Equation for Domain Wall Motion in
  Antiferromagnets},\ \href
  {https://link.aps.org/doi/10.1103/PhysRevB.99.214436} {\bibfield  {journal}
  {\bibinfo  {journal} {Phys. Rev. B}\ }\textbf {\bibinfo {volume} {99}},\
  \bibinfo {pages} {214436} (\bibinfo {year} {2019})}\BibitemShut {NoStop}%
\end{thebibliography}%

\end{document}